\documentclass[a4paper,fleqn,usenatbib]{mnras}

\usepackage{newtxtext,newtxmath}

\usepackage[T1]{fontenc}
\usepackage{ae,aecompl}

\usepackage{graphicx}	
\newsavebox\CBox

\usepackage{amsmath}	
\usepackage{amssymb}	

\usepackage{amsfonts}
\usepackage{bm}
\usepackage{hyperref}
\usepackage{threeparttable}
\usepackage{array}
\usepackage{multirow}
\usepackage{subcaption}
\usepackage{framed}
\usepackage{rotating}

\title[Megaparsec-scale structure around SPT2349$-$56]
{Megaparsec-scale structure around the proto-cluster core SPT2349$-$56 at $z\,{=}\,4.3$}

\author[Hill et al.]
{Ryley Hill,$^{1}$
Scott Chapman,$^{1,2,3}$
Douglas Scott,$^{1}$
Yordanka Apostolovski,$^{4,5}$ \newauthor
Manuel Aravena,$^{6}$
Matthieu B{\'e}thermin,$^{7}$
C.~M.~Bradford,$^{8}$
Rebecca E.~A.~Canning,$^{9}$ \newauthor
Carlos De Breuck,$^{10}$ 
Chenxing Dong,$^{11}$
Anthony Gonzalez,$^{11}$ 
Thomas R.~ Greve,$^{12,13}$ \newauthor
Christopher C.~Hayward,$^{14}$ 
Yashar Hezaveh,$^{14,15}$ 
Katrina Litke,$^{16}$ 
Matt Malkan,$^{17}$ \newauthor
Daniel P.~Marrone,$^{16}$
Kedar Phadke,$^{18}$ 
Cassie Reuter,$^{18}$ 
Kaja Rotermund,$^{3}$ \newauthor
Justin Spilker,$^{19}$ 
Joaquin D.~Vieira,$^{18}$
Axel Wei{\ss}$^{20}$
\\
$^{1}$Department of Physics and Astronomy, University of British Columbia, 6225 Agricultural Road, Vancouver, V6T 1Z1, Canada\\
$^{2}$National Research Council, Herzberg Astronomy and Astrophysics, 5071 West Saanich Road, Victoria, V9E 2E7, Canada\\
$^{3}$Department of Physics and Atmospheric Science, Dalhousie University, 6310 Coburg Road, B3H 4R2, Halifax, Canada\\
$^{4}$Departamento de Ciencias Fisicas, Universidad Andres Bello, Fernandez Concha 700, Santiago, 7591538, Chile\\
$^{5}$Millennium Institute of Astrophysics (MAS), Nuncio Monse{\~n}or Sotero Sanz 100, Santiago, 7500000, Chile\\
$^{6}$N\'ucleo de Astronom\'{\i}a, Facultad de Ingenier\'{\i}a y Ciencias, Universidad Diego Portales, Av. Ej{\'e}rcito 441, Santiago, 8320000, Chile\\
$^{7}$Laboratoire d'Astrophysique de Marseille, 38 rue Fr{\'e}d{\'e}ric Joliot-Curie, Marseille, 13013, France\\
$^{8}$Harvard-Smithsonian Center for Astrophysics, 60 Garden Street, Cambridge, MA 02138, USA\\
$^{9}$Kavli Institute for Particle Astrophysics and Cosmology, 452 Lomita Mall, Stanford, CA 94305-4085, USA\\
$^{10}$European Southern Observatory, Karl Schwarzschild Stra{\ss}e 2, Garching, D-85748, Germany\\
$^{11}$Department of Astronomy, University of Florida, 211 Bryant Space Science Center, Gainesville, FL 32611-2055, USA\\
$^{12}$Department of Physics and Astronomy, University College London, Gower Street, London, WC1E 6BT, UK\\
$^{13}$Cosmic Dawn Center, Holbergsgade 14, Copenhagen, DK-1057, Denmark\\
$^{14}$Center for Computational Astrophysics, Flatiron Institute, 162 Fifth Avenue, New York, NY 10010, USA\\
$^{15}$D{\'e}partement de Physique, Universit{\'e} de Montr{\'e}al, 1375 Avenue Th{\'e}r{\`e}se-Lavoie-Roux, Montr{\'e}al, H2V 0B3, Canada\\
$^{16}$Steward Observatory, University of Arizona, 933 North Cherry Avenue, Tucson, AZ 85721, USA\\
$^{17}$Department of Physics and Astronomy, University of California, 900 University Avenue, Riverside, CA 90095-1547, USA\\
$^{18}$Department of Astronomy, University of Illinois, 1002 West Green Street, Urbana, IL 61801, USA\\
$^{19}$Department of Astronomy, University of Texas at Austin, 2515 Speedway, Stop C1400, Austin, TX 78712, USA\\
$^{20}$Max-Planck-Institut f{\"u}r Radioastronomie, Auf dem H{\"u}gel 69, Bonn, D-53121, Germany
}

\date{2 April 2020}

\pubyear{2020}

\begin{document}
\label{firstpage}
\pagerange{\pageref{firstpage}--\pageref{lastpage}}
\maketitle

\begin{abstract}
\noindent We present an extensive ALMA spectroscopic follow-up programme of the $z\,{=}\,4.3$ structure SPT2349$-$56, one of the most actively star-forming proto-cluster cores known, to identify additional members using their [C{\sc ii}] 158\,$\mu$m and \mbox{CO(4--3)} lines. In addition to robustly detecting the 14 previously published galaxies in this structure, we identify a further 15 associated galaxies at $z\,{=}\,4.3$, resolving 55$\,{\pm}\,$5\,per cent of the 870-$\mu$m flux density at 0.5\,arcsec resolution compared to 21\,arcsec single-dish data. These galaxies are distributed into a central core containing 23 galaxies extending out to 300\,kpc in diameter, and a northern extension, offset from the core by 400\,kpc, containing three galaxies. We discovered three additional galaxies in a red {\it Herschel\/}-SPIRE source 1.5\,Mpc from the main structure, suggesting the existence of many other sources at the same redshift as SPT2349$-$56 that are not yet detected in the limited coverage of our data. An analysis of the velocity distribution of the central galaxies indicates that this region may be virialized with a mass of (9$\pm$5)$\,{\times}\,$10$^{12}$\,M$_{\odot}$, while the two offset galaxy groups are about 30 and 60\,per cent less massive and show significant velocity offsets from the central group. We calculate the [C{\sc ii}] and far-infrared number counts, and find evidence for a break in the [C{\sc ii}] luminosity function. We estimate the average SFR density within the region of SPT2349$-$56 containing single-dish emission (a proper diametre of 720\,kpc), assuming spherical symmetry, to be roughly 4$\,{\times}\,10^4$\,M$_{\odot}$\,yr$^{-1}$\,Mpc$^{-3}$; this may be an order of magnitude greater than the most extreme examples seen in simulations.
\end{abstract}

\begin{keywords}
galaxies -- formation: galaxies -- evolution: submm -- galaxies
\end{keywords}

\section{Introduction}
\label{introduction}

The largest gravitationally-bound objects in the Universe are galaxy clusters, which have evolved from the largest overdensities seeded in the very early Universe into Mpc-sized structures presently containing thousands of galaxies. Cosmological simulations and observations indicate that these structures are built up hierarchically, where small overdensities initially collapsed and later merged to form large overdensities; however, the details of this process are far from understood, and in particular, we don't yet know how cluster formation affects galaxy evolution, and what roles may be played by active galactic nuclei (AGN) feedback \citep[e.g.,][]{mcnamara2012,pike2014,smolcic2017}, or by star-formation downsizing \citep[e.g.,][]{magliocchetti2013,miller2015,wilkinson2017}.

One way to investigate these issues is to look for clues in local, fully-formed clusters. Local clusters are dominated by elliptical galaxies \citep[e.g.][]{dressler1980} that are much more red than their field counterparts \citep[e.g.,][]{wake2005,stott2007}, and similarly show very little star-formation activity \citep[e.g.,][]{balogh1998,lewis2002,tanaka2004}. These observations suggest that the bulk of the star-formation activity in galaxy clusters occurred before redshifts of 2 \citep[e.g.,][]{snyder2012,willis2020}.

A more direct way to investigate galaxy cluster formation is to observe galaxy clusters at high redshifts. We now find galaxy clusters out to redshift 2 by looking for observational signatures such as X-rays emitted by hot inter-cluster gas \citep[e.g.,][]{rosati2009,gobat2011,andreon2014,wang2016,mantz2018}, the Sunyaev-Zeldovich effect \citep[e.g.,][]{planck2014-a36,bleem2015,huang2019}, and galaxy-based searches \citep[e.g.,][]{andreon2009,papovich2010,zeimann2012,stanford2012,muzzin2013}. However, beyond this epoch, these observational signatures become much less defined as these structures have not yet virialized. With this in mind, following \citet{overzier2016}, we adopt the definition that a `galaxy cluster' is a virialized object with $M\,{>}\,10^{14}\,$M$_{\odot}$, and a `proto-cluster' is a structure that will one day become a galaxy cluster. Proto-clusters may have high merger rates and correspondingly high star-formation rates \citep[SFRs; e.g.][]{casey2016}, thus containing a large number of dusty galaxies. These galaxies would then be more easily observed as luminous starbursts at millimetre/submillimetre (mm/submm) wavelengths and as AGN at radio wavelengths \citep[e.g.,][]{miley2008,galametz2013,rigby2014}, motivating searches in these regimes.

Distant proto-clusters are excellent laboratories for studying not only the details of cluster formation, but also galaxy evolution and star formation, since these processes are likely undergoing their most active phase at this epoch. A number of proto-clusters have been discovered beyond redshifts of 2, typically through their rest-frame optical emission, which traces unobscured stellar light \citep[e.g.,][]{steidel2000,shimasaku2003,steidel2005,venemans2007,chiang2015,dey2016,harikane2019}, or as overdensities of submm galaxies (SMGs), which probes their rest-frame far-infrared emission and traces star formation \citep[e.g.,][]{tamura2009,chapman2009,dannerbauer2014,chiang2015,flores-cacho2015,umehata2015,casey2015,hung2016,oteo2018,lacaille2019,kneissl2019}. However, comparing these systems to current simulations is challenging due to their very low number density, which requires large simulated cosmological volumes, and because they contain very massive galaxies with high gas and stellar densities that require significant resolution to simulate accurately.

Recently, one such structure, SPT2349$-$56, was identified as an incredibly luminous 870-$\mu$m flux density source at redshift 4.3 ($S_{870\,\mu{\rm m}}\,{=}\,110\,{\pm}\,10\,$mJy, corresponding to a SFR ${>}\,$10$^4$\,M$_{\odot}\,$yr$^{-1}$) within which 14 SMGs were spectroscopically confirmed in the core region, making it potentially one of the highest density proto-clusters known at this epoch \citep{miller2018}. However, only the central component of the structure was probed, and 36\,per cent of the single-dish flux density resolved, leaving open the possibility that the remaining flux density could be due to chance alignments along the line of sight \citep[e.g.,][]{hayward2018}. Establishing more proto-cluster members through further spectroscopic observations would provide further evidence that this system is the progenitor to a rich galaxy cluster (perhaps even as large as the Coma Cluster), as opposed to a starved core that evolves into a much smaller galaxy group (as seen in some systems, see e.g. \citealt{lovell2018}). Additionally, redshift 4.3 SMGs in the field are known to be quite rare \citep[the median redshift being about 2.5, see][]{chapman2005,simpson2014}, making this a particularly interesting and statistically robust sample of galaxies undergoing accelerated evolution from which we can learn about the complex interplay between star formation, galaxy formation, and cluster formation.

In this paper, we report the results from an extensive follow-up programme of SPT2349$-$56 using the Atacama Large Millimeter/submillimeter Array \citep[ALMA;][]{wootten2009}, which aimed to spectroscopically confirm new proto-cluster members and spatially resolve the galaxies responsible for the intense star-formation observed. In Section \ref{observations}, we outline how SPT2349$-$56 was selected, summarize previous observations of this proto-cluster, and describe our new ALMA follow-up efforts. In Section \ref{analysis} we present our data analysis methods, including our search for new galaxies, and in Section \ref{results} we present our results. Section \ref{discussion} discusses our findings, and the paper is summarized and concluded in Section \ref{conclusion} . We assume a $\Lambda$CDM model with parameters from \citet{planck2014-a15} throughout.

\section{Observations}
\label{observations}

\subsection{Selection from the South Pole Telescope survey}
\label{spt_survey}

SPT2349$-$56 was initially discovered as part of the South Pole Telescope (SPT) extragalactic mm-wave point-source catalogue \citep{vieira2010,mocanu2013,everett2020}, a collection of bright ($S_{2\,\mathrm{mm}}\,{>}\,5\,$mJy at ${>}\,4.5\,\sigma$) sources found in the SPT 2500\,deg$^2$ survey that are unresolved by SPT's 1\,arcmin beam. From a total sample of over 1000 objects, roughly 200 were classified as dusty star-forming galaxies based on their spectral indices. Of these, the brightest were followed up with the Atacama Pathfinder Experiment (APEX) telescope's Large APEX BOlometer CAmera \citep[LABOCA;][]{kreysa2003,siringo2009} instrument at 870\,$\mu$m, and a flux selection was made at $S_{870\,\mu{\rm m}}\,{>}\,25\,$mJy, resulting in a final sample of 81 SMGs.

A dedicated follow-up campaign using a number of optical-through-mm wavelength facilities, including ALMA, the Spectral and Photometric Imaging REceiver \citep[SPIRE;][]{griffin2010} on board the {\it Herschel\/} satellite, and {\it Spitzer\/}'s Infrared Array Camera \citep[IRAC;][]{fazio2004}, was subsequently undertaken to determine the nature of these incredibly bright star-forming galaxies \citep{vieira2013,weiss2013}; some reached flux densities of 100\,mJy at 870\,$\mu$m, meaning that they could only be strong gravitational lenses or collections of galaxies densely packed within SPT's 1\,arcmin beam. It was found that about 90\,per cent of the sources are indeed strong gravitational lenses with magnification factors reaching up to about 30, and that the remaining 10\,per cent show no evidence for lensing and are instead likely to be intrinsically ultra-luminous galaxies or collections of galaxies \citep{hezaveh2013,spilker2016}. 

SPT2349$-$56 is the brightest of these unlensed sources, with $S_{1.4\,{\rm mm}}\,{=}\,23.3\,$mJy. Further follow-up with ALMA in Cycle 1 at 3\,mm and with the APEX telescope's First Light APEX Submillimeter Heterodyne \citep[FLASH;][]{heyminck2006} instrument revealed that the structure is composed of a bright central component at a redshift of 4.3 \citep{greve2012,strandet2016}, and a fainter northern extension (see Fig.~\ref{coverage}).

\subsection{Follow-up ALMA observations}
\label{alma_observations}

Since its discovery and redshift determination, SPT2349$-$56 has been the subject of numerous ALMA follow-up studies. High-resolution spectroscopy targeting the \mbox{CO(4--3)}, [C{\sc ii}] 158\,$\mu$m, and [N{\sc ii}] 205\,$\mu$m transitions in the core region of SPT2349$-$56 were carried out in Cycles 3 and 4 and used to securely identify 14 central galaxies \citep{miller2018}; for reference, the depths of these observations were 0.2\,mJy\,beam$^{-1}$ for the \mbox{CO(4--3)} transition, 1.1\,mJy\,beam$^{-1}$ for the [C{\sc ii}] transition, and 0.4\,mJy\,beam$^{-1}$ for the [N{\sc ii}] transition. Here we report on a suite of new ALMA observations undertaken during Cycles 5 and 6, covering a much larger area with greater depth.

The Cycle 5 observations used in this paper targeted two of the lines observed in the previous cycles, [C{\sc ii}] in Band 7 ($\nu_{\rm rest}\,{=}\,1900.537\,$GHz) and \mbox{CO(4--3)} in Band 3 ($\nu_{\rm rest}\,{=}\,461.041\,$GHz). Our [C{\sc ii}] coverage included a three-pointing mosaic of the brighter central component of SPT2349$-$56, covering a much larger area than the existing data, and a single pointing of the previously unobserved northern component, both down to a depth of about 0.3\,mJy\,beam$^{-1}$ per 13\,km\,s$^{-1}$ channel. Our \mbox{CO(4--3)} pointings covered the entire 870-$\mu$m emission region, including the previously unobserved northern component, down to 0.07\,mJy\,beam$^{-1}$ per 54\,km\,s$^{-1}$ channel. The [C{\sc ii}] observations were tuned to place the line in the centre of the upper sideband, while for the \mbox{CO(4--3)} observations the tuning was set to place the line in the lower sideband.

The Cycle 6 observations presented in this paper also targeted the [C{\sc ii}] and \mbox{CO(4--3)} transitions in Bands 7 and 3, respectively. The goal of the [C{\sc ii}] observations was to cover most of the central component of SPT2349$-$56 with a frequency setup similar to that of the Cycle 5 data, but with higher angular resolution in order to resolve morphologies. The goal of the \mbox{CO(4--3)} observations was to provide coverage of the outskirts of the structure. The setup of these Band 3 observations was also chosen to be similar to that of the Cycle 5 \mbox{CO(4--3)} observations in order to allow the data to be combined into a single deep \mbox{CO(4--3)} map. The depth of the Band 7 data was approximately 0.4\,mJy\,beam$^{-1}$ per 13\,km\,s$^{-1}$ channel, and the depth of the Band 3 data was approximately 0.06\,mJy\,beam$^{-1}$ per 54\,km\,s$^{-1}$ channel. We note that the Cycle 5 and 6 depths quoted above do not include existing observations from previous cycles.

\begin{figure*}
\includegraphics[width=\textwidth]{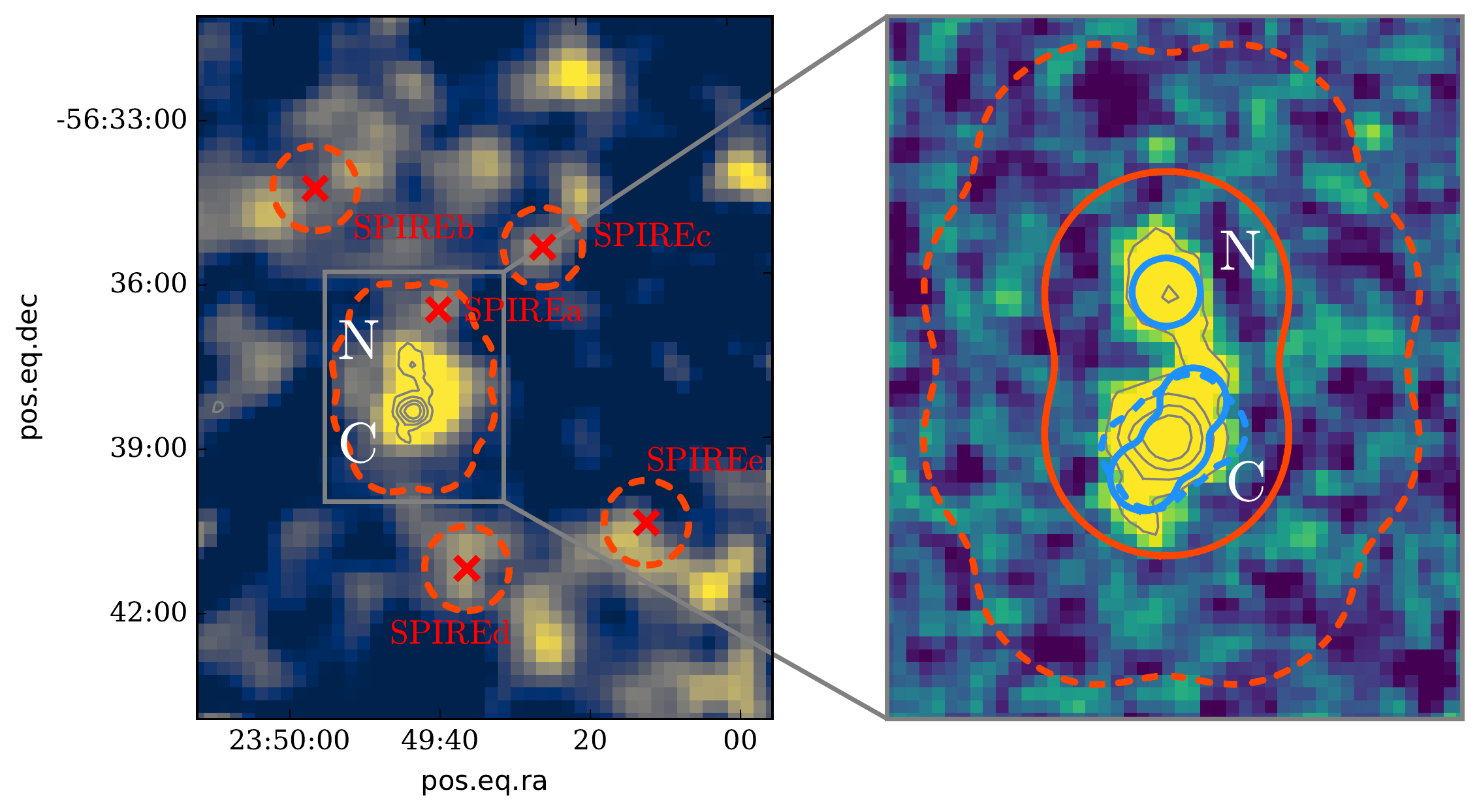}
\caption{Summary of the ALMA data presented in this paper. {\it Left:} The background image shows the {\it Herschel\/}-SPIRE image at 500\,$\mu$m, with our Band 3 \mbox{CO(4--3)} coverage outlined in red dashed contours; the five previously-selected red {\it Herschel\/} sources are indicated by red crosses, and the grey contours outline the LABOCA 870-$\mu$m emission. The red {\it Herschel} sources are obtained using 250\,$\mu$m positions as priors where the angular resolution is best, so in this image some sources (particularly SPIREb) are blended. {\it Right:} Expanded view of the left panel with LABOCA 870-$\mu$m data shown in the background. The green solid and dashed contours show our Band 7 [C{\sc ii}] coverage from Cycles 5 and 6, respectively, while the red solid and dashed contours show our Band 3 \mbox{CO(4--3)} coverage, also from Cycles 5 and 6, respectively.}
\label{coverage}
\end{figure*}

In addition to targeting the main structure of SPT2349$-$56, we also used existing {\it Herschel\/}-SPIRE data to identify five red {\it Herschel\/} sources surrounding SPT2349$-$56 using 250\,$\mu$m positions as priors; here `red' is defined as $S_{500\,\mu{\rm m}}\,{>}\,S_{350\,\mu{\rm m}}\,{>}\,S_{250\,\mu{\rm m}}$, with a signal-to-noise above 3 at both 250 and 500\,$\mu$m (see \citealt{miller2018} for details). One of these five sources lies close to the central structure and was covered by our extended \mbox{CO(4--3)} mosaic; the remaining four sources were targeted in dedicated Band 3 observations in Cycle 6, with the expected \mbox{CO(4--3)} transition centred in the lower sideband, and these pointings reached depths of 0.1--0.2\,mJy\,beam$^{-1}$ per 54\,km\,s$^{-1}$ channel. An overview of the observations is shown in Fig.~\ref{coverage} and summarized in Table \ref{table:observations}. It is also important to recall that the sensitivity of these observations to source-detection depend on the synthesized beams, which in turn depend on the array configurations -- thus in Table \ref{table:observations} we also provide beamsizes for each dataset.

\setlength\tabcolsep{3pt}
\setlength\extrarowheight{2pt}
\begin{table*}
\centering
\caption{Summary of ALMA data presented in this paper. Rows in bold indicate maps that were independently searched for [C{\sc ii}] or \mbox{CO(4--3)} lines.}
\label{table:observations}
\begin{threeparttable}
\begin{tabular}{llccccc}
\hline
Cycle & Description & Line transition & Channel width & RMS per channel & Synthesized beamsize & Area \\
& & & [km\,s$^{-1}$] & [mJy\,beam$^{-1}$] & major/minor [arcsec] & [arcmin$^2$] \\
\hline
5 & 3-point mosaic of central LABOCA source & [C{\sc ii}] & 13 & 0.27 & 0.56/0.49 & 0.33 \\
{\bf 5} & {\bf Single pointing of northern LABOCA source} & {\bf [C{\sc ii}]} & {\bf 13} & {\bf 0.32} & {\bf 0.51/0.38} & {\bf 0.13} \\
5 & 2-point mosaic of entire LABOCA source & \mbox{CO(4--3)} & 54 & 0.072 & 1.01/0.84 & 2.8 \\
6 & 6-point mosaic of central LABOCA source & [C{\sc ii}] & 13 & 0.41 & 0.23/0.17 & 0.45 \\
6 & 8-point mosaic of outer LABOCA region & \mbox{CO(4--3)} & 54 & 0.14 & 0.52/0.45 & 9.2 \\
{\bf 6} & {\bf Single pointing of SPIREb} & {\bf \mbox{CO(4--3)}} & {\bf 54} & {\bf 0.13} & {\bf 1.03/0.81} & {\bf 1.9} \\
{\bf 6} & {\bf Single pointing of SPIREc} & {\bf \mbox{CO(4--3)}} & {\bf 54} & {\bf 0.17} & {\bf 0.71/0.61} & {\bf 1.7} \\
{\bf 6} & {\bf Single pointing of SPIREd} & {\bf \mbox{CO(4--3)}} & {\bf 54} & {\bf 0.12} & {\bf 0.97/0.83} & {\bf 1.9} \\
{\bf 6} & {\bf Single pointing of SPIREe} & {\bf \mbox{CO(4--3)}} & {\bf 54} & {\bf 0.14} & {\bf 0.94/0.82} & {\bf 1.9} \\
\hline
{\bf 5+6} & {\bf Combined map of central LABOCA source}$^{\rm a}$ & {\bf [C{\sc ii}]} & {\bf 13} & {\bf 0.22} & {\bf 0.35/0.29} & {\bf 0.43} \\
{\bf 5+6} & {\bf Combined map of outer LABOCA region}$^{\rm b}$ & {\bf \mbox{CO(4--3)}} & {\bf 54} & {\bf 0.064} & {\bf 0.85/0.72} & {\bf 7.2} \\
\hline
\end{tabular}
\begin{tablenotes}
\item $^{\rm a}$Combination of the 3-point mosaic of the central LABOCA source from Cycle 5 and the 6-point mosaic of the central LABOCA source from Cycle 6.
\item $^{\rm b}$Combination of the 2-point mosaic of the entire LABOCA source from Cycle 5 and the 8-point mosaic of the outer LABOCA region from Cycle 6.
\end{tablenotes}
\end{threeparttable}
\end{table*}

\section{Data analysis}
\label{analysis}

\subsection{Data reduction}
\label{reduction}

The ALMA data were calibrated using {\sc casa}\footnote{\url{https://casa.nrao.edu}} \citep{mcmullin2007} and the observatory-provided calibration scripts. Dirty and cleaned data cubes for each of the observations reported in Table \ref{table:observations} were produced using the {\sc casa} function {\tt tclean} with Briggs weighting and a robust parameter of 0.5. Continuum images were also produced using {\tt tclean} for the sidebands that did not contain any line emission using multi-frequency synthesis (MFS).

Additionally, we combined the $uv$ data from the Cycle 5 and 6 observations of [C{\sc ii}] in the core region (`3-point mosaic of central LABOCA source' and `6-point mosaic of central LABOCA source' in Table \ref{table:observations}) to produce a single, deep [C{\sc ii}] data cube, and similarly we combined the $uv$ tables from the Cycle 5 and 6 observations of \mbox{CO(4--3)} around the entire structure (`2-point mosaic of entire LABOCA source' and `8-point mosaic of outer LABOCA region' in Table \ref{table:observations}) to produce a deep data cube of \mbox{CO(4--3)}. The maximum depth of the deep [C{\sc ii}] map was 0.2\,mJy\,beam$^{-1}$ per 13\,km\,s$^{-1}$ channel, and the synthesized beam major/minor full width at half maximum (FWHM) was 0.35/0.29\,arcsec. For the deep \mbox{CO(4--3)} map, the maximum depth achieved was 0.06\,mJy\,beam$^{-1}$ per 54\,km\,s$^{-1}$ channel and the synthesized beam major/minor FWHM was 0.85/0.72\,arcsec.

\subsection{Source extraction}
\label{source_search}

Given the significantly deeper data for the core of SPT2349$-$56 compared to that reported in \citet{miller2018} (which was 0.9\,mJy\,beam$^{-1}$ per 13\,km\,s$^{-1}$ channel at the observed [C{\sc ii}] frequency), we expect to discover a number of new sources in both line emission and in the continuum. However, lacking any knowledge of where these new sources might be in our data cubes, it was necessary to perform a source search over the entire surveyed area around the frequencies of the expected lines. In addition, since some of the fainter sources in \citet{miller2018} were not significantly detected in [C{\sc ii}] but instead were derived from a joint analysis of [C{\sc ii}] and \mbox{CO(4--3)}, we would like to confirm their [C{\sc ii}] properties.

To accomplish this, we used the publicly-available code {\sc LineSeeker} \citep[see][for details]{gonzalez-lopez2017,gonzalez-lopez2019}. Briefly, {\sc LineSeeker} convolves a primary beam-uncorrected data cube with a number of Gaussians of varying width along the spectral axis. The noise per channel is assessed iteratively by computing the standard deviation of all the pixels in a given channel, then re-computing the standard deviation of all the pixels whose absolute values are lower than 5 times the initial noise estimate. Signal-to-noise ratio (S/N) peaks are then located in both positive and negative flux density pixels and returned to the user.

We ran {\sc LineSeeker} on a total of seven data cubes: 1) the combined, extra-deep [C{\sc ii}] map of the main core region; 2) the combined, deep \mbox{CO(4--3)} map of the entire structure plus the outskirts; 3) the [C{\sc ii}] map of the fainter northern region; and 4--7) the four pointings of surrounding red {\it Herschel\/} sources. Here we chose to search for sources through the dirty data cubes in order to minimize the possibility of picking up artefacts introduced by the cleaning. We searched for [C{\sc ii}] peaks ranging from a single channel to 1000\,km\,s$^{-1}$ in FWHM, over a velocity range encompassing $\pm$1500\,km\,s$^{-1}$ relative to the mean redshift of 4.304 reported by \citet{miller2018}. This velocity range corresponds to the total bandwidth available in the sideband containing the expected [C{\sc ii}] emission. Within each data cube we took all positive-pixel line peaks with a S/N greater than the most significant negative-pixel line peak to be detections.

In the [C{\sc ii}] map of the core, we found negative peaks down to a S/N of 6.2. Using this as our threshold, we identified the 14 known sources found by \citet{miller2018}, and nine new sources. In the [C{\sc ii}] pointing of the northern region there were negative peaks down to a S/N of 5.9, but only one bright source was found to be more significant than this. Across our deep \mbox{CO(4--3)} map the most significant negative peak was at a S/N of 5.9, and in addition to finding 11 of the above [C{\sc ii}] sources, two additional sources were found just outside of our [C{\sc ii}] coverage in the northern region. Out of our four red {\it Herschel\/} targets, sources were only found in one pointing, SPIREc. Here, the most negative peak was at a S/N of 5.6, and three sources were more significant than this. In the other three pointings, SPIREb, SPIREd, and SPIREe, negative peaks were seen down to 5.9, 6.1, and 6.3, respectively. There were no correspondingly more significant positive peaks. Despite the fact that these {\it Herschel\/} follow-up pointings did not turn up any sources, we note that the targets were all quite low S/N in the SPIRE data, and none showed significant 870-$\mu$m emission in our LABOCA map. This means that there could be other red sources at the same redshift as SPT2349$-$56 with low S/N in the {\it Herschel\/} data that we have not yet targeted.

We also searched our maps for continuum sources (i.e.~interloping foreground/background sources not associated with the structure of SPT2349$-$56) by averaging over all channels. The noise levels of these continuum maps were estimated on a pixel-by-pixel basis by calculating the local rms within circles of 6\,arcsec (after masking all of the sources detected by their line emission), and then we ran a peak-finding algorithm on the resulting maps, looking for both positive flux density and negative flux density peaks. Similar to our line search, we looked for positive flux density peaks with a higher significance than the most significant negative flux density peak. This time we chose one S/N cutoff for all Band 3 data and another for all Band 7 data, based on the most significant negative peaks across all five and two maps, respectively.

This continuum search found negative peaks down to a S/N of 5.0 in the Band 3 data and 5.6 in the Band 7 data. While no new sources were found in the Band 3 data that are more significant than a S/N of 5.0, two were found in the Band 7 data in the central region and one in the northern region. These 29 line detections and three continuum detections constitute our current sample of SPT2349$-$56 galaxies, and are summarized in Fig.~\ref{all_sources}. Sources are named firstly according to the region where they are located, where `C' refers to the core, `N' refers to the northern component, and `SPIREc' refers to the red {\it Herschel\/} source SPIREc, and secondly in order of decreasing [C{\sc ii}] line strength for the core, and decreasing \mbox{CO(4--3)} line strength for the northern component and SPIREc (see Section \ref{source_properties} for details about line strength measurements). Continuum-only sources are designated as `NL' (no line) and ordered by decreasing 850\,$\mu$m flux density.

\subsection{Extended [C{\sc ii}] emission}
\label{cii_arc}

\begin{figure*}
\includegraphics[width=\textwidth]{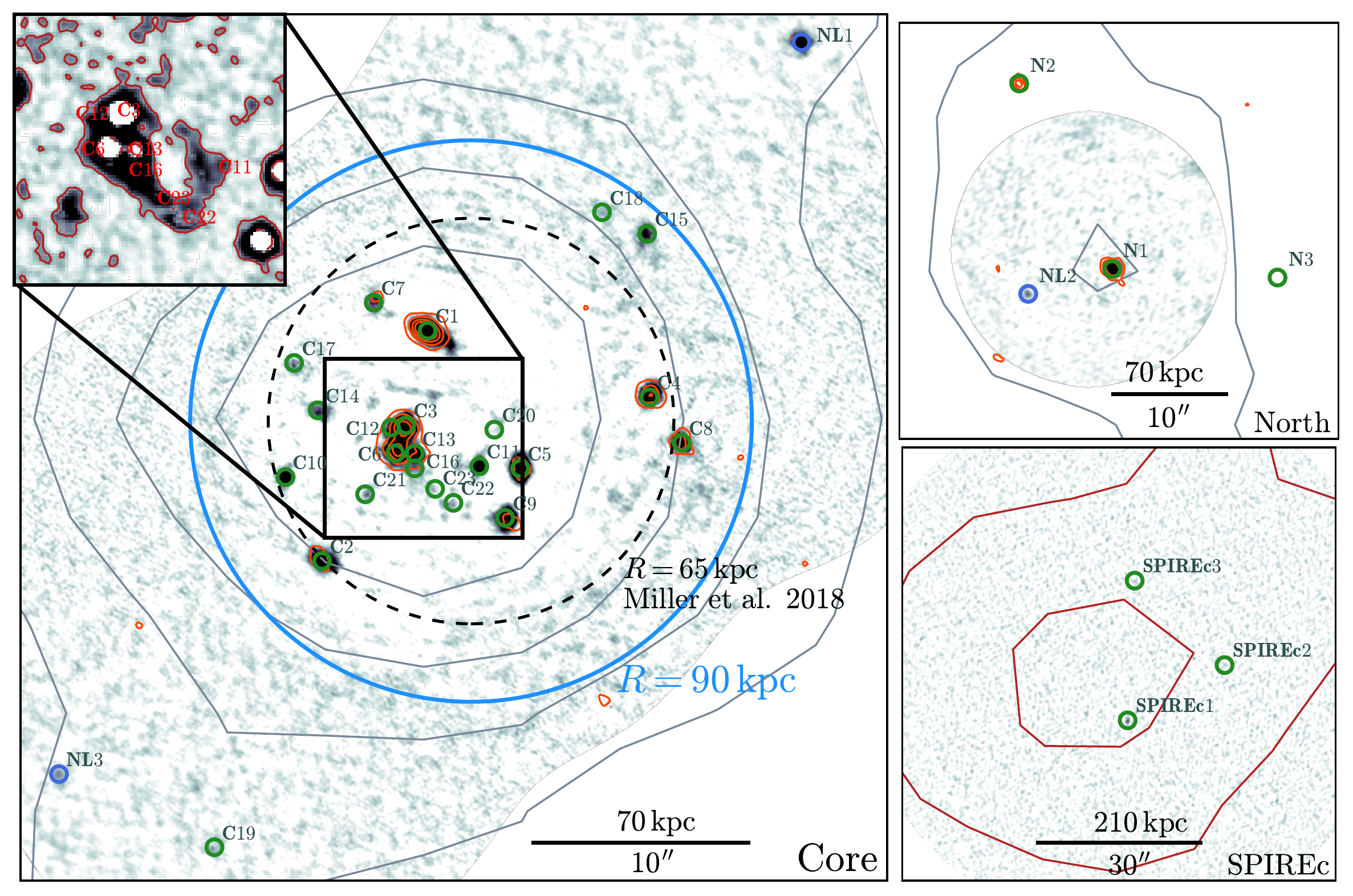}
\caption{Summary of source detections. {\it Left:} The [C{\sc ii}] data of the central component are shown in the background, averaged over $\pm1500\,$km\,s$^{-1}$, with LABOCA 870-$\mu$m contour levels shown in grey. Red contours show our \mbox{CO(4--3)} data obtained from the combined map, also averaged over $\pm1500\,$km\,s$^{-1}$. Sources detected via line emission are circled in green, and those detected via continuum emission are circled in blue. A blue circle of radius 90\,kpc (proper distance) has been draw around the 850-$\mu$m flux-weighted centre, which is used here as the nominal radius of the core (see Section \ref{virialization}). For reference, a black dashed circle of radius 65\,kpc (also proper distance) is shown, which was used by \citet{miller2018} when our wider coverage data were not available. The inset panel shows the extended [C{\sc ii}] emission found near the centre of the field after averaging channels between 130 and 420\,km\,s$^{-1}$, with pixels fainter than 1$\sigma$ and brighter than 10$\sigma$ masked (where $\sigma$ is the local rms); the contours are 1.25$\sigma$, showing that the extended emission rises above the 1$\sigma$ level in an arc around galaxies C16, C23, and C11. {\it Right, top:} Same as the left panel, but for the northern component of SPT2349$-$56. {\it Right, bottom:} The background image shows our Band 3 observation of the {\it Herschel\/} source SPIREc, averaged from $-$500 to $-$1500\,km\,s$^{-1}$. The dark red contours are 3.5 and 4.5$\sigma$ contours of the {\it Herschel\/} 500-$\mu$m flux density.}
\label{all_sources}
\end{figure*}

Around the bright central galaxies C3, C6, and C13 (see Fig.~\ref{all_sources}) an arc of extended [C{\sc ii}] gas is seen between 357.98 and 358.33\,GHz, or between 130 and 420\,km\,s$^{-1}$ relative to the mean redshift of the structure. The typical specific intensity of this arc is about 1\,mJy\,arcsec$^{-2}$ at a single channel, with peaks at the 2\,mJy\,arcsec$^{-2}$ level, and the local noise was estimated to be 0.3\,mJy\,arcsec$^{-2}$ by taking the rms within a 6\,arcsec-diametre circular aperture just outside of the extended emission region after masking all known sources. The diametre of this arc is roughly 2.8\,arcsec, or about 19\,kpc in proper distance. To investigate this extended emission, we average over these frequencies (i.e.~compute a moment 0 map) while masking pixels below 1$\sigma$ and above 10$\sigma$ (where $\sigma$ is the local rms, calculated in the same way as above). The resulting map is shown as the inset in Fig.~\ref{all_sources}, along with 1.25$\sigma$-level contours. It can be seen that there is extended emission above the 1.25$\sigma$ level encircling a number of central galaxies, and in particular sources C16, C22, and C23 are embedded within it (source C11 is spatially close to this emission as well, but has a significantly different velocity offset). We note that there is no clear sign of this extended gas in the \mbox{CO(4--3)} data.

There are many ways to explain the nature of the extended emission seen here, including tidal tails resulting from gravitational interactions amongst galaxies, in particular through major mergers \citep[e.g.,][]{toomre1972,barnes1988}, or an expanding shell of ionized gas (as seen for example in local gas-rich galaxies, e.g., \citealt{heiles1979,mcclure-griffiths2002}). Extended gas emission has been detected in some proto-clusters through the \mbox{CO(1--0)} transition \citep{emonts2016,dannerbauer2017}, the \mbox{CO(3--4)} transition \citep{ginolfi2017}, and in [C\textsc{i}] \citep{emonts2018}, and it has also been statistically detected surrounding $z\,{=}\,$4--7 galaxies through stacking analyses \citep{fujimoto2019,ginolfi2019}. The extended gas in these systems could affect the evolution of the embedded galaxies \citep[see e.g.][]{dannerbauer2017}, although the number of sources available for such an analysis still remains low. However, a more thorough analysis is required to understand what we are seeing here, and will be done in future work.

\subsection{Source properties}
\label{source_properties}

The statistical properties of our un-combined maps, where the array was in a single configuration, are more homogenous and easy to estimate (i.e. the depth is more uniform throughout most of the surveyed area), and additionally have lower angular resolution, reducing the amount of resolved-out flux, thus we estimated the continuum strengths and line properties of our sources using the Cycle 5 3-pointing mosaic for the central sources, and the Cycle 5 single-pointing for the northern sources. Similarly, for Band 3 we used the Cycle 5 2-pointing mosaic to measure line and continuum properties. Lastly, the high-resolution Band 7 imaging from Cycle 6 was used to estimate the sizes of our sources. All line and continuum measurements were made on primary beam-corrected maps.

\subsubsection{Line and continuum emission}

There are many ways to measure line strengths and continuum flux densities of galaxies in inteferometric data. For example, one could model the galaxies directly in the $uv$ plane; however, some of our data cubes contain over 20 galaxies, many of which are somewhat resolved, as well as extended emission, making such a procedure very computationally expensive. Instead, we choose to obtain these properties through aperture photometry, and then to compare the results to peak flux photometry to assess if our sources are indeed resolved. Elliptical apertures were designed for each source manually using the [C{\sc ii}] data cubes (or otherwise the \mbox{CO(4--3)} data cubes when no [C{\sc ii}] imaging was available) as follows. 

First, 2\,arcsec-diametre circular apertures were placed on each source, and spectra were obtained by integrating over the pixels and binning the channels by a factor of 4 in order to reduce the noise. From the resulting spectra we tried fitting a constant, a single Gaussian and a double Gaussian model to the line profiles, and took the model with the reduced $\chi^2$ closest to 1 to be the best fit. We then stacked the channels from $-3\sigma$ to $3\sigma$ (where $\sigma$ is the standard deviation of the best-fitting linewidth), or for cases where two Gaussians was a better fit, from $-3\sigma_{\rm L}$ to $+3\sigma_{\rm R}$, where $\sigma_{\rm L}$ and $\sigma_{\rm R}$ are from the left and right Gaussian fits, respectively. We also stacked the channels that did not contain line emission to produce continuum images. We plotted our apertures overtop of these line and continuum images and adjusted the apertures to enclose each source out to about 2$\sigma$ in the image plane, then repeated the procedure until the apertures enclosed each source in our best-fitting line and continuum stacks out to 2$\sigma$. These same apertures were then used on the \mbox{CO(4--3)} maps, and for our \mbox{CO(4--3)} spectra the noise was reduced by binning by a factor of 2.

For the three galaxies detected only in the continuum (NL1, NL2, and NL3) the best-fitting models are a constant; these sources could be chance line-of-sight alignments of galaxies at other redshifts, or sources with small line/continuum ratios. We note that NL1 and NL3 also lie close to the edge of our [C{\sc ii}] maps where the primary beam response is low, making it possible that their [C{\sc ii}] lines are simply undetected due to the noise. These sources might also be outside of the velocity range probed by our [C{\sc ii}] observations (${\pm}\,$1500\,km\,s$^{-1}$) and too faint in \mbox{CO(4--3)}; this interpretation is consistent with NL2 and NL3, whose 850\,$\mu$m continuum flux densities are the same magnitude as other sources with no (or low S/N) \mbox{CO(4--3)} detections, but NL1 is much brighter at 850\,$\mu$m, and sources of similar brightness have well-detected \mbox{CO(4--3)} lines.

Redshifts were determined from the best-fitting Gaussian means, and line strengths were obtained by integrating the spectra from $-3\sigma$ to $3\sigma$. In the case of a two-peaked fit, we calculated the best-fitting amplitude-weighted average of the two best-fitting means to calculate a redshift, and the integration bound for the line strength was $-3\sigma_{\rm L}$ to $+3\sigma_{\rm R}$. Continuum flux densities were measured in the line-free continuum stacks using the same photometry apertures. Positions were calculated from the brightest pixel in the stacked line images except for NL1, NL2, and NL3, where we used the brightest pixel in the stacked continuum images.

We checked that our apertures were not missing flux by comparing line strengths and continuum flux densities computed with larger aperture sizes. We looked at 10 bright and isolated sources, then determined their line strengths by integrating over the same frequency range as above but increasing the aperture size by 1 pixel up to 6 pixels, and similarly determined their continuum flux densities by stacking the remaining line-free channels. We found that over this aperture range half of the resulting line strengths increased by less than 2\,per cent relative to the apertures we used for our measurements, and the remaining half increased by less than 15\,per cent, while seven continuum flux densities strictly decreased and the remaining three increased by less than 2\,per cent. Therefore, while increasing our aperture sizes may capture more line emission, it would also on average result in a loss in continuum flux density, thus our apertures should be striking a good balance between these two measurements.

We also made line strength and continuum flux density measurements assuming that our sources are unresolved. To do this, for each source we took the stacked line emission maps and obtained spectra at the positions of the brightest pixel, then integrated the spectra across the same channels as in our aperture photometry analysis. Similarly, continuum flux densities were obtained from the brightest pixels in the stacked continuum maps. For the 10 brightest isolated sources, we found that the median ratio of peak-pixel to aperture-photometry line strength in the [C\textsc{ii}] cubes was 0.52, and 0.80 for the continuum flux density ratio. These results show that for our higher frequency data, some galaxies are indeed somewhat resolved. For the \mbox{CO(4--3)} data cubes, these ratios are 1.08 and 1.09, respectively, showing that here our data are consistent with point sources, but that we are not missing much flux in our apertures. For the remainder of this work, we use only measurements obtained from aperture photometry.

We found that the weighted mean redshift of the sources located within the core is 4.30280$\pm$0.00002, in agreement with the findings of \citet{miller2018}, who determined the mean redshift of the core sources to be 4.304$\pm$0.002. Meanwhile, the weighted mean redshift of the northern component is 4.31309$\pm$0.00016 (a velocity offset of 580\,km\,s$^{-1}$ relative to the core), and that of SPIREc is 4.28171$\pm$0.00018 (a velocity offset of -1200\,km\,s$^{-1}$ relative to the core). For the remainder of this paper, we report all line-of-sight velocities relative to the mean redshift of the main central component.

In Table \ref{table:cont} we provide the positions, relative velocities, and continuum flux densities of all 32 galaxies in our sample, and for completeness we also give continuum flux densities at 1.1\,mm independently derived from the Band 7 imaging described in \citet{miller2018} using the same method and apertures outlined above. Here, the positions and relative velocities are from our [C\textsc{ii}] detections when possible, but for galaxies N2, N3, SPIREc1, SPIREc2, and SPIREc3, which only have \mbox{CO(4--3)} observations, we report positions and relative velocities measured from the \mbox{CO(4--3)} line. For galaxies NL1, NL2, and NL3, where no line was detected, we provide the positions of the peak pixels in the continuum maps. For cases where the continuum flux density S/N was less than 2 we give 1$\sigma$ upper limits. We note that sources C3 and C12, as well as sources C13 and C16, are completely blended in the \mbox{CO(4--3)} data cubes. Since sources C3 and C13 are much brighter than sources C12 and C16, respectively, we simply report C12 and C16 as non-detections in Band 3 and provide line and continuum measurements of C3 and C13 only.

In Table \ref{table:line} we give the integrated line strengths, line luminosities, and linewidths obtained from the fits. Where the line profile is best fit by a double Gaussian, we report both FWHM. For cases where no line is detected we provide 1$\sigma$ upper limits on line strengths and luminosities calculated as the average uncertainty from our measurements. Cutouts of all these sources, showing both stacked continuum and line emission contours, alongside their corresponding spectra with best-fitting Gaussian profiles, are shown in in Appendix \ref{appendix1}.

\subsubsection{Far-infrared luminosities}
\label{fir_lums}

We model the spectral energy distributions (SEDs) of our dusty submm galaxies following the greybody function outlined in \citet{greve2012}:
\begin{equation}\label{mbb}
S_{\nu} (T_{\rm d}) \propto \left( \frac{\nu}{\nu_0} \right)^{\beta} \frac{\nu^3}{\exp(h \nu / k_{\rm B} T_{\rm d})-1},
\end{equation}
\noindent
where $\nu_0$ is the critical frequency (or wavelength), $\beta$ is the emissivity index of the dust, $T_{\rm d}$ is the dust temperature, and $h$ and $k_{\rm B}$ are the Planck and Boltzmann constants, respectively. The continuum flux density measurements presented in this paper do not reach the peak of this SED (about 400\,$\mu$m in the observed frame assuming fiducial model parameters), but instead lie in the Rayleigh-Jeans (long wavelength) regime, meaning that we can explore the range of model parameters that are consistent with the colours (i.e.~slopes) seen in our data -- a more detailed SED analysis using additional mid-infrared photometry will be presented in a future paper.

\begin{figure}
\includegraphics[width=\columnwidth]{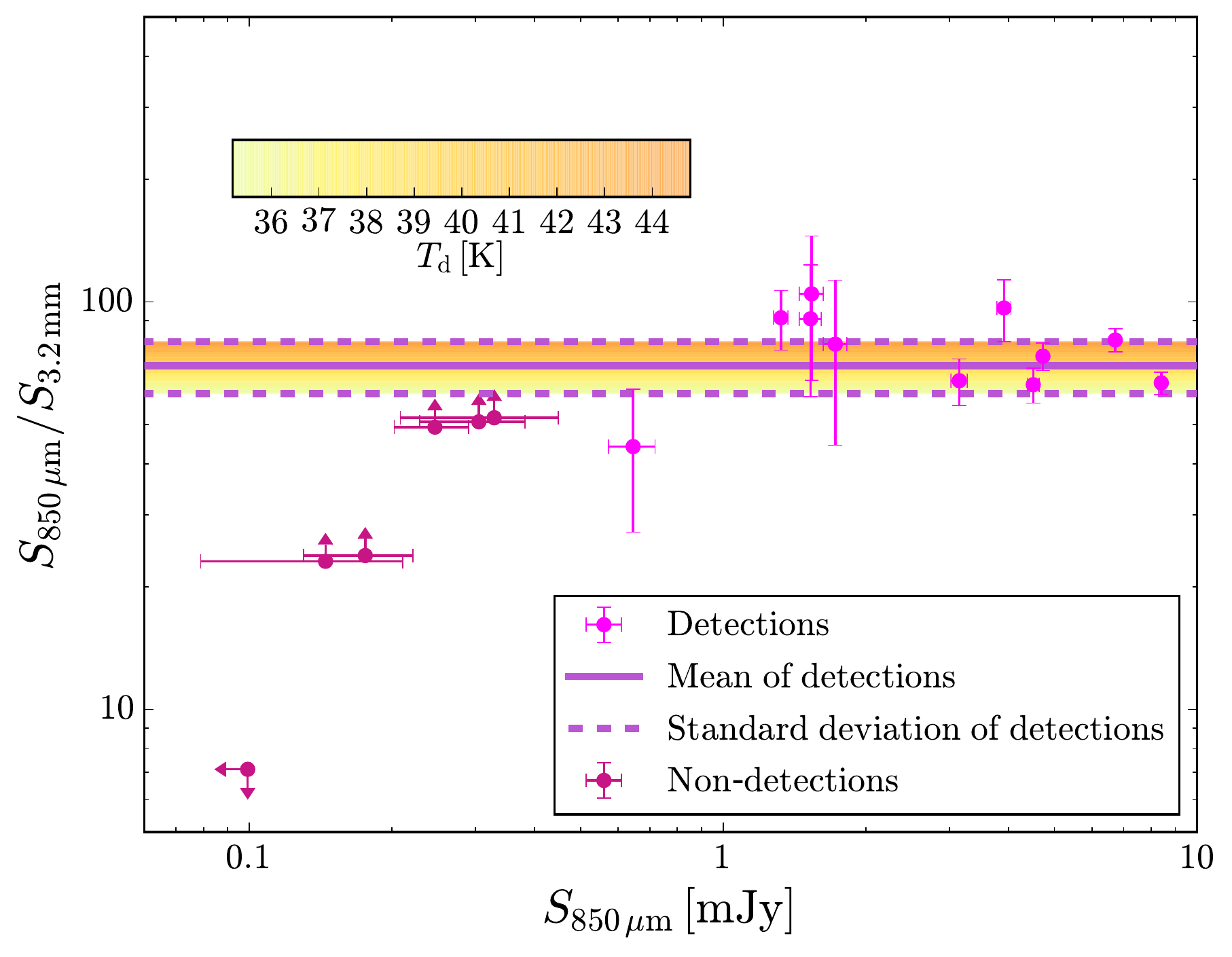}
\caption{$S_{850\,\mu{\rm m}}$ over $S_{3.2\,{\rm mm}}$ as a function of $S_{850\,\mu{\rm m}}$ for all central galaxies with a detection in at least one of these wavelengths, with non-detections shown as upper limits. The weighted mean ratio for sources with detections in both bands is 70 with a weighted standard deviation of 10, as indicated by the solid and dashed purple lines, respectively. Sources with only upper limits available are consistent with this ratio as well, except for one outlier, C17, which has strong 3.2\,mm emission but is undetected at 850\,$\mu$m. The range of dust temperatures that provide $S_{850\,\mu{\rm m}}/S_{3.2\,{\rm mm}}$ colours within this spread are indicated by the colourbar and range from 35.2 to 44.8\,K, with 39.6\,K providing the closest match to the observed weighted mean colour.}
\label{flux_ratio}
\end{figure}

In Fig.~\ref{flux_ratio} we have plotted the continuum flux density ratio $S_{850\,\mu{\rm m}}/S_{3.2\,{\rm mm}}$ versus $S_{850\,\mu{\rm m}}$ for all central sources with a detection in at least one of these wavelengths, and we show upper limits for the non-detections. We see that the sources with available $S_{850\,\mu{\rm m}}$ and $S_{3.2\,{\rm mm}}$ flux density measurements (11 in total) have similar colours (albeit with some spread), and we calculate the weighted mean flux density ratio to be 70 with a weighted standard deviation of 10. We note that the upper limits provided by sources lacking one of these continuum measurements are also consistent with this mean colour, except for one outlying source, C17, which has strong 3.2\,mm emission but is undetected at 850\,$\mu$m.

We next investigate the range of dust temperatures consistent with the observed spread in $S_{850\,\mu{\rm m}}/S_{3.2\,{\rm mm}}$ by fixing the emissivity index to 2, the critical wavelength to 100\,$\mu$m, and adopting the mean cluster redshift of 4.303. We then calculate the $S_{850\,\mu{\rm m}}/S_{3.2\,{\rm mm}}$ colours predicted by Eq.~\ref{mbb} for a range of finely-gridded dust temperatures, and check to see if the colours fall within the observed weighted standard deviation. In Fig.~\ref{flux_ratio} we show our results, finding that consistent dust temperatures range from 35.2 to 44.8\,K and that $T_{\rm d}\,{=}\,39.6$\,K provides the closest match to the observed weighted mean colour. This is in agreement with results from \citet{strandet2016}, who found the median dust temperature of SPT SMGs to be 39$\,{\pm}\,10\,$K, and in particular estimated the SPT2349$-$56 dust temperature to be 46.7$\,{\pm}\,$2.8\,K.

We next fit the amplitude of the greybody SED of Eq.~\ref{mbb} to each source with one or more continuum detections (see Table \ref{table:cont}) using the same parameters as above and with $T_{\rm d}\,{=}\,39.6$\,K; for sources with one flux density detection, this simply amounts to scaling the greybody function. The best-fitting models were integrated from 42 to 500\,$\mu$m in order to obtain far-infrared luminosities ($L_{\rm FIR}$), consistent with the integration range used in similar studies, and uncertainties were estimated by adding in quadrature the uncertainty from the fit and the difference in the far-infrared luminosity obtained from using the lower and upper limits for $T_{\rm d}$ (35.2\,K and 44.8\,K, respectively) found above. For sources with no continuum detections we scale our $S_{850\,\mu{\rm m}}$ upper limits (except for source N3, where we scale our $S_{3.2\,{\rm mm}}$ upper limit) to obtain upper limits on $L_{\rm FIR}$. We lastly use a conversion factor of 0.95$\,{\times}\,10^{-10}$\,M$_{\odot}$\,yr$^{-1}$/L$_{\odot}$ (from \citealt{kennicutt1998}, modified for a Chabrier initial mass function; see \citealt{chabrier2003}) to convert the far-infrared luminosities into SFRs; these values are all provided in Table \ref{table:cont}.

\subsubsection{Half-light radii}

Lastly, half-light radii were measured using the high-resolution Band 7 dust continuum data obtained in Cycle 6. As shown in Table \ref{table:observations}, the resolution of this data is about 0.2\,arcsec, sufficient to resolve many of the galaxies in our sample, but on the other hand the depth is about 25\,per cent worse than the Band 7 data covering the same region from Cycle 5, thus we only expect to see resolved images of our brightest sources.

Based on our best-fitting Gaussian frequency profiles described above, we created 3$\,{\times}\,$3\,arcsec cutouts of each source by stacking the frequency channels in our high-resolution data containing no line emission in order to produce high-resolution continuum images. We then modelled these images as elliptical S{\'e}rsic profiles convolved with the best-fitting elliptical Gaussian synthesized beam produced by {\tt tclean} during the imaging process, and we allowed the S{\'e}rsic index to vary. Since many sources are not detected in the continuum in this data, fits were only done for sources brighter than 3$\sigma$ in the continuum stacks. We also note that source C7 resolves into a complicated structure, possibly a pair of merging galaxies, and we do not attempt to fit a S{\'e}rsic profile and measure a half-light radius for it. The best-fitting half-light radii, which in our model indicates the sizes of the major axes, are provided in Table \ref{table:cont}, and in Appendix \ref{appendix2} we show the high-resolution continuum images, best-fitting models, and residuals. We find that the half-light radii range from 0.7 to 2.1\,kpc, with a mean of 1.5\,kpc and a dispersion of 0.5\,kpc.

\setlength\tabcolsep{2pt}
\setlength\extrarowheight{2pt}
\begin{table*}
\centering
\caption{Observed properties and derived quantities of sources found in the SPT2349$-$56 proto-cluster field. In this table dashes indicate non-detections, and dots represent cases where no data are available. All upper limits are 1$\sigma$. Sources are named firstly according to the region where they are located (where `C' refers to the core, `N' refers to the northern component, and `SPIREc' refers to the red {\it Herschel\/} source SPIREc), and secondly in order of decreasing [C{\sc ii}] line strength for the core, and decreasing \mbox{CO(4--3)} line strength for the northern component and SPIREc. Continuum-only sources are designated as `NL' and ordered by decreasing 850-$\mu$m flux density. Names given in \citet{miller2018} are provided in brackets.}
\label{table:cont}
\begin{threeparttable}
\begin{tabular}{lcccccccccc}
\hline
Name & Ra Dec & $\Delta v^{\rm a}$ & $R_{1/2}^{\rm b}$ & $D^{\rm c}$ & $S_{3.2\,{\rm mm}}^{\rm d}$ & $S_{1.1\,{\rm mm}}^{\rm e}$ & $S_{850\,\mu{\rm m}}^{\rm f}$ & $L_{\rm FIR}^{\rm g}$ & SFR$^{\rm h}$ \\
& [J2000] & [km\,s$^{-1}$] & [kpc] & [kpc] & [$\mu$Jy] & [mJy] & [mJy] & [$10^{10}$\,L$_{\odot}$] & [M$_{\odot}$\,yr$^{-1}$] \\
\hline
\multicolumn{9}{c}{Core} \\
\hline
C1 (A) & 23:49:42.65 $-$56:38:19.4 & 15$\pm$4 & 2.01$\pm$0.01 & 32$\pm$3 & 133$\pm$8 & 6.21$\pm$0.22 & 8.41$\pm$0.14 & 1030$_{-300}^{+420}$ & 980$_{-290}^{+400}$ \\
C2 (J) & 23:49:43.25 $-$56:38:30.1 & -544$\pm$2 & 2.10$\pm$0.06 & 66$\pm$3 & 22$\pm$9 & 1.08$\pm$0.15 & 1.72$\pm$0.10 & 210$_{-60}^{+80}$ & 200$_{-60}^{+80}$ \\
C3 (B) & 23:49:42.78 $-$56:38:23.8 & -86$\pm$11 & 0.91$\pm$0.02 & 21$\pm$2 & 83$\pm$5 & 5.17$\pm$0.05 & 6.72$\pm$0.07 & 890$_{-260}^{+370}$ & 840$_{-250}^{+350}$ \\
C4 (D) & 23:49:41.41 $-$56:38:22.5 & -83$\pm$13 & 1.33$\pm$0.02 & 58$\pm$2 & 72$\pm$7 & 2.90$\pm$0.09 & 4.51$\pm$0.13 & 570$_{-170}^{+240}$ & 540$_{-160}^{+230}$ \\
C5 (F) & 23:49:42.13 $-$56:38:25.8 & 235$\pm$6 & 1.96$\pm$0.02 & 22$\pm$3 & 49$\pm$6 & 2.24$\pm$0.07 & 3.15$\pm$0.12 & 420$_{-130}^{+180}$ & 400$_{-120}^{+170}$ \\
C6 (C) & 23:49:42.84 $-$56:38:25.1 & 471$\pm$3 & 0.96$\pm$0.05 & 26$\pm$2 & 64$\pm$5 & 3.41$\pm$0.05 & 4.73$\pm$0.07 & 620$_{-180}^{+260}$ & 590$_{-170}^{+240}$ \\
C7 (K) & 23:49:42.96 $-$56:38:18.1 & 631$\pm$3 & -- & 49$\pm$3 & 15$\pm$5 & 0.54$\pm$0.11 & 0.65$\pm$0.07 & 82$_{-25}^{+34}$ & 78$_{-24}^{+32}$ \\
C8 (E) & 23:49:41.22 $-$56:38:24.6 & -22$\pm$4 & 0.72$\pm$0.02 & 68$\pm$2 & 41$\pm$7 & 2.55$\pm$0.10 & 3.91$\pm$0.13 & 490$_{-150}^{+210}$ & 460$_{-140}^{+200}$ \\
C9 (I) & 23:49:42.22 $-$56:38:28.1 & 153$\pm$6 & 1.73$\pm$0.08 & 33$\pm$3 & 15$\pm$6 & 0.85$\pm$0.08 & 1.54$\pm$0.09 & 180$_{-50}^{+80}$ & 170$_{-50}^{+70}$ \\
C10 (H) & 23:49:43.45 $-$56:38:26.2 & -819$\pm$6 & 0.91$\pm$0.02 & 62$\pm$2 & 17$\pm$6 & 0.95$\pm$0.08 & 1.53$\pm$0.08 & 180$_{-60}^{+80}$ & 180$_{-50}^{+70}$ \\
C11 (L) & 23:49:42.36 $-$56:38:25.7 & -409$\pm$6 & 1.01$\pm$0.04 & 15$\pm$3 & $<$6 & $<$0.19 & 0.33$\pm$0.12 & 39$_{-18}^{+21}$ & 37$_{-17}^{+20}$ \\
C12 & 23:49:42.86 $-$56:38:23.9 & -615$\pm$7 & -- & 26$\pm$2 & $<$5 & $<$0.10 & 0.25$\pm$0.04 & 29$_{-10}^{+13}$ & 28$_{-9}^{+12}$ \\
C13 (G) & 23:49:42.73 $-$56:38:25.1 & 264$\pm$17 & 0.73$\pm$0.12 & 21$\pm$2 & 14$\pm$2 & 1.18$\pm$0.04 & 1.32$\pm$0.05 & 180$_{-50}^{+80}$ & 170$_{-50}^{+70}$ \\
C14 (N) & 23:49:43.27 $-$56:38:23.1 & 101$\pm$5 & -- & 49$\pm$2 & $<$5 & 0.24$\pm$0.10 & $<$0.18 & 51$_{-27}^{+31}$ & 49$_{-25}^{+30}$ \\
C15 & 23:49:41.42 $-$56:38:14.9 & -107$\pm$8 & -- & 82$\pm$3 & $<$6 & \dots & 0.30$\pm$0.08 & 36$_{-14}^{+17}$ & 34$_{-13}^{+16}$ \\
C16 & 23:49:42.73 $-$56:38:25.8 & 231$\pm$16 & -- & 24$\pm$3 & $<$5 & $<$0.10 & $<$0.11 & $<$5 & $<$5 \\
C17 (M) & 23:49:43.41 $-$56:38:20.9 & 49$\pm$7 & -- & 60$\pm$2 & 14$\pm$5 & $<$0.23 & $<$0.10 & 110$_{-60}^{+80}$ & 110$_{-60}^{+80}$ \\
C18 & 23:49:41.67 $-$56:38:13.9 & -328$\pm$13 & -- & 79$\pm$3 & $<$6 & \dots & 0.14$\pm$0.07 & 17$_{-9}^{+10}$ & 16$_{-9}^{+10}$ \\
C19 & 23:49:43.86 $-$56:38:43.4 & -441$\pm$8 & -- & 159$\pm$3 & $<$7 & \dots & 0.18$\pm$0.05 & 21$_{-8}^{+10}$ & 20$_{-8}^{+9}$ \\
C20 & 23:49:42.28 $-$56:38:24.0 & -557$\pm$8 & -- & 8$\pm$2 & $<$5 & $<$0.05 & $<$0.08 & $<$9 & $<$9 \\
C21 & 23:49:43.01 $-$56:38:27.0 & 124$\pm$7 & -- & 41$\pm$3 & $<$4 & $<$0.07 & $<$0.04 & $<$5 & $<$5 \\
C22 & 23:49:42.51 $-$56:38:27.4 & 163$\pm$3 & -- & 27$\pm$3 & $<$4 & $<$0.05 & $<$0.04 & $<$5 & $<$5 \\
C23 & 23:49:42.61 $-$56:38:26.7 & 112$\pm$6 & -- & 25$\pm$3 & $<$2 & $<$0.04 & $<$0.04 & $<$5 & $<$5 \\
NL1 & 23:49:40.55 $-$56:38:06.0 & -- & -- & 161$\pm$3 & 37$\pm$10 & \dots & 4.10$\pm$0.09 & -- & -- \\
NL3 & 23:49:44.73 $-$56:38:40.0 & -- & -- & 174$\pm$3 & $<$17 & \dots & 0.56$\pm$0.05 & -- & -- \\
\hline
\multicolumn{9}{c}{North} \\
\hline
N1 & 23:49:42.53 $-$56:37:33.2 & 480$\pm$10 & \dots & 348$\pm$4 & 156$\pm$7 & \dots & 14.97$\pm$0.15 & 1730$_{-510}^{+720}$ & 1640$_{-490}^{+680}$ \\
N2 & 23:49:43.53 $-$56:37:16.7 & 493$\pm$22 & \dots & 466$\pm$4 & 79$\pm$11 & \dots & \dots & 650$_{-270}^{+400}$ & 620$_{-250}^{+380}$ \\
N3 & 23:49:40.75 $-$56:37:33.9 & 1532$\pm$29 & \dots & 355$\pm$4 & $<$8 & \dots & \dots & $<$64 & $<$60 \\
NL2 & 23:49:43.43 $-$56:37:35.4 & -- & \dots & 338$\pm$3 & $<$8 & \dots & 1.11$\pm$0.08 & -- & -- \\
\hline
\multicolumn{9}{c}{SPIREc} \\
\hline
SPIREc1 & 23:49:24.95 $-$56:35:38.6 & -1148$\pm$12 & \dots & 1512$\pm$3 & 39$\pm$12 & \dots & \dots & 330$_{-160}^{+220}$ & 310$_{-150}^{+210}$ \\
SPIREc2 & 23:49:22.78 $-$56:35:28.4 & -1326$\pm$35 & \dots & 1647$\pm$3 & $<$10 & \dots & \dots & $<$86 & $<$82 \\
SPIREc3 & 23:49:24.79 $-$56:35:12.8 & -1274$\pm$21 & \dots & 1656$\pm$3 & $<$23 & \dots & \dots & $<$80 & $<$76 \\
\hline
\end{tabular}
\begin{tablenotes}
\item $^{\rm a}$ Line-of-sight velocity relative to the mean redshift of the main central component of SPT2349, $z=4.303$.
\item $^{\rm b}$ Half-light radius (as proper distance) from fitting S{\'e}rsic profiles to the high-resolution Band 7 continuum data obtained in Cycle 6.
\item $^{\rm c}$ Proper distance from 850-$\mu$m flux-weighted centre at 23:49:42.41 $-$56:38:23.6.
\item $^{\rm d}$ Continuum flux density at 3.2\,mm, measured from the \mbox{CO(4--3)} map channels with no line emission.
\item $^{\rm e}$ Continuum flux density at 1.1\,mm, independently measured from the Band 7 imaging described in \citet{miller2018}.
\item $^{\rm f}$ Continuum flux density measured at 850\,$\mu$m from the [C{\sc ii}] map channels with no line emission.
\item $^{\rm g}$ Far-infrared luminosity, obtained by fitting a greybody function to the available continuum flux density measurements \citep[see][]{greve2012} and integrating from 42 to 500\,$\mu$m. In the fitting process $\beta$ was fixed to 2, $\lambda_0$ was fixed to 100\,$\mu$m, and the dust temperature, $T_{\rm d}$, was fixed to 39.6\,K, while uncertainties were estimated by varying the dust temperature between 35.2 and 44.8\,K (see Section \ref{fir_lums}). Where only continuum flux density upper limits are available, we scale the 850-$\mu$m flux density upper limit to estimate an upper limit on $L_{\rm FIR}$, except for source N3, where we scale the 3.2-mm flux density upper limit.
\item $^{\rm h}$ Using a conversion factor from $L_{\rm FIR}$ to SFR of 0.95$\,{\times}\,10^{-10}$\,M$_{\odot}$\,yr$^{-1}$/L$_{\odot}$.
\end{tablenotes}
\end{threeparttable}
\end{table*}

\setlength\tabcolsep{2pt}
\setlength\extrarowheight{2pt}
\begin{table*}
\centering
\caption{Line characterization and derived quantities of sources found in the SPT2349$-$56 proto-cluster field. In this table dashes indicate non-detections, and dots represent cases where no data are available. All upper limits are 1$\sigma$, and the limits on line strengths and luminosities were derived as the average uncertainty from our measurements. Sources are named firstly according to the region where they are located (where `C' refers to the core, `N' refers to the northern component, and `SPIREc' refers to the red {\it Herschel\/} source SPIREc), and secondly in order of decreasing [C{\sc ii}] line strength for the core, and decreasing \mbox{CO(4--3)} line strength for the northern component and SPIREc. Continuum-only sources are designated as `NL' and ordered by decreasing 850-$\mu$m flux density. Names given in \citet{miller2018} are provided in brackets.}
\label{table:line}
\begin{threeparttable}
\begin{tabular}{lccccccccc}
\hline
Name & $F_{[\rm C\textsc{ii}]}^{\rm a}$ & $L^{\rm b}_{[\rm C\textsc{ii}]}$ & FWHM$_{[\rm C\textsc{ii}]}^{\rm c}$ & $F_{\rm CO(4-3)}^{\rm a}$ & $L^{\rm b}_{\rm CO(4-3)}$ & FWHM$_{\rm CO(4-3)}^{\rm c}$ & $M_{\rm dyn}^{\rm d}$ & $M_{\rm gas}^{\rm e}$ & $M_{\rm halo}^{\rm f}$ \\
& [Jy\,km\,s$^{-1}$] & [10$^{8}$\,L$_{\odot}$] & [km\,s$^{-1}$] & [Jy\,km\,s$^{-1}$] & [10$^{7}$\,L$_{\odot}$] & [km\,s$^{-1}$] & [10$^{10}$\,M$_\odot$] & [10$^{10}$\,M$_\odot$] & [10$^{10}$\,M$_\odot$] \\
\hline
\multicolumn{10}{c}{Core} \\
\hline
C1 (A) & 16.86$\pm$0.20 & 100.9$\pm$1.2 & 486$\pm$14/493$\pm$13 & 0.98$\pm$0.03 & 14.2$\pm$0.5 & 399$\pm$42/850$\pm$108 & 27.0$\pm$1.1 & 7.5$\pm$0.7 & 320$\pm$30 \\
C2 (J) & 8.82$\pm$0.13 & 52.6$\pm$0.8 & 343$\pm$5 & 0.27$\pm$0.02 & 3.9$\pm$0.3 & 326$\pm$28 & 6.9$\pm$0.3 & 2.1$\pm$0.2 & 88$\pm$10 \\
C3 (B) & 7.89$\pm$0.12 & 47.2$\pm$0.7 & 152$\pm$13/606$\pm$31 & 0.56$\pm$0.02 & 8.1$\pm$0.3 & 281$\pm$33/494$\pm$41 & 9.9$\pm$1.0 & 4.3$\pm$0.4 & 180$\pm$20 \\
C4 (D) & 5.90$\pm$0.15 & 35.3$\pm$0.9 & 394$\pm$31/465$\pm$41 & 0.38$\pm$0.02 & 5.6$\pm$0.3 & 273$\pm$43/624$\pm$127 & 13.8$\pm$1.7 & 3.0$\pm$0.3 & 130$\pm$10 \\
C5 (F) & 5.19$\pm$0.15 & 31.1$\pm$0.9 & 270$\pm$13/506$\pm$26 & 0.14$\pm$0.02 & 2.0$\pm$0.3 & 290$\pm$72/304$\pm$68 & 18.0$\pm$1.5 & 1.1$\pm$0.2 & 45$\pm$8 \\
C6 (C) & 5.16$\pm$0.07 & 31.0$\pm$0.4 & 372$\pm$6 & 0.44$\pm$0.02 & 6.5$\pm$0.2 & 402$\pm$17 & 3.7$\pm$0.2 & 3.4$\pm$0.3 & 150$\pm$10 \\
C7 (K) & 3.76$\pm$0.09 & 22.6$\pm$0.5 & 312$\pm$7 & 0.13$\pm$0.01 & 1.9$\pm$0.2 & 280$\pm$27 & 3.6$\pm$1.4 & 1.0$\pm$0.1 & 42$\pm$5 \\
C8 (E) & 3.68$\pm$0.12 & 22.0$\pm$0.7 & 181$\pm$17/264$\pm$16 & 0.31$\pm$0.02 & 4.5$\pm$0.3 & 489$\pm$36 & 2.1$\pm$0.2 & 2.4$\pm$0.3 & 100$\pm$10 \\
C9 (I) & 3.35$\pm$0.11 & 20.1$\pm$0.7 & 473$\pm$14 & 0.12$\pm$0.02 & 1.7$\pm$0.2 & 491$\pm$64 & 10.8$\pm$0.8 & 0.9$\pm$0.1 & 39$\pm$6 \\
C10 (H) & 2.96$\pm$0.10 & 17.6$\pm$0.6 & 197$\pm$17/320$\pm$22 & 0.14$\pm$0.02 & 2.1$\pm$0.2 & 460$\pm$51 & 3.6$\pm$0.4 & 1.1$\pm$0.2 & 47$\pm$7 \\
C11 (L) & 2.70$\pm$0.11 & 16.2$\pm$0.6 & 362$\pm$13 & 0.04$\pm$0.01 & 0.7$\pm$0.2 & 269$\pm$82 & 3.7$\pm$0.3 & 0.3$\pm$0.1 & 15$\pm$5 \\
C12 & 2.62$\pm$0.09 & 15.6$\pm$0.5 & 285$\pm$18 & <0.02 & <0.2 & -- & 3.0$\pm$1.2 & $<$0.2 & $<$7 \\
C13 (G) & 1.84$\pm$0.08 & 11.1$\pm$0.5 & 226$\pm$37/315$\pm$46 & 0.10$\pm$0.01 & 1.5$\pm$0.1 & 305$\pm$80/426$\pm$51 & 3.1$\pm$0.9 & 0.8$\pm$0.1 & 34$\pm$4 \\
C14 (N) & 1.70$\pm$0.08 & 10.2$\pm$0.5 & 241$\pm$12 & 0.03$\pm$0.01 & 0.4$\pm$0.2 & 239$\pm$88 & 2.1$\pm$0.9 & 0.2$\pm$0.1 & 9$\pm$4 \\
C15 & 1.65$\pm$0.09 & 9.9$\pm$0.5 & 339$\pm$19 & <0.02 & <0.2 & -- & 4.2$\pm$1.7 & $<$0.2 & $<$7 \\
C16 & 1.11$\pm$0.06 & 6.7$\pm$0.3 & 243$\pm$46 & <0.02 & <0.2 & -- & 2.2$\pm$1.2 & $<$0.2 & $<$7 \\
C17 (M) & 0.93$\pm$0.09 & 5.6$\pm$0.5 & 191$\pm$17 & <0.02 & <0.2 & -- & 1.3$\pm$0.6 & $<$0.2 & $<$7 \\
C18 & 0.86$\pm$0.09 & 5.1$\pm$0.5 & 318$\pm$30 & <0.02 & <0.2 & -- & 3.7$\pm$1.6 & $<$0.2 & $<$7 \\
C19 & 0.85$\pm$0.07 & 5.1$\pm$0.4 & 192$\pm$17 & <0.02 & <0.2 & -- & 1.4$\pm$0.6 & $<$0.2 & $<$7 \\
C20 & 0.51$\pm$0.06 & 3.0$\pm$0.4 & 170$\pm$21 & <0.02 & <0.2 & -- & 1.1$\pm$0.5 & $<$0.2 & $<$7 \\
C21 & 0.35$\pm$0.04 & 2.1$\pm$0.2 & 153$\pm$17 & <0.02 & <0.2 & -- & 0.9$\pm$0.4 & $<$0.2 & $<$7 \\
C22 & 0.33$\pm$0.03 & 2.0$\pm$0.1 & 79$\pm$5 & <0.02 & <0.2 & -- & 0.2$\pm$0.1 & $<$0.2 & $<$7 \\
C23 & 0.28$\pm$0.03 & 1.6$\pm$0.2 & 119$\pm$13 & <0.02 & <0.2 & -- & 0.5$\pm$0.2 & $<$0.2 & $<$7 \\
NL1 & <0.09 & <0.5 & -- & <0.02 & <0.2 & -- & -- & -- & -- \\
NL3 & <0.09 & <0.5 & -- & <0.02 & <0.2 & -- & -- & -- & -- \\
\hline
\multicolumn{10}{c}{North} \\
\hline
N1 & 18.87$\pm$0.20 & 113.3$\pm$1.2 & 414$\pm$27/520$\pm$18 & 1.55$\pm$0.03 & 22.6$\pm$0.4 & 672$\pm$12 & 16.2$\pm$6.5 & 12.0$\pm$1.0 & 510$\pm$40 \\
N2 & \dots & \dots & \dots & 0.65$\pm$0.04 & 9.4$\pm$0.5 & 307$\pm$61/645$\pm$86 & 18.7$\pm$8.5 & 5.0$\pm$0.5 & 210$\pm$20 \\
N3 & \dots & \dots & \dots & 0.12$\pm$0.02 & 1.8$\pm$0.3 & 421$\pm$65 & 6.5$\pm$3.3 & 0.9$\pm$0.2 & 40$\pm$7 \\
NL2 & <0.09 & <0.5 & -- & <0.02 & <0.2 & -- & -- & -- & -- \\
\hline
\multicolumn{10}{c}{SPIREc} \\
\hline
SPIREc1 & \dots & \dots & \dots & 0.72$\pm$0.05 & 10.4$\pm$0.7 & 326$\pm$45/333$\pm$42 & 7.9$\pm$3.5 & 5.5$\pm$0.6 & 240$\pm$30 \\
SPIREc2 & \dots & \dots & \dots & 0.37$\pm$0.05 & 5.4$\pm$0.7 & 409$\pm$148/410$\pm$93 & 12.3$\pm$7.2 & 2.9$\pm$0.5 & 120$\pm$20 \\
SPIREc3 & \dots & \dots & \dots & 0.25$\pm$0.04 & 3.7$\pm$0.6 & 354$\pm$51 & 4.6$\pm$2.2 & 1.9$\pm$0.3 & 83$\pm$15 \\
\hline
\end{tabular}
\begin{tablenotes}
\item $^{\rm a}$ Line strength, obtained by integrating the spectra over $[-3\sigma,3\sigma]$, where $\sigma$ is the linewidth (in units of standard deviation) as determined from the best fit. In the case of a two-peaked fit, the integration bounds are $[-3\sigma_{\rm L},3\sigma_{\rm R}]$, where $\sigma_{\rm L}$ and $\sigma_{\rm R}$ are from the left and right Gaussian fits, respectively.
\item $^{\rm b}$ Line luminosity, calculated using $L=4 \pi D_{\rm L}^2 F$, where $D_{\rm L}$ is the luminosity distance and $F$ is the line strength.
\item $^{\rm c}$ FWHM from the best-fitting single Gaussian models for the emission lines; where two peaks are detected, we provide both FWHM.
\item $^{\rm d}$ Dynamical mass, calculated using the isotropic virial estimator, $M_{\rm dyn}\,{=}\,2.8\,{\times}\,10^5\,{\rm FWHM}^2\,R_{1/2}$\,M$_{\odot}$ \citep[e.g.][]{spitzer1987}, where FWHM is the linewidth in km\,s$^{-1}$ and $R_{1/2}$ is the half-light radius in kpc. Here we used [C{\sc ii}] linewidth measurements wherever possible, otherwise \mbox{CO(4--3)} linewidth measurements were used. For sources best fit by double Gaussians, we treated each Gaussian separatlely and summed the results. For galaxies where we could not measure $R_{1/2}$ from our high-resolution data, we adopted $R_{1/2}\,{=}\,(1.5\,{\pm}\,0.5)\,$kpc, corresponding to the mean half-light radius of our sample.
\item $^{\rm e}$ Gas mass derived from our observations of CO: \mbox{CO(4--3)} luminosities were converted to \mbox{CO(1--0)} luminosities using a conversion factor of $r_{4,1}\,{=}\,0.60\,{\pm}0.05$, the average line strength ratio of SPT SMGs \citep{spilker2014}, and then we applied a scaling factor of $\alpha_{\rm CO}\,{=}\,1$\,M$_{\odot}$/(K\,km\,s$^{-1}$\,pc$^2$).
\item $^{\rm f}$ Halo mass, obtained from gas mass using a scaling factor of 42.8 \citep{rennehan2019}, appropriate for galaxies with gas mass fractions of 0.7 and stellar mass-to-halo mass fractions of 0.01.
\end{tablenotes}
\end{threeparttable}
\end{table*}

\section{Results}
\label{results}

\subsection{Velocity distribution}
\label{velocities}

\begin{figure*}
\includegraphics[width=\textwidth]{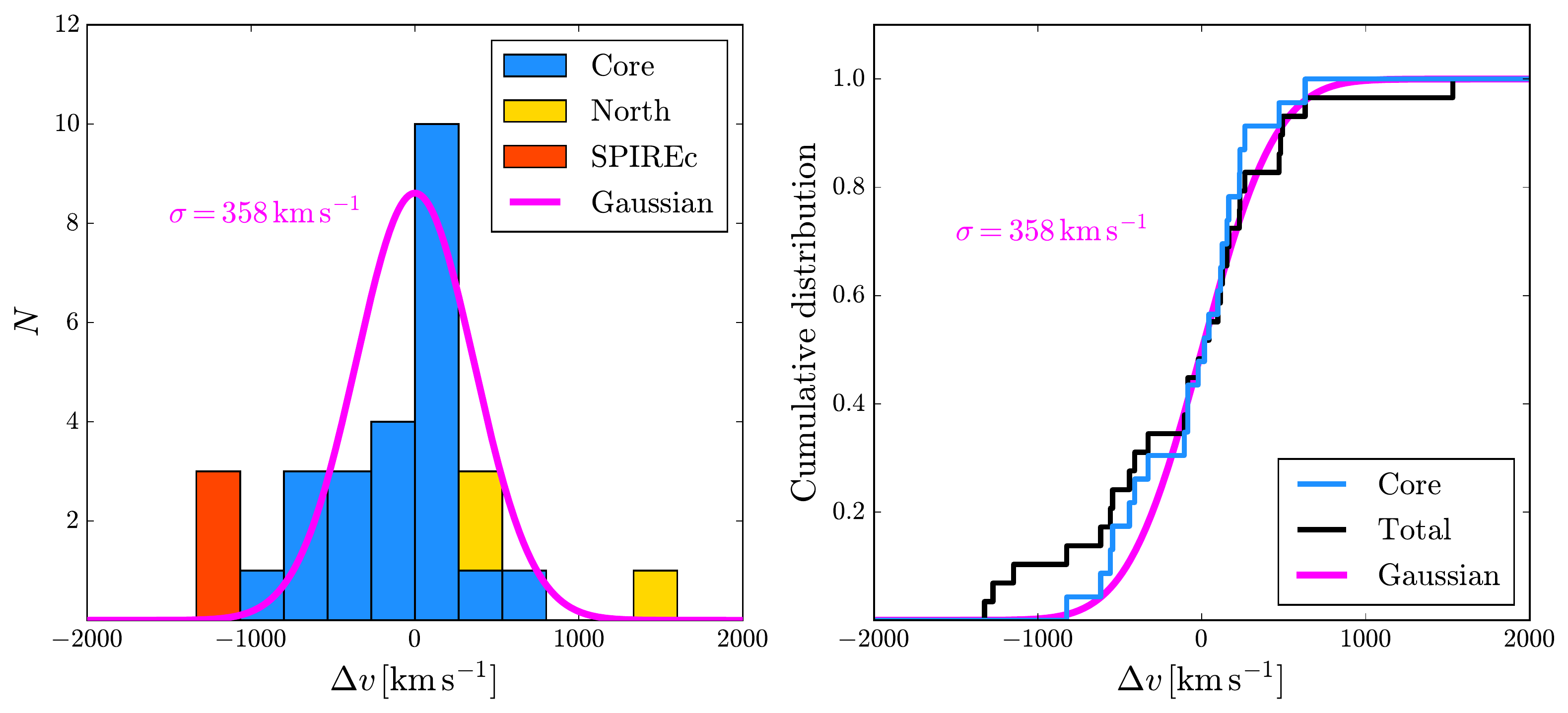}
\caption{Histogram ({\it left}) and cumulative distribution function ({\it right}) of proto-cluster member velocities relative to the mean redshift of 4.303. Main core members are shown in blue and compared to a Gaussian distribution with the same standard deviation shown in magenta. Additionally, the three galaxies found in the northern region are shown added to the core histogram in yellow, and the three galaxies found in SPIREc are shown added to the core histogram in red. The cumulative distribution of all sources is shown in black.}
\label{v_dist}
\end{figure*}

The spatial distribution of sources, shown in Fig.~\ref{coverage} and Fig.~\ref{all_sources}, is suggestive of three main groups of galaxies: a core group; a northern group; and a SPIREc group. To investigate this, in Fig.~\ref{v_dist} we show the velocity histogram of our sources, with core sources coloured in blue, northern sources coloured in yellow, and SPIREc sources coloured in red. The distribution of galaxies in the northern and SPIREc components are indeed largely offset from the core by about ${\pm}\,$1000\,km\,s$^{-1}$, indicative of surrounding galaxy groups or filaments of a central structure, however small number statistics means that we cannot draw any strong conclusions here. In Fig.~\ref{v_dist} we also show the cumulative distribution of the core sources, and compare this to the cumulative distribution of the whole sample.

The standard deviation of relative radial velocities within the central region was calculated to be 358\,km\,s$^{-1}$, and in Fig.~\ref{v_dist} we show a Gaussian distribution with a standard deviation of 358\,km\,s$^{-1}$ for comparison. We can see that the core galaxies do exhibit a nearly Gaussian distribution, as expected for relaxed systems that are gravitationally bound, but there may be some hint of interesting substructure within the distribution. To test the Gaussianity of our sample, we perform a Lilliefors test \citep{lilliefors1967}, which is based on the commonly-used Kolmogorov-Smirnov test for cases where the underlying mean and variance are not known, but instead estimated by the sample mean and variance. Our null hypothesis is that the relative radial velocities in the core follow a Gaussian distribution. First, we find the maximum distance between the sample cumulative distribution and the Gaussian cumulative distribution with our sample mean and variance (i.e.~the maximum distance between the blue and magenta curves shown in Fig.~\ref{v_dist}) to be 0.15, and second, we find the corresponding $p$-value from the Lilliefors distribution (which takes into account a range of underlying means and variances) to be 0.20, meaning that the probability of obtaining a maximum distance of 0.15 or greater is 20\,per cent if the relative radial velocities are Gaussian-distributed. We therefore do not find strong evidence for rejecting the null hypothesis that the relative radial velocities in the core are Gaussian-distributed.

\subsection{Mass estimates}
\label{mass}

\subsubsection{Mass from velocity dispersion}
\label{virialization}

Under the assumption that the core of SPT2349$-$56 has virialized, we can ask what its total mass is based on the velocity dispersion of the constituent galaxies. While it is probably not entirely true that the core has completely virialized as it may still be in the process of collapsing, our analysis of the distribution of relative velocities suggests that this is not a bad approximation. 

Before estimating the mass we must first deal with completeness, since we want to calculate the velocity dispersion within a region that has been approximately uniformly sampled. We first compute the 850-$\mu$m flux-weighted centre of the core, and we find that the smallest proper distance between this centre and a point where our primary-beam response drops to 0.5 is about 90\,kpc (or about 13\,arcsec). We therefore take this to be the effective radius of the core region where our data is approximately complete; this removes galaxy C19 from the analysis. We note that \citet{miller2018} used a smaller radius of 65\,kpc in their analysis, since the newer more extended data were not available at the time. For reference, we have plotted the circle used in this analysis in Fig.~\ref{all_sources}, as well as the circle used by \citet{miller2018}.

The velocity dispersion of the galaxies within this region is $\sigma_r\,{=}\,$370$_{-89}^{+67}$\,km\,s$^{-1}$ using the biweight estimator \citep{beers1990}, which is a more robust estimator for the underlying standard deviation when the sample size is not large, with the uncertainty estimated using bootstrap resampling. We then convert this velocity dispersion into a mass using the scaling relation found by \citet{evrard2008}, which is based on a suite of $N$-body simulations run with different cosmologies:
\begin{equation}
M_{200}=\frac{1}{h(z)}\left(\frac{\sigma_r}{1082.9\,\mathrm{km\,s^{-1}}}\right)^{1/0.3361} 10^{15}\,\mathrm{M}_{\odot},
\end{equation}
\noindent
where $h(z)\,{=}\,H(z)/100\,$km\,s$^{-1}$\,Mpc$^{-1}$. The resulting mass is (9$\pm$5)$\,{\times}\,$10$^{12}$\,M$_{\odot}$; we note that \citet{miller2018} estimated a core mass (using the same scaling relation) of (12$\pm$7)$\,{\times}\,$10$^{12}$\,M$_{\odot}$, based on the velocity dispersion of 14 sources. If we restrict our calculation to the same sources we find a velocity dispersion of 371$_{-138}^{+93}$\,km\,s$^{-1}$ and a mass of (9$\pm$7)$\,{\times}\,$10$^{12}$\,M$_{\odot}$, which is entirely consistent with their result.

\begin{figure*}
\includegraphics[width=\textwidth]{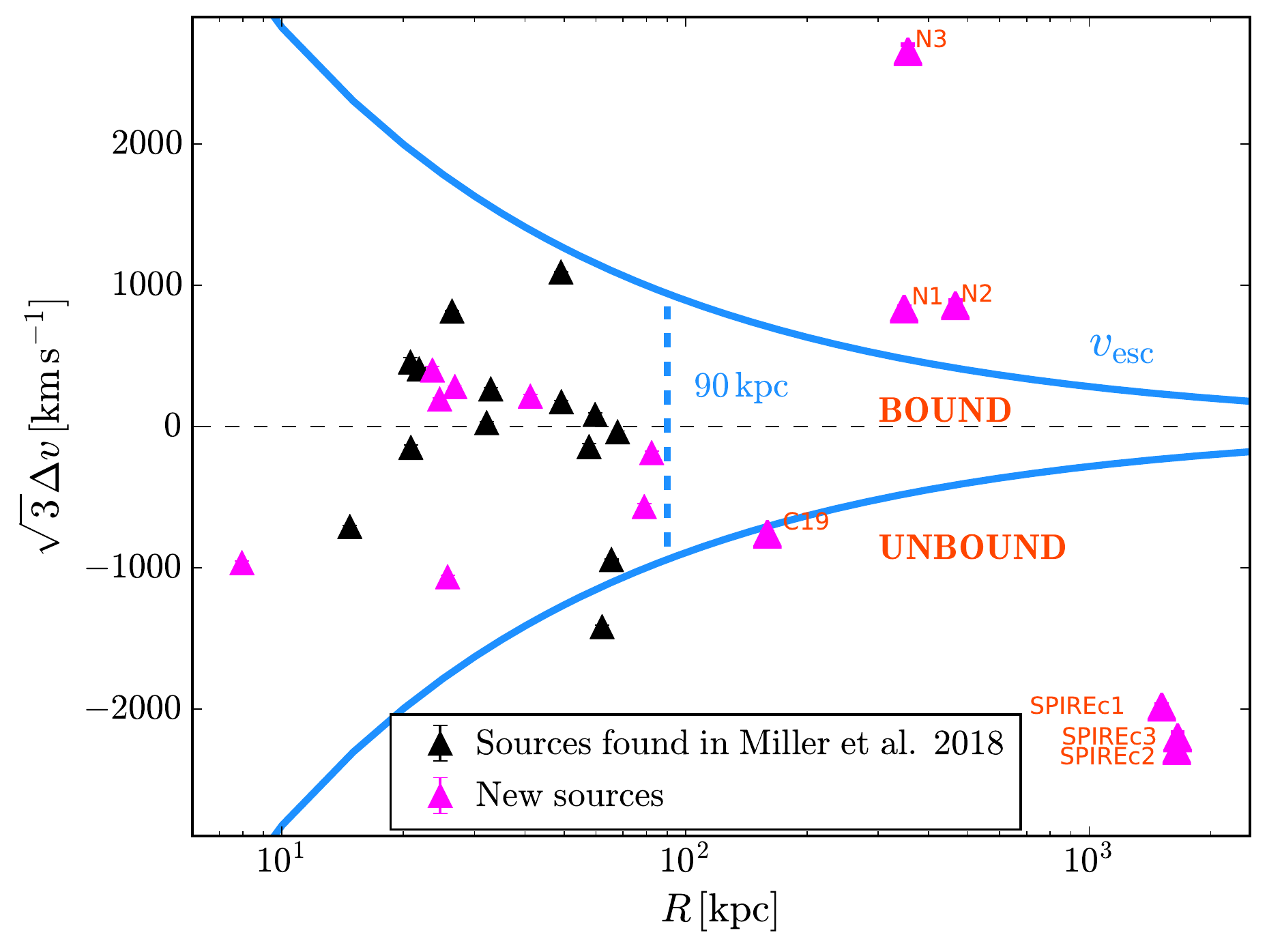}
\caption{Line-of-sight velocities relative to the mean redshift of SPT2349$-$56 ($z=4.303$) as a function of distance from the 850-$\mu$m flux-weighted centre of the proto-cluster, with the 14 galaxies originally detected by \citet{miller2018} shown in black and our new galaxies shown in magenta. Here the velocities are scaled by $\sqrt{3}$ as an estimate of the true 3-dimensional velocity. We note that statistical error bars are plotted here, but they are smaller than the symbol sizes. The estimated mass contained within 90\,kpc (proper distance) is (9$\pm$5)$\,{\times}\,$10$^{12}$\,M$_{\odot}$, and the two blue curves show the escape velocity (positive and negative) as a function of distance from this mass. The galaxies found in the northern component of the structure and in SPIREc have velocities outside of the region enclosed by these two curves are not expected to be gravitationally bound to the central mass.}
\label{r_v}
\end{figure*}

Another question we can ask is, given the presence of a 9$\,{\times}\,$10$^{12}$\,M$_{\odot}$ central object, are the surrounding components gravitationally bound. In Fig.~\ref{r_v} we show the measured relative velocities of the galaxies in the proto-cluster system (scaled by a factor of $\sqrt{3}$, since we have only measured line-of-sight velocities). We also plot the escape speed for a point mass of 9$\,{\times}\,$10$^{12}$\,M$_{\odot}$ as a function of distance, given by $v_{\rm esc}\,{=}\,\sqrt{2GM/R}$; galaxies with velocities that fall within the envelope are expected to be gravitationally bound, and those outside the envelope have velocities that exceed the escape speed of the central mass and are expected to leave the system on hyperbolic orbits. We find that the galaxies found in the northern extent of SPT2349$-$56 and in SPIREc show velocities greater than the escape velocity. While this suggests that the northern and SPIREc galaxies will not end up falling into the potential well of the core, it is still very likely that they will remain bound within the entire proto-cluster system.

\subsubsection{Galaxy dynamical masses}
\label{dynamical}

An interesting comparison to make with the total proto-cluster core mass evaluated above is to look at the sum of the masses of the individual galaxies within the proto-cluster core. First, we probe the dynamical masses of the galaxies in our sample through their measured linewidths. To do this, we have adopted the isotropic virial estimator, which relates linewidths to radii using the equation
\begin{equation}
\label{mdyn}
M_{\rm dyn} = 2.8 \times 10^5 \, {\rm FWHM}^2 \, R\,\mathrm{M}_{\odot},
\end{equation}
\noindent
where FWHM is the linewidth in km\,s$^{-1}$ and $R$ is the half-light radius in kpc \citep[e.g.][]{spitzer1987}. We note that this equation assumes the gas to be dispersion-dominated, whereas the constant of proportionality can be smaller by a factor of a few for gas rotating in a disc inclined relative to the line-of-sight.

In order to estimate dynamical masses we thus need linewidths and radii. For linewidths we use the results from our best-fitting Gaussian profiles of [C{\sc ii}] emission (since most of the galaxies are detected in [C{\sc ii}]), and for the galaxies for which we have no [C{\sc ii}] data we use linewidths determined from fits to the \mbox{CO(4--3)} profiles. Next, we used the half-light radii measured from our continuum high-resolution data when possible, and otherwise adopted the mean half-light radius of sample, $R_{1/2}\,{=}\,1.5\,{\pm}\,0.5\,$kpc (where the uncertainty is the standard deviation). Lastly, for sources best fit by double-Gaussian profiles, we treat each Gaussian separately and sum the two resulting masses to obtain the total dynamical mass -- this assumes that double-Gaussian profiles are a result of two unresolved sources, as opposed to for example a rotating disc, consistent with the use of Eq.~\ref{mdyn}. Table \ref{table:line} shows the results of this calculation.

We find that the sum of the dynamical masses of the galaxies within the core is 1.3$\,{\times}\,10^{12}\,$M$_{\odot}$; this is less than the total core mass estimated through the velocity dispersion of 9$\,{\times}\,10^{12}\,$M$_{\odot}$, which is to be expected since we are only measuring the cores of the galaxies here, and there could be much more mass between the individual galaxies than we have detected. For comparison, if we use \mbox{CO(4--3)} linewidths for just those galaxies in the core where this line is detected, we find a total core mass of 1.3$\,{\times}\,10^{12}\,$M$_{\odot}$. However, only half the core galaxies are actually detected in \mbox{CO(4--3)}, so the total core mass from this diagnostic should be about a factor of two larger; our dynamical mass estimates should therefore be treated as uncertain to within a factor of a few. We note that this total mass is in line with what is expected for brightest cluster galaxies (BCGs), and in Section \ref{discussion_spatial} we discuss in more detail the possibility that these core galaxies could merge into a BCG.

For comparison, the total dynamical mass of the galaxies found in the northern extension of the proto-cluster is 4.1$\,{\times}\,10^{11}\,$M$_{\odot}$, or approximately 3 times smaller than the core, and similarly we find the total mass of the galaxies in the SPIREc component is 2.5$\,{\times}\,10^{11}\,$M$_{\odot}$, or approximately 5 times smaller than the core. 

\subsubsection{Galaxy gas and halo masses}
\label{halo}

Another useful property we can derive is the gas masses of the individual galaxies within the proto-cluster structure. We calculate this property as follows: firstly, line luminosities were calculated in units of K\,km\,s$^{-1}$\,pc$^2$ (i.e. $L^{\prime}$ units, see e.g. \citealt{solomon1997}) using the equation
\begin{equation}\label{Lprime}
L^{\prime}_{\mathrm{CO(4-3)}} = \frac{c^2}{2 k_{\mathrm{B}}} \nu_{\mathrm{obs}}^{-2} D_{\mathrm{L}}^2 F_{\mathrm{CO(4-3)}} (1+z)^{-3}.
\end{equation}
\noindent
These were then converted to $L^{\prime}_{\rm CO(1-0)}$ luminosities using a conversion factor of $r_{4,1}\,{=}\,0.60\,{\pm}0.05$, the average line strength ratio of SPT SMGs in \citet{spilker2014}; lastly, to convert to gas mass, we need to adopt a value of $\alpha_{\rm CO}$. There are many values of $\alpha_{\rm CO}$ reported in the literature, ranging from 1 for starbursting galaxies to 4 for more quiescently star-forming galaxies \citep[e.g.,][]{tacconi2008,daddi2010,carilli2013,aravena2016a,bothwell2017}. Our sample spans quite a large dynamical range, and probably includes both typical star-forming galaxies and rare starbursting galaxies. However, for simplicity, we adopt a conversion factor of $\alpha_{\rm CO}\,{=}\,1$\,M$_{\odot}$/(K\,km\,s$^{-1}$\,pc$^2$) to obtain gas masses, and note that these quantities should be interpreted primarily as order-of-magnitude estimates, while multi-line analyses of these galaxies will follow in future work. Table \ref{table:line} shows the resulting gas masses estimated using the procedure outlined above.

\citet{miller2018} present observations of \mbox{CO(2--1)} within the core of SPT2349$-$56, and uses the results to independently derive the gas masses of these sources. These observations do not resolve individual sources, but instead should be representative of the total core gas mass, and can be used to verify the CO line strength conversion factor we have used. They find a total core gas mass of 1.8\,${\times}\,10^{11}\,$M$_{\odot}$, and meanwhile we find that the sum of the gas masses within the core reach a total value of 2.9\,${\times}\,10^{11}\,$M$_{\odot}$. However, the \mbox{CO(2--1)} data of \citet{miller2018} is much shallower than our \mbox{CO(4--3)} data and may be missing a significant fraction of the total flux; nonetheless, the two measurements do provide consistent results to within a factor of a few. Similarly, the total gas mass of the northern component is found to be 1.8\,${\times}\,10^{11}\,$M$_{\odot}$, and that of SPIREc is 1.0\,${\times}\,10^{11}\,$M$_{\odot}$.

The gas masses of galaxies should in turn be proportional to their stellar masses and corresponding halo masses. \citet{rennehan2019} derive a scaling factor of 42.8 between gas mass and halo mass, assuming a gas mass fraction of 0.7 and a stellar mass-to-halo mass fraction of 0.01. We have applied this factor to the galaxies in our sample and provide the results in Table \ref{table:line}. For reference, the total halo mass of the core found here is 1.3\,${\times}\,10^{13}\,$M$_{\odot}$, while for the northern component the sum is 7.7\,${\times}\,10^{12}\,$M$_{\odot}$, and for the SPIREc component the sum is 4.4\,${\times}\,10^{12}\,$M$_{\odot}$. We note that this procedure treats the underlying dark matter distribution as the sum of individual halos, while the velocity dispersion method used above assumes that the individual galaxies are on orbits inside of a larger host halo. 

\subsection{Far-infrared properties}

\begin{figure*}
\includegraphics[width=0.45\textwidth,height=6.2cm]{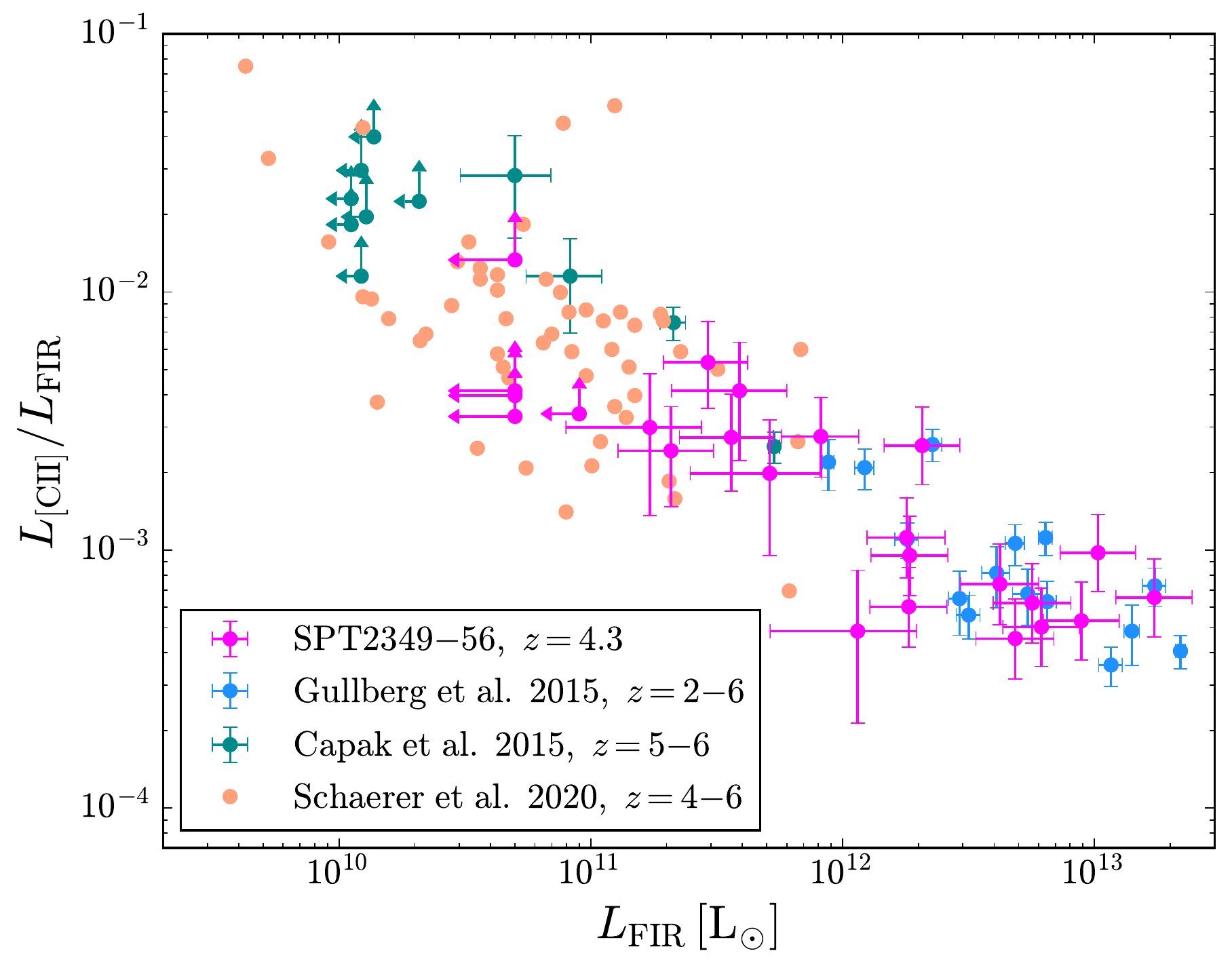}
\includegraphics[width=0.45\textwidth,height=6.1cm]{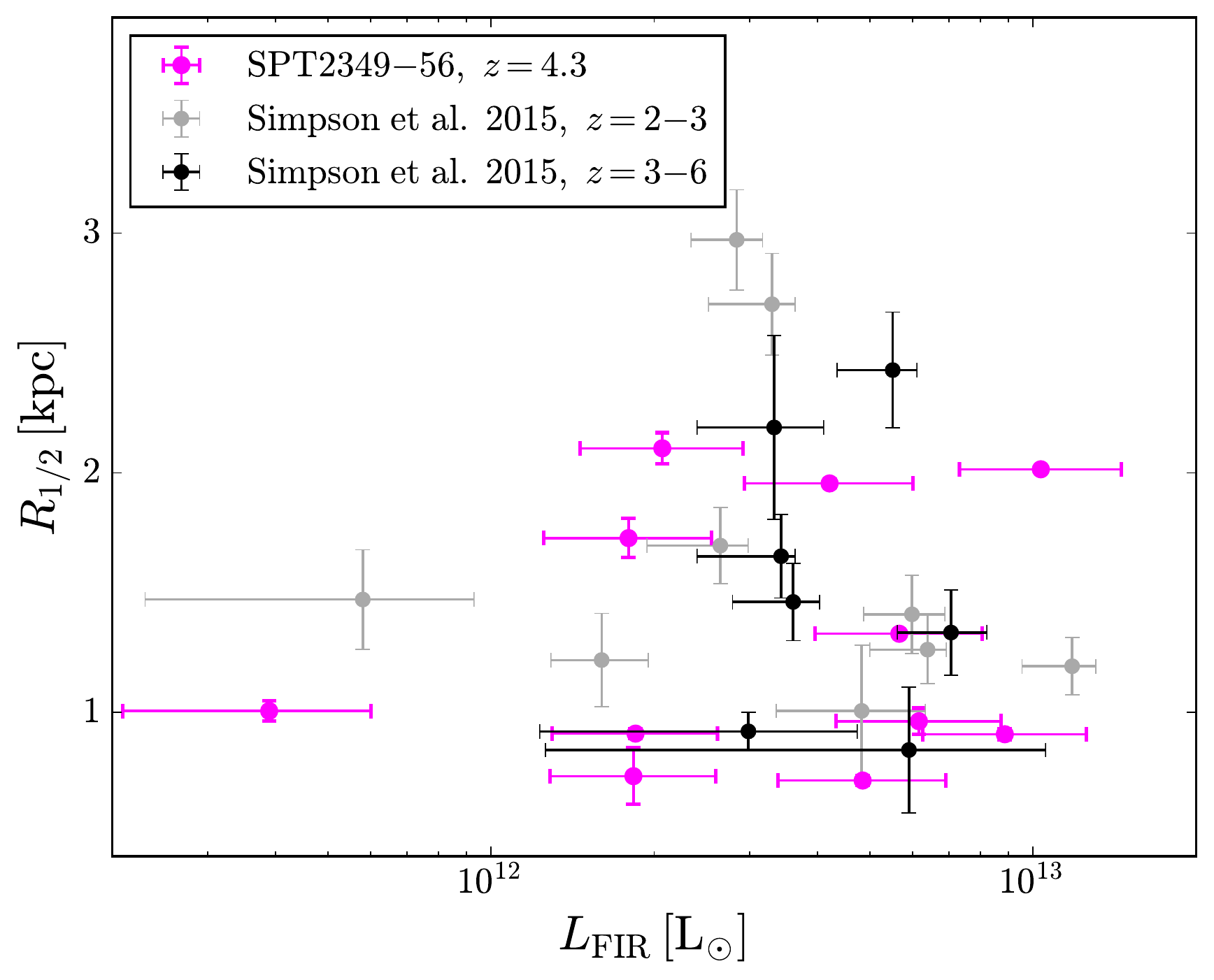}
\caption{{\it Left:} $L_{[\rm C\textsc{ii}]}$/$L_{\rm FIR}$ as a function of $L_{\rm FIR}$. Also shown are $z\,{=}\,5$--6 galaxies from \citet{capak2015}, $z\,{=}\,4$--6 galaxies from the ALPINE survey \citep{schaerer2020}, and a sample of SPT lensed SMGs from \citet{gullberg2015}, with de-magnification corrections from the lensing models of \citet{spilker2016}; in this plot, literature values of $L_{\rm IR}$ have been converted to $L_{\rm FIR}$ using a conversion factor of 1.628, the mean $L_{\rm IR}$/$L_{\rm FIR}$ ratio of the ALPINE galaxies. {\it Right:} Half-light radii of the galaxies in our sample that have high-resolution Band 7 dust continuum emission brighter than 3$\sigma$, plotted against far-infrared luminosity. These galaxies were fit with elliptical S{\'e}rsic profiles with a variable S{\'e}rsic index (see Appendix \ref{appendix2}). Also shown are half-light radii measurements of field SMGs from \citet{simpson2015a}, split into a low-$z$ group ($z\,{=}\,2$--3), and a high-$z$ group ($z\,{=}\,3$--6). The mean size of our galaxies is found to be 1.5\,kpc, with a dispersion of 0.5\,kpc, while the galaxies from \citet{simpson2015a} have a median value of 1.2$\,{\pm}\,$0.1\,kpc; we thus find no siginificant difference between these samples, and moreover there appears to be no trend with far-infrared luminosity.}
\label{l_fir}
\end{figure*}

In Fig.~\ref{l_fir} we show the ratio $L_{[\rm C\textsc{ii}]}$/$L_{\rm FIR}$ as a function of $L_{\rm FIR}$ for the galaxies in SPT2349$-$56. In this plot we compare our sample to galaxies between $z\,{=}\,5$ and 6 from \citet{capak2015}, between $z\,{=}\,4$ and 6 from the ALPINE survey \citep{schaerer2020}, and between $z\,{=}\,2$ and 6 from the catalogue of SPT lensed galaxies of \citet{gullberg2015} for which good lensing models have been derived by \citet{spilker2016}. These are amongst the largest samples of sources where similar far-infrared observations around $z\,{=}\,4$ exist, and they also span an interesting luminosity range; the galaxies from \citet{capak2015} and \citet{schaerer2020} are expected to be `typical' star-forming galaxies and have lower luminosities, and meanwhile the SPT SMGs should represent a homogenous sample of bright and rare star-forming galaxies, having been found in a large sky survey. For the ALPINE galaxies in this comparison, \citet{schaerer2020} provide infrared luminosities, $L_{\rm IR}$, which is the integral of the SED evaluated between 8 and 1000\,$\mu$m; however, they report that the mean ratio of $L_{\rm IR}/L_{\rm FIR}$ for their sample is 1.628, thus we have divided their infrared luminosities by this ratio in Fig.~\ref{l_fir}. Similarly, \citet{capak2015} provide $L_{\rm IR}$ as opposed to $L_{\rm FIR}$, so we have simply divided their values by the same factor.

A decreasing trend of $L_{[\rm C\textsc{ii}]}$/$L_{\rm FIR}$ with increasing $L_{\rm FIR}$ beyond $L_{\rm FIR}\,{=}\,10^{11}$L$_{\odot}$ has been found in many surveys \citep[e.g.,][]{luhman1998,gracia-carpio2011,farrah2013,diaz-santos2013}, including the surveys that we compare our sample to. There is a similar trend in $L_{[\rm C\textsc{ii}]}$ versus $L_{\rm FIR}$ in our sources as well, which are located not only at the same redshift but also within the same environment.

The $L_{[\rm C\textsc{ii}]}$/$L_{\rm FIR}$ ratio has been considered a diagnostic for the identification of AGN candidates \citep[e.g.,][]{stacey2010}, where low $L_{[\rm C\textsc{ii}]}$/$L_{\rm FIR}$ ratios (${\lesssim}\,10^{-3}$) are potentially correlated with AGN-powered systems. In the context of SPT2349$-$56, our data are uniquely sensitive to SMGs, and it has been proposed that these types of galaxies are tightly linked to AGN \citep[e.g.,][]{sanders1988,tacconi2002,veilleux2009,simpson2012}. Perhaps some of the galaxies we have discovered so-far in this proto-cluster system contain an AGN, and it would be interesting to conduct future studies to investigate whether or not the fraction of AGN in this system is greater than what is observed in other galaxy cluster environments at lower ($z\,{\lesssim}\,3$) redshifts \citep[e.g.,][]{alexander2005,laird2010,hart2011,martini2013,macuga2019}; this could be done with future observations targeting for example high-$J$ CO lines and X-rays.

In Fig.~\ref{l_fir} we also plot half-light radii as a function of far-infrared luminosity. We compare our results to field SMG size measurements presented by \citet{simpson2015a}. These measurements were taken by ALMA at 870\,$\mu$m with 0.3\,arcsec resolution, comparable to our data. The authors were able to fit elliptical Gaussian profiles to a subset of 23 SMGs (17 of which have sufficient photometry to constrain photometric redshifts, see \citealt{simpson2017}), finding a median half-light diametre (radius) of 2.4$\,{\pm}\,$0.2\,kpc (1.2$\,{\pm}\,$0.1\,kpc), comparable with out results. We note that in our model, a S{\'e}rsic index of 0.5 is equivalent to a Gaussian profile, and that the half-light radius of our model has the same interpretation as the half-light diametres measured by \citet{simpson2015a} (modulo a factor of exactly 2), namely that they are radii enclosing half the integrated intensity of the model. We additionally split this field sample in two, based on the photometric redshifts provided by \citet{simpson2017}: a low-$z$ sample ($z\,{=}\,2$--3), and a high-$z$ sample ($z\,{=}\,3$--6). \citet{simpson2015a} report that they do not find any trend in SMG size with luminosity or redshift, and this seems to be in agreement with our sample.

\begin{figure*}
\includegraphics[width=0.45\textwidth,height=6.2cm]{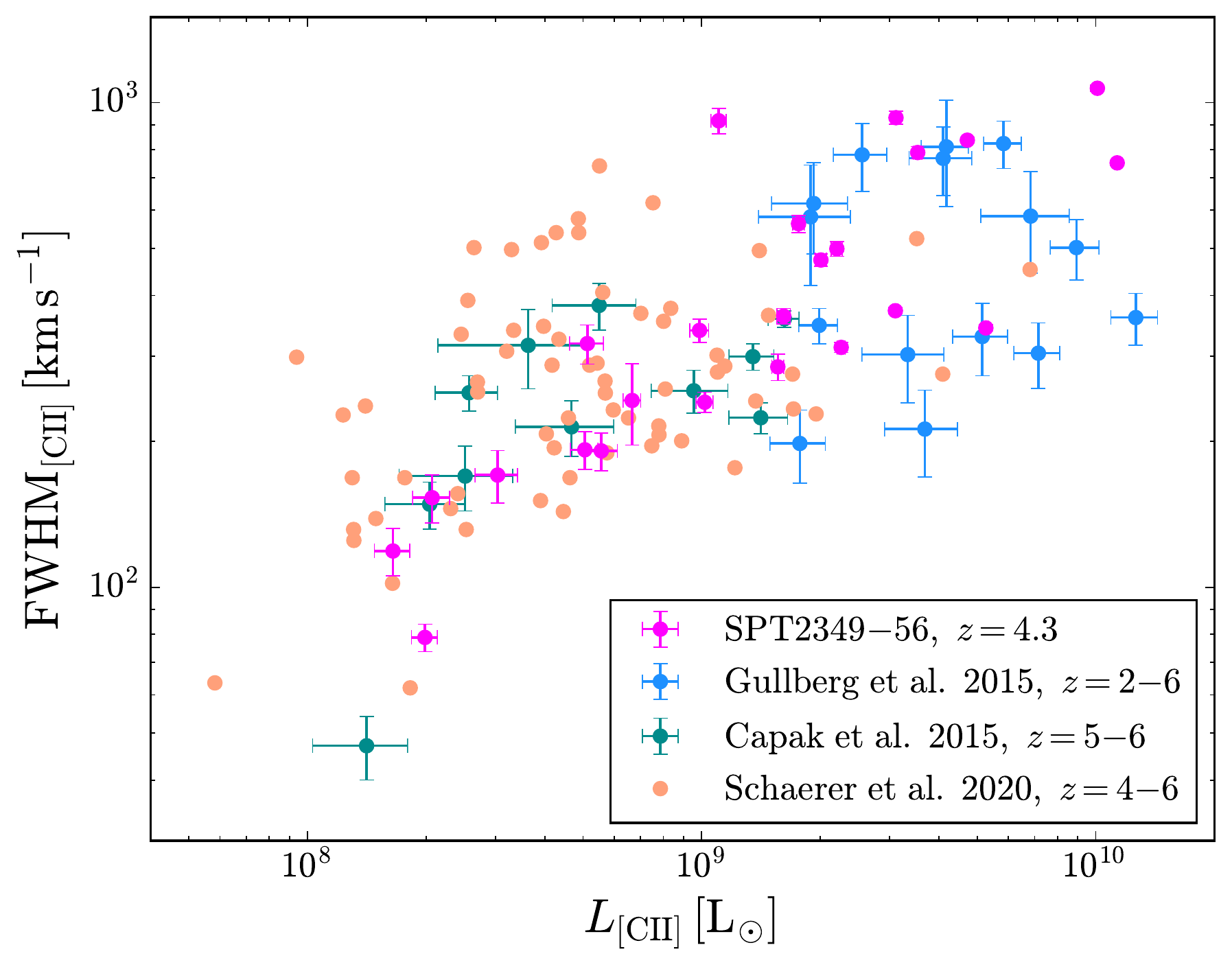}
\includegraphics[width=0.45\textwidth,height=6.2cm]{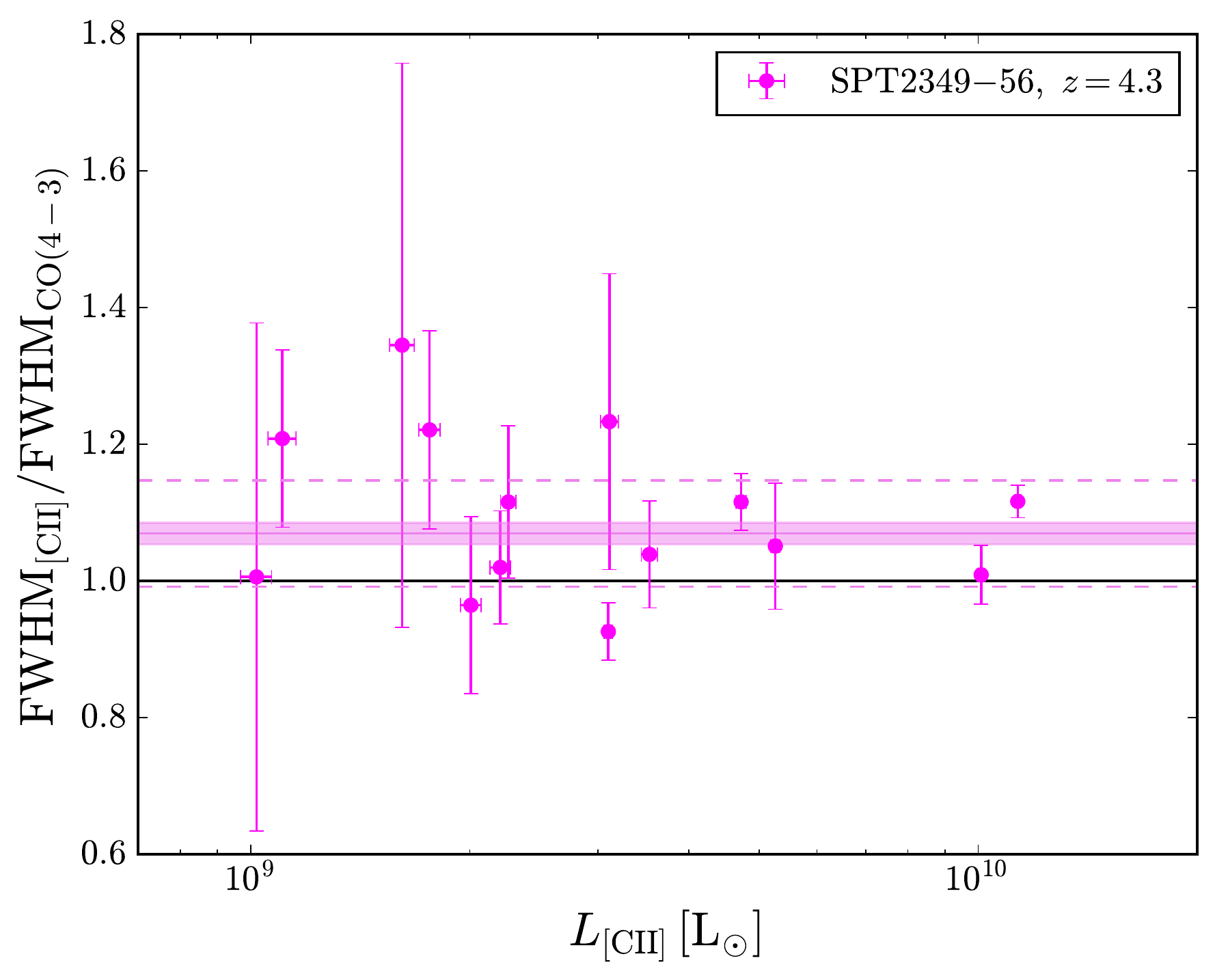}
\caption{{\it Left:} [C{\sc ii}] FWHM as a function of $L_{\rm [CII]}$, compared to three high-$z$ samples from the literature. For sources best fit by a double Gaussian, we show the FWHM obtained from forcing a single-Gaussian fit. {\it Right:} The ratio of [C{\sc ii}] FWHM to \mbox{CO(4--3)} FWHM as a functino of $L_{[\rm C\textsc{ii}]}$. The weighted mean ratio is found to be 1.07$\,{\pm}\,0.02$ (shown by the pink shaded region), with a weighted standard deviation of 0.08 (shown by the pink dotted lines), suggesting that the [C{\sc ii}] linewidths may on average be slightly larger than the \mbox{CO(4--3)} linewidths, but there is no trend with increasing $L_{[\rm C\textsc{ii}]}$.}
\label{fwhm}
\end{figure*}

Figure \ref{fwhm} shows the [C{\sc ii}] FWHM of our sources as a function of $L_{\rm [CII]}$. For sources best fit by a double Gaussian, we show the FWHM obtained from forcing a single-Gaussian fit. Galaxies N2, N3, SPIREc1, SPIREc2, and SPIREc3 do not have Band 7 [C{\sc ii}] data, and are excluded from this plot. On this plot we also show the three high-$z$ samples from the literature described above. We find that the FWHM generally increases with increasing $L_{\rm [CII]}$, although with considerable spread, and that there is no discernible difference between the samples. This lack of difference between proto-cluster galaxies and field galaxies has also been observed in other systems, for example through \mbox{CO(1--0)} linewidths in the Spiderweb system \citep{dannerbauer2017}.

Figure \ref{fwhm} also shows the ratio of [C{\sc ii}] FWHM to \mbox{CO(4--3)} FWHM for all of the sources in our sample that have both of these lines detected. We find that the weighted mean ratio of these linewidths is 1.07$\,{\pm}\,0.02$ (with a weighted standard deviation of 0.08), suggesting that the [C{\sc ii}] linewidths in our sample are slightly larger than the \mbox{CO(4--3)} linewidths, although the difference is small. We point out that the resolution of our \mbox{CO(4--3)} data is about twice that of our [C{\sc ii}] data, so this is likely not due to the difference in beamsizes between the measurements, but instead could be a physical result of the galaxies themselves (e.g., difference in emission extension or optical depth). In Fig.~\ref{fwhm} these ratios are shown plotted against $L_{[\rm C\textsc{ii}]}$; however, there is no evidence for a correlation between the quantities.

\subsection{Number counts and luminosity functions}

The SPT2349$-$56 region represents a significant overdensity in [C{\sc ii}] emitters, so we do not expect the line luminosity function to be comparable to those estimated for the average Universe. Nevertheless, to understand how overdense this region is compared to field regions, we estimate our space density of [C{\sc ii}] emitters as follows. First, we restrict our analysis to include only sources with both Band 3 and Band 7 coverage, thus removing N2, N3, and all SPIREc sources. Second, we have estimate the normalization volume by calculating the area of our [C{\sc ii}] maps (both the deep map of the core and the single pointing of the northern section) where the primary beam response was greater than 0.5, and for the depth we have used 1.8\,Mpc (proper distance), which is the diametre of the sphere used to estimate the virial mass of the core in Section \ref{virialization}. The resulting proper volume is 0.1\,Mpc$^3$, and we use this to normalize our number counts. 

\begin{figure*}
\includegraphics[width=0.45\textwidth]{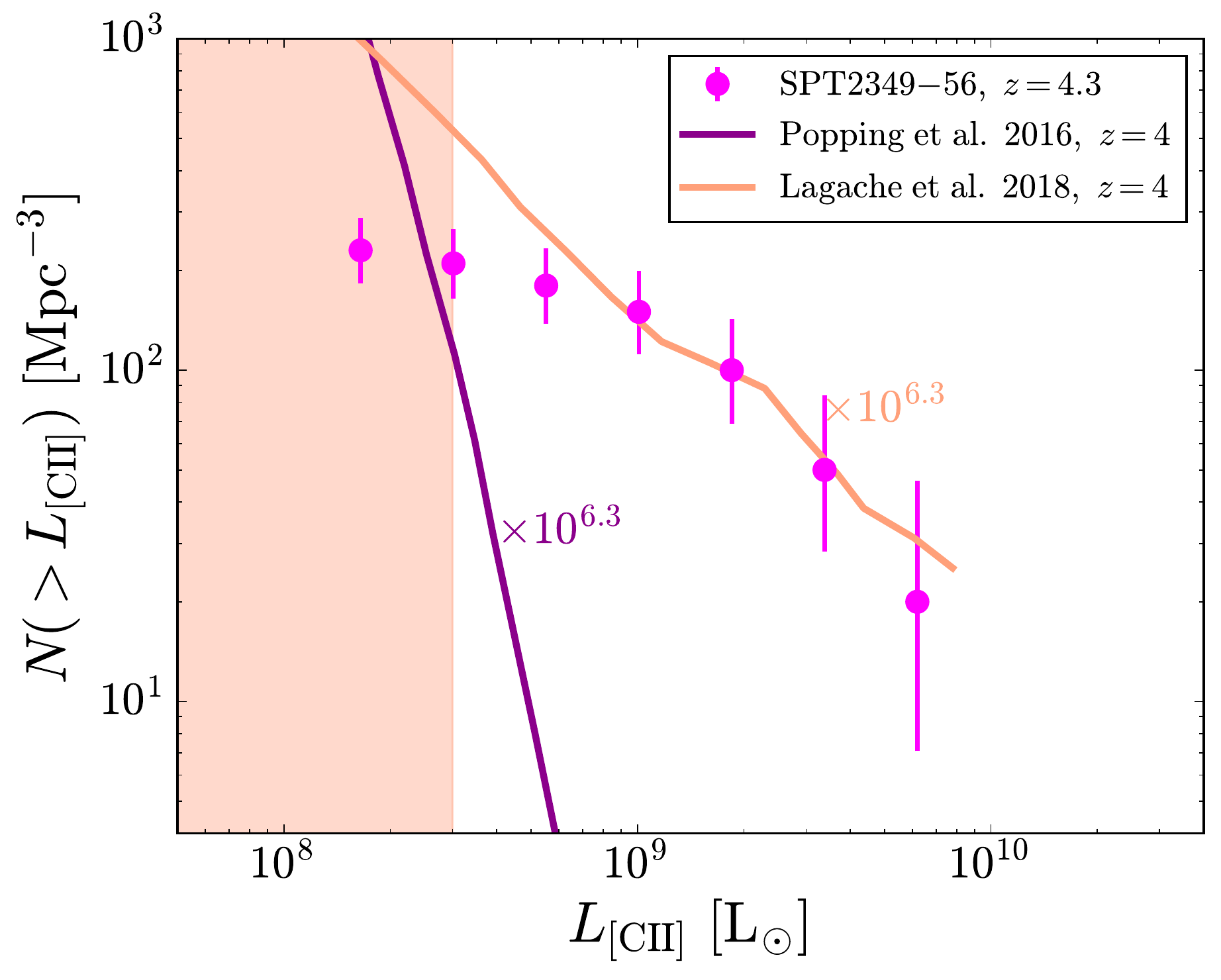}
\includegraphics[width=0.45\textwidth]{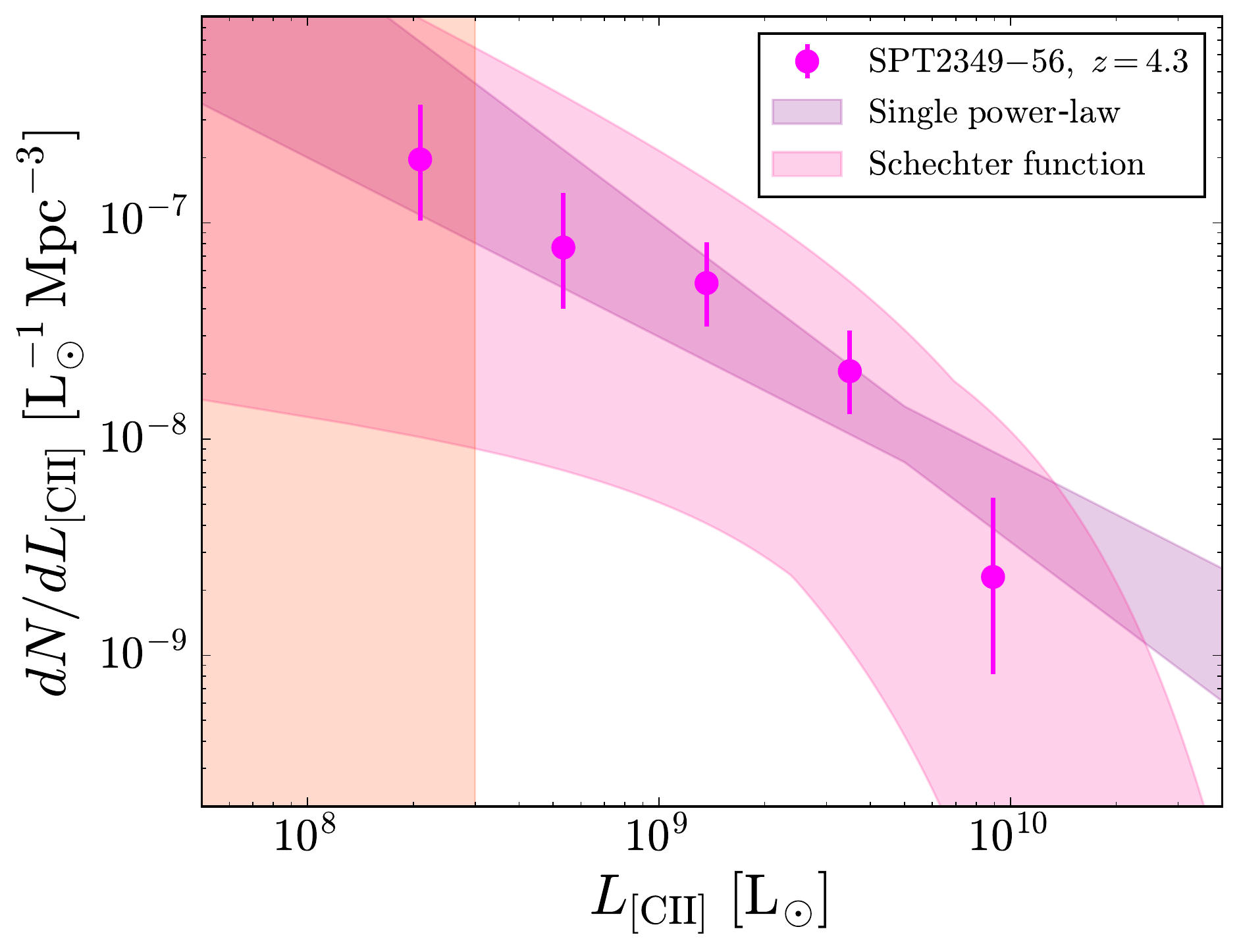}
\includegraphics[width=0.45\textwidth]{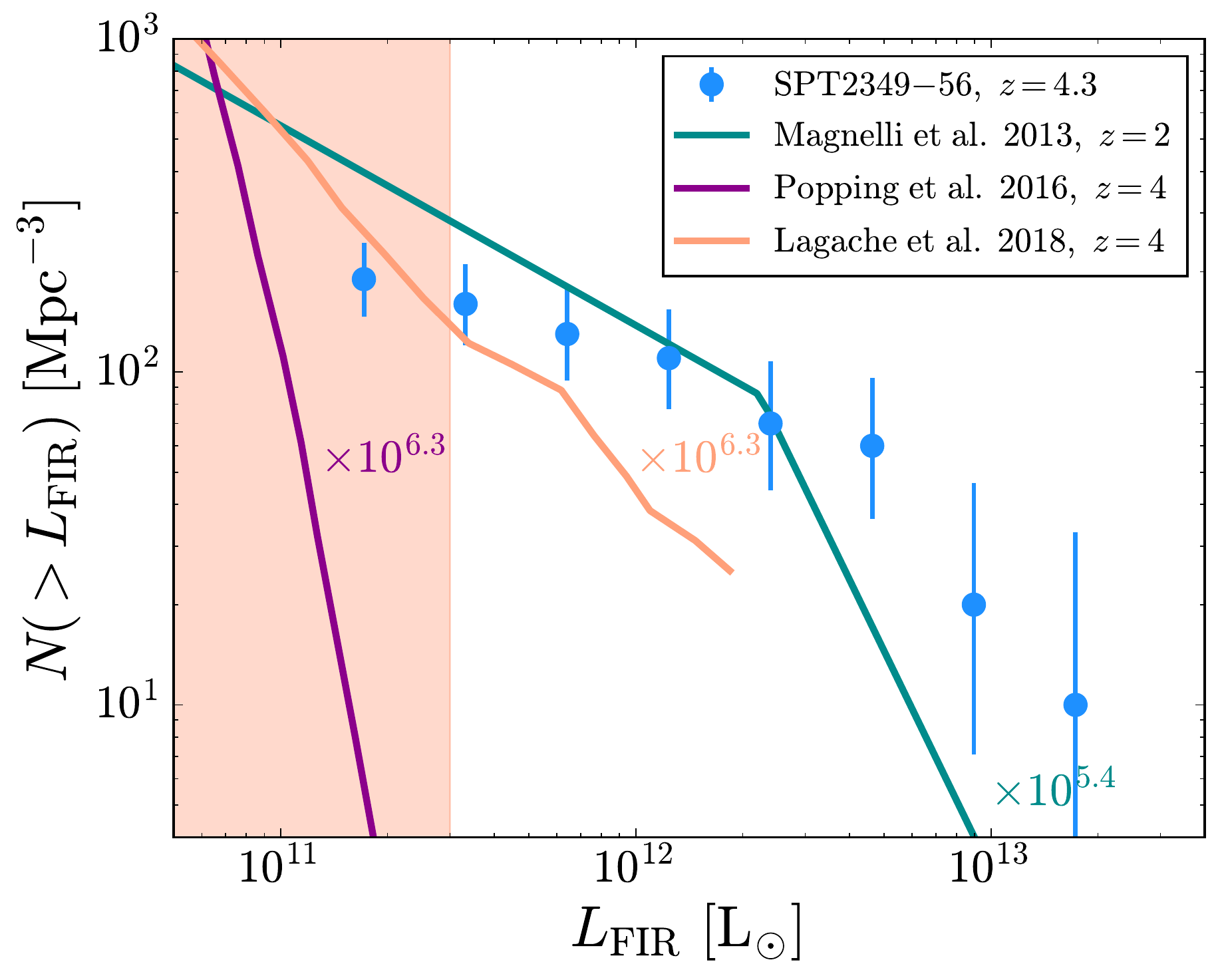}
\includegraphics[width=0.45\textwidth]{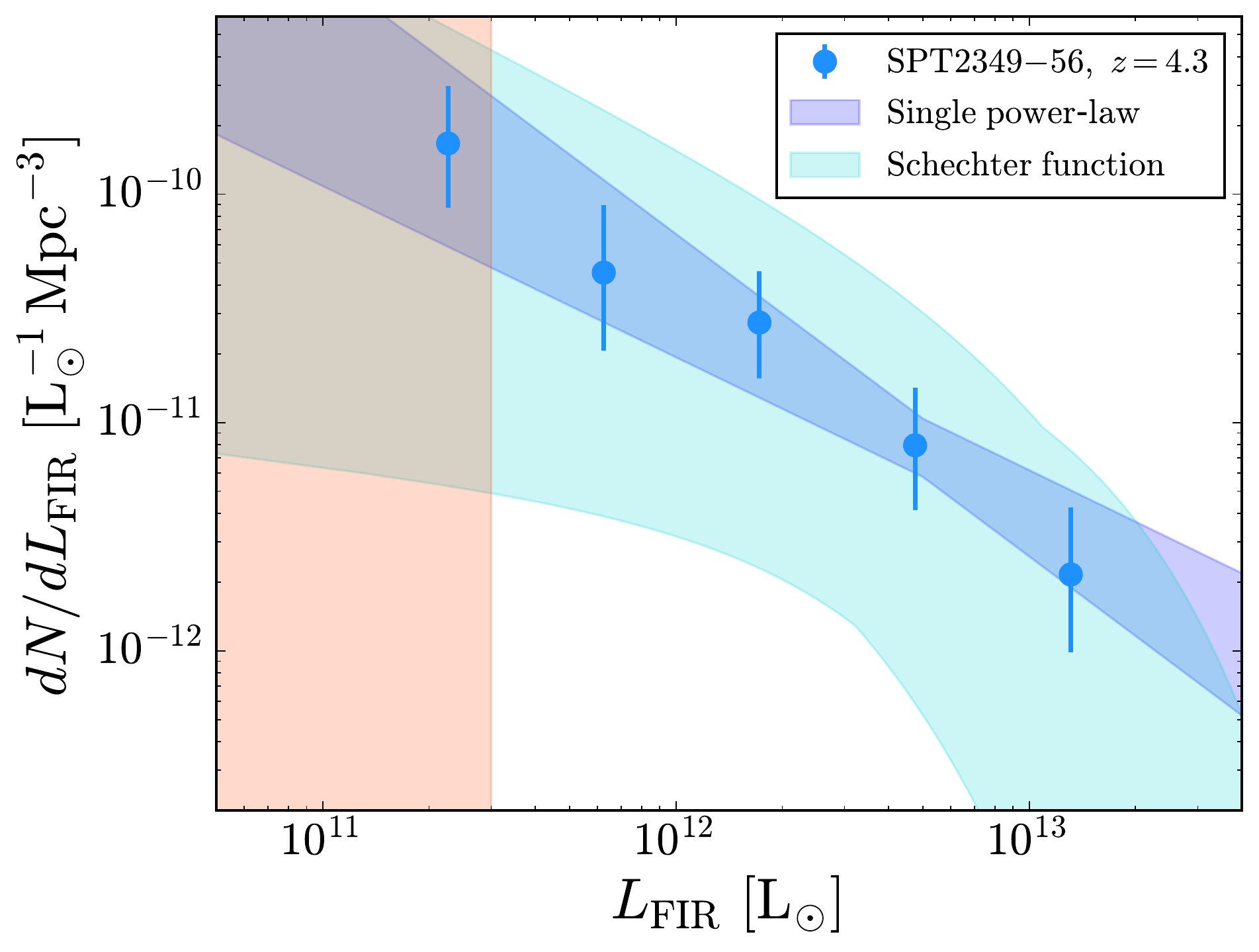}
\caption{{\it Left:} $L_{[\rm C\textsc{ii}]}$ and $L_{\rm FIR}$ cumulative number counts for all sources where have both Band 3 and Band 7 data. The red shaded region indicates where our data is no longer complete. Models for the cosmological $L_{[\rm C\textsc{ii}]}$ luminosity function at $z\,{=}\,4$ are shown from \citet{popping2016} and \citet{lagache2018}; an estimate for the $L_{\rm FIR}$ luminosity function at $z\,{=}\,2$ is shown from \citet{magnelli2013}, and the models from \citet{popping2016} and \citet{lagache2018} have been scaled using the relation from \citet{lagache2018}. Since our survey volume is quite small and centred on a known proto-cluster, the volume normalization factors between these models and our observations are very different. We have thus scaled the $L_{[\rm C\textsc{ii}]}$ models by a factor of 10$^{6.3}$ in order to line up the model from \citet{lagache2018} to our data at about 10$^9$\,L$_{\odot}$, and we have scaled the $L_{\rm FIR}$ luminosity function from \citet{magnelli2013} by a factor of 10$^{5.4}$ to line it up with our data at about 10$^{12}$\,L$_{\odot}$. {\it Right:} Corresponding $L_{[\rm C\textsc{ii}]}$ and $L_{\rm FIR}$ differential number counts for the same sources. The coloured shaded regions are single power-law and Schechter functions with parameters optimized to our data by minimizing the negative log-likelihood given by Eq.~\ref{likelihood}, encompassing lower and upper 68\,per cent confidence interval limits of the optimal parametres.}
\label{counts}
\end{figure*}

Fig.~\ref{counts} shows the $L_{[\rm C\textsc{ii}]}$ and $L_{\rm FIR}$ cumulative and differential number counts of our sources in bins logarithmically spaced between $10^{8}$\,L$_{\odot}$ and $10^{10}$\,L$_{\odot}$ for $L_{[\rm C\textsc{ii}]}$, and between $10^{11}$\,L$_{\odot}$ and $10^{13}$\,L$_{\odot}$ for $L_{\rm FIR}$. In this figure we compare our $L_{[\rm C\textsc{ii}]}$ cumulative number counts to models for the $z\,{=}\,4$ $L_{[\rm C\textsc{ii}]}$ luminosity function from \citet{popping2016} and \citet{lagache2018}. The volume normalizations from these models are about 6 orders of magnitude below our results, and we have scaled their counts by a factor of 10$^{6.3}$ in order to line up the model from \citet{lagache2018} to our data at about 10$^9$\,L$_{\odot}$. \citet{lagache2018} find a scaling relation between $L_{[\rm C\textsc{ii}]}$ and SFR (see Eq.~10 in their paper), which can be combined with the $L_{\rm FIR}$-SFR scale factor adopted here to obtain predictions for the $L_{\rm FIR}$ luminosity function. We show these predictions alongside a measurement of the cumulative number count at $z\,{=}\,2$ from a {\it Herschel\/}-PACS survey of the GOODS field \citep{magnelli2013}, scaled by a factor of 10$^{5.4}$ to again match our counts at about 10$^{12}$\,L$_{\odot}$. 

We can estimate the extent to which our counts are complete by looking at the detection thresholds from Section \ref{source_search}. The mean threshold from our search of the deep core data cube and our single pointing of the northern component is a S/N of 6, and the mean uncertainty in our $L_{[\rm C\textsc{ii}]}$ measurements is about 0.5$\,{\times}\,10^{8}\,$L$_{\odot}$. By multiplying these two values we find that we expect to be complete down to $L_{[\rm C\textsc{ii}]}\,{\approx}\,$3$\,{\times}\,10^{8}\,$L$_{\odot}$; we show this completeness limit as a red shaded region, and we note that this calculation assumes that all of our sources were selected from data with uniform sensitivity. The $L_{\rm FIR}$ number count limit is roughly set by continuum detection; at 850\,$\mu$m, the mean uncertainty is 0.09\,mJy, and we set a S/N threshold of 2, meaning that we are sensitive to continuum emission down to about 0.2\,mJy; scaling this flux density to a far-infrared luminosity gives 3$\,{\times}\,10^{11}\,$L$_{\odot}$.

Unfortunately, a proper comparison between these models and our data is quite difficult owing to the fact that we have surveyed a very small volume containing a known proto-cluster field, where we expect to find many sources, whereas the other works are for galaxies in the field. Nonetheless, our data provide the first statistically significant measurements of these luminosity functions at $z\,{>}\,4$, and we can qualitatively compare their shapes to what is expected in the field; we find that the slope of the $L_{[\rm C\textsc{ii}]}$ model given by \citet{lagache2018} is consistent with our results, while the slope from \citet{popping2016} is much steeper, and for the $L_{\rm FIR}$ counts both of these models and the luminosity function from \citet{magnelli2013} at $z\,{=}\,2$ are steeper than what we see in SPT2349$-$54. 

This observation, that the $L_{\rm FIR}$ luminosity function is biased towards bright galaxies compared to what is predicted for field galaxies, may be a result of ongoing mergers within this proto-cluster. Mergers between galaxies are expected to trigger bursts of star formation \citep[e.g.,][]{tacconi2008,engel2010,luo2014,chen2015}, and hence increase far-infrared luminosities, and there is growing evidence that mergers are more common in overdense regions such as proto-clusters \citep{lotz2013,hine2016}. Additionally, $N$-body simulations using SPT2349$-$56 as the approximate initial conditions show that most of the core galaxies within this proto-cluster field should indeed merge within a timescale of about 100\,Myr \citep{rennehan2019}, consistent with this interpretation.

Looking at the differential number counts in Fig.~\ref{counts}, it appears as though we might be seeing a change in slope between low- and high-luminosity sources at a characteristic $L_{\star}$. In order to investigate this, we use the maximum-likelihood approach to fit for model parameters, first described by \citet{marshall1983} and later adopted by e.g. \citet{wall2008}, modified to describe luminosity functions at a fixed redshift, and we use the approximation that all sources were selected from maps of uniform depth. As opposed to the typical least-squares approach, which depends on a choice of number-count bins, the maximum-likelihood approach uses all of the available data, and is therefore desirable when the sample size is not large. Following Eq.~2 of \citet{marshall1983}, we remove the redshift dependence of the model luminosity function, and minimize the following negative log-likelihood:
\begin{equation}
\label{likelihood}
S = -2 \ln \mathcal{L} = -2 \sum_{i=1}^N \ln \phi (L_i) + 2 V \int _{L_\mathrm{a}}^{L_\mathrm{b}} \phi (L) dL + C.
\end{equation}
\noindent
In this equation $N$ is the sample size (here the number of sources above our completeness limits), $\phi(L)$ is the model luminosity function (in units of L$_{\odot}^{-1}$\,Mpc$^{-3}$), $V$ is the volume of the survey (here set to be 0.1\,Mpc$^3$), $L_{\mathrm{a}}$ and $L_{\mathrm{b}}$ are the luminosity limits of the sample (which we take to be between the completeness limits and the maximum luminosities in the sample), and $C$ is a constant independent of the model.

We test two models, a single power-law and a Schechter function, where the Schechter function contains a break at a characteristic $L_{\star}$. We minimize the negative log-likelihood and obtain marginalized probability distributions for the model parameters using a Markov chain Monte Carlo (MCMC), allowing the normalization, slope, and characteristic luminosity (if present) to vary. The models are then compared by calculating the final likelihood ratios. Lastly, parameters are estimated from the means of the marginalized posterior probability distributions, and uncertainties are estimated by calculating the 68\,per cent confidence intervals.

The results of this analysis are shown in Table \ref{table:lf}, and in Fig.~\ref{counts} the functions are shown as shaded regions encompassing lower and upper limits of the optimal parametres. We find that for our $L_{[\rm C\textsc{ii}]}$ number counts, a Schechter function is slightly favoured over the single-power law, while for the $L_{\rm FIR}$ number counts a Schechter function does not improve the likelihood compared to the single power-law, although we emphasize that the ratio of the $L_{[\rm C\textsc{ii}]}$ likelihood functions is still quite close to 1, meaning that the models are quite similar.

\setlength\tabcolsep{2pt}
\begin{table*}
\centering
\caption{Optimal model parameters for the luminosity functions shown in Fig.~\ref{counts}. The models tested were a single power-law and a Schechter function, and the optimal parameters were obtained by minimizing the negative-log likelihood function given by Eq.~\ref{likelihood}; parameter values are the means of the marginalized posterior distributions, and uncertainties are the 68\,per cent confidence intervals. The left half of the table shows results for the $L_{[\rm C\textsc{ii}]}$ luminosity function, and the right half of the table shows results for the $L_{\rm FIR}$ luminosity function. The models were compared by calculating the final likelihood ratios.}
\label{table:lf}
\begin{threeparttable}
\begin{tabular}{lcccc|lcccc}
\hline
\multicolumn{5}{c|}{$L_{[\rm C\textsc{ii}]}$} & \multicolumn{5}{c}{$L_{\rm FIR}$} \\
\hline
$\phi(L)$ & $\phi_{\star}$ & $L_{\star}$ & $\alpha$ & \multirow{2}{*}{$\frac{\mathcal{L}_{\rm model}}{\mathcal{L}_{\rm single}}$} & $\phi(L)$ & $\phi_{\star}$ & $L_{\star}$ & $\alpha$ & \multirow{2}{*}{$\frac{\mathcal{L}_{\rm model}}{\mathcal{L}_{\rm single}}$} \\
& [$10^{-9}$\,L$_{\odot}^{-1}$\,Mpc$^{-3}$] & [$10^{9}$\,L$_{\odot}$] & & & & [$10^{-12}$\,L$_{\odot}^{-1}$\,Mpc$^{-3}$] & [$10^{12}$\,L$_{\odot}$] & & \\
\hline
$\phi_{\star}\left(\frac{L}{L_{\star}}\right)^{\alpha}$ & 12$_{-4}^{+3}$ & 5$^{\rm a}$ & -1.0$_{-0.2}^{+0.2}$ & 1.0 & $\phi_{\star}\left(\frac{L}{L_{\star}}\right)^{\alpha}$ & 8$_{-3}^{+2}$ & 5$^{\rm a}$ & -1.0$_{-0.2}^{+0.2}$ & 1.0 \\
\multicolumn{5}{c|}{} & \multicolumn{5}{c}{} \\
$\phi_{\star}\left(\frac{L}{L_{\star}}\right)^{\alpha}e^{L/L_{\star}}$ & 39$_{-32}^{+12}$ & 6$_{-3}^{+1}$ & -0.5$_{-0.3}^{+0.3}$ & 1.3 & $\phi_{\star}\left(\frac{L}{L_{\star}}\right)^{\alpha}e^{L/L_{\star}}$ & 22$_{-18}^{+4}$ & 9$_{-5}^{+2}$ & -0.4$_{-0.4}^{+0.2}$ & 1.0 \\
\hline
\end{tabular}
\begin{tablenotes}
\item $^{\rm a}$ $L_{\star}$ was fixed to 5 and is not a free parameter of this model.
\end{tablenotes}
\end{threeparttable}
\end{table*}

\section{Discussion}
\label{discussion}

\subsection{Spatial distribution}
\label{discussion_spatial}

Based on our observations, we see three distinct components in SPT2349$-$56: the main central core, consisting of 23 gravitationally bound galaxies; a northern component containing one very bright galaxy and two satellites; and another small group of galaxies located about 1.5\,Mpc in proper distance from the core. The central galaxies likely represent the early phases of BCG formation; indeed, $N$-body simulations with initial conditions approximately matching SPT2349$-$56 predict almost complete assimilation on a timescale of 100\,Myr. This indicates that by redshift 3 a massive, BCG-like elliptical galaxy will already be in place at the core of this proto-cluster \citep{rennehan2019}. Our analysis suggests that the remaining two components will not merge with the core, but will instead become very massive galaxies embedded within the overall cluster.

There are several features that are immediately apparent regarding the spatial distribution of galaxies in SPT2349$-$56. First, despite the fact that our search of red {\it Herschel\/} sources only turned up one surrounding halo, the initial targets were very low S/N sources in the SPIRE data to begin with, and in fact did not show any significant 870-$\mu$m emission in our LABOCA map. It is therefore possible that there could be other low S/N {\it Herschel\/} sources in this field that we have not yet targeted, but that are also at the same redshift as SPT2349$-$56 if this system is indeed a proto-cluster. Additionally, it should be emphasized that these mm/submm observations are only probing dusty, star-forming galaxies, and that many of the galaxies within the structure are probably much brighter at rest-frame optical wavelengths. Indeed, Lyman-$\alpha$ emitters have been discovered in a dedicated MUSE follow-up of SPT2349$-$56, mostly outside the core region where we measure the strong SMG overdensity (Apostolovski et al.~in prep.).

Second, our large \mbox{CO(4--3)} survey of the region surrounding the main 870\,$\mu$m LABOCA emission area did not uncover any new sources, despite the fact that sources were found in SPIREc. Our large \mbox{CO(4--3)} map spanned roughly 300--400\,kpc in proper distance beyond this LABOCA emission area, corresponding to a surrounding proper volume of order 1\,Mpc$^3$, while the SPIREc group is located about 1.5\,Mpc from the LABOCA region. The sensitivity to \mbox{CO(4--3)} line emission outside of the core LABOCA emission area is about 0.05\,Jy\,km\,s$^{-1}$ (at the 1\,$\sigma$ level), or in terms of gas mass, 0.4$\,{\times}\,10^{10}$\,M$_{\odot}$. In this case the S/N threshold of our line search was 5.9, meaning that we can constrain the gas masses of potential sources surrounding SPT2349$-$56 to be ${<}\,2\,{\times}\,10^{10}$\,M$_{\odot}$.

It is interesting to compare this gas mass limit to what one would expect from the galaxy main-sequence (MS). Using a gas-fraction of 0.7 (from \citealt{rennehan2019}, using the results of \citealt{narayanan2012} and \citealt{tadaki2019}, and used throughout this paper), our gas-mass sensitivity corresponds to a stellar-mass sensitivity of $M_{\star}\,{<}\,1\,{\times}\,10^{10}$\,M$_{\odot}$. Next, using the $z\,{=}\,3.8$--4.9 MS relation from \citet{pearson2018}, which is derived from a {\it Herschel\/} survey of the COSMOS field, the above stellar-mass limit corresponds to a SFR limit of ${<}\,$100\,M$_{\odot}$\,yr$^{-1}$. For comparison, the sensitivity of the large \mbox{CO(4--3)} map to 3.2\,mm continuum emission outside of the core LABOCA emission area is roughly 10\,$\mu$Jy at the 1$\sigma$ level, corresponding to a SFR of about 80\,M$_{\odot}$\,yr$^{-1}$; since we applied a continuum-detection S/N threshold of 5.0, this limits continuum sources to have SFRs ${<}\,$400\,M$_{\odot}$\,yr$^{-1}$. 

\subsection{Cluster mass of SPT2349$-$56}

Next, we turn to the total mass of the proto-cluster SPT2349$-$56 as inferred from our new observations. We have estimated the mass of the core using the velocity dispersion of the core galaxies, and found a value of 9$\,{\times}\,$10$^{12}$\,M$_{\odot}$. This mass does not include the northern component of the proto-cluster nor SPIREc, and based on our total dynamical and gas/halo mass estimates, these components should be 30--60 and 20--30\,per cent the mass of the core, respectively. Thus, to within a factor of a few, we would expect the total mass of SPT2349$-$56 to be roughly 1--2\,${\times}\,10^{13}$\,M$_{\odot}$. Similarly, the halo mass of the core, as derived by scaling the total gas mass, is 1.3\,${\times}\,10^{13}\,$M$_{\odot}$, while for the northern component the mass is 7.7\,${\times}\,10^{12}\,$M$_{\odot}$, and for the SPIREc component the mass is 4.4\,${\times}\,10^{12}\,$M$_{\odot}$, making the total mass about 2.5$\,{\times}\,$10$^{13}$\,M$_{\odot}$; while there are large systematic uncertainties which have not been incorporated into this estimation, the results are consistent with what we get using the velocity dispersion.

So-far we have only probed the mass of SPT2349$-$56 within the central proper ${\sim}\,$500\,kpc of the structure (except for SPIREc) and down to a certain galaxy-mass limit, but we can estimate how much mass there is left out to about a Mpc using the gas mass limit of our large \mbox{CO(4--3)} mosaic (${<}\,2\,{\times}\,10^{10}$\,M$_{\odot}$) and assuming a gas mass fraction of 0.7 (constraining stellar masses to be ${<}\,1\,{\times}\,10^{10}$\,M$_{\odot}$). \citet{vanderburg2013} and \citet{nantais2016} give best-fitting stellar mass functions for clusters at $z=1$ and 1.5, respectively, within a circle of proper radius 1\,Mpc, which closely matches our survey area. We estimate the total amount of stellar mass undetected in our proto-cluster using these two models by integrating them from 0 to $1\,{\times}\,10^{10}$\,M$_{\odot}$; since \citet{vanderburg2013} and \citet{nantais2016} have not normalized their models, we divide the integrals by the number of clusters used to fit each model (10 and four, respectively) to get the mass for one cluster. This gives $5\,{\times}\,10^{10}$\,M$_{\odot}$ for the $z=1$ stellar mass function, and $0.5\,{\times}\,10^{10}$\,M$_{\odot}$ for the $z=1.5$ stellar mass function. Converting this back to gas mass using the conversion factor of 0.7, these masses correspond to (1--10)$\,{\times}\,10^{10}$\,M$_{\odot}$. Our total gas mass of the core and the northern section of SPT2349$-$56 is 4.7$\,{\times}\,10^{11}$\,M$_{\odot}$ (SPIREc is further than 1\,Mpc), meaning that roughly 2--20\,per cent of the gas mass might remain undetected in less luminous \mbox{CO(4--3)} sources in this large surrounding area. We note that even if all this mass were contained within the central several hundred kpc of SPT2349$-$56, it would not dramatically increase the sum of the halo masses and become inconsistent with the mass from the velocity dispersion.

In Fig.~\ref{mass} we show the total halo mass estimates we have derived from our \mbox{CO(4--3)} observations for the central, northern, and SPIREc components, and also summed over all three components. Each source is shown as a range of plausible halo masses (denoted by arrows), where the tip of the arrow assumes the gas mass-to-halo mass conversion factor of 48.3. The base of the arrow is a conservative lower limit assuming that the gas mass encompases all of the baryonic mass (neglecting for instance stars), and using the cosmic baryon fraction $f_{\rm b}\,{=}\,\Omega_{\rm b}\,{/}\,\Omega_{\rm c}\,{=}\,0.19$ \citep{planck2014-a15} to convert gas mass to halo mass (the factor is $1\,{/}\,f_{\rm b}\,{=}\,5.3$); we note that the gas mass-to-halo mass conversion factor used in this paper corresponds to a baryon fraction of $f_{\rm b}\,{=}\,0.033$. The arrows are used to emphasize that the results are strictly lower limits, both as the halo masses could be higher than our fiducial conversion \citep{rennehan2019}, and also as there may be additional gas mass in sources that we have not yet detected.

Figure \ref{mass} compares our results to $z\,{>}\,4$, SMG-rich proto-clusters found in the literature with similar CO observations as a function of redshift, namely the Distant Red Core at $z\,{=}\,4.00$, where 10 proto-cluster members have detections in \mbox{CO(6--5)} \citep{oteo2018}, and GN20 at $z\,{=}\,4.05$, where three proto-cluster members have detections in \mbox{CO(4--3)} \citep{daddi2009,hodge2013}. For these literature proto-clusters, we first converted the reported total line strengths into luminosities (in units of K\,km\,s$^{-1}$\,pc$^2$) using Eq.~\ref{Lprime}. We next used the appropriate conversion factors from \citet{spilker2014} to obtain \mbox{CO(1--0)} line luminosities ($r_{6,1}\,{=}\,0.46$ for \mbox{CO(6--5)}, and $r_{4,1}\,{=}\,0.60$ for \mbox{CO(4--3)}, the same factor used on our data). We then applied an $\alpha_{\rm CO}$ factor of 1\,M$_{\odot}$/(K\,km\,s$^{-1}$\,pc$^2$) to obtain gas masses, and lastly converted these gas masses to halo masses using the scaling factor of 42.8 from \citet{rennehan2019}; these are the same scaling relations applied to our data throughout this paper. We find that these systems are similar in mass, and do not differ by more than a factor of a few. Also shown on this plot are a sample of $z\,{>}\,4$ SMGs with gas masses derived from CO detections \citep{marrone2018}, converted to halo mass as described above. These sources are all shown as lower-limit arrows, with the base of the arrow representing the halo mass one would obtain assuming the cosmic baryon fraction, as described above.

Lastly, we show predictions for the largest cluster halos expected in a 25\,deg$^2$ survey, a 2500\,deg$^2$ survey, and a full-sky survey \citep{harrison2013,marrone2018}. We find that SPT23494$-$56, having been selected from a 2500\,deg$^2$ survey, is a factor of about 4 smaller than the largest expected halo at $z\,{=}\,4.3$, suggesting that it may be amongst the largest overdensities to be found in the survey at that redshift. On the other hand, assuming the lower limit on this structure's halo mass to be true, this factor is about 40, and thus there should exist of a population of comparable (and larger) halos in SPT's 2500\,deg$^2$ survey. However, as SPT2349$-$56 was selected as the brightest unlensed point source in this survey, these remaining halos must either be in a quiescent phase of evolution, implying that the star-formation burst seen in SPT2349$-$56 has a very short comparative timescale, or that they do not contain enough material to follow a typical Coma cluster-like trajectory \citep[as in e.g.][]{chiang2013}.

With this in mind, one important question to ask is what will be the final mass of SPT2349$-$56 at redshift 0, and whether we can locate examples of this final state to learn about the formation and growth of the most massive galaxy clusters. A study was carried out looking for massive merger events in the MultiDark Planck 2 \citep[MDPL2;][]{riebe2013,klypin2016} simulation, specifically, events where more than five halos of mass greater than 2$\,{\times}\,10^{11}\,$M$_{\odot}$ entered the virial radius of a halo that is less than 20 times the mass of their sum, and where the final cluster mass was greater than 5$\,{\times}\,10^{14}\,$M$_{\odot}$ \citep{rennehan2019}. It was found that about 10\,per cent of clusters with present-day masses ${>}\,10^{15}$\,M$_{\odot}$ formed through such a massive merger event at high redshift, and such an event is consistent with what we are observing with SPT2349$-$56. Another important feature of SPT2349$-$56 is it's rarity, being the brightest unlensed point source in the 2500\,deg$^2$ of sky surveyed by SPT. It seems plausible that it will remain a rare object up until redshift 0, in correspondence with ${>}\,10^{15}$\,M$_{\odot}$ galaxy clusters such as the Coma Cluster, continually accreting mass through mergers similar to what we are currently seeing with the northern component and SPIREc. 

\begin{figure*}
\includegraphics{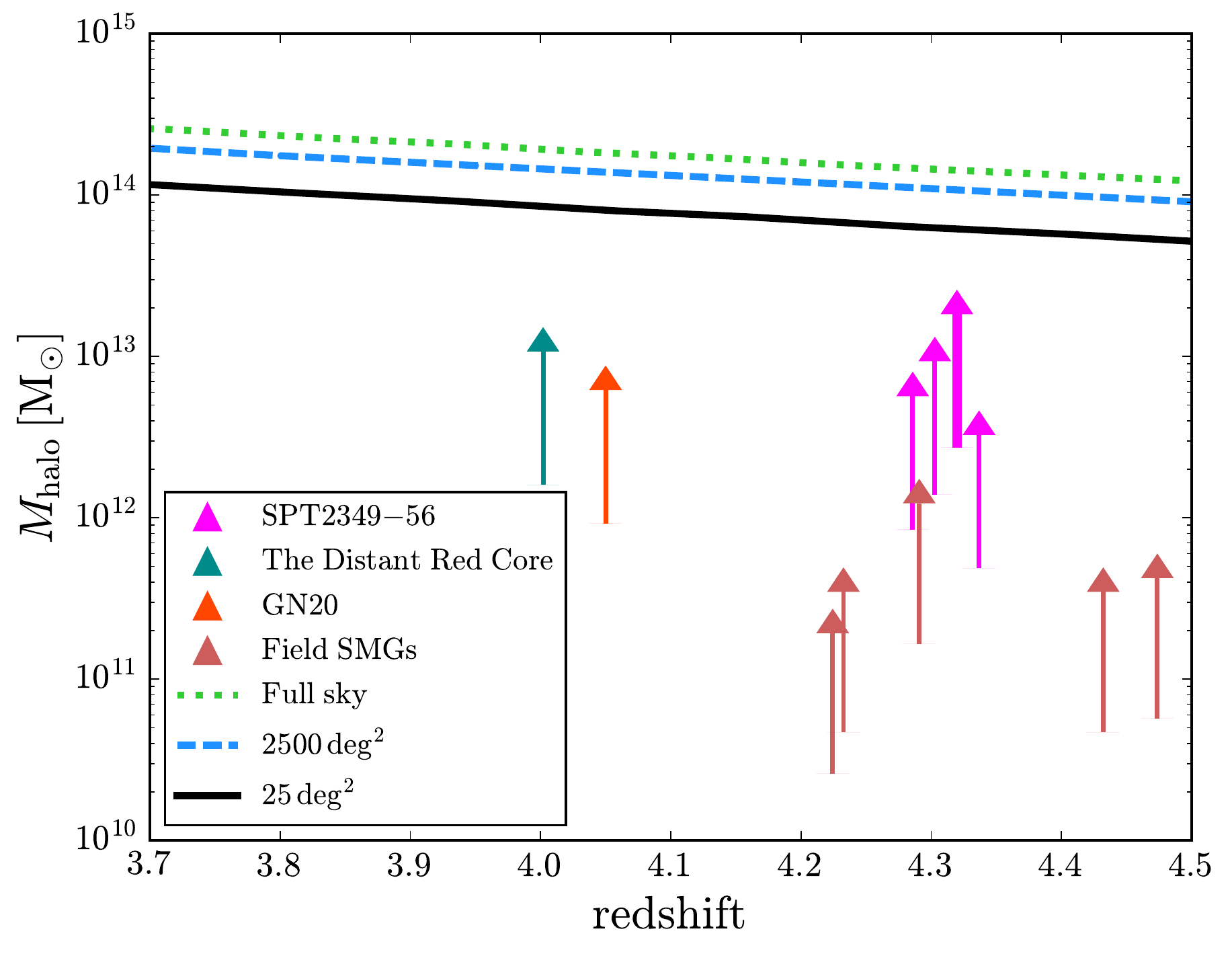}
\caption{Halo mass estimates for SPT2349$-$56, as derived from our observations of \mbox{CO(4--3)} (see Section \ref{halo}); the thin arrows indicate the halo masses of the central, northern, and SPIREc components, while the thick arrow indicates their summed mass. The tip of the arrow assumes a gas mass-to-halo mass conversion factor of 42.8 \citep{rennehan2019}, while the bottom of the arrow is a conservative lower limit assuming the gas mass encompases all of the baryonic mass, and using the cosmic baryon fraction to convert to halo mass (the factor is $1\,{/}\,f_{\rm b}\,{=}\,5.3$). The arrows emphasize that these estimates are lower limits, both because the gas mass-to-halo mass conversion could be higher, and because there may be additional gas mass in sources that we have not yet detected. Also shown are halo mass estimates for $z\,{>}\,4$, SMG-rich proto-clusters in the literature with similar CO observations, converted to halo mass following the same procedure: the Distant Red Core \citep{oteo2018,long2020}, and GN20 \citep{daddi2009,hodge2013}. The solid, dashed, and dotted tracks show the largest expected halos in a 25\,deg$^2$ survey, a 2500\,deg$^2$ survey, and a full-sky survey, respectively \citep{harrison2013,marrone2018}. A sample of $z\,{>}\,4$ SMGs with CO-derived gas masses (also converted to halo mass as above) are shown as brown arrows \citep{marrone2018}.}
\label{mass}
\end{figure*}

\subsection{Star formation}

We can determine the extent to which our observations have resolved the star-formation within SPT2349$-$56 by comparing the total 850\,$\mu$m continuum flux density of our sources seen at 0.5\,arcsec resolution to the total 870\,$\mu$m flux density measured by LABOCA and seen at 21\,arcsec resolution. Our sources total 60.3$\,{\pm}\,$0.4\,mJy at 850\,$\mu$m, compared to 110.0$\,{\pm}\,$9.5\,mJy at 870\,$\mu$m as seen by LABOCA, meaning that we have resolved only 55$\pm$5\,per cent of the total star-formation. This is consistent with the fact that our \mbox{CO(4--3)} data covering the entire LABOCA 870\,$\mu$m emission region is not sensitive to galaxies with SFRs below about 100\,M$_{\odot}$\,yr$^{-1}$, and there could still be many more star-forming galaxies under this limit.

\begin{figure*}
\includegraphics{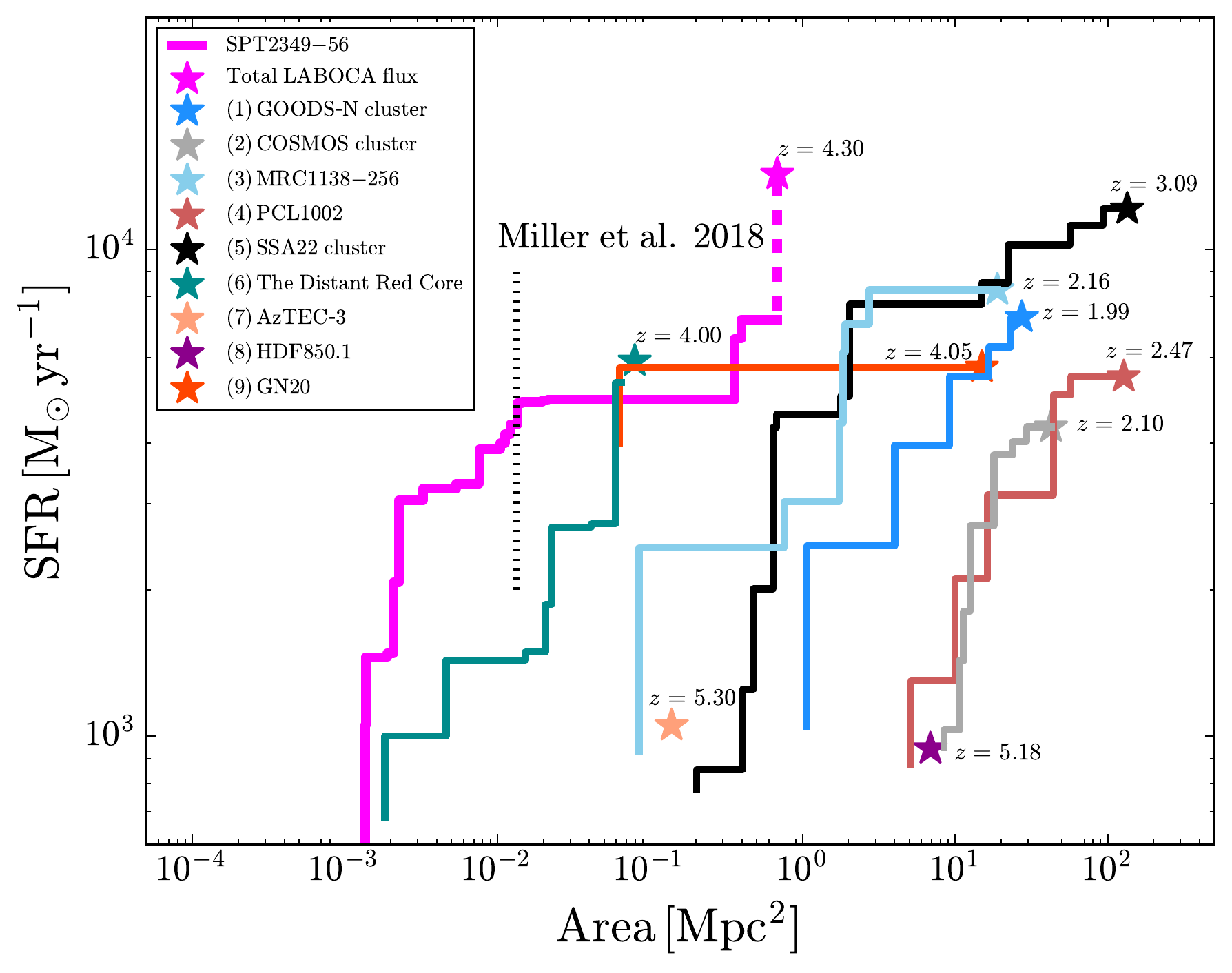}
\caption{The cumulative SFR as a function of circular projected area from the centre of SPT2349$-$56. The SFR derived from the total LABOCA flux is shown as the magenta star, and the spatial extent of the data presented in \citet{miller2018} is shown as the dashed black line (although it is important to keep in mind that many new galaxies have now been found even below this line). Also shown are curves for other proto-clusters from the literature (1 - \citealt{chapman2009}; 2 - \citealt{yuan2014}; 3 - \citealt{dannerbauer2014}; 4 - \citealt{casey2015}; 5 - \citealt{umehata2015}; 6 - \citealt{oteo2018,long2020}; 7 - \citealt{capak2011}; 8 - \citealt{walter2012}; 9 - \citealt{daddi2009,hodge2013}). SPT2349$-$56 not only contains the highest total SFR, but is also much more dense than the other proto-clusters shown here.}
\label{curve-of-growth}
\end{figure*}

It is important to recall that SPT2349$-$56 was initially selected because of its bright mm-wavelength flux density, and for objects selected in this way it is the density of star formation (rather than stars or hot gas) that makes them stand out. To put this into context, in Fig.~\ref{curve-of-growth} we show the integrated SFR as a function of projected area from the centre of the proto-cluster, along with other proto-clusters containing spectroscopically-detected SMGs reported in the literature: the GOODS-N proto-cluster at $z\,{=}\,1.99$ \citep[e.g.][]{chapman2009}; the COSMOS proto-cluster at $z\,{=}\,2.10$ \citep{yuan2014}; MRC1138$-$256 at $z\,{=}\,2.16$ \citep[e.g.][]{dannerbauer2014}; PCLL1002 at $z\,{=}\,2.47$ \citep[e.g.][]{casey2015}; the SSA22 proto-cluster at $z\,{=}\,3.09$ \citep[e.g.][]{umehata2015}; the Distant Red Core from the $H$-ATLAS survey at $z\,{=}\,4.00$ \citep{oteo2018,long2020}; and the concentration of SMGs around AzTEC-3 at $z\,{=}\,5.30$ \citep[][]{capak2011}, around HDF850.1 at $z\,{=}\,5.18$ \citep[][]{walter2012}, and around GN20 at $z\,{=}\,4.05$ \citep[e.g.][]{hodge2013} -- see \citet{casey2016} for more details. Here, for the other proto-clusters we have converted observed 870-$\mu$m flux densities to SFRs assuming a modified blackbody with $T_{\rm d}\,{=}\,39.6\,$K and an emissivity index of $\beta\,{=}\,2$ (the same model applied to our sample), fixed at the redshifts reported in the papers. SPT2349$-$56 not only contains the largest integrated SFR seen in a proto-cluster field so far, but is also significantly more dense in terms of projected area. 

We can convert points on this plot into SFR densities by assuming spherical geometry. The core of SPT2349$-$56 contains 23 galaxies within a projected proper radius of 150\,kpc, and summing up the SFRs of these galaxies gives an SFR density of 3.5$\,{\times}\,10^5$\,M$_{\odot}$\,yr$^{-1}$\,Mpc$^{-3}$. Similarly, the three galaxies in the northern extent of SPT2349$-$56 lie within a projected proper radius of approximately 130\,kpc and have a total SFR density of 2.5$\,{\times}\,10^5$\,M$_{\odot}$\,yr$^{-1}$\,Mpc$^{-3}$, while the three SPIREc galaxies are contained within a proper radius of 110\,kpc and reach 7.4$\,{\times}\,10^4$\,M$_{\odot}$\,yr$^{-1}$\,Mpc$^{-3}$. The total LABOCA emission region (i.e.~including the core and the northern clump) extends about 720\,kpc in the North-South direction. Using a sphere of proper radius 360\,kpc we can thus estimate a total SFR density of 3.7$\,{\times}\,10^4$\,M$_{\odot}$\,yr$^{-1}$\,Mpc$^{-3}$.

While the interpretation of this number is unclear, since the volume specified is somewhat arbitrary, this calculation at least provides a simple way to compare the star-formation densities we are observing to simulations. Indeed, it appears that total SFR densities are roughly an order of magnitude larger than what is seen in current simulations \citep[e.g.,][]{saro2009,granato2015}, motivating further work in both comparing the data to the simulations and in modelling star formation at high redshift in the most massive galaxy clusters.

\section{Summary and conclusion}
\label{conclusion}

We have used ALMA to resolve galaxies within SPT2349$-$56, the core of a massive galaxy proto-cluster at redshift 4.3. Initial single-dish submm observations revealed an extended structure covering several hundred kpc and containing over 10$^4$\,M$_{\odot}$\,yr$^{-1}$ in star formation, making this an incredibly active and rare structure at this epoch.

Our observations included 850-$\mu$m pointings targeting the [C{\sc ii}] line and 3.2-mm pointings targeting the \mbox{CO(4--3)} transition. Our data cover the entire single-dish flux density region detected by previous observations, and include the surrounding area out to about 400\,kpc in proper distance and four nearby red {\it Herschel\/}-SPIRE sources. A line and continuum source search revealed a total of 29 galaxies at redshift 4.303 and three continuum-only galaxies that are potentially line-of-sight interlopers. From the line profiles we measured line strengths, linewidths, and line luminosities, and we used continuum flux-density measurements to constrain far-infrared luminosities and SFRs. 

SPT2349$-$56 resolves into a large central core containing 23 galaxies and a northern group located 400 proper kiloparsecs away containing three galaxies, while one of the red {\it Herschel\/}-SPIRE sources resolves into another group of three galaxies located about 1.5\,Mpc (in proper distance) from the central region. Given the low S/N of the red {\it Herschel\/}-SPIRE sources selected for follow-up here, and given that none of these targets showed significant 870-$\mu$m emission in our LABOCA map, we argue that there could be many other sources at the same redshift as SPT2349$-$56 that are not yet detected in our current data. $N$-body simulations predict that the core galaxies will merge into a BCG, while an analysis of the distribution of line-of-sight velocities within the central region suggests that the remaining two groups are not gravitationally bound to the core system and will remain distributed within the overall galaxy cluster.

We compare the far-infrared properties of our proto-cluster galaxies to samples of field galaxies found around the same redshift. We find good agreement between the far-infrared luminosities, line strengths, line widths, and physical radii, despite the fact that our proto-cluster galaxies are found within such a dense environment.

$L_{[\rm C\textsc{ii}]}$ and $L_{\rm FIR}$ cumulative and differential number counts were computed and compared to models and measurements of these luminosity functions for field galaxies at similar redshifts from the literature. As our observations target a known proto-cluster with many more sources compared to the field, the normalization factors of our number counts are many orders of magnitude larger than the previous works, but we can compare their shapes. We find that our $L_{\rm FIR}$ number counts are biased towards the bright end compared to current models of high-$z$ field galaxies, suggesting a possible increase in SFR during the proto-cluster assembly process. We also looked for evidence of a break in these luminosity functions by comparing single power-law models to Schechter functions using a maximum-likelihood approach. We found a Schechter function provided the highest likelihood to our $L_{[\rm C\textsc{ii}]}$ number counts, while our $L_{\rm FIR}$ number counts are best described by a single power-law.

We have estimated the mass of SPT2349$-$56 using several techniques. First, we looked at the velocity dispersion of the core galaxies, which suggests a mass of (9$\,{\pm}\,5$)$\,{\times}\,$10$^{12}$\,M$_{\odot}$, consistent with the mass of BCGs that inhabit the cores present-day galaxy clusters. Next, we estimated the total mass of SPT2349$-$56 resolved so-far by scaling the gas masses of the constituent galaxies to halo masses, and we found that the total halo mass summed over the core region, the northern region, and the SPIREc region is 2.5$\,{\times}\,$10$^{13}$\,M$_{\odot}$. While there are large systematic uncertainties in this quantity, we find the results to be comparable to other massive proto-cluster systems in the literature.

SPT2349$-$56 reaches a total SFR density of around 4$\,{\times}\,10^4$\,M$_{\odot}$\,yr$^{-1}$\,Mpc$^{-3}$, something that simulations may be currently incapable of generating at high redshift. Future studies are therefore needed to compare the data to the simulations in more detail, and to understand the mechanisms responsible for producing such copious amounts of star formation at such an early epoch.

Our results suggest that SPT2349$-$56 is the progenitor of a $10^{13}$\,M$_{\odot}$ Coma-like cluster core, surrounded by a number of groups of galxies, each only a factor of a few times less massive than the core. This is clearly a unique proto-cluster system, and is an example of one of the most active large-scale environments seen during the peak of its star formation. 

\section*{ACKNOWLEDGMENTS}

This paper makes use of the following ALMA data: ADS/JAO.ALMA\#2017.1.00273.S, and ADS/JAO.ALMA\#2018.1.00058.S. ALMA is a partnership of ESO (representing its member states), NSF (USA) and NINS (Japan), together with NRC (Canada), MOST and ASIAA (Taiwan), and KASI (Republic of Korea), in cooperation with the Republic of Chile. The Joint ALMA Observatory is operated by ESO, AUI/NRAO and NAOJ. {\it Herschel\/} is an ESA space observatory with science instruments provided by European-led Principal Investigator consortia and with important participation from NASA. Based on observations made with ESO Telescopes at the La Silla Paranal Observatory under programme ID 299.A-5045(A). The National Radio Astronomy Observatory is a facility of the National Science Foundation operated under cooperative agreement by Associated Universities, Inc. The SPT is supported by the National Science Foundation through grant PLR-1248097, with partial support through PHY-1125897, the Kavli Foundation and the Gordon and Betty Moore Foundation grant GBMF 947. This work was supported by the Natural Sciences and Engineering Research Council of Canada. The Flatiron Institute is supported by the Simons Foundation. MA has been supported by the grant ``CONICYT+PCI+REDES 190194''. D.P.M., J.D.V., K.C.L., and K.P. acknowledge support from the US NSF under grants AST-1715213 and AST-1716127. K.C.L acknowledges support from the US NSF NRAO under grants SOSPA5-001 and SOSPA4-007, respectively. J.D.V. acknowledges support from an A. P. Sloan Foundation Fellowship.

\bibliographystyle{mnras}
\bibliography{spt2349_redshifts}

\appendix

\section{Spectra}
\label{appendix1}

Here we show all 32 sources detected in our [C{\sc ii}] and \mbox{CO(4--3)} observations of SPT2349$-$56. For each source, in the left-hand panel we show 3$\,{\times}\,3\,$arcsec cutouts, except for source C1, where we show a 4$\,{\times}\,4\,$arcsec cutout. Continuum images obtained by stacking all channels containing no line emission are shown in the background, and overlaid are corresponding continuum contours starting at 2$\sigma$ and increasing in steps of 3$\sigma$. We also show line emission contours from stacking all channels between $-3\sigma$ and $3\sigma$ (where $\sigma$ is the standard deviation of the best-fitting linewidth), or for cases where two Gaussians were a better fit, between $-3\sigma_{\rm L}$ and $+3\sigma_{\rm R}$, where $\sigma_{\rm L}$ and $\sigma_{\rm R}$ are from the left and right Gaussian fits, respectively. These contours also start at 2$\sigma$ and increase in steps of 3$\sigma$.

In the righthand panels we show our continuum-subtracted spectra. Plotted overtop of the spectra is shown the best-fitting single or double Gaussian functions, or for the three cases where no line is detected, a constant function with 0 amplitude. The shaded region ranges from $-3\sigma$ to $3\sigma$ (or from $-3\sigma_{\rm L}$ and $+3\sigma_{\rm R}$ for double-Gaussian fits), corresponding to the range used to calculate line strengths. \mbox{CO(4--3)} spectra are not shown for sources C12 and C16 since they are completely blended with sources C3 and C13, respectively.

\begin{figure*}
\begin{subfigure}{.45\textwidth}
\begin{framed}
\includegraphics[width=\textwidth]{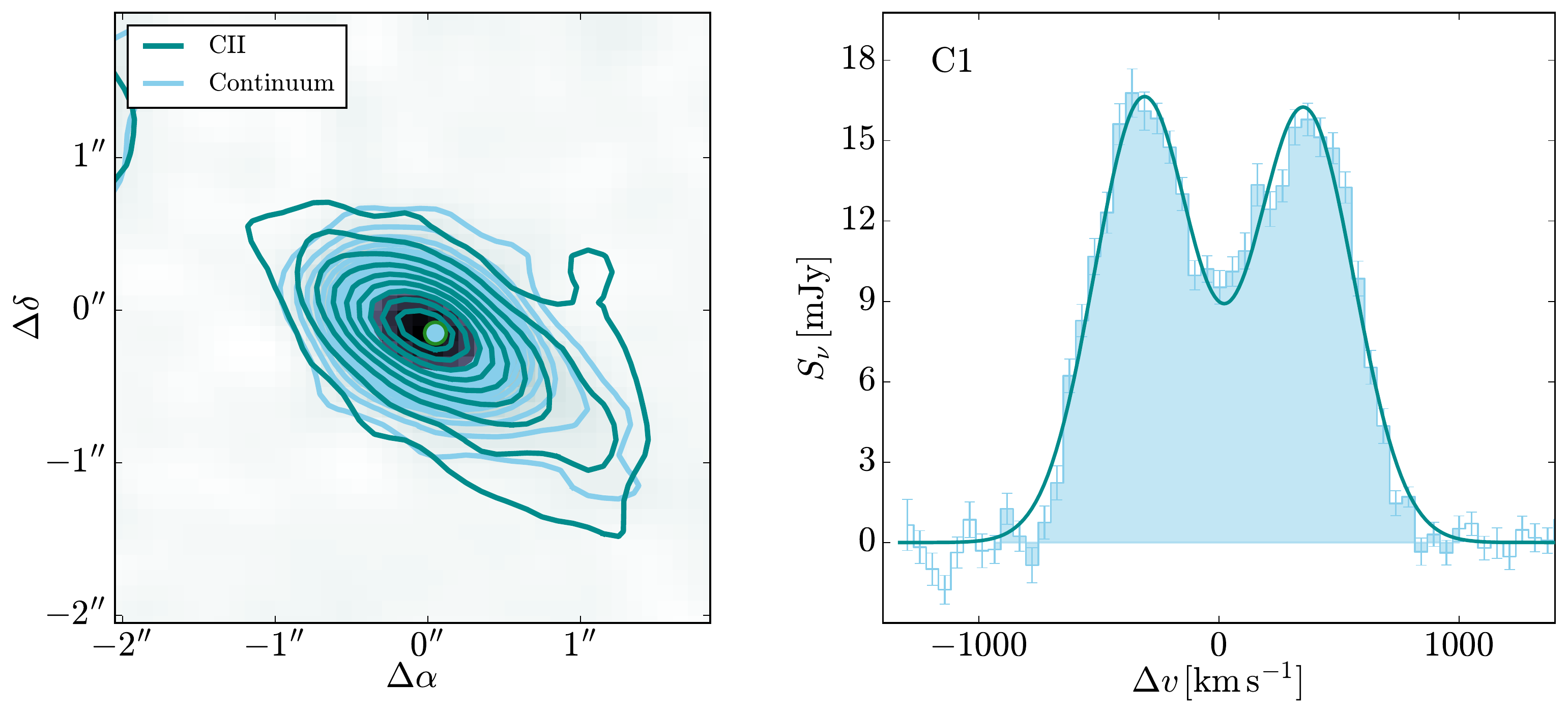}
\includegraphics[width=\textwidth]{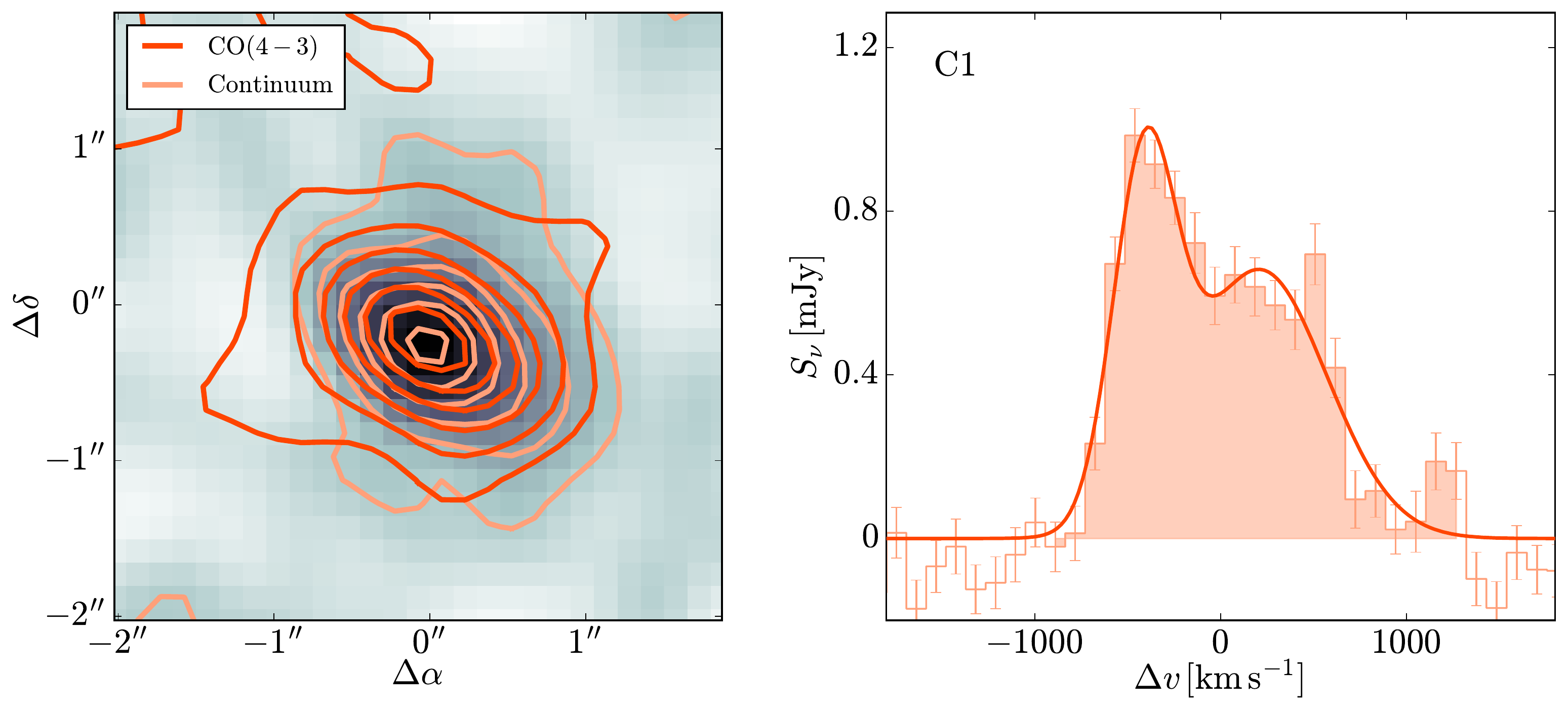}
\end{framed}
\end{subfigure}
\begin{subfigure}{.45\textwidth}
\begin{framed}
\includegraphics[width=\textwidth]{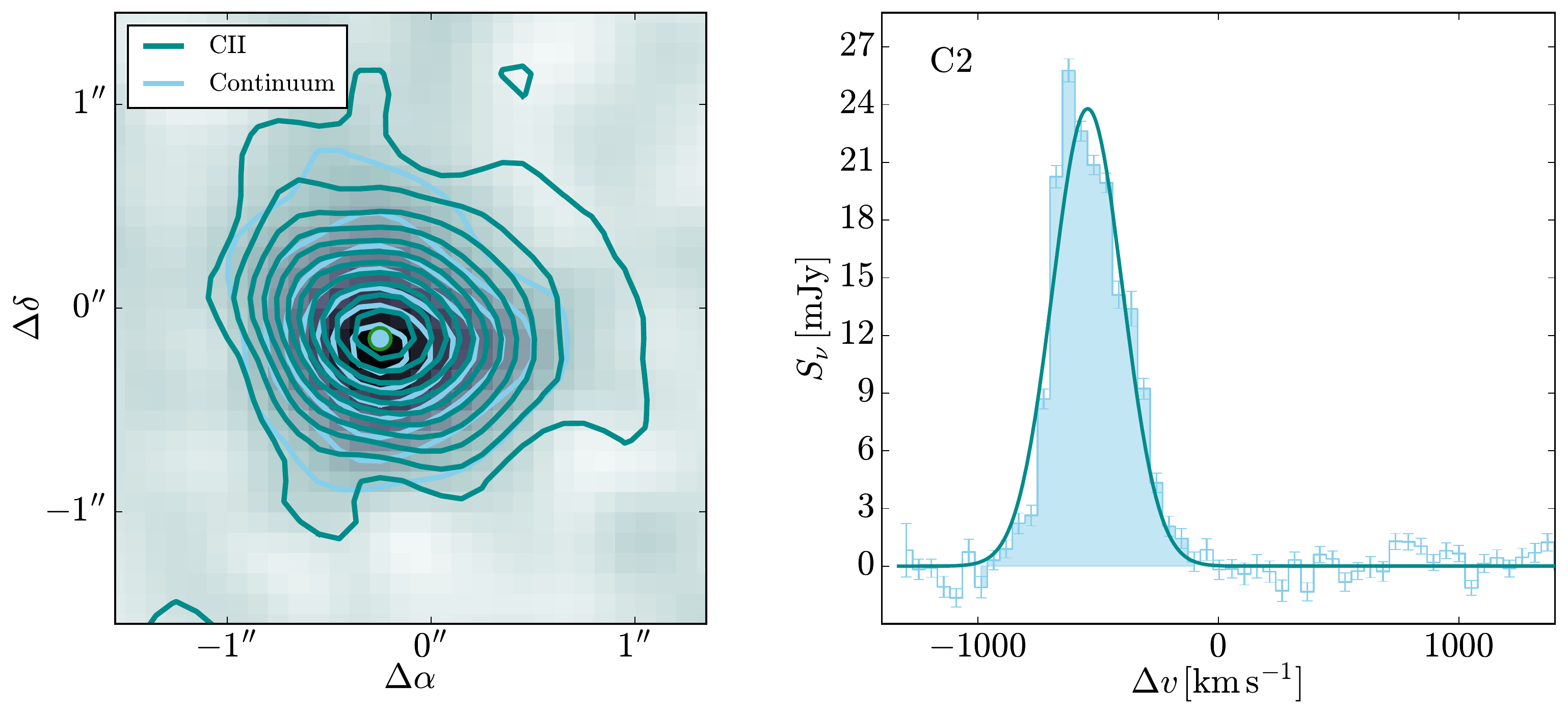}
\includegraphics[width=\textwidth]{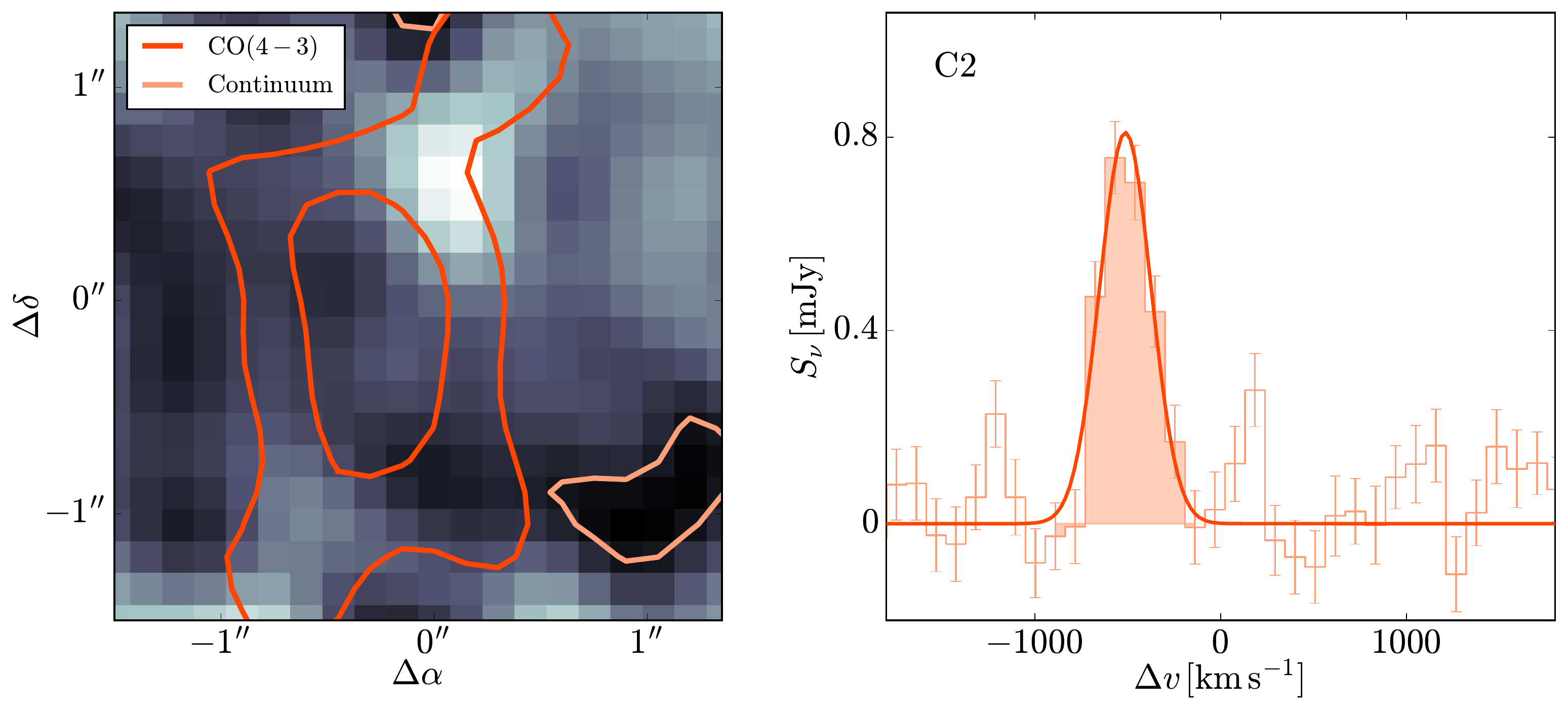}
\end{framed}
\end{subfigure}
\begin{subfigure}{.45\textwidth}
\begin{framed}
\includegraphics[width=\textwidth]{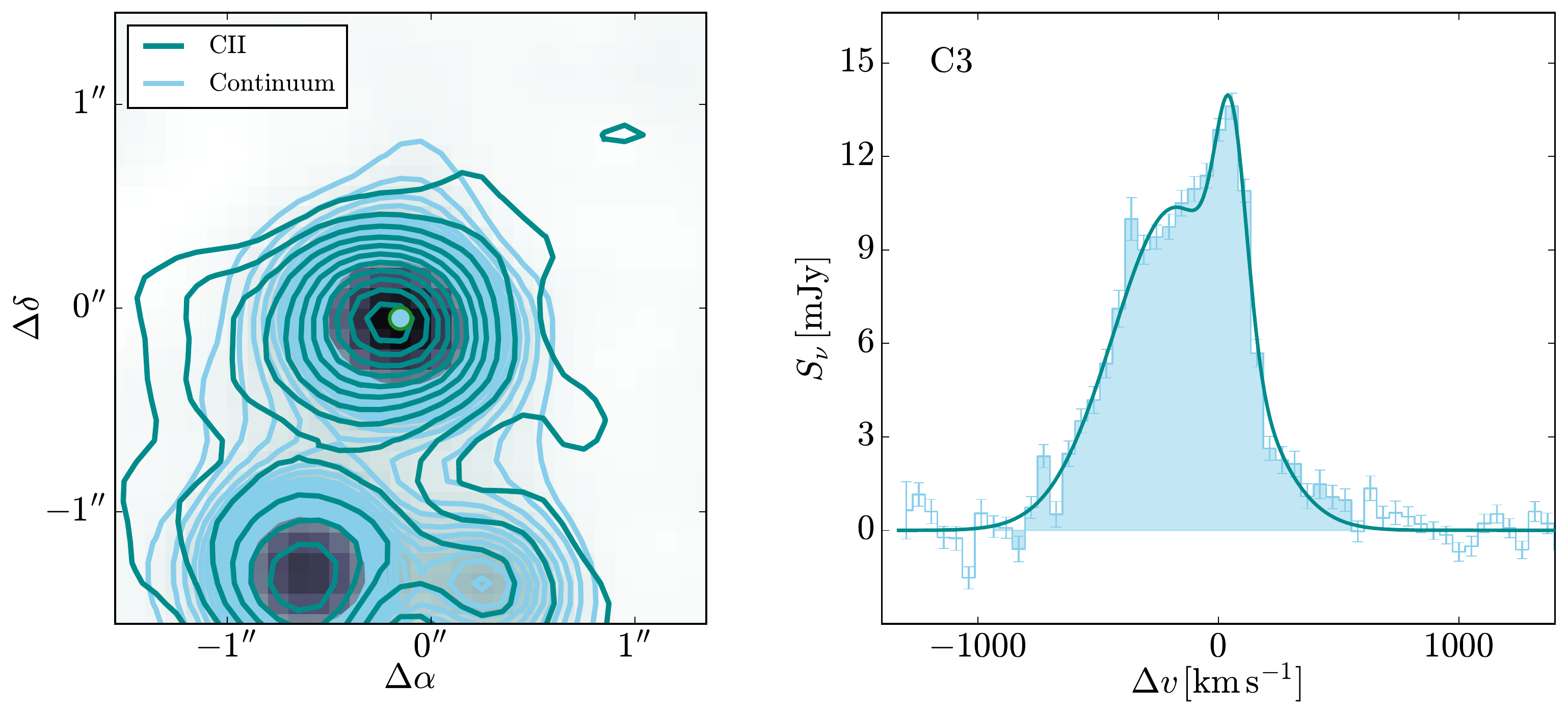}
\includegraphics[width=\textwidth]{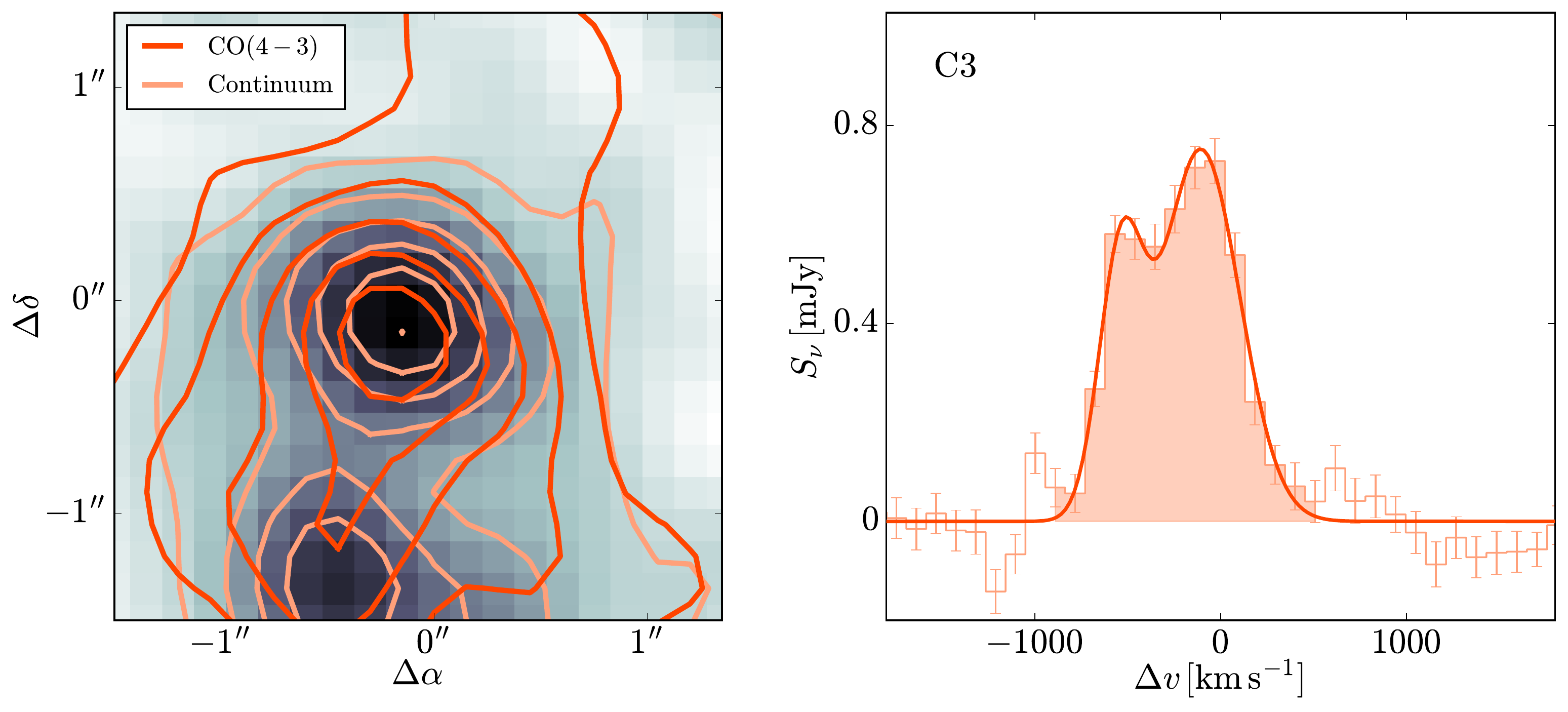}
\end{framed}
\end{subfigure}
\begin{subfigure}{.45\textwidth}
\begin{framed}
\includegraphics[width=\textwidth]{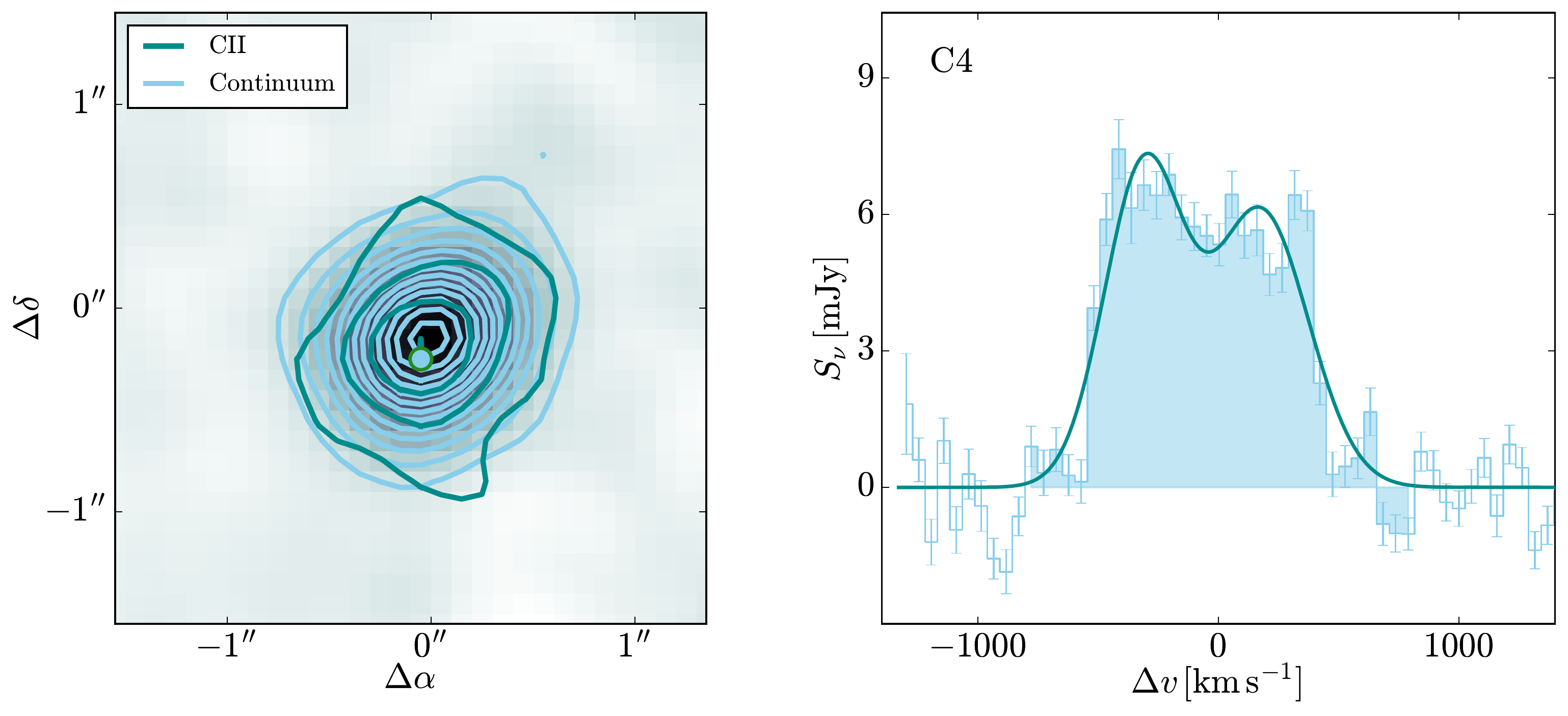}
\includegraphics[width=\textwidth]{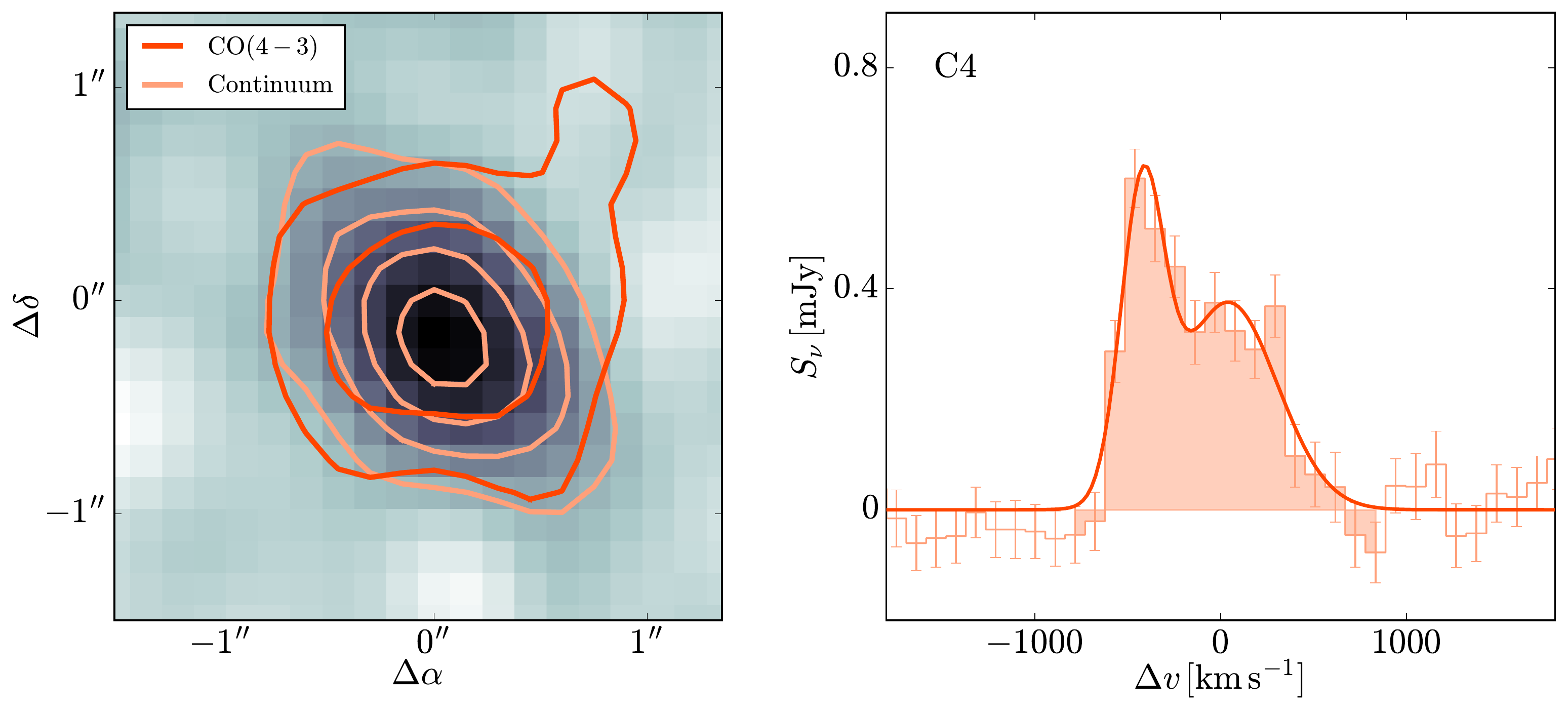}
\end{framed}
\end{subfigure}
\begin{subfigure}{.45\textwidth}
\begin{framed}
\includegraphics[width=\textwidth]{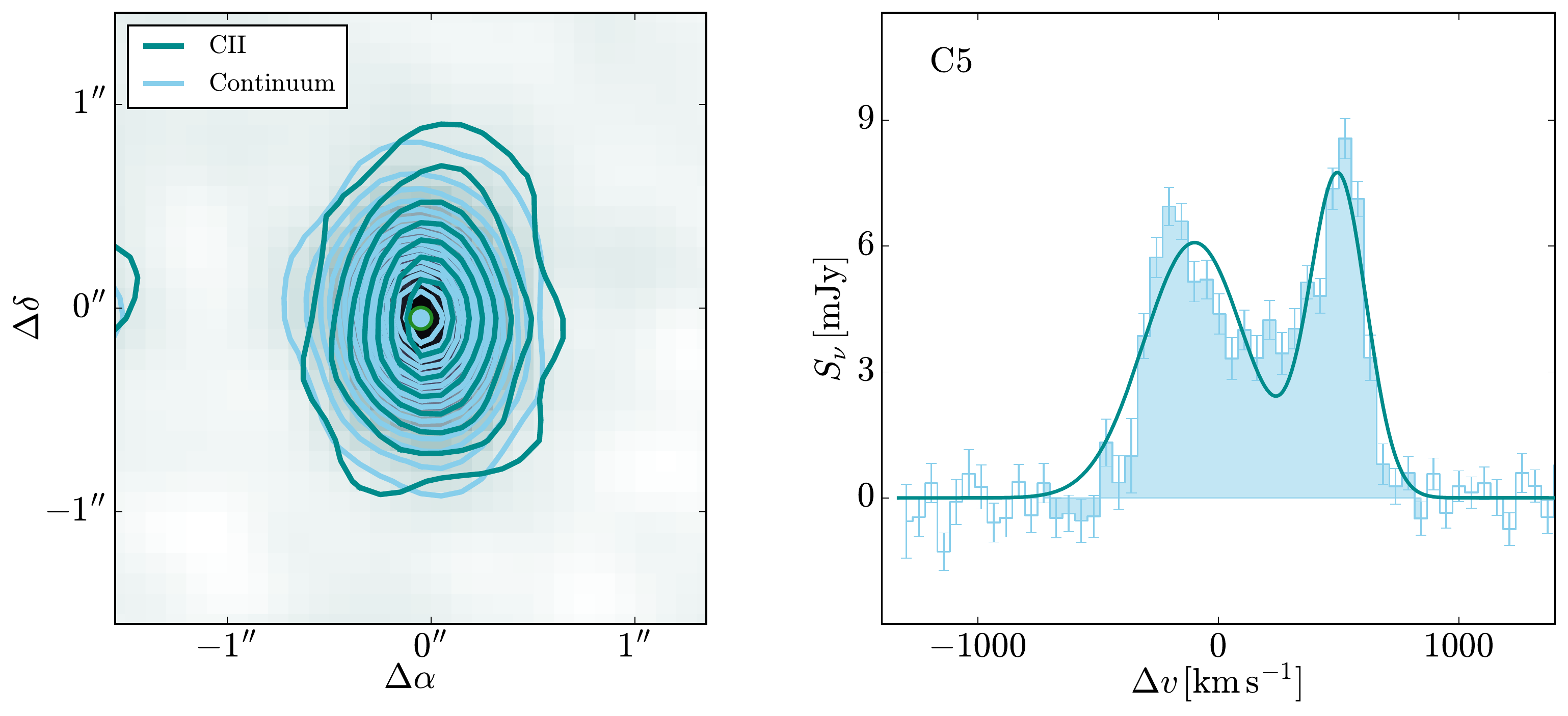}
\includegraphics[width=\textwidth]{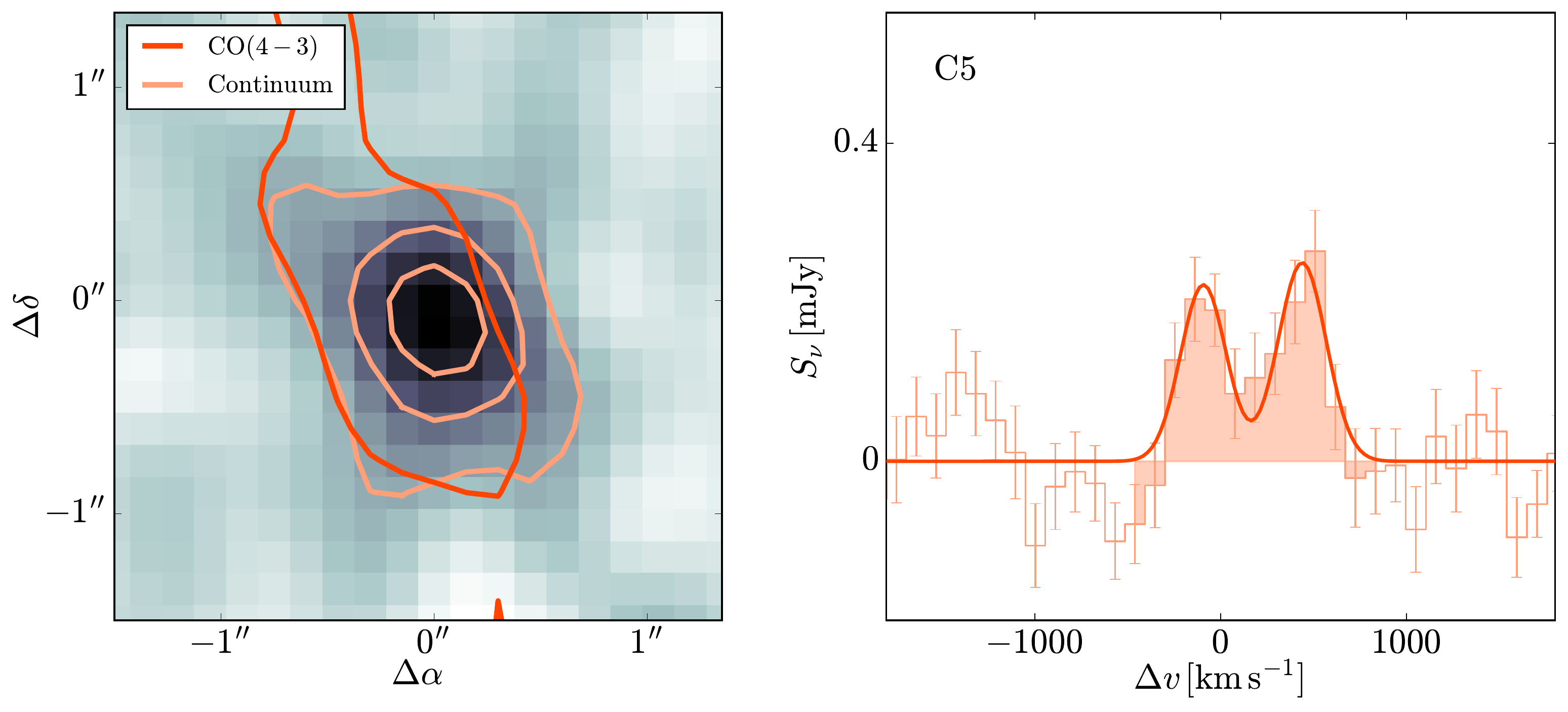}
\end{framed}
\end{subfigure}
\begin{subfigure}{.45\textwidth}
\begin{framed}
\includegraphics[width=\textwidth]{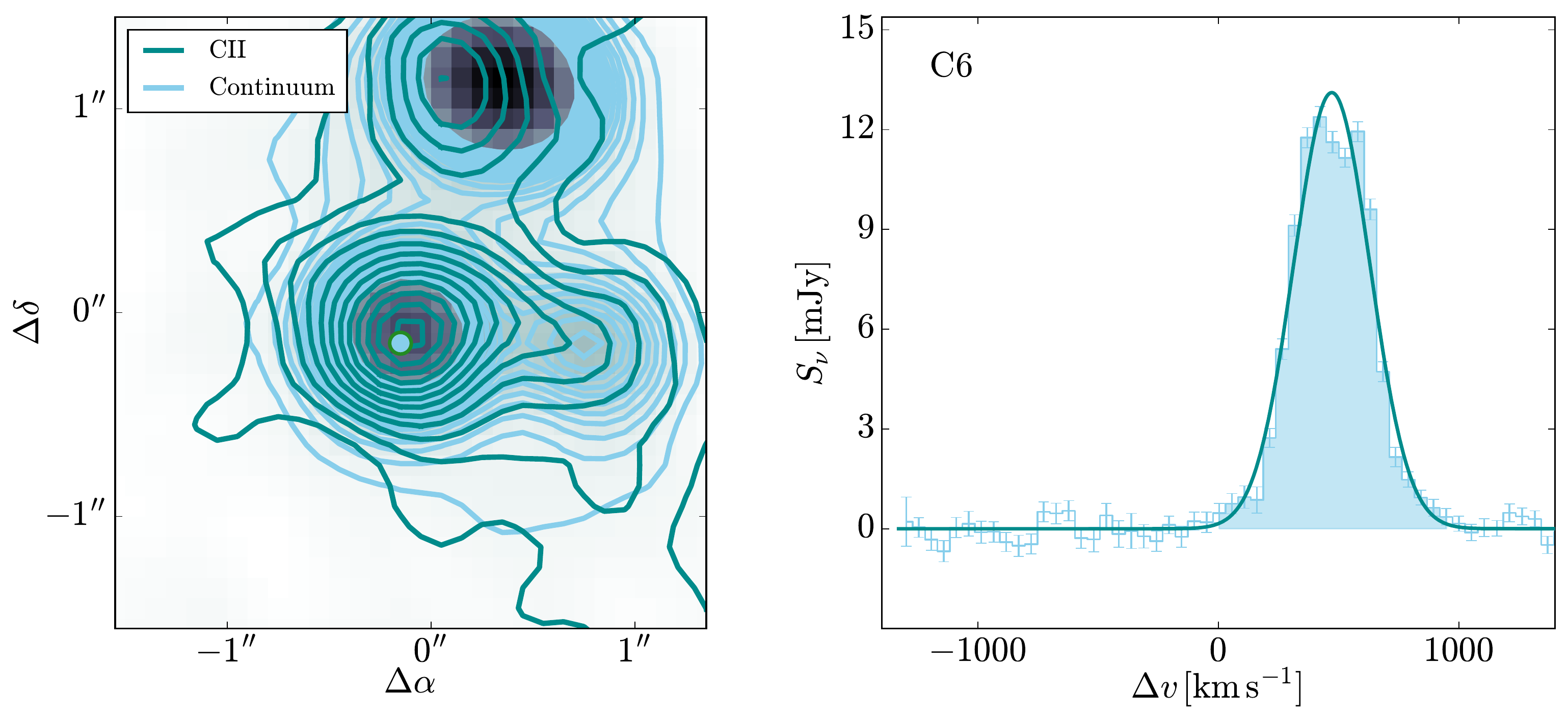}
\includegraphics[width=\textwidth]{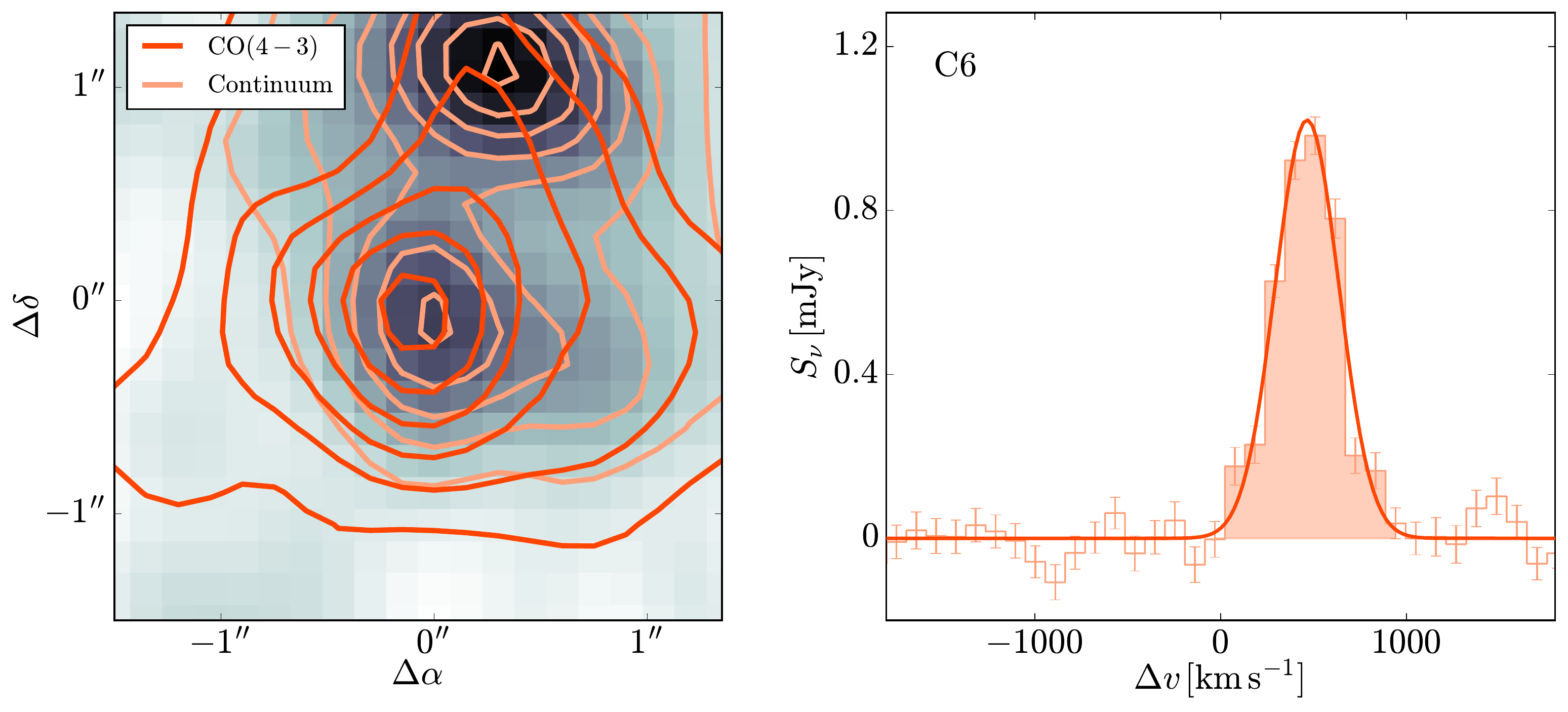}
\end{framed}
\end{subfigure}
\caption{{\it Left:} The background images are 3$\,{\times}\,3\,$arcsec cutouts (or 4$\,{\times}\,4\,$arcsec cutouts for source C1) of continuum images obtained by stacking all channels containing no line emission. Overlaid are continuum contours starting at 2$\sigma$ and increasing in steps of 3$\sigma$, and line emission contours obtained by stacking best-fitting line emission channels (i.e. moment 0 maps), also starting at 2$\sigma$ and increasing in steps of 3$\sigma$. {\it Right:} Continuum-subtracted spectra, with corresponding best-fitting constant, single Gaussian, or double Gaussian functions. The shaded region ranges from $-3\sigma$ to $3\sigma$ (or from $-3\sigma_{\rm L}$ and $+3\sigma_{\rm R}$ for double-Gaussian fits, where $\sigma_{\rm L}$ and $\sigma_{\rm L}$ are the left and right Gaussians, respectively), which corresponds to the range used to calculate line strengths.}
\label{cutouts}
\end{figure*}

\renewcommand{\thefigure}{A\arabic{figure} (Cont.)}
\addtocounter{figure}{-1}
\begin{figure*}
\begin{subfigure}{.45\textwidth}
\begin{framed}
\includegraphics[width=\textwidth]{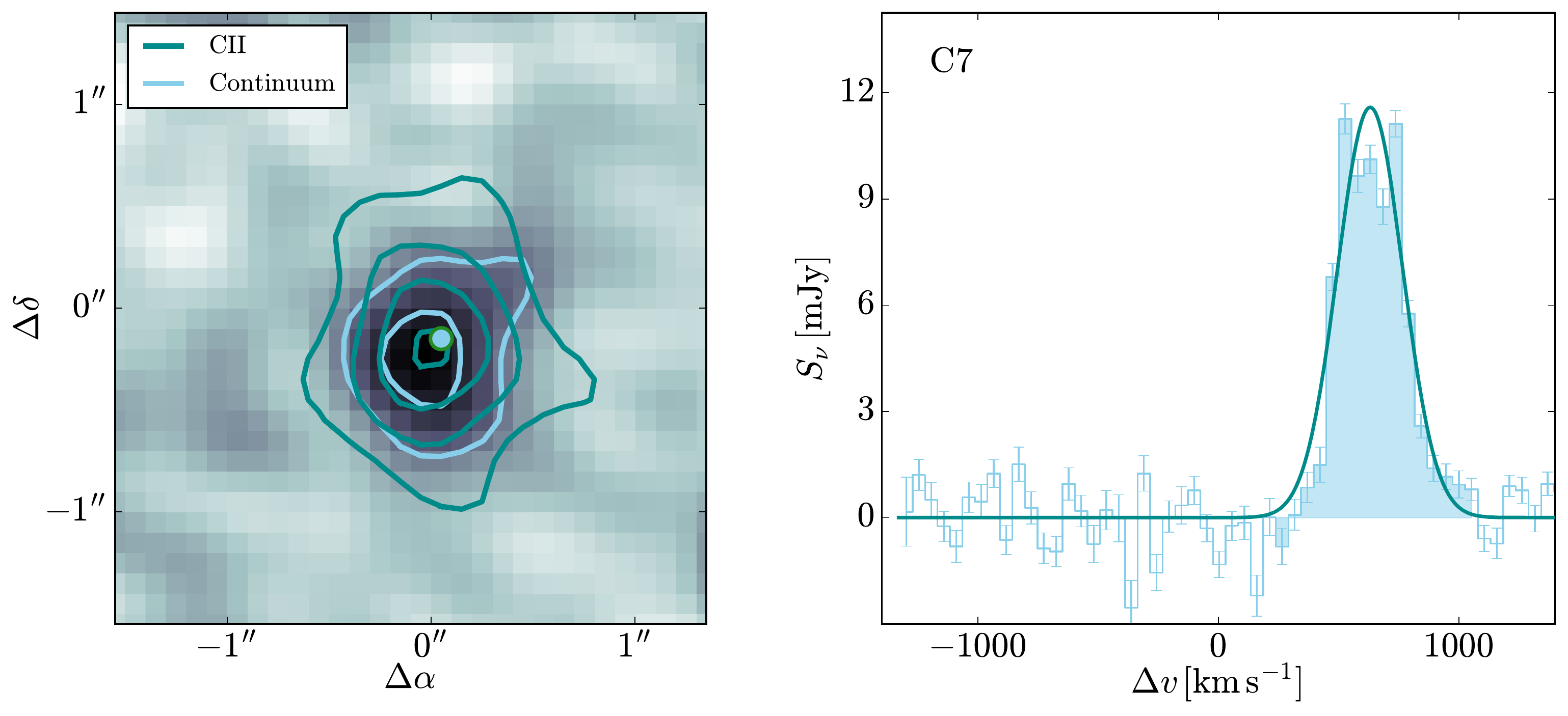}
\includegraphics[width=\textwidth]{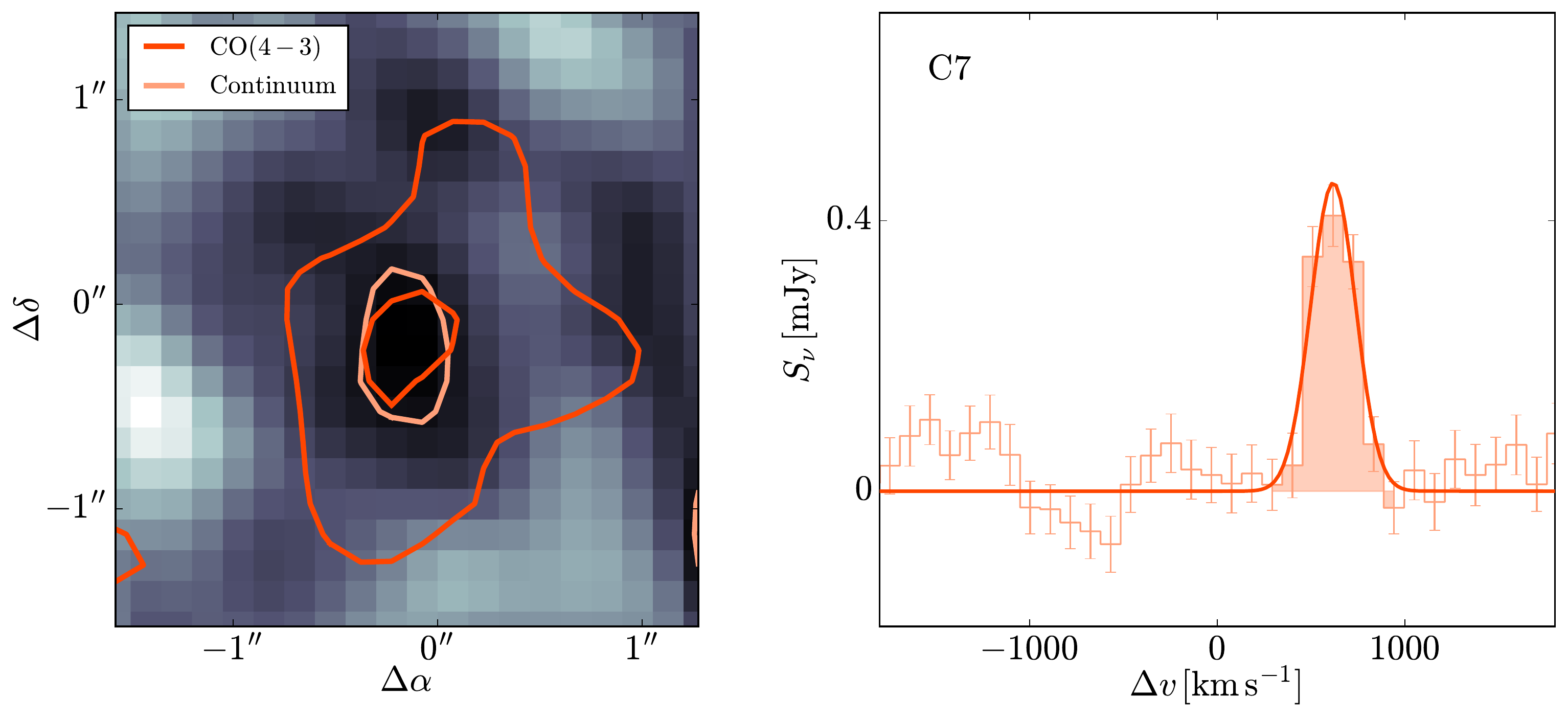}
\end{framed}
\end{subfigure}
\begin{subfigure}{.45\textwidth}
\begin{framed}
\includegraphics[width=\textwidth]{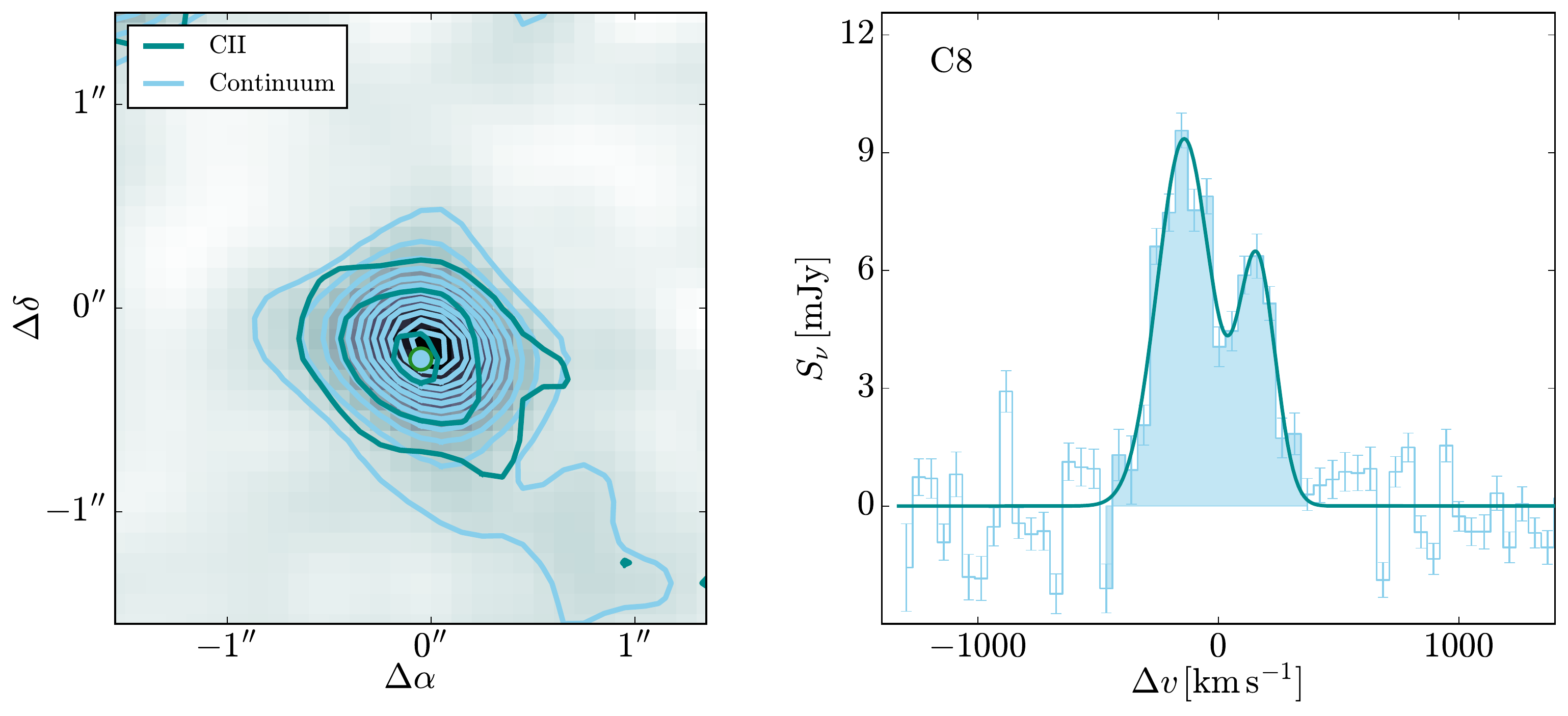}
\includegraphics[width=\textwidth]{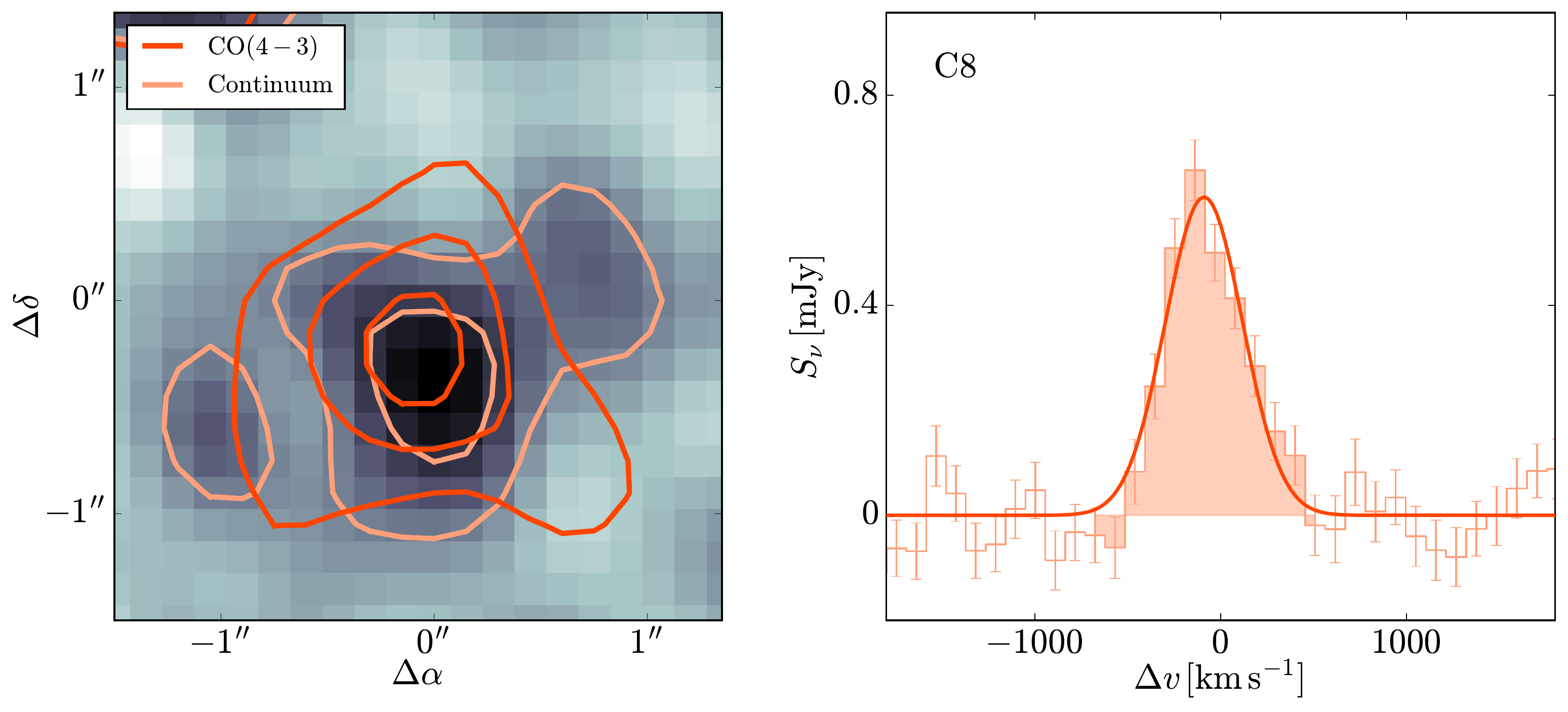}
\end{framed}
\end{subfigure}
\begin{subfigure}{.45\textwidth}
\begin{framed}
\includegraphics[width=\textwidth]{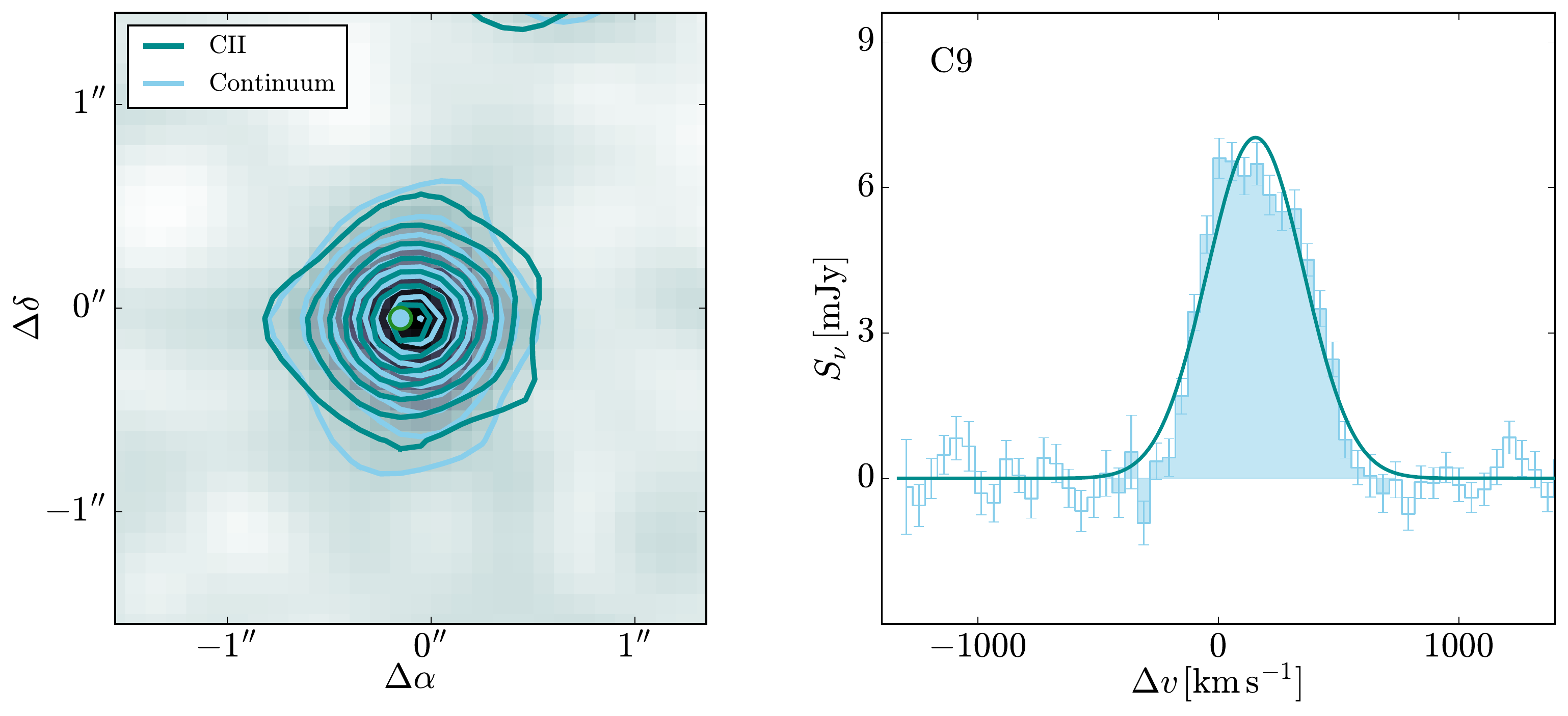}
\includegraphics[width=\textwidth]{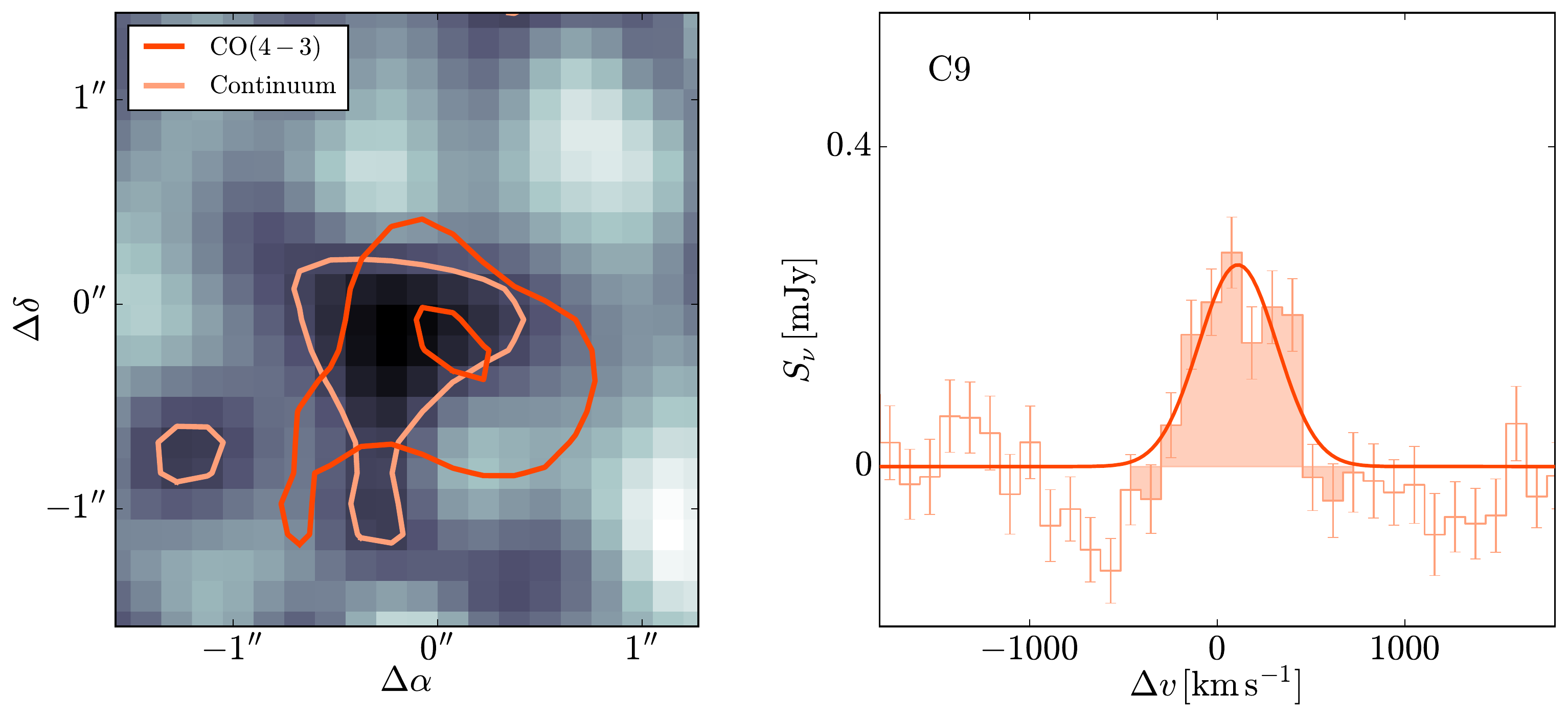}
\end{framed}
\end{subfigure}
\begin{subfigure}{.45\textwidth}
\begin{framed}
\includegraphics[width=\textwidth]{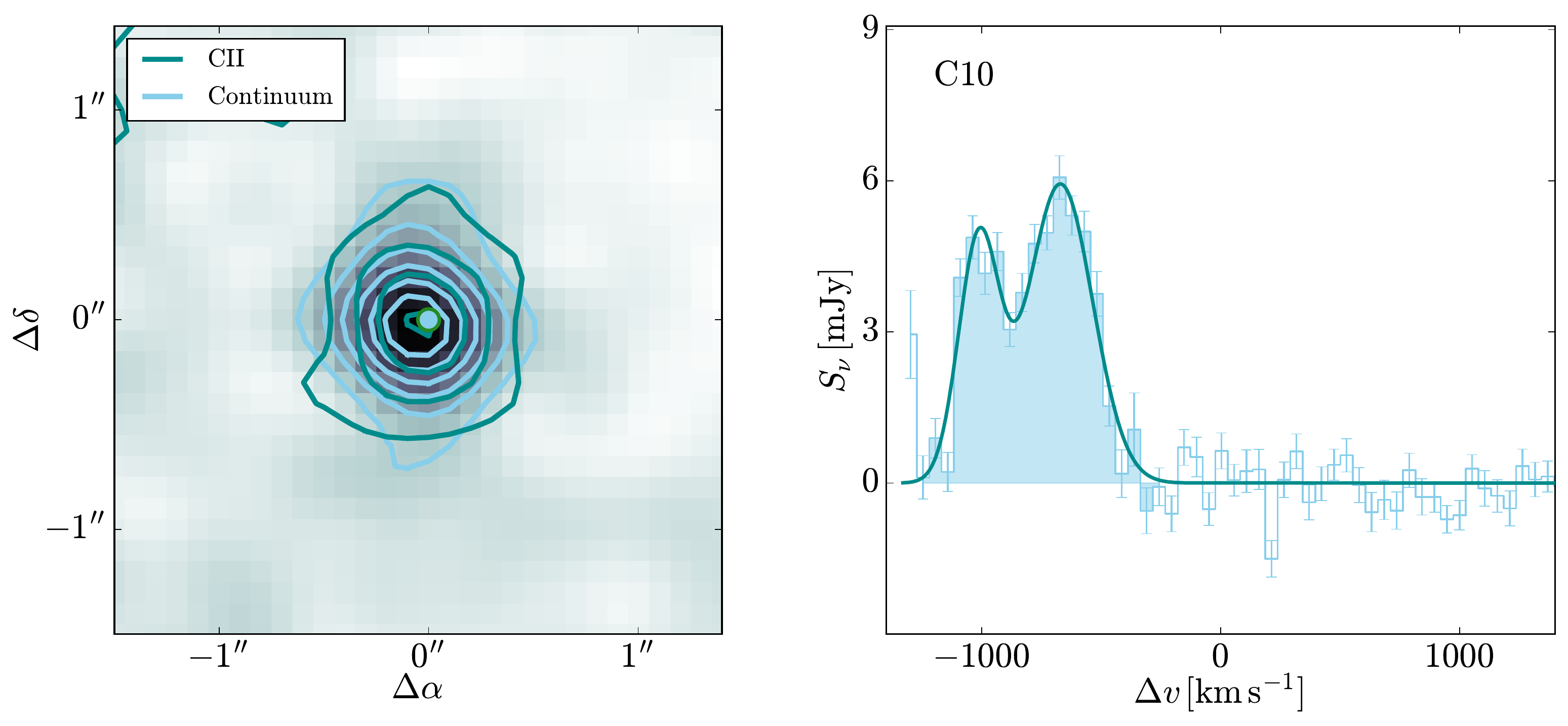}
\includegraphics[width=\textwidth]{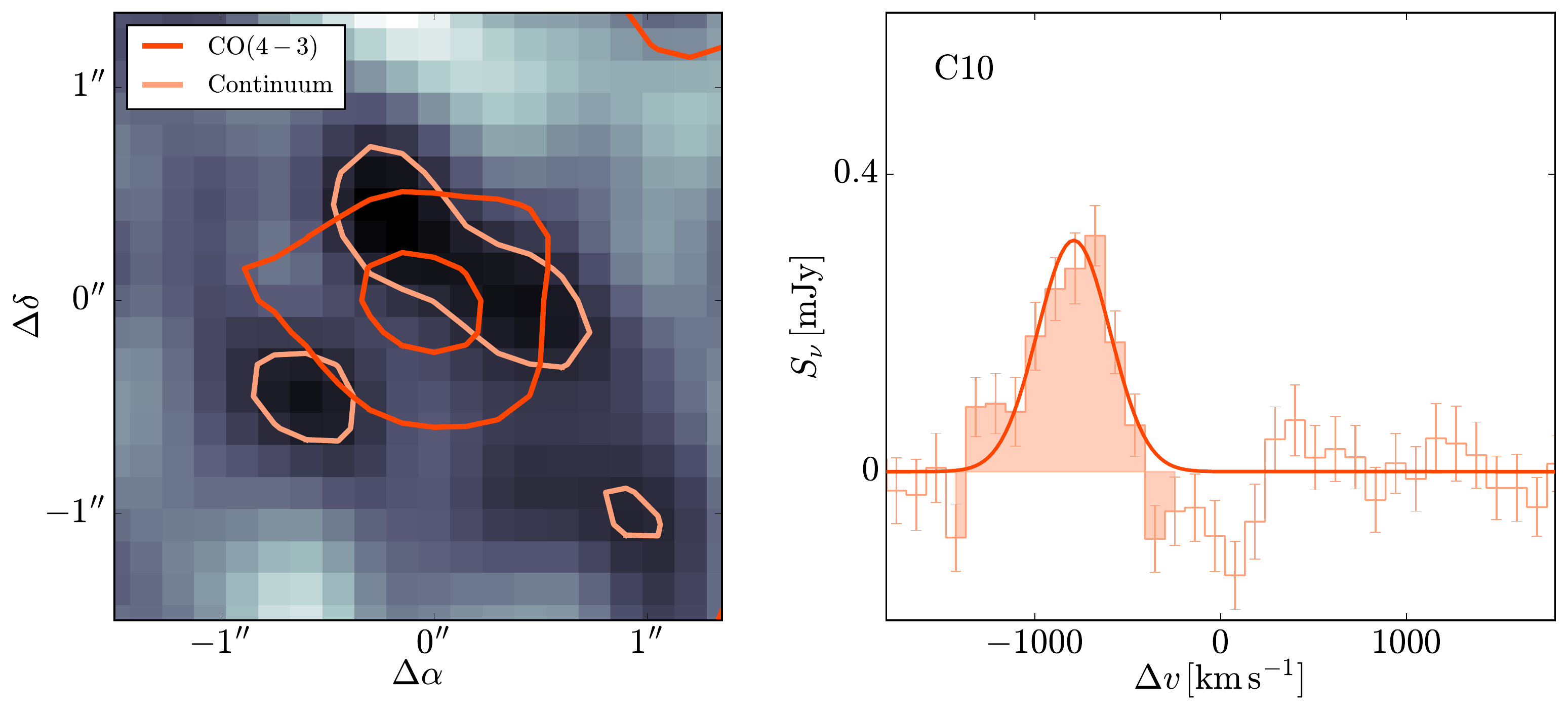}
\end{framed}
\end{subfigure}
\begin{subfigure}{.45\textwidth}
\begin{framed}
\includegraphics[width=\textwidth]{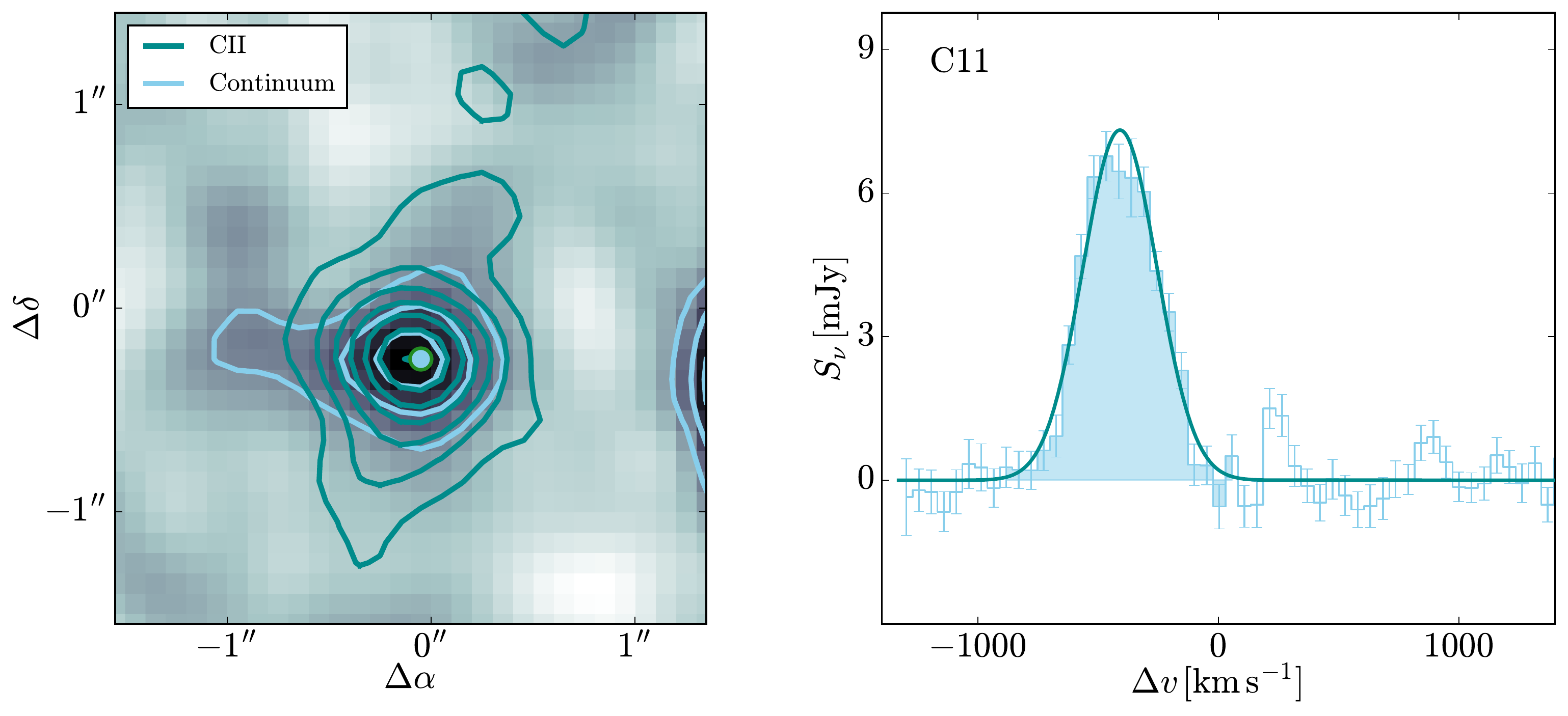}
\includegraphics[width=\textwidth]{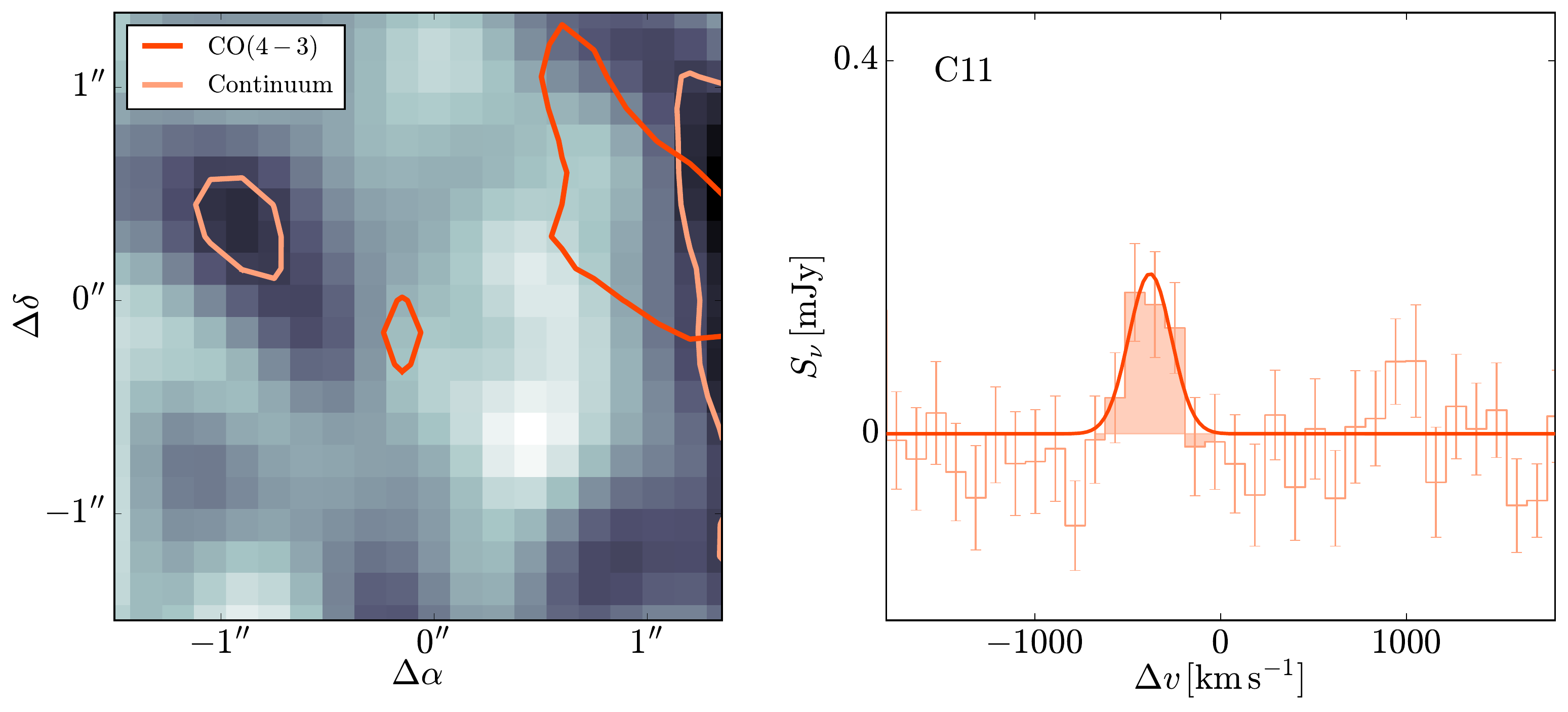}
\end{framed}
\end{subfigure}
\begin{subfigure}{.45\textwidth}
\begin{framed}
\includegraphics[width=\textwidth]{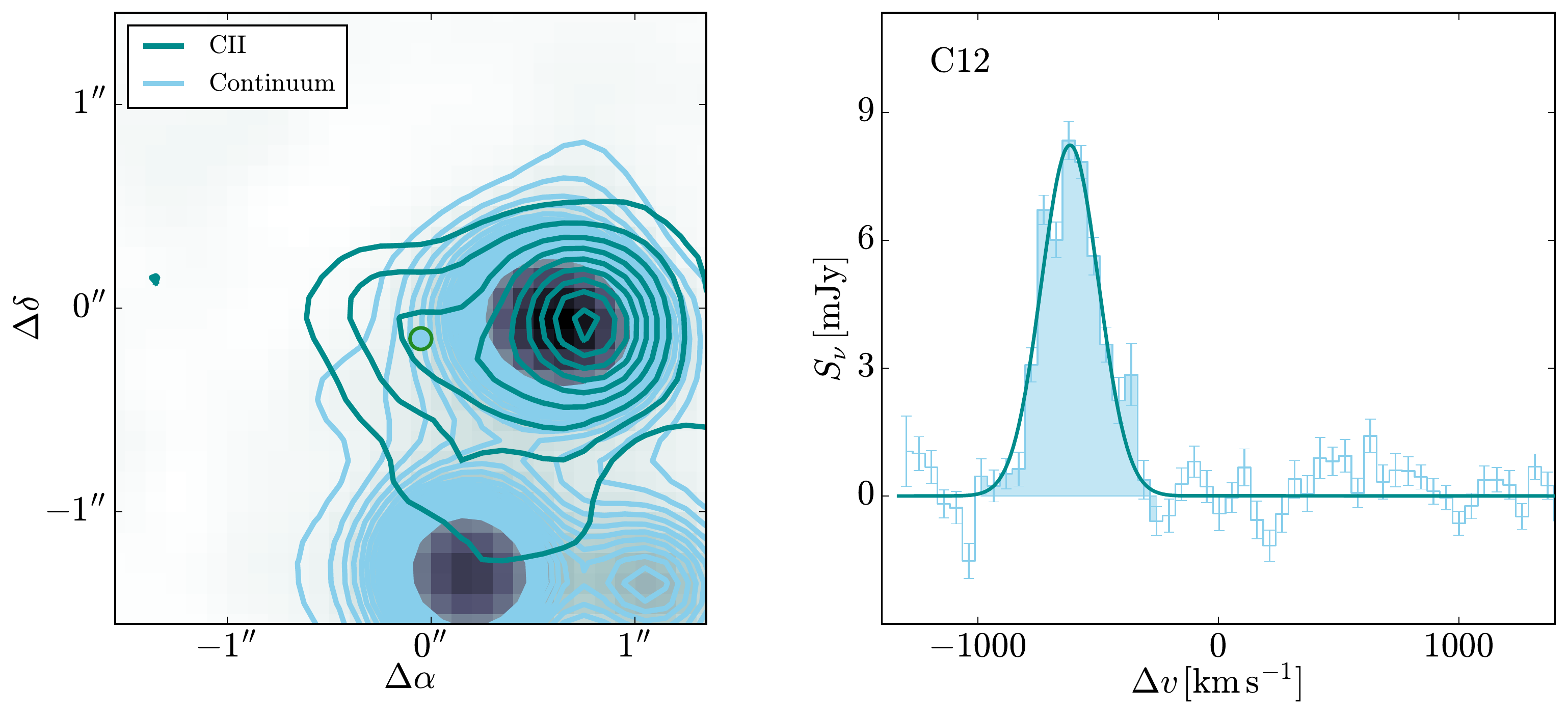}
\includegraphics[width=\textwidth]{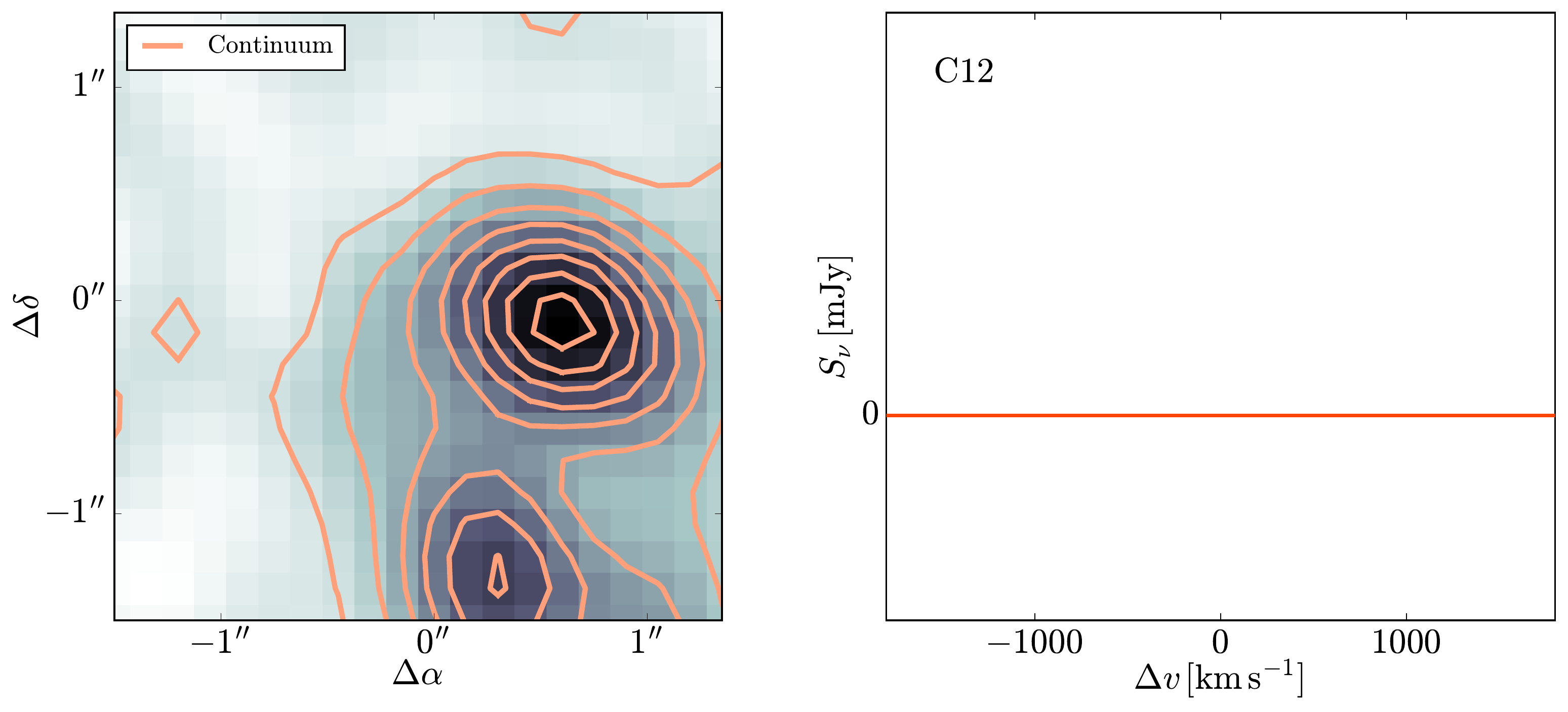}
\end{framed}
\end{subfigure}
\caption{}
\end{figure*}
\renewcommand{\thefigure}{\arabic{figure}}

\renewcommand{\thefigure}{A\arabic{figure} (Cont.)}
\addtocounter{figure}{-1}
\begin{figure*}
\begin{subfigure}{.45\textwidth}
\begin{framed}
\includegraphics[width=\textwidth]{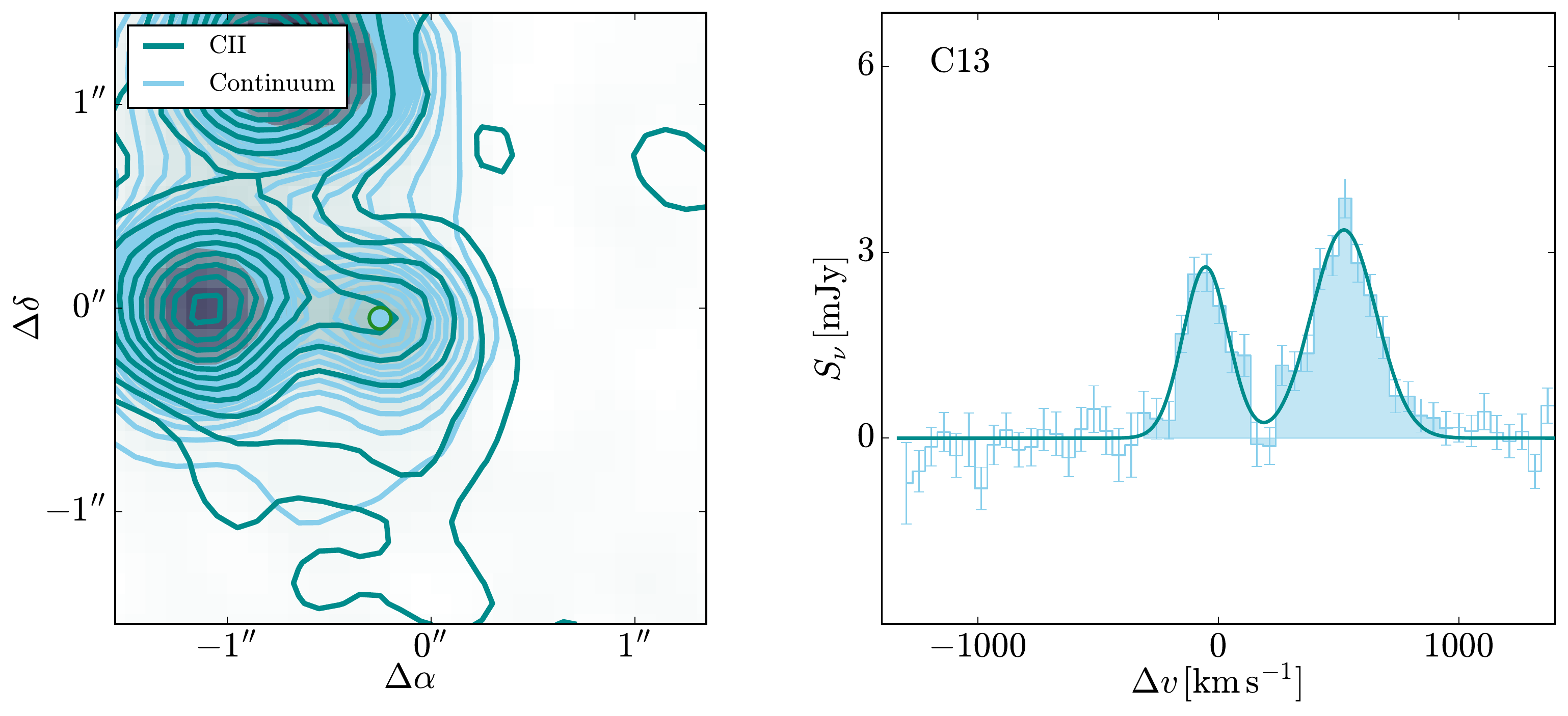}
\includegraphics[width=\textwidth]{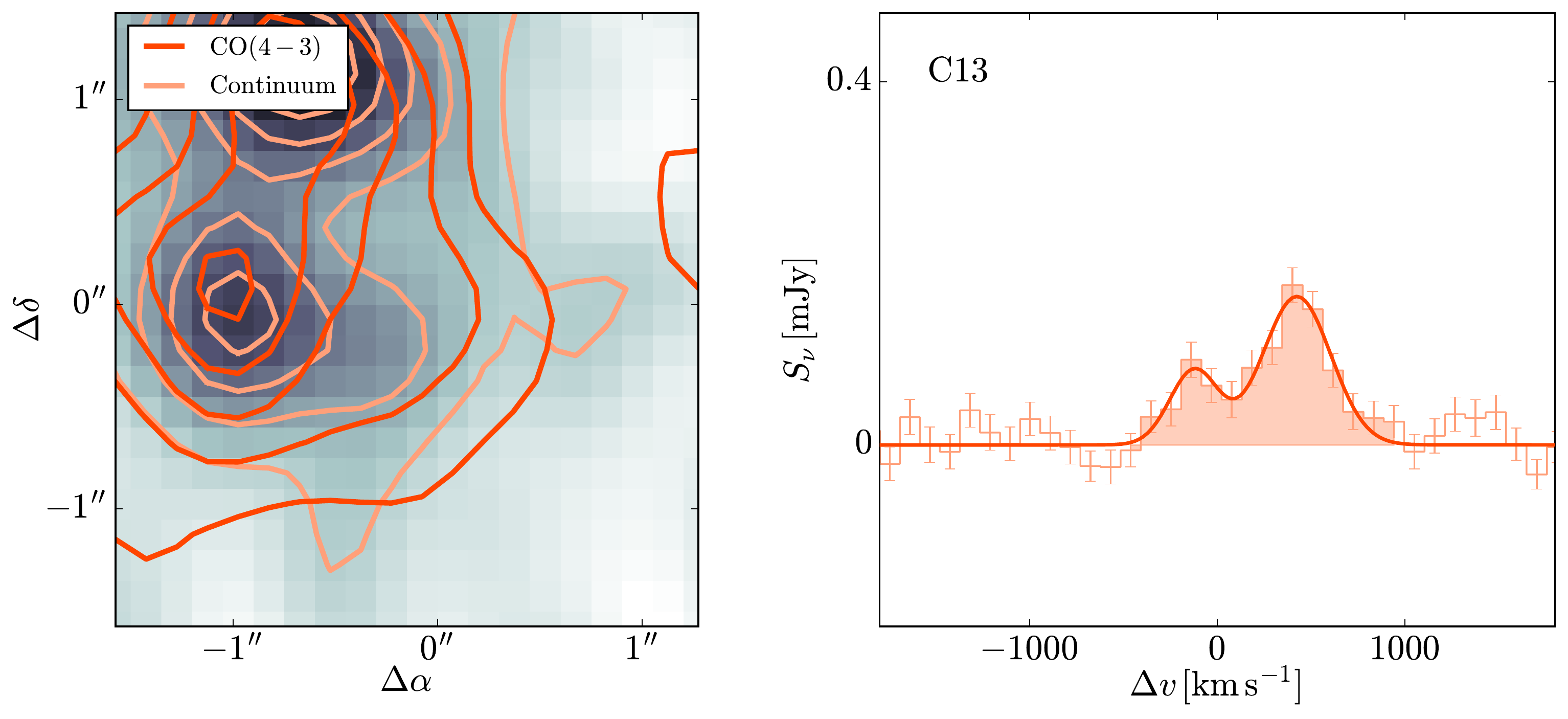}
\end{framed}
\end{subfigure}
\begin{subfigure}{.45\textwidth}
\begin{framed}
\includegraphics[width=\textwidth]{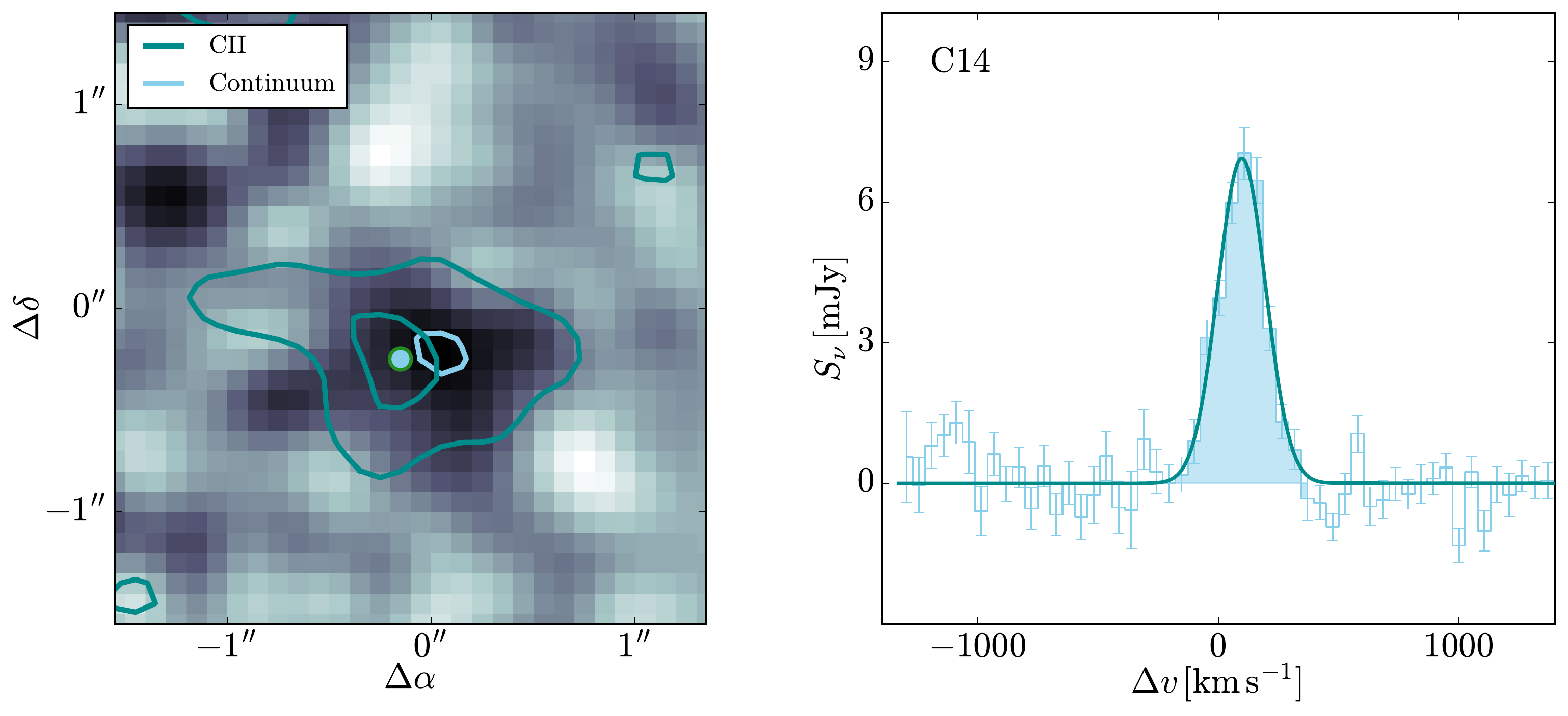}
\includegraphics[width=\textwidth]{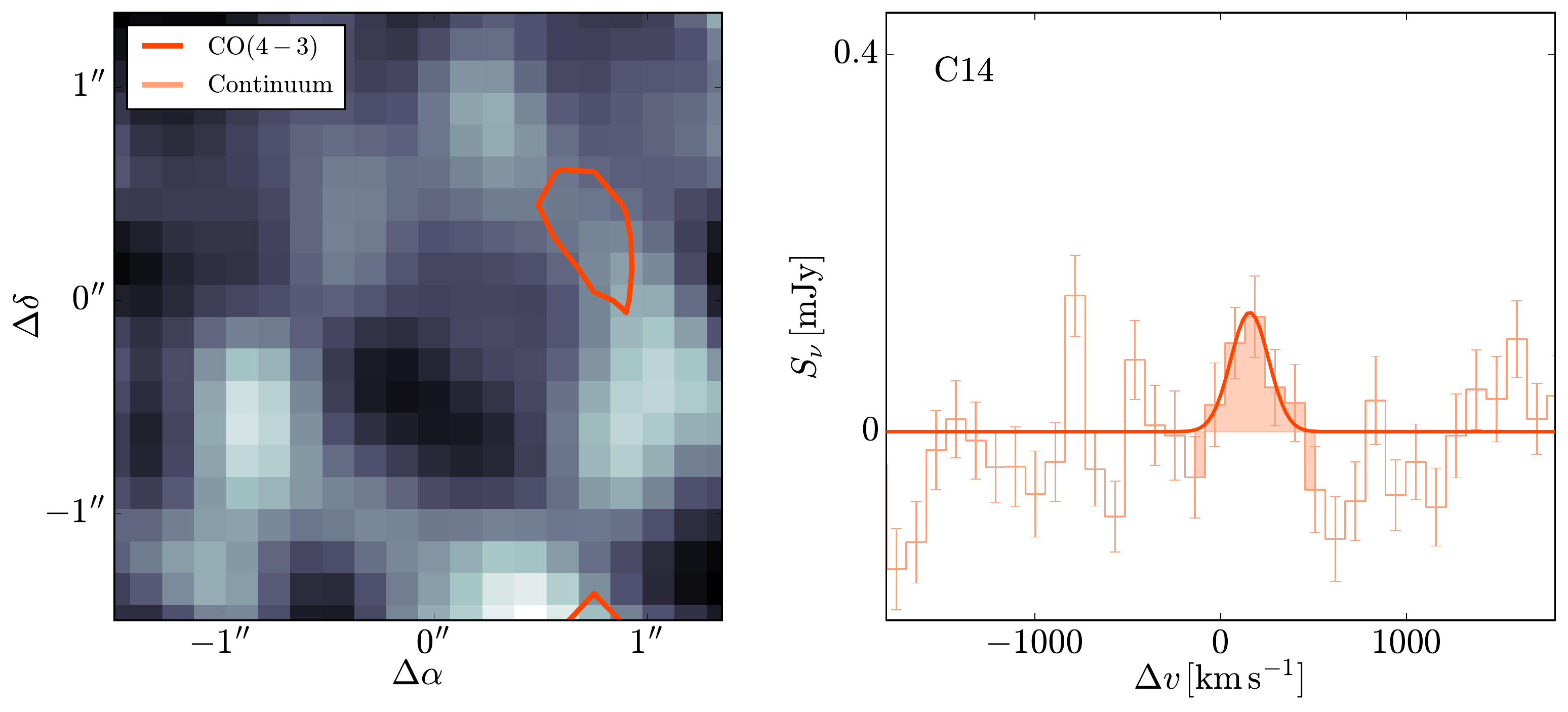}
\end{framed}
\end{subfigure}
\begin{subfigure}{.45\textwidth}
\begin{framed}
\includegraphics[width=\textwidth]{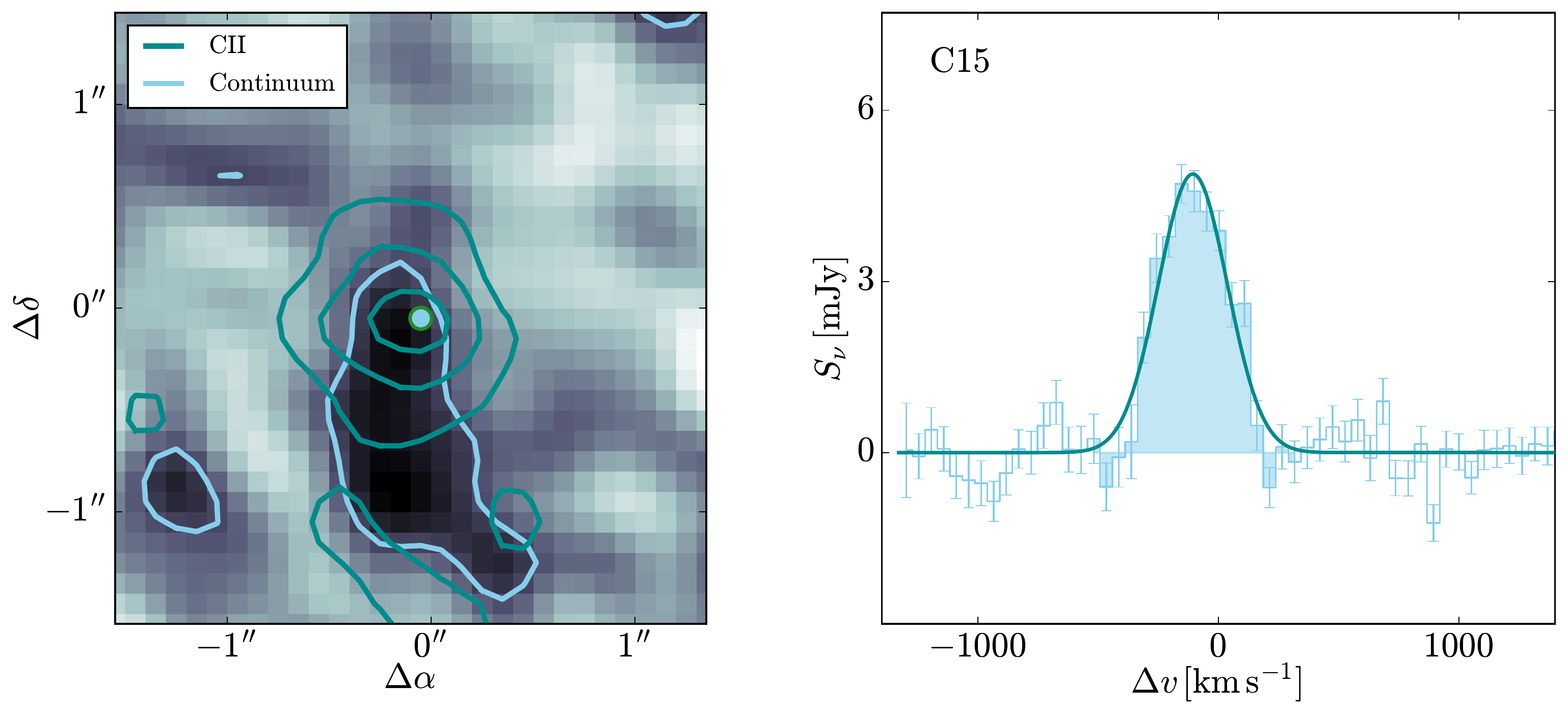}
\includegraphics[width=\textwidth]{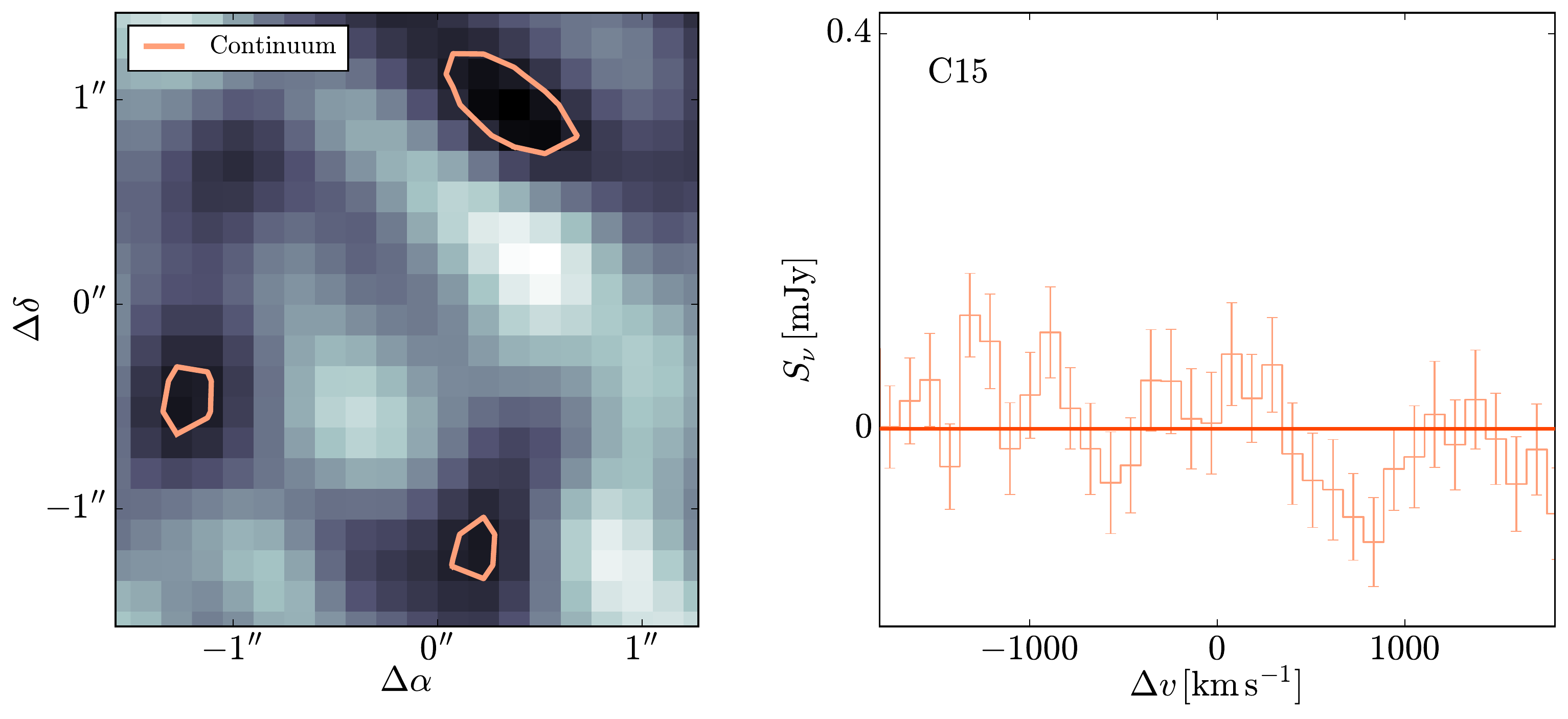}
\end{framed}
\end{subfigure}
\begin{subfigure}{.45\textwidth}
\begin{framed}
\includegraphics[width=\textwidth]{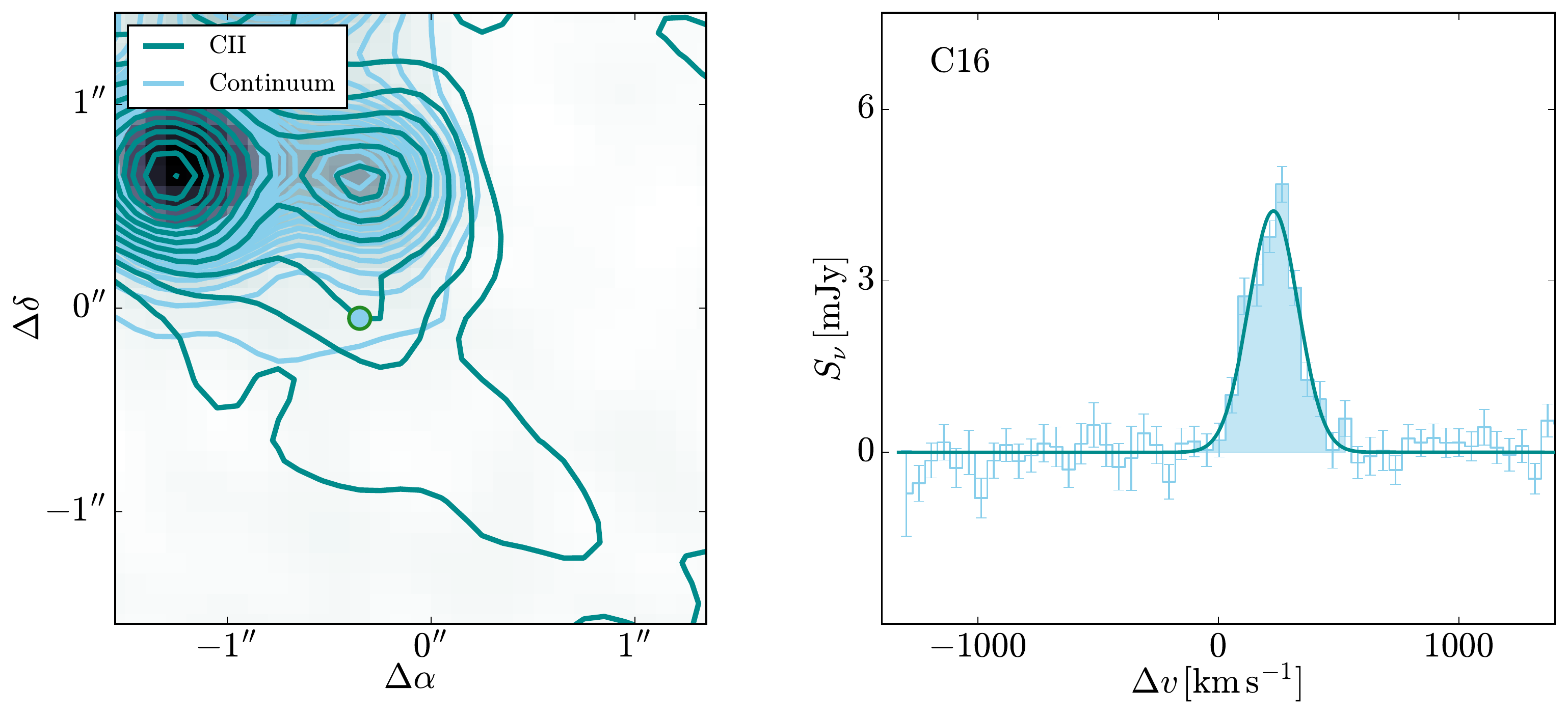}
\includegraphics[width=\textwidth]{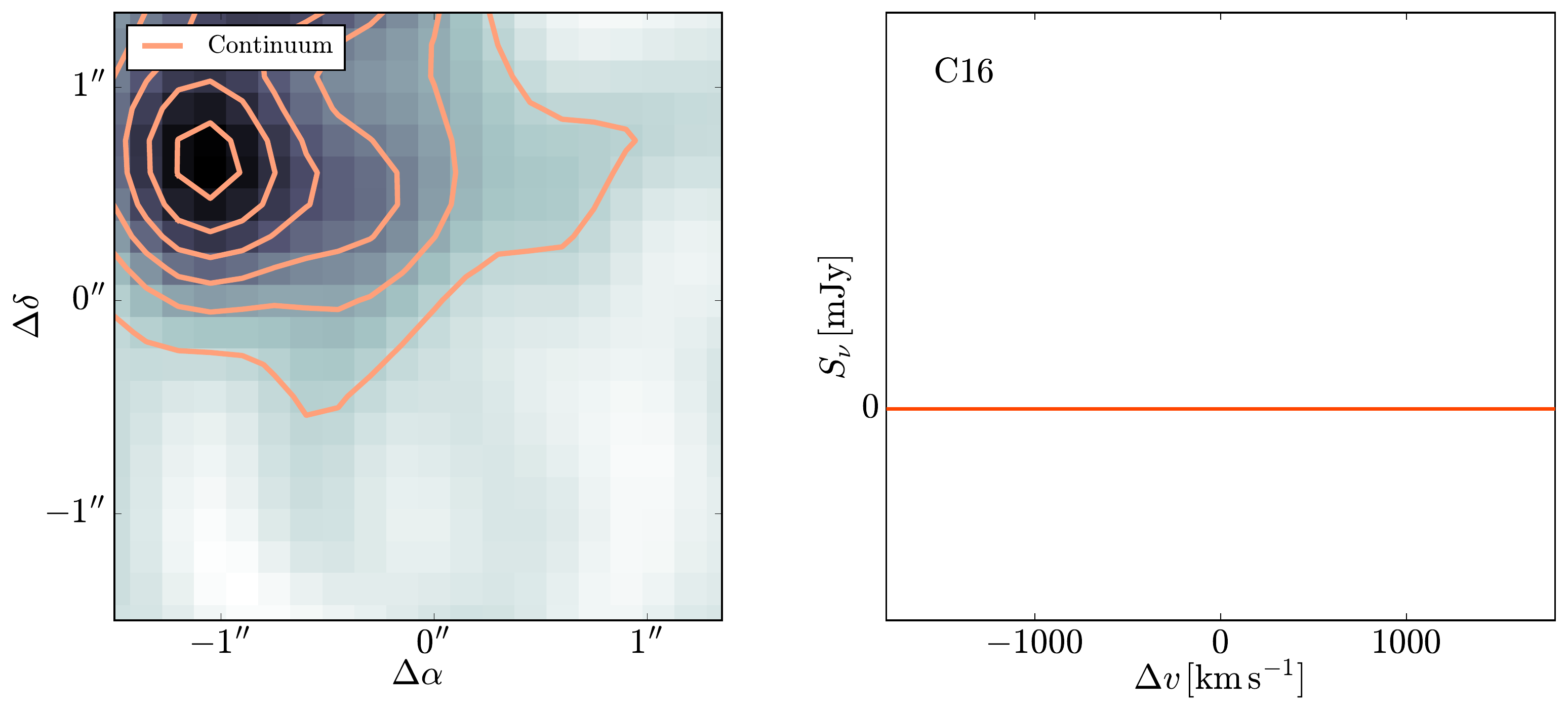}
\end{framed}
\end{subfigure}
\begin{subfigure}{.45\textwidth}
\begin{framed}
\includegraphics[width=\textwidth]{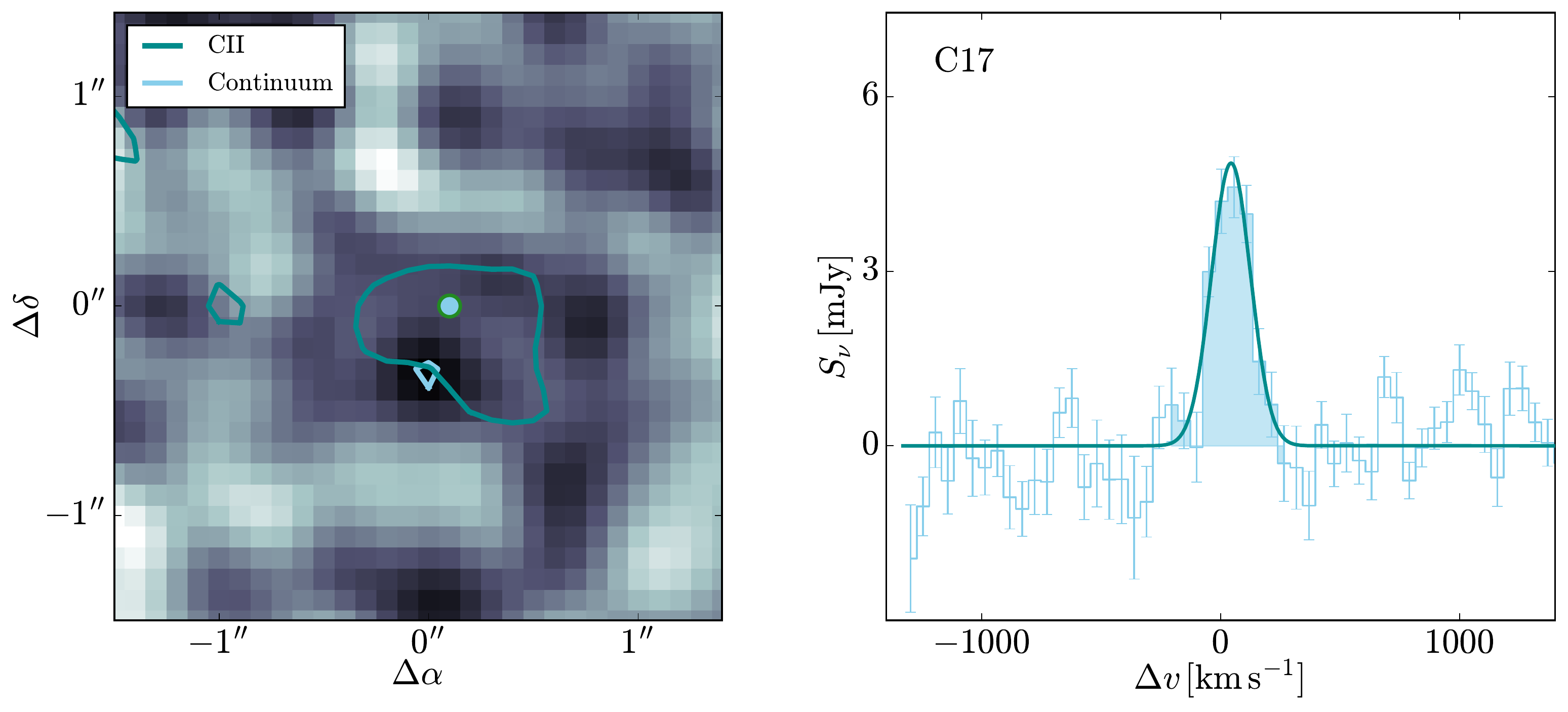}
\includegraphics[width=\textwidth]{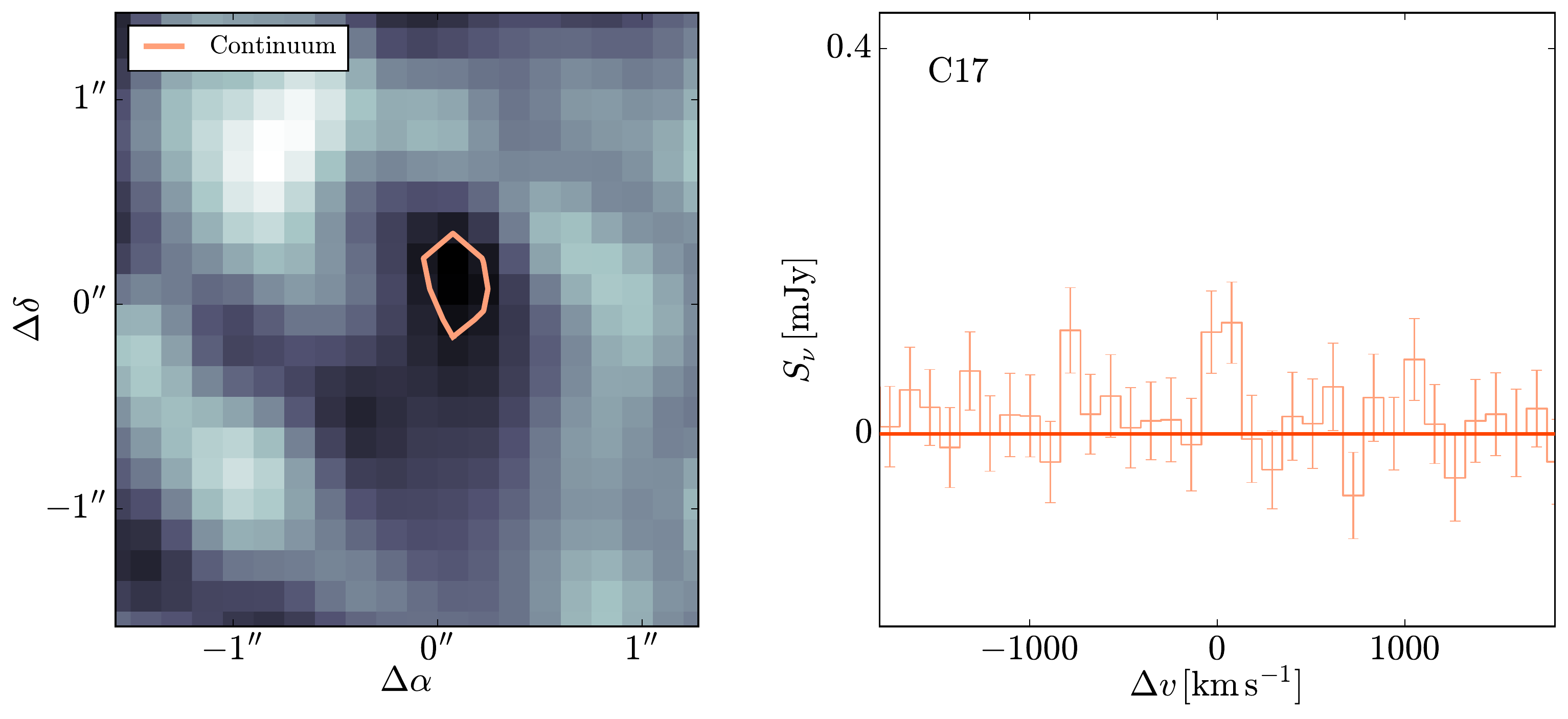}
\end{framed}
\end{subfigure}
\begin{subfigure}{.45\textwidth}
\begin{framed}
\includegraphics[width=\textwidth]{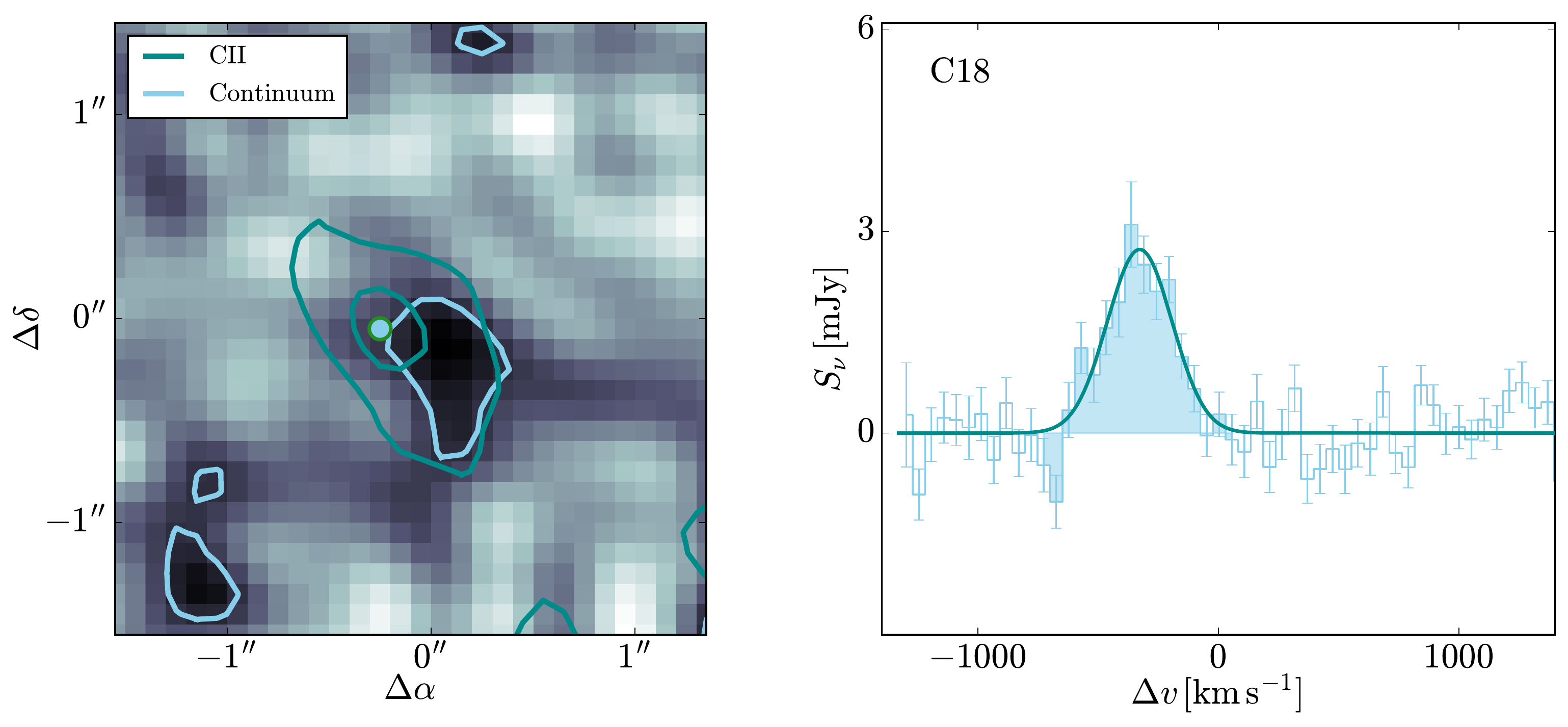}
\includegraphics[width=\textwidth]{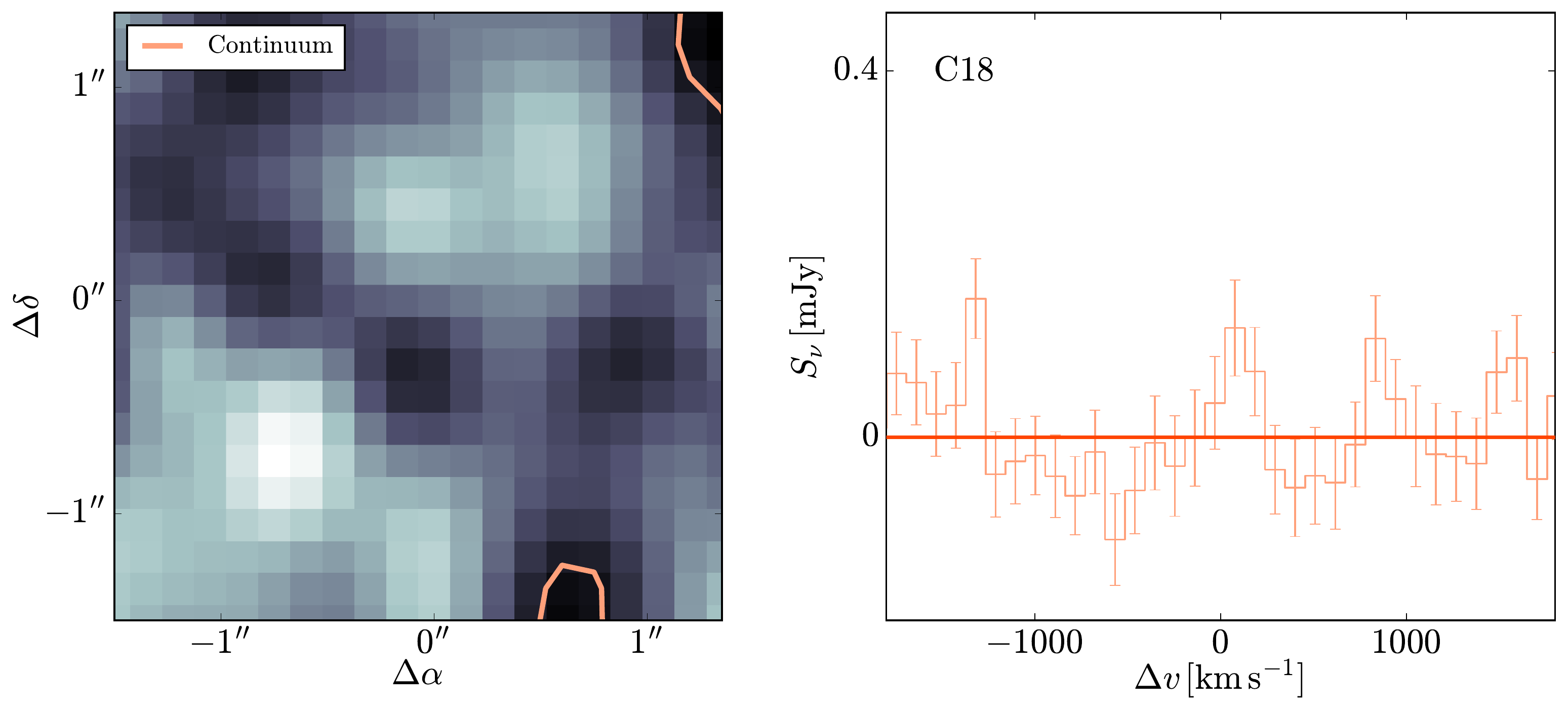}
\end{framed}
\end{subfigure}
\caption{}
\end{figure*}
\renewcommand{\thefigure}{\arabic{figure}}

\renewcommand{\thefigure}{A\arabic{figure} (Cont.)}
\addtocounter{figure}{-1}
\begin{figure*}
\begin{subfigure}{.45\textwidth}
\begin{framed}
\includegraphics[width=\textwidth]{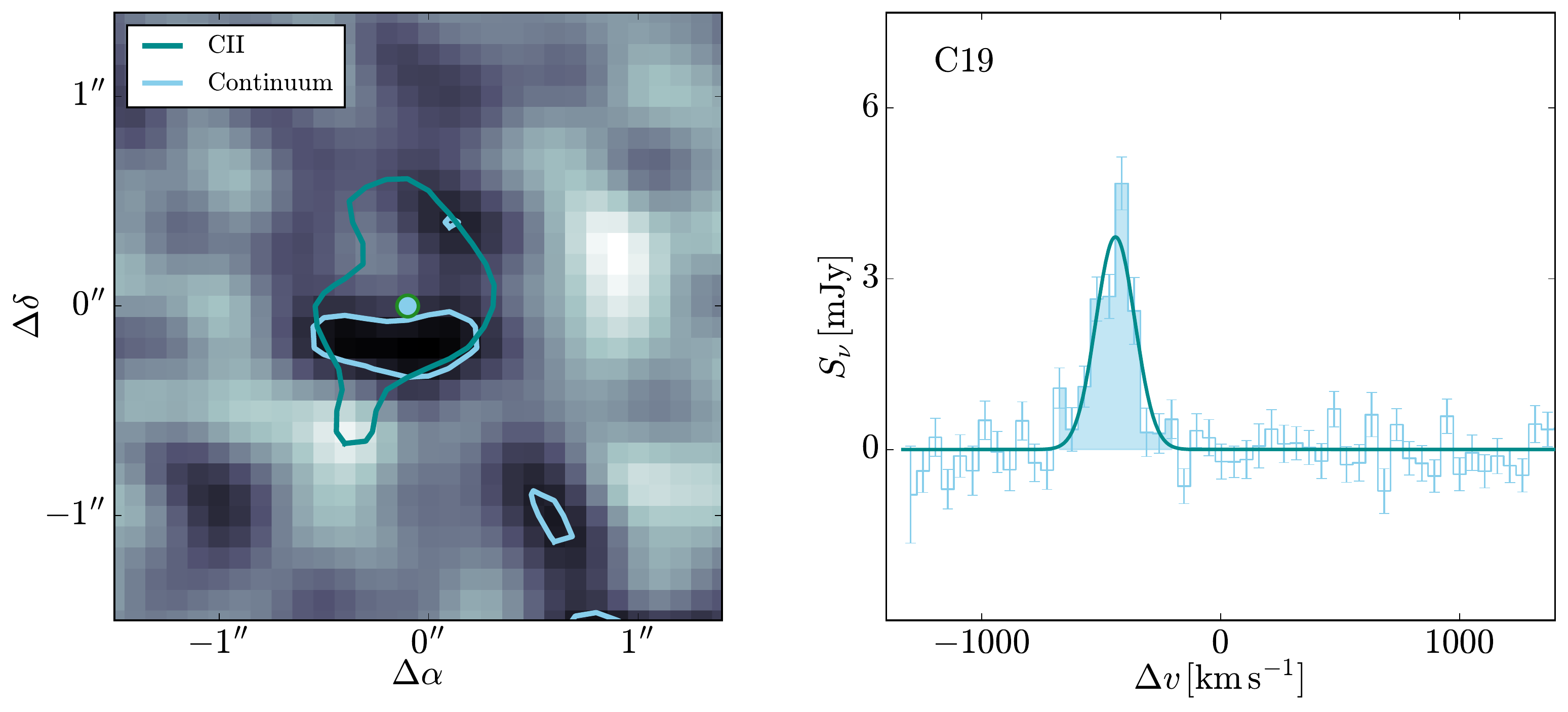}
\includegraphics[width=\textwidth]{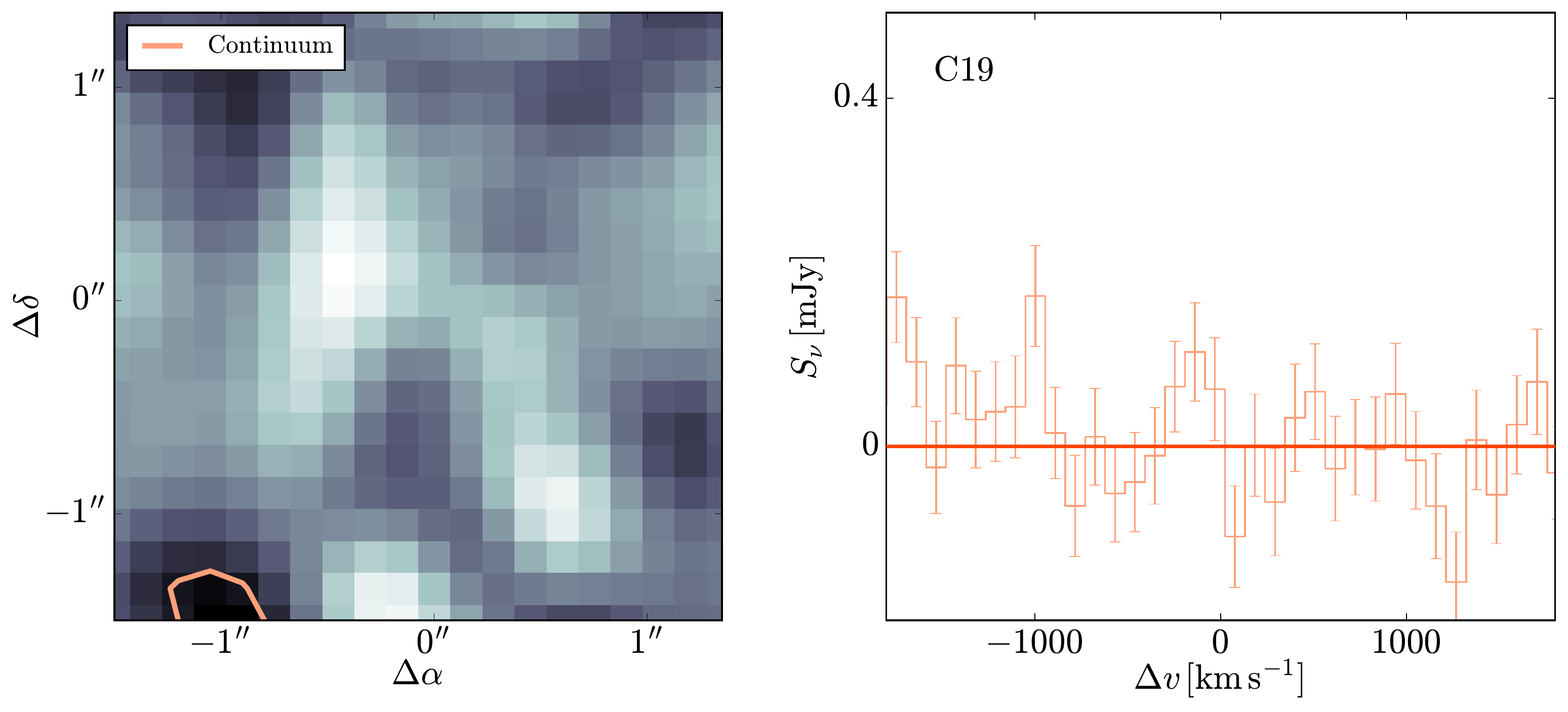}
\end{framed}
\end{subfigure}
\begin{subfigure}{.45\textwidth}
\begin{framed}
\includegraphics[width=\textwidth]{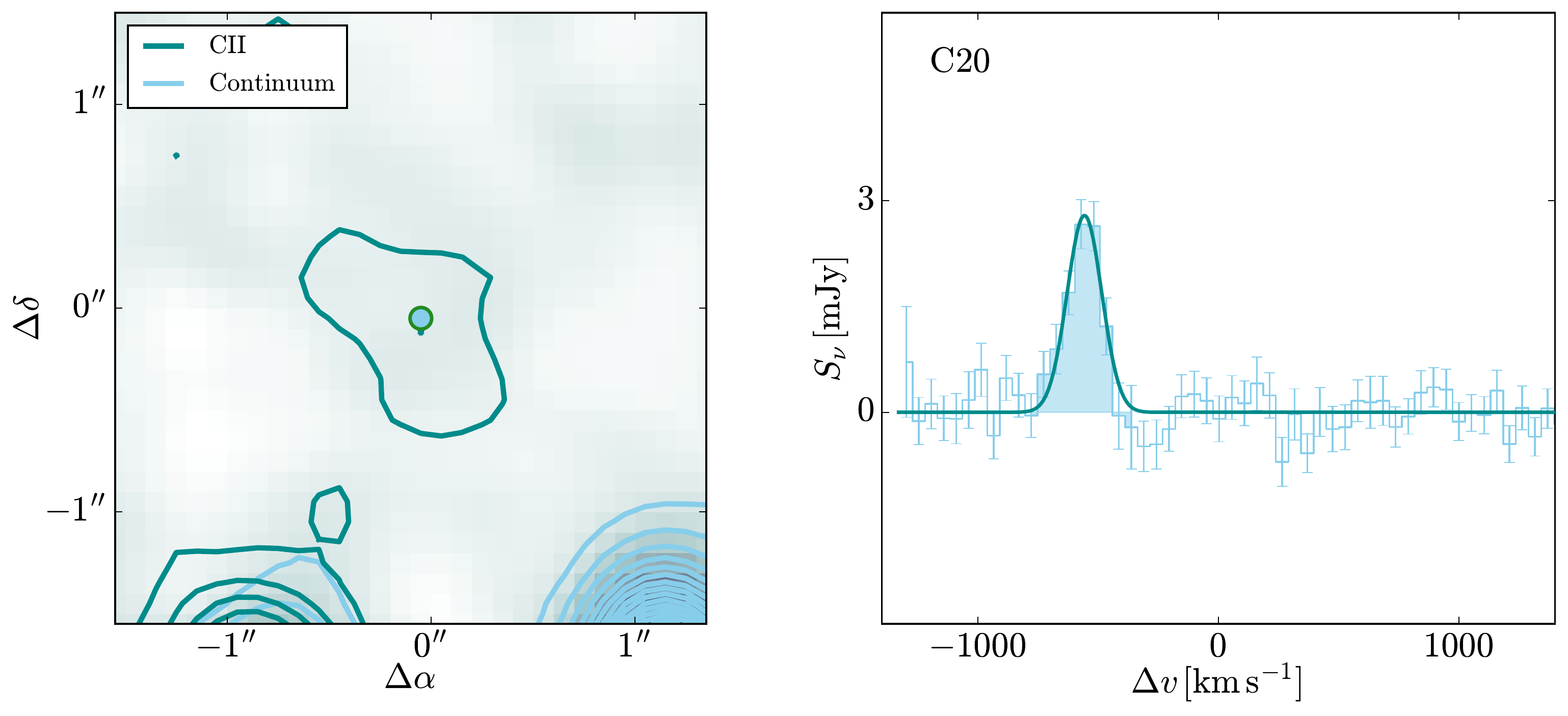}
\includegraphics[width=\textwidth]{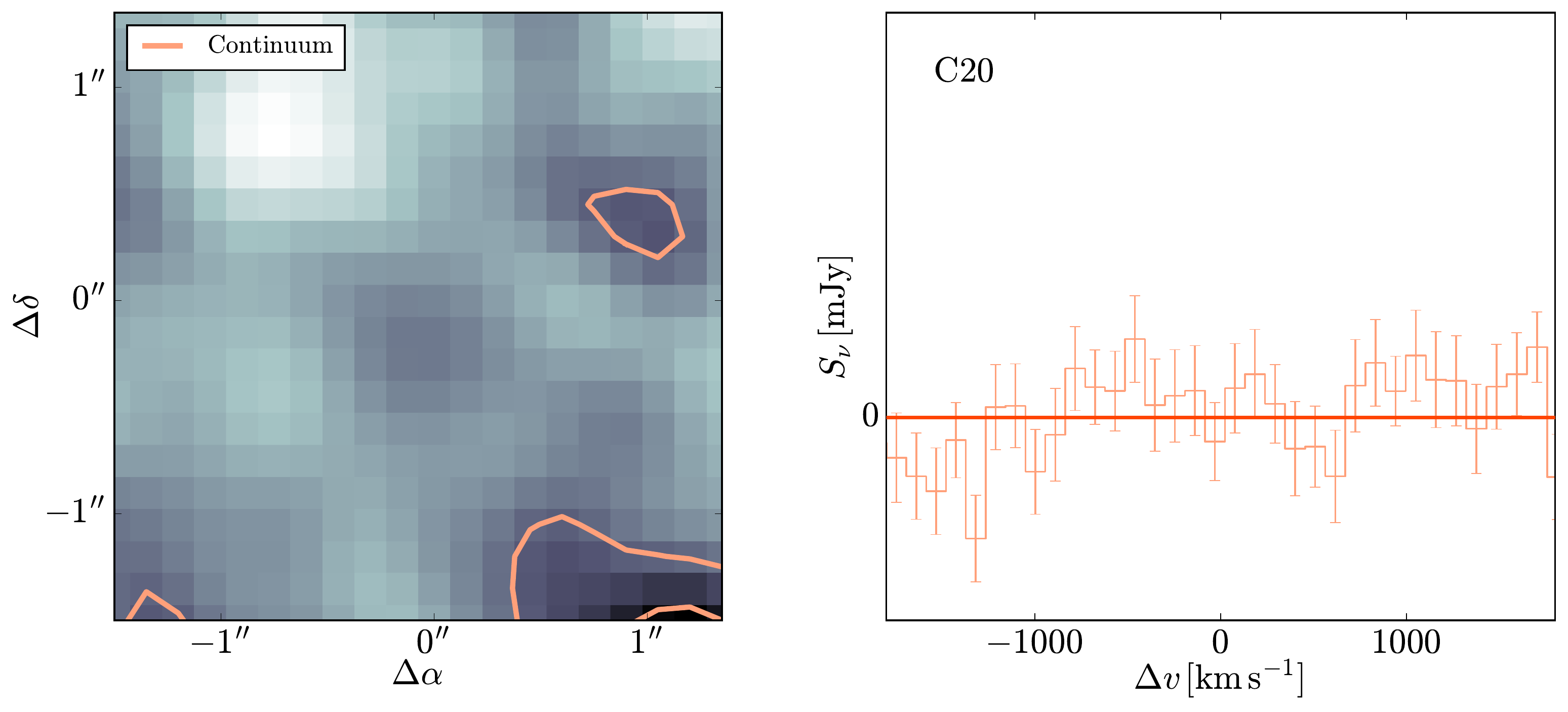}
\end{framed}
\end{subfigure}
\begin{subfigure}{.45\textwidth}
\begin{framed}
\includegraphics[width=\textwidth]{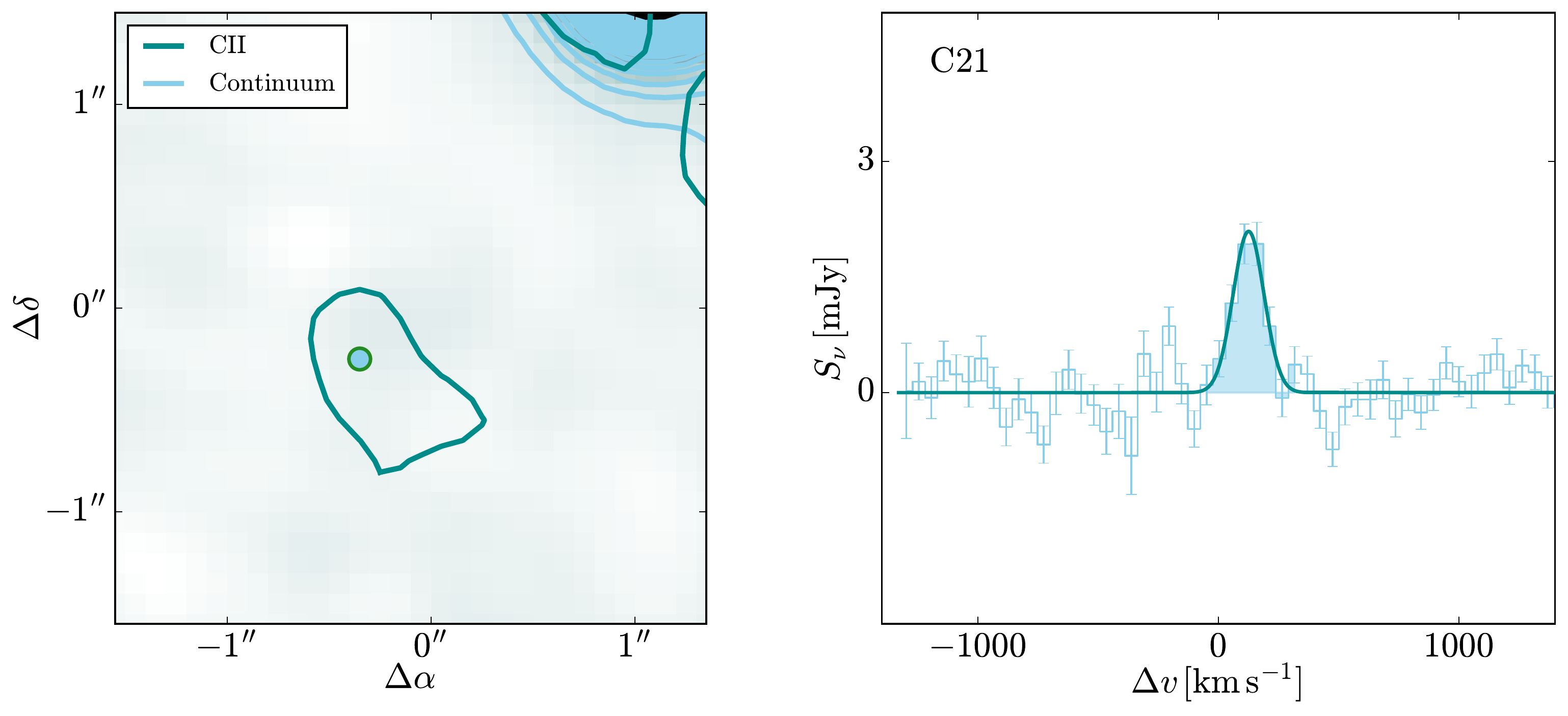}
\includegraphics[width=\textwidth]{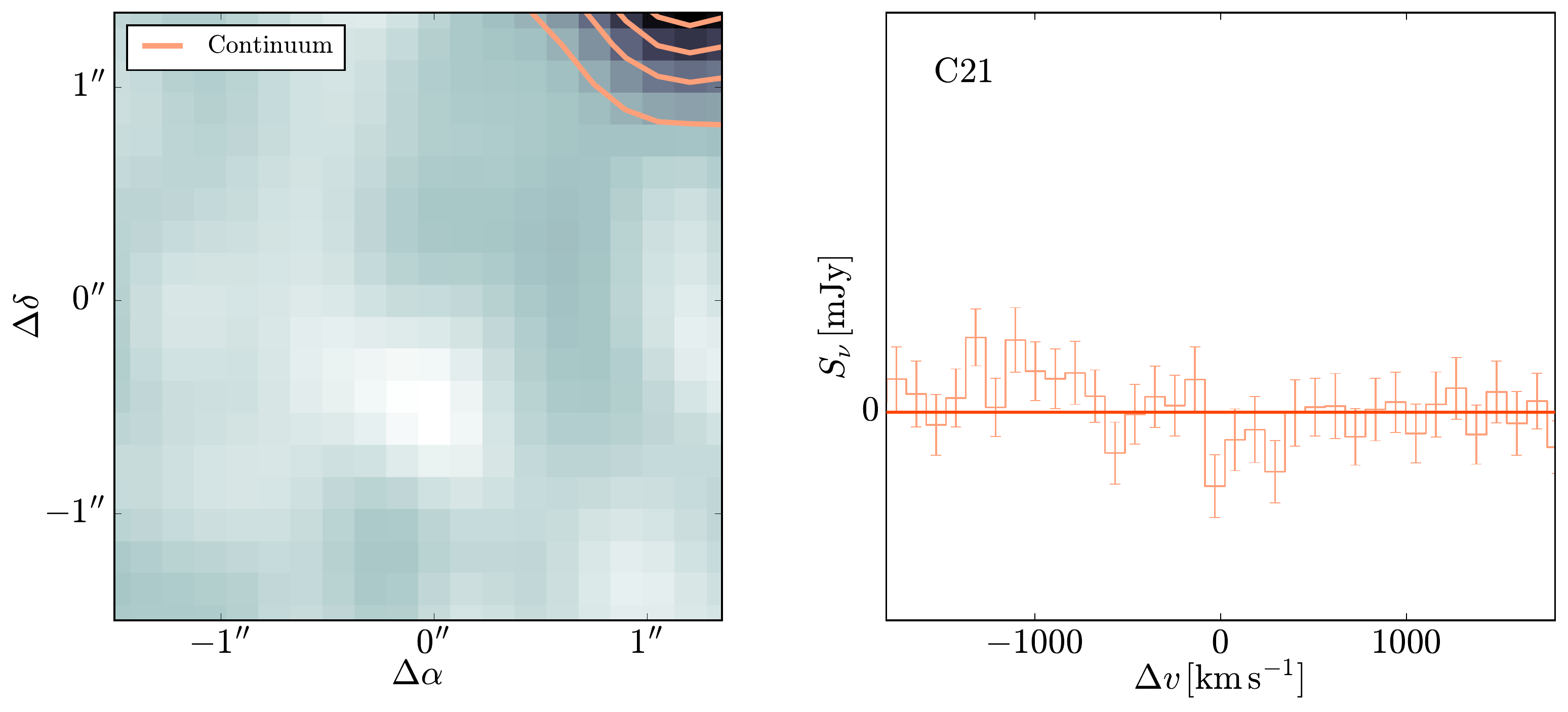}
\end{framed}
\end{subfigure}
\begin{subfigure}{.45\textwidth}
\begin{framed}
\includegraphics[width=\textwidth]{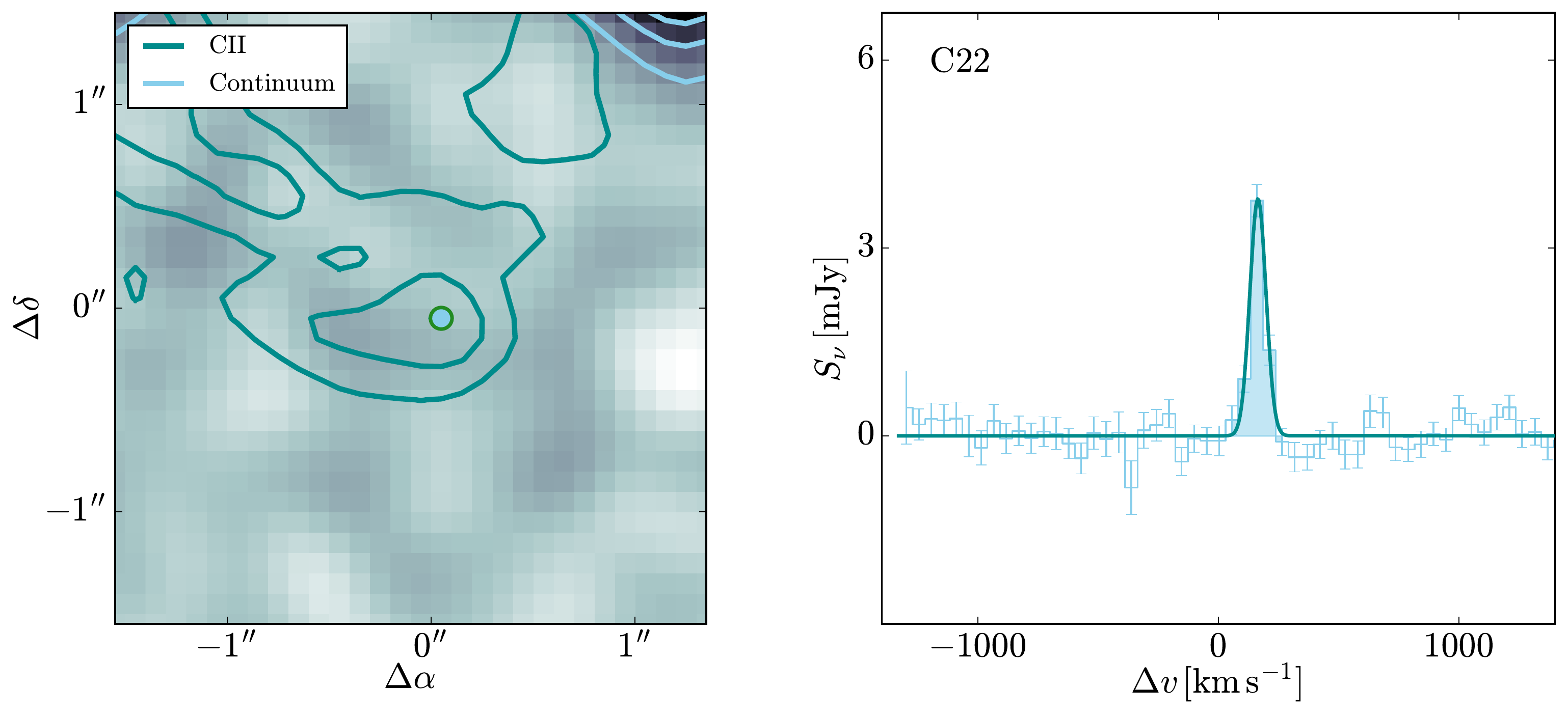}
\includegraphics[width=\textwidth]{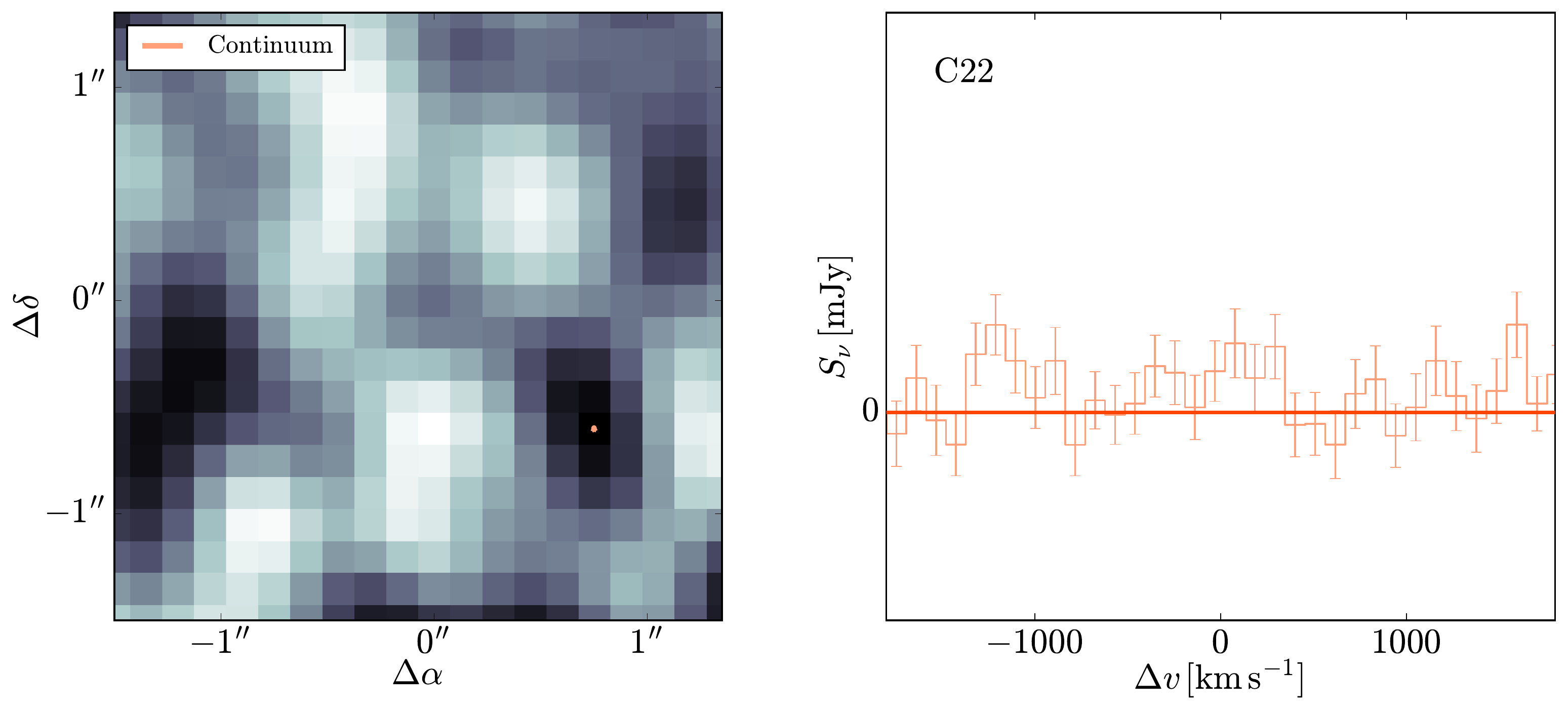}
\end{framed}
\end{subfigure}
\begin{subfigure}{.45\textwidth}
\begin{framed}
\includegraphics[width=\textwidth]{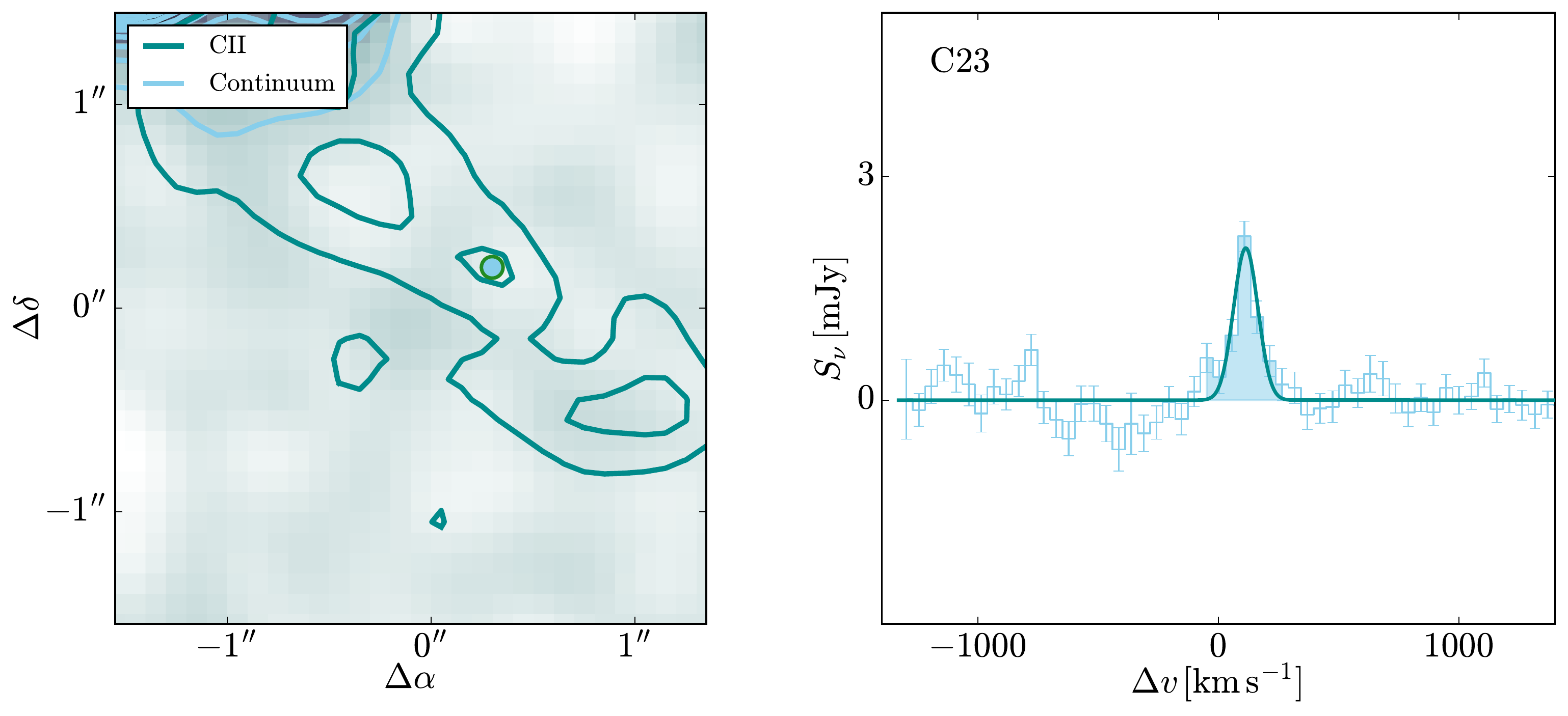}
\includegraphics[width=\textwidth]{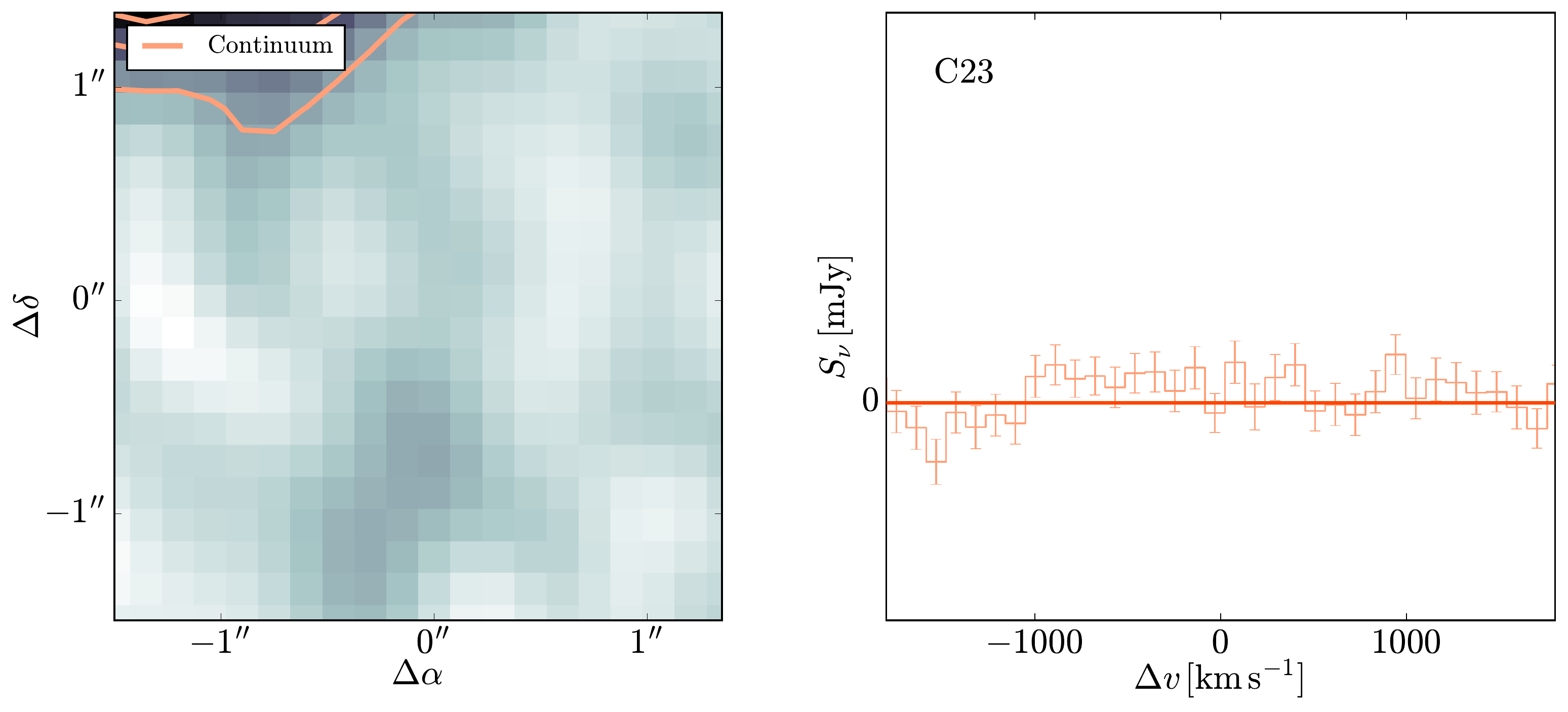}
\end{framed}
\end{subfigure}
\begin{subfigure}{.45\textwidth}
\begin{framed}
\includegraphics[width=\textwidth]{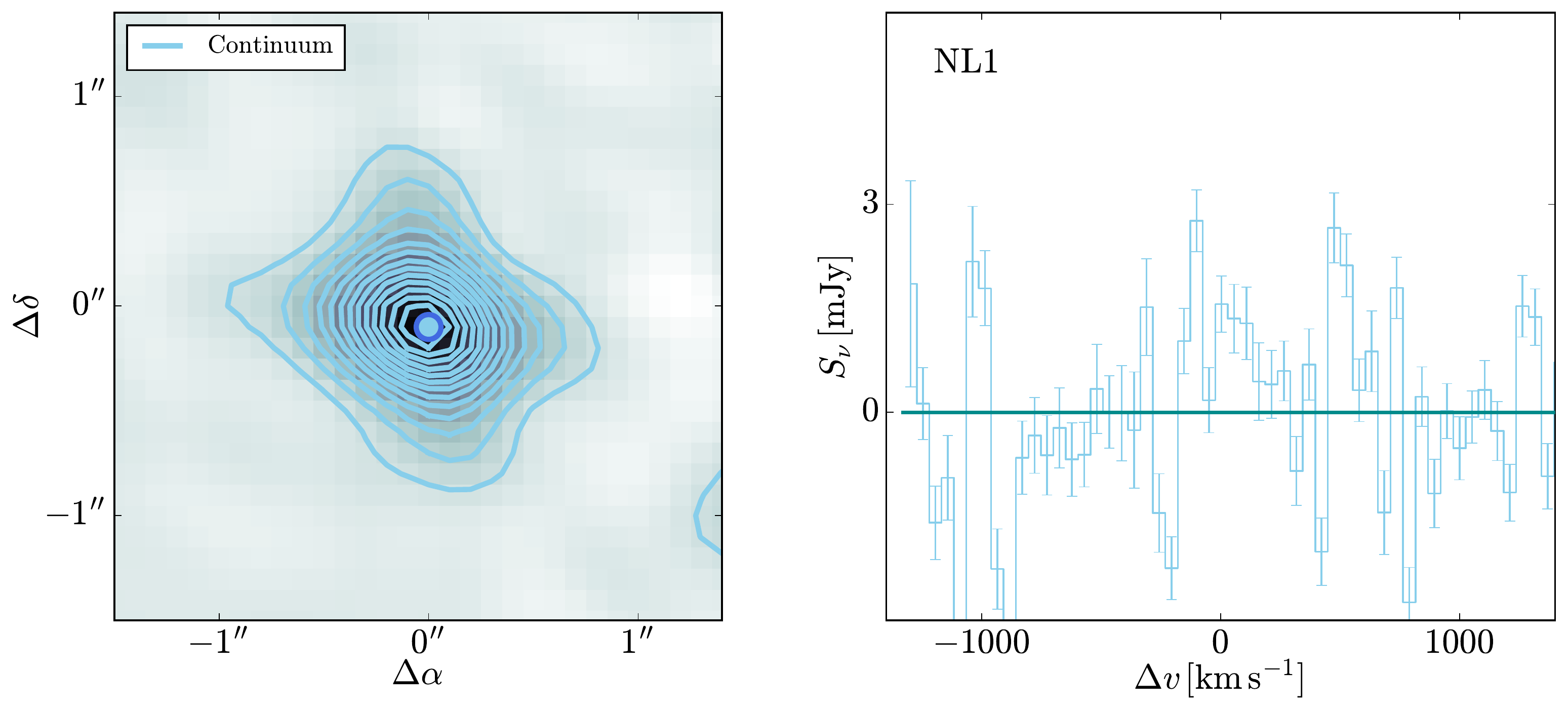}
\includegraphics[width=\textwidth]{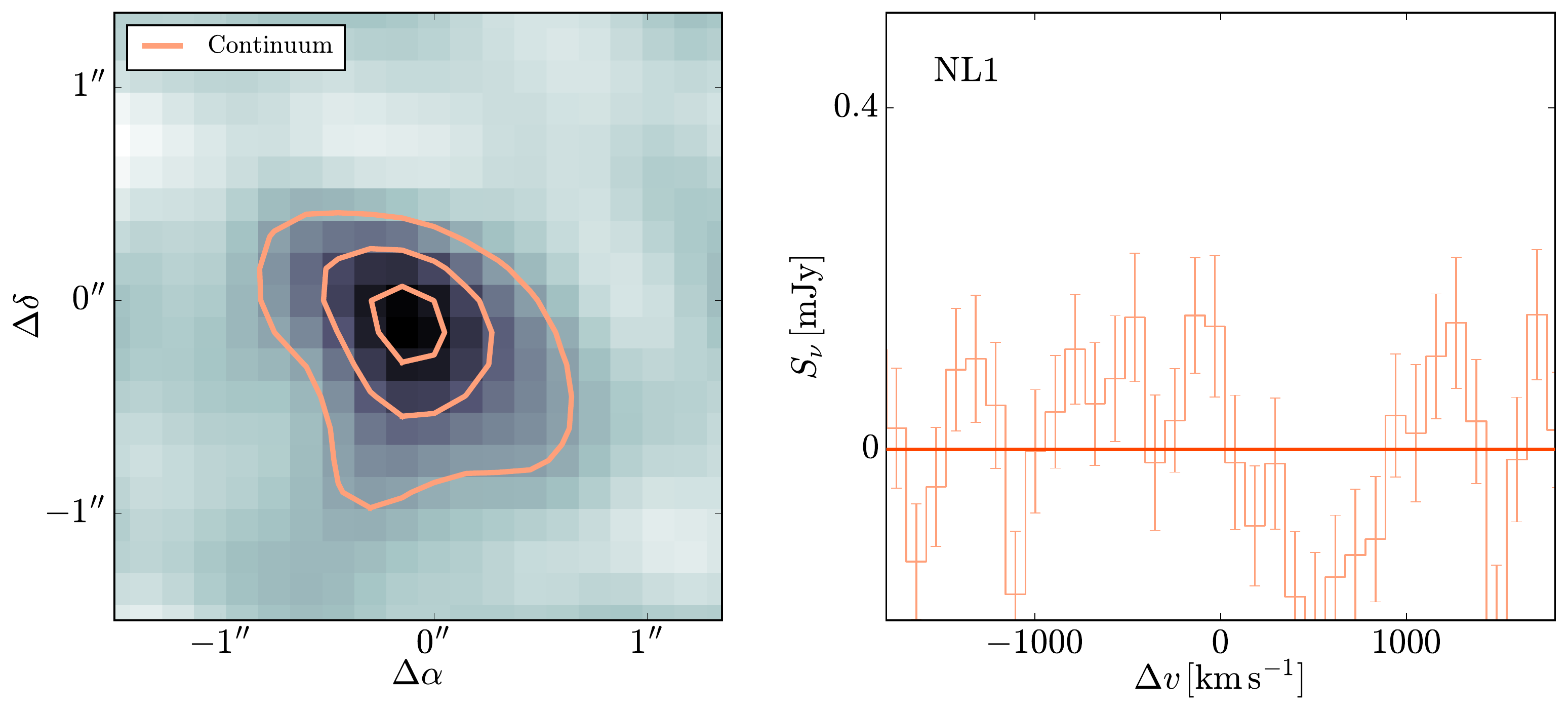}
\end{framed}
\end{subfigure}
\caption{}
\end{figure*}
\renewcommand{\thefigure}{\arabic{figure}}

\renewcommand{\thefigure}{A\arabic{figure} (Cont.)}
\addtocounter{figure}{-1}
\begin{figure*}
\begin{subfigure}{.45\textwidth}
\begin{framed}
\includegraphics[width=\textwidth]{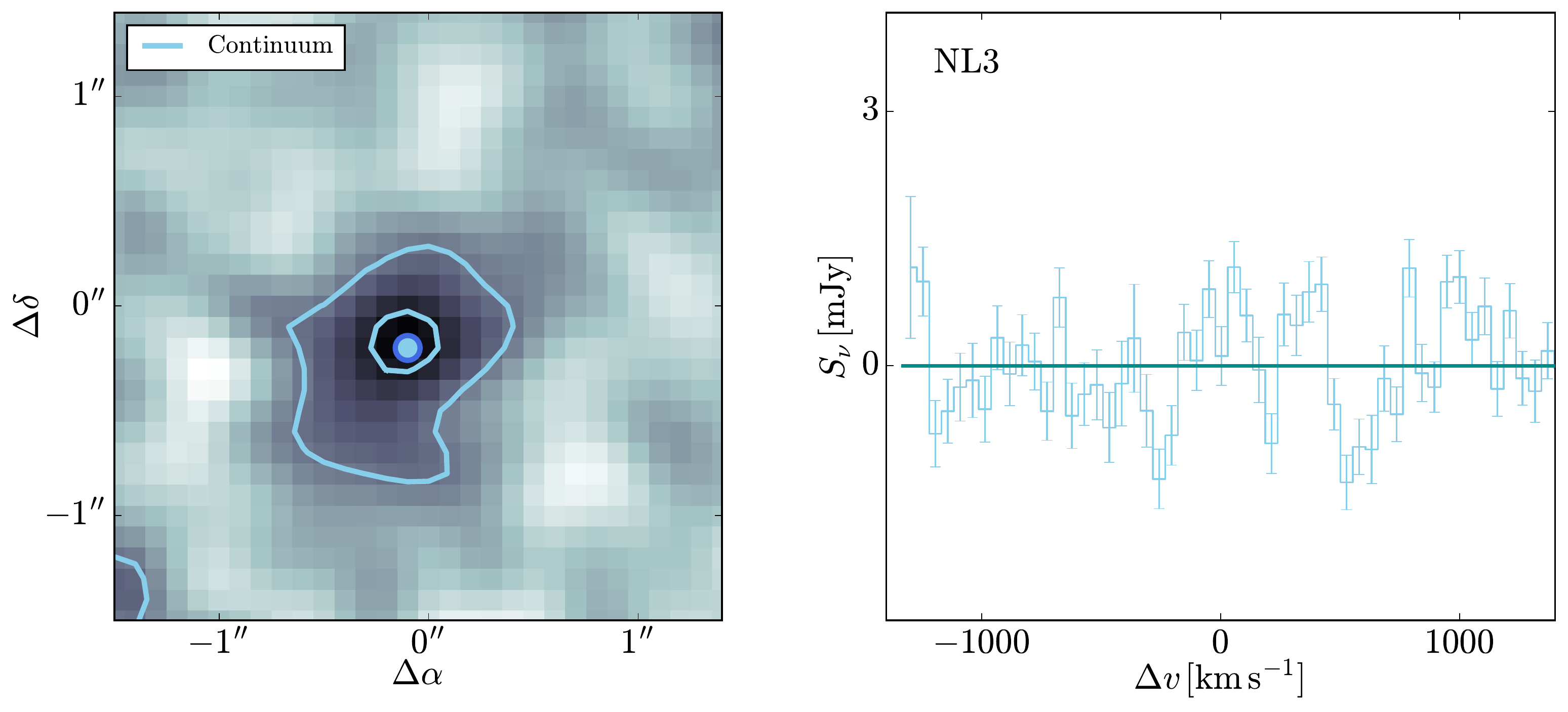}
\includegraphics[width=\textwidth]{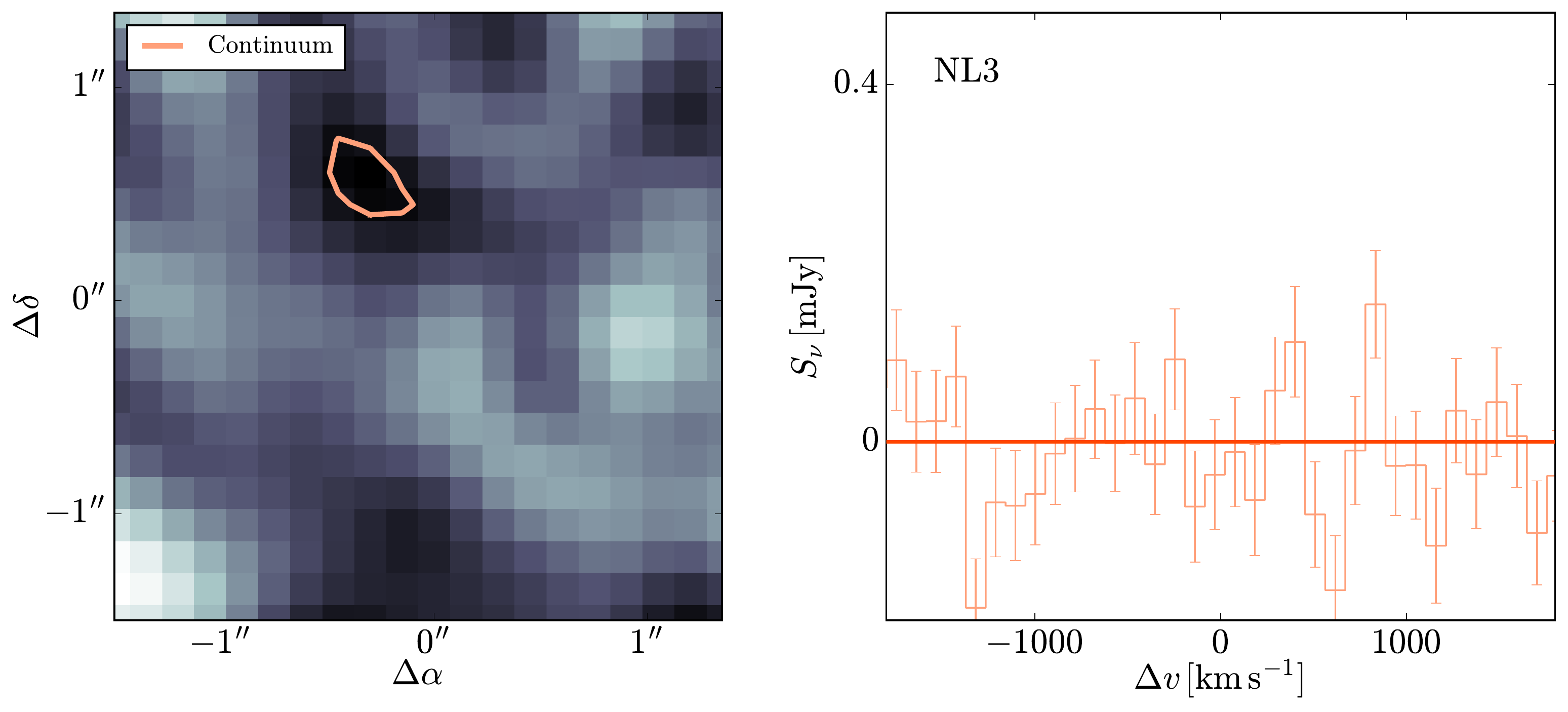}
\end{framed}
\end{subfigure}
\begin{subfigure}{.45\textwidth}
\begin{framed}
\includegraphics[width=\textwidth]{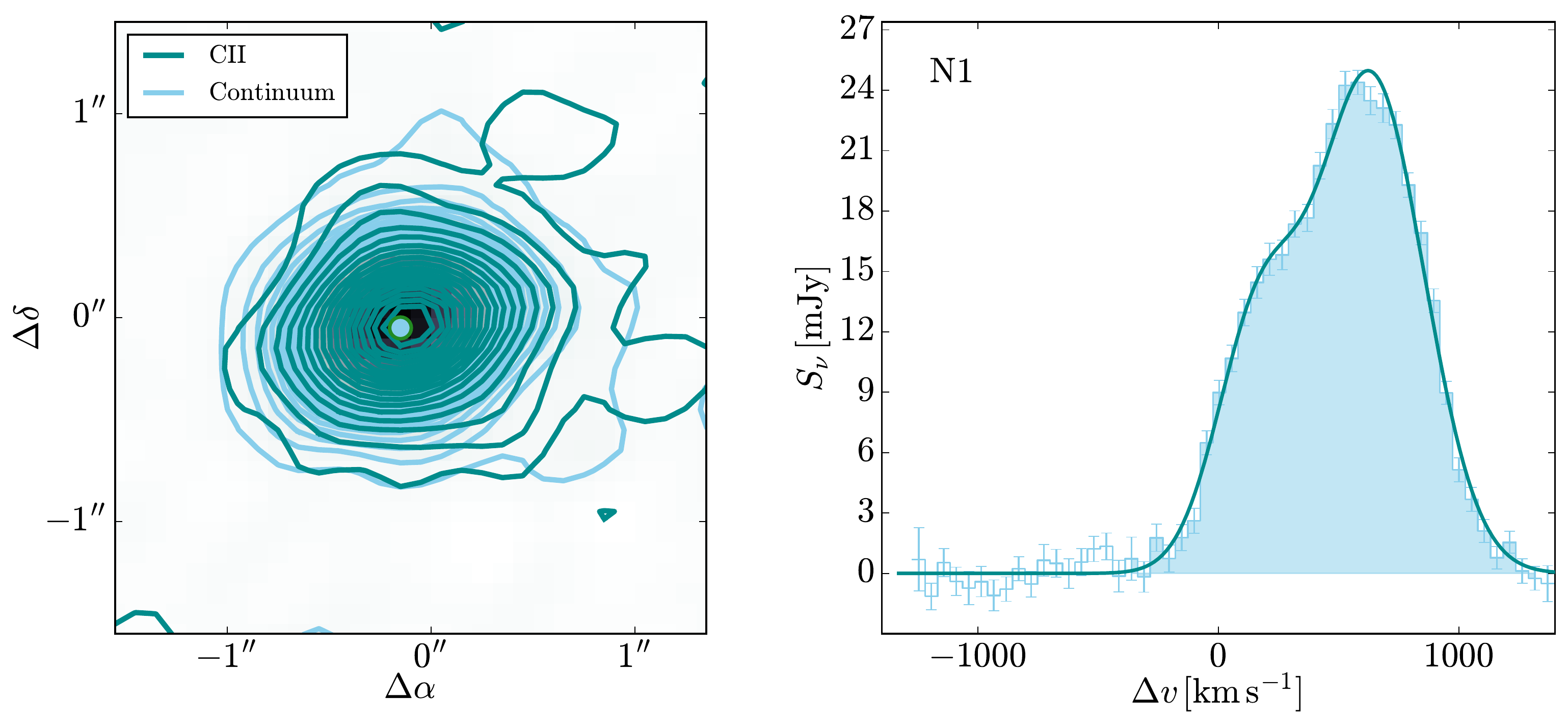}
\includegraphics[width=\textwidth]{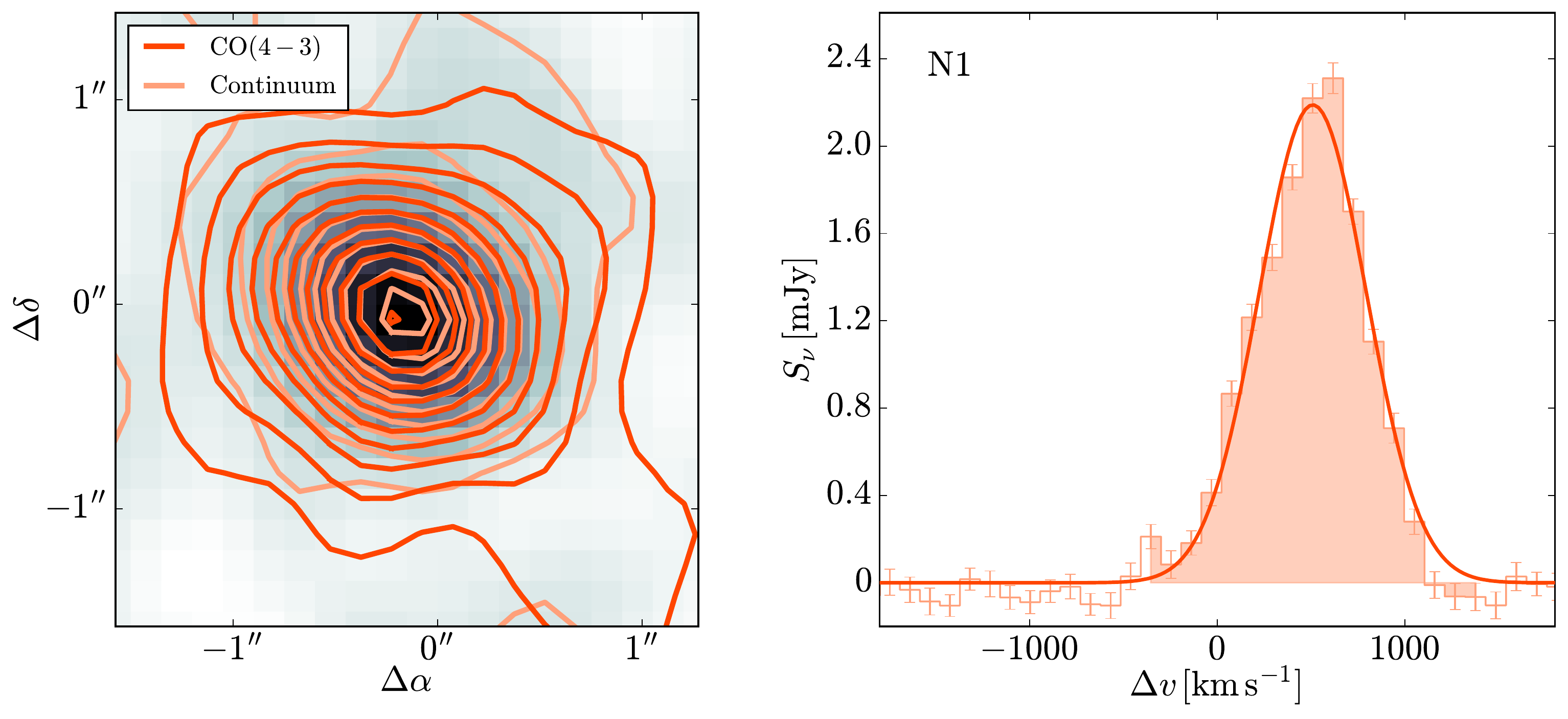}
\end{framed}
\end{subfigure}
\begin{subfigure}{.45\textwidth}
\begin{framed}
\includegraphics[width=\textwidth]{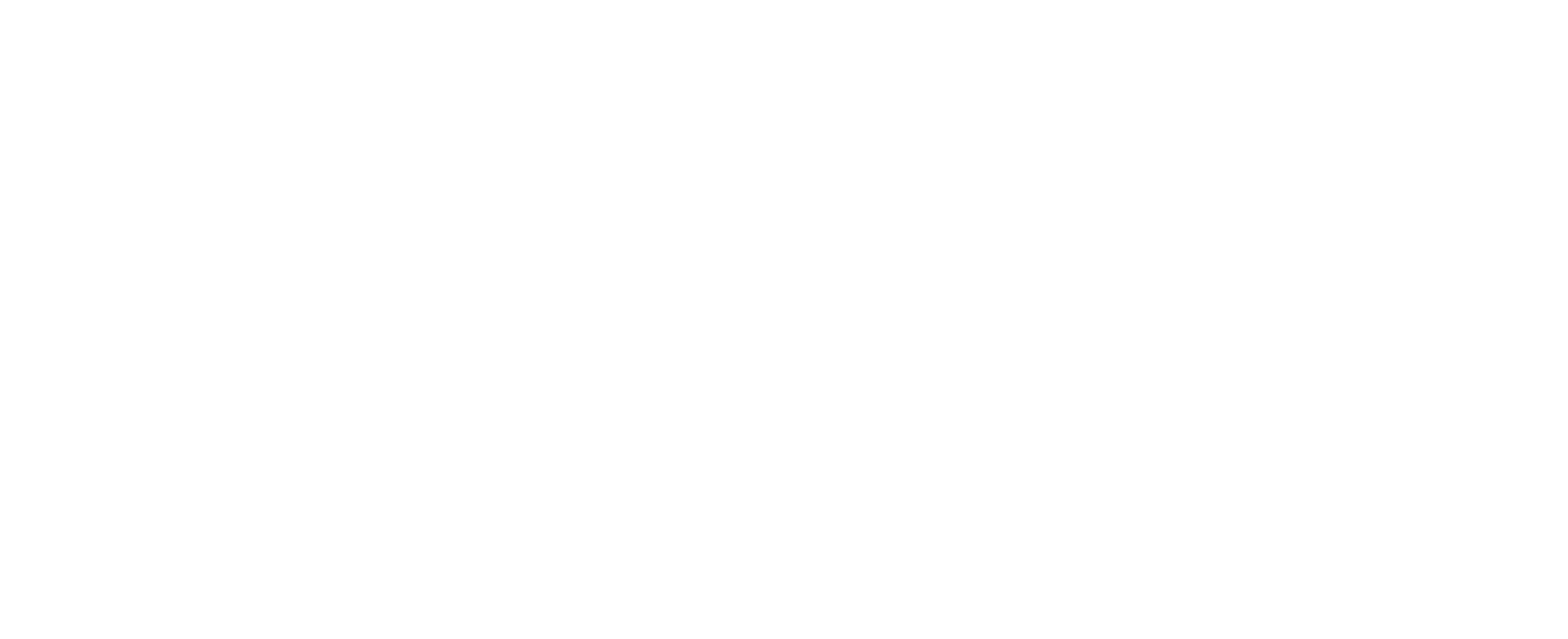}
\includegraphics[width=\textwidth]{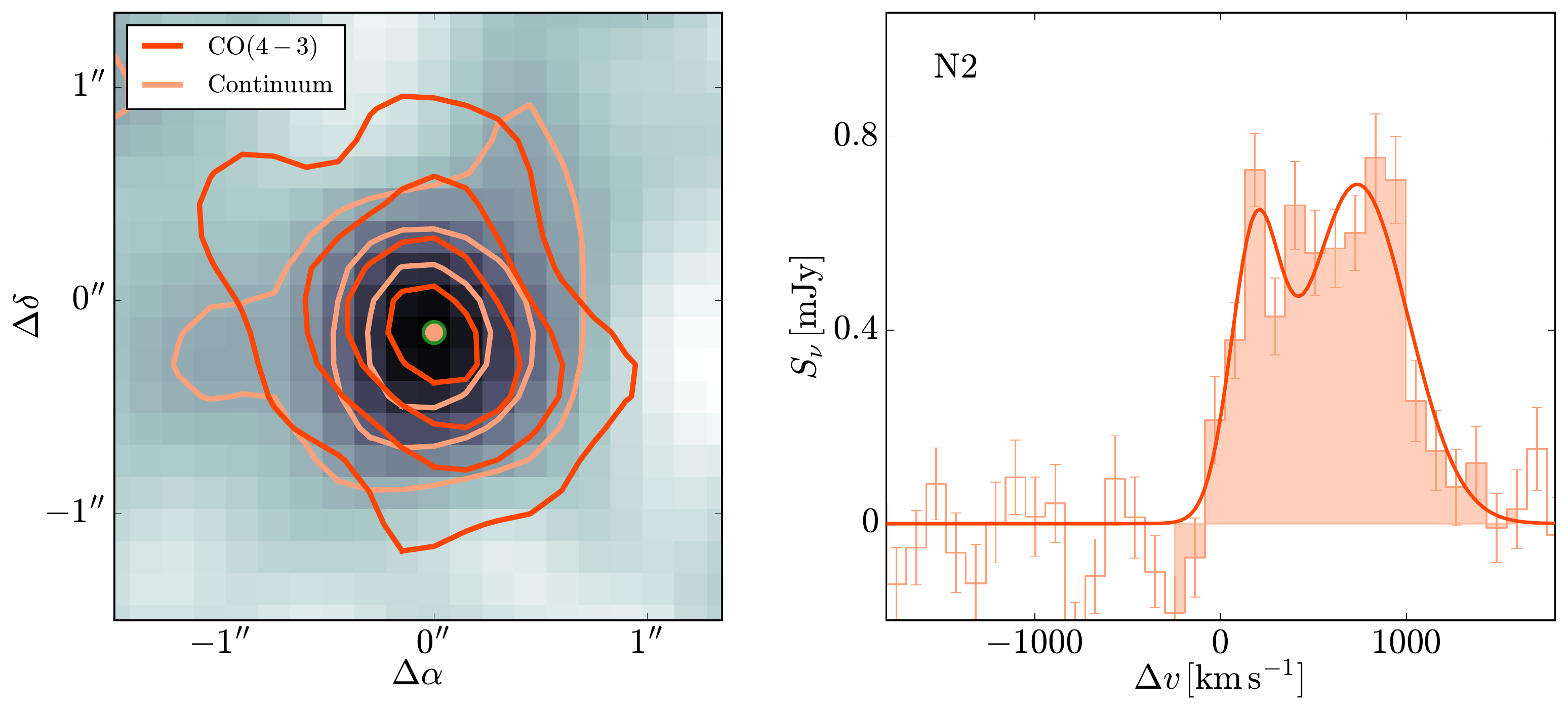}
\end{framed}
\end{subfigure}
\begin{subfigure}{.45\textwidth}
\begin{framed}
\includegraphics[width=\textwidth]{CII/blank.pdf}
\includegraphics[width=\textwidth]{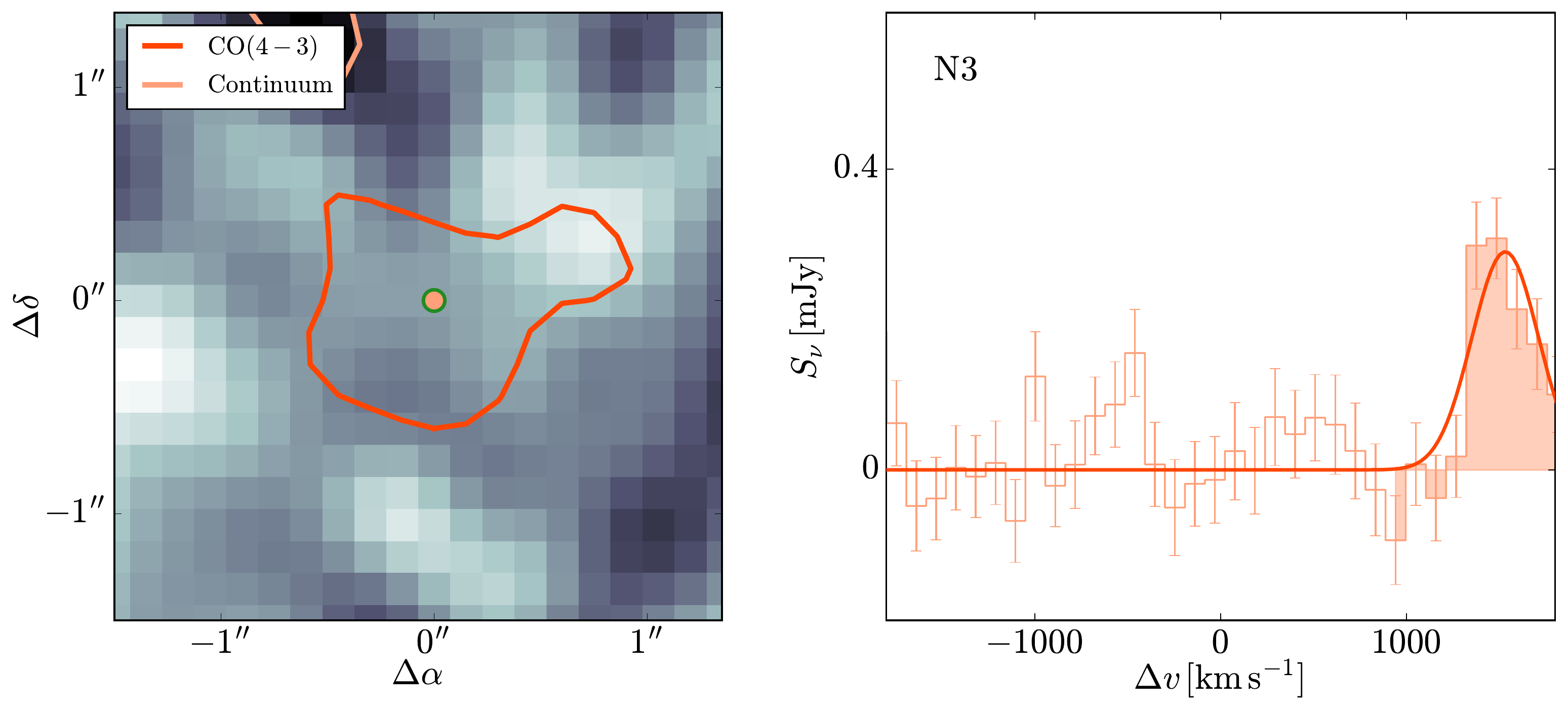}
\end{framed}
\end{subfigure}
\begin{subfigure}{.45\textwidth}
\begin{framed}
\includegraphics[width=\textwidth]{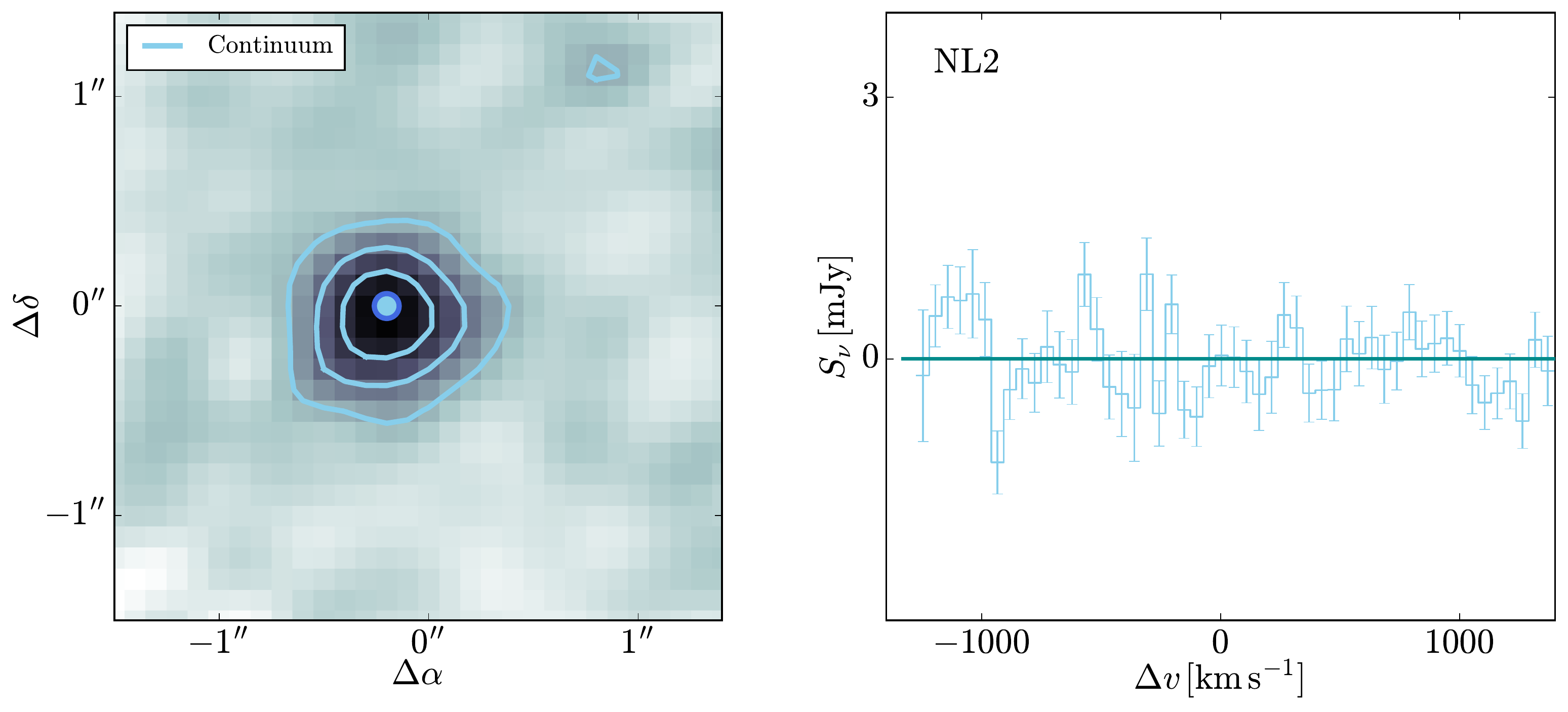}
\includegraphics[width=\textwidth]{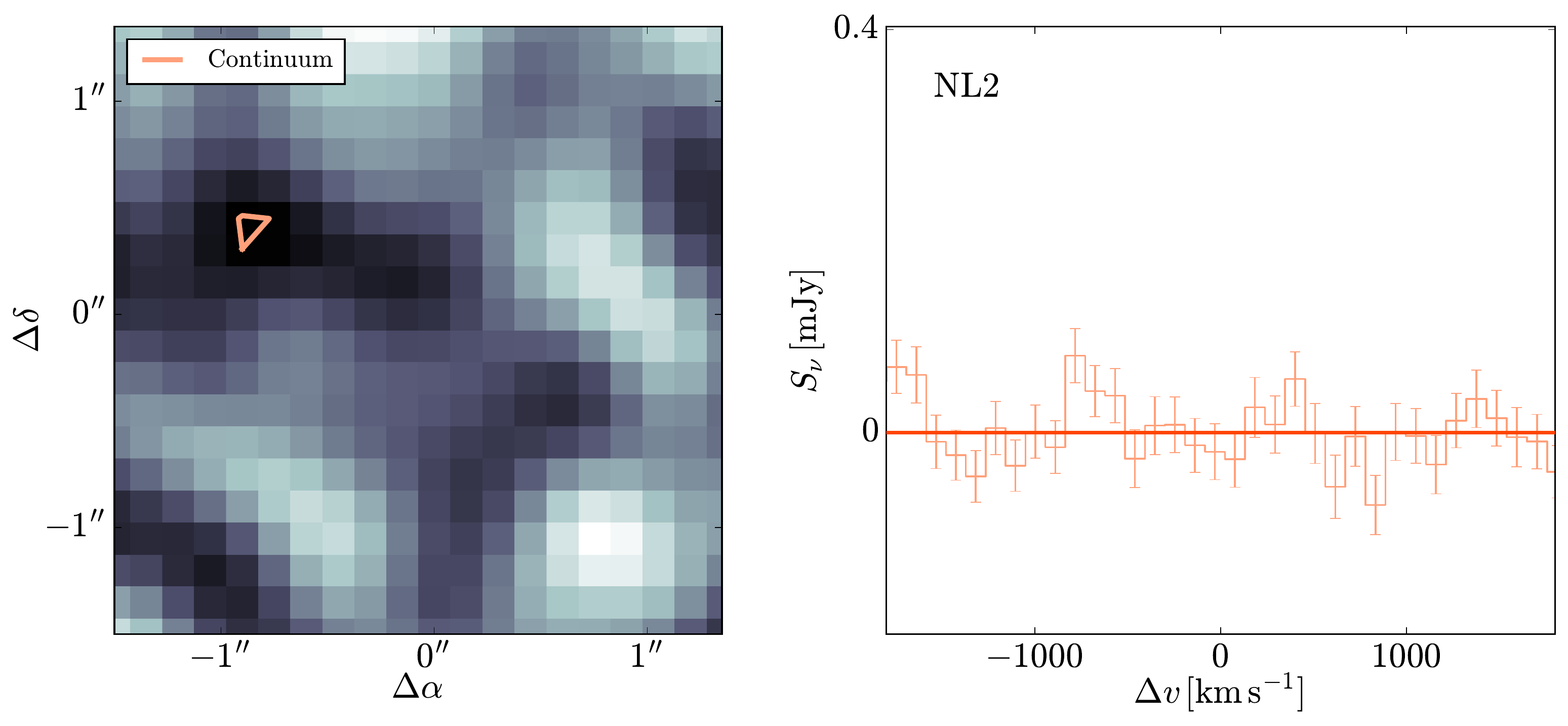}
\end{framed}
\end{subfigure}
\begin{subfigure}{.45\textwidth}
\begin{framed}
\includegraphics[width=\textwidth]{CII/blank.pdf}
\includegraphics[width=\textwidth]{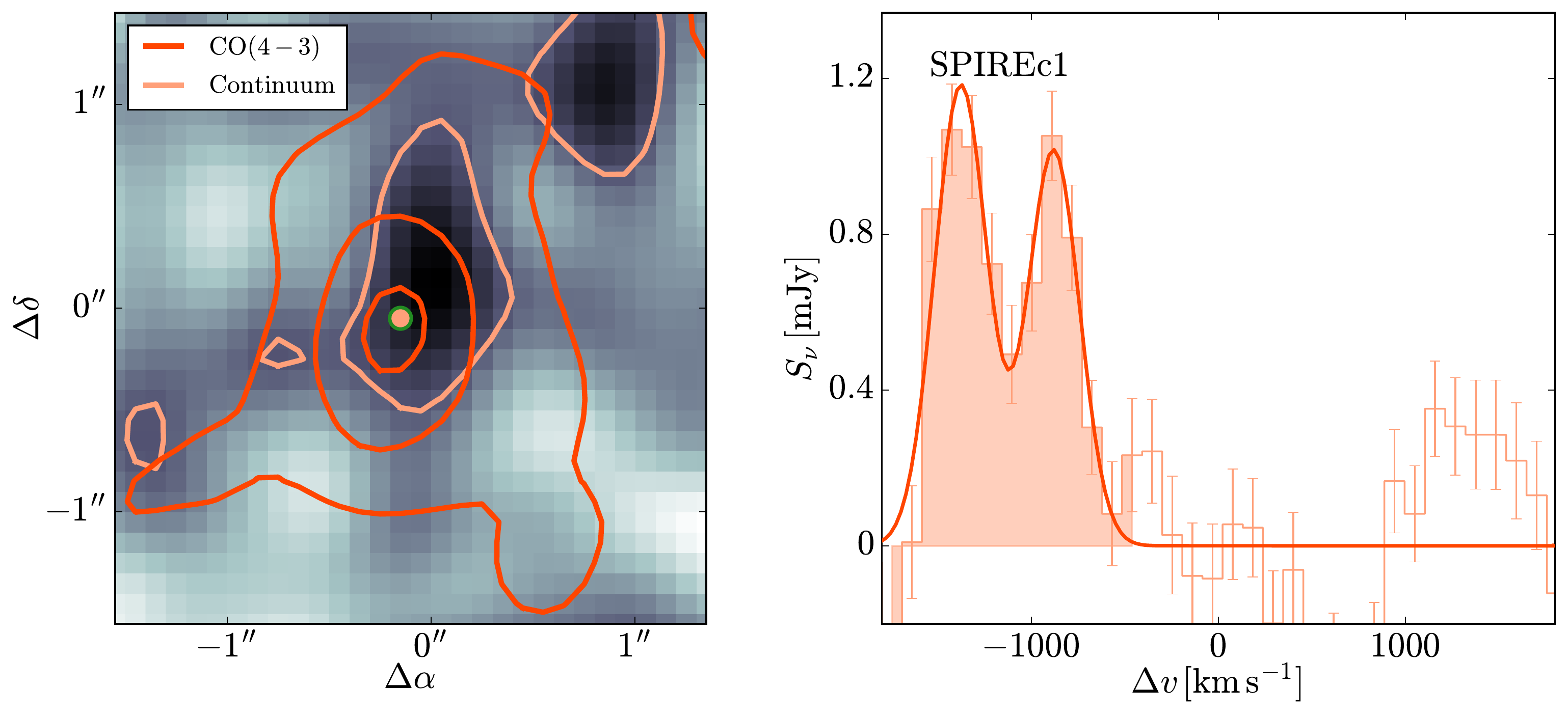}
\end{framed}
\end{subfigure}
\caption{}
\end{figure*}
\renewcommand{\thefigure}{\arabic{figure}}

\renewcommand{\thefigure}{A\arabic{figure} (Cont.)}
\addtocounter{figure}{-1}
\begin{figure*}
\begin{subfigure}{.45\textwidth}
\begin{framed}
\includegraphics[width=\textwidth]{CII/blank.pdf}
\includegraphics[width=\textwidth]{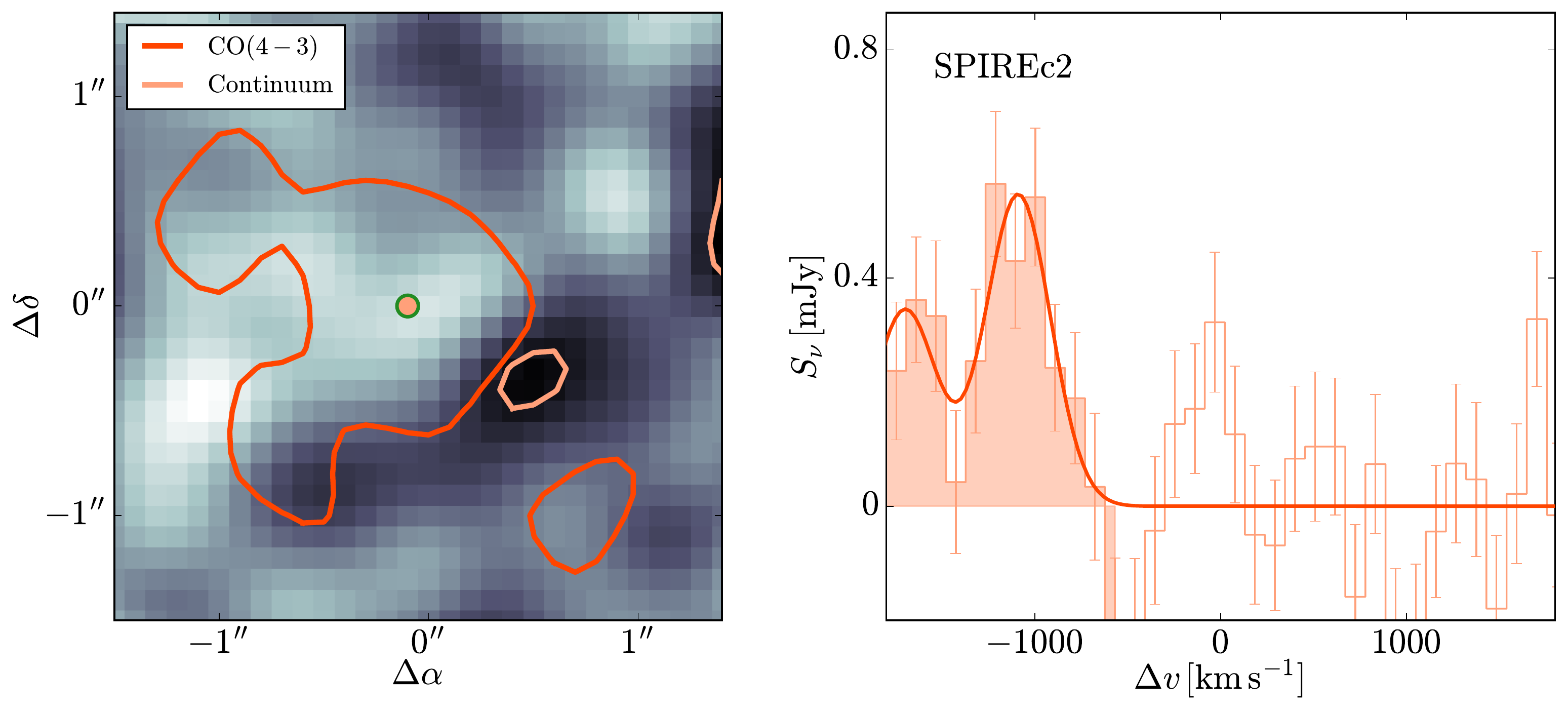}
\end{framed}
\end{subfigure}
\begin{subfigure}{.45\textwidth}
\begin{framed}
\includegraphics[width=\textwidth]{CII/blank.pdf}
\includegraphics[width=\textwidth]{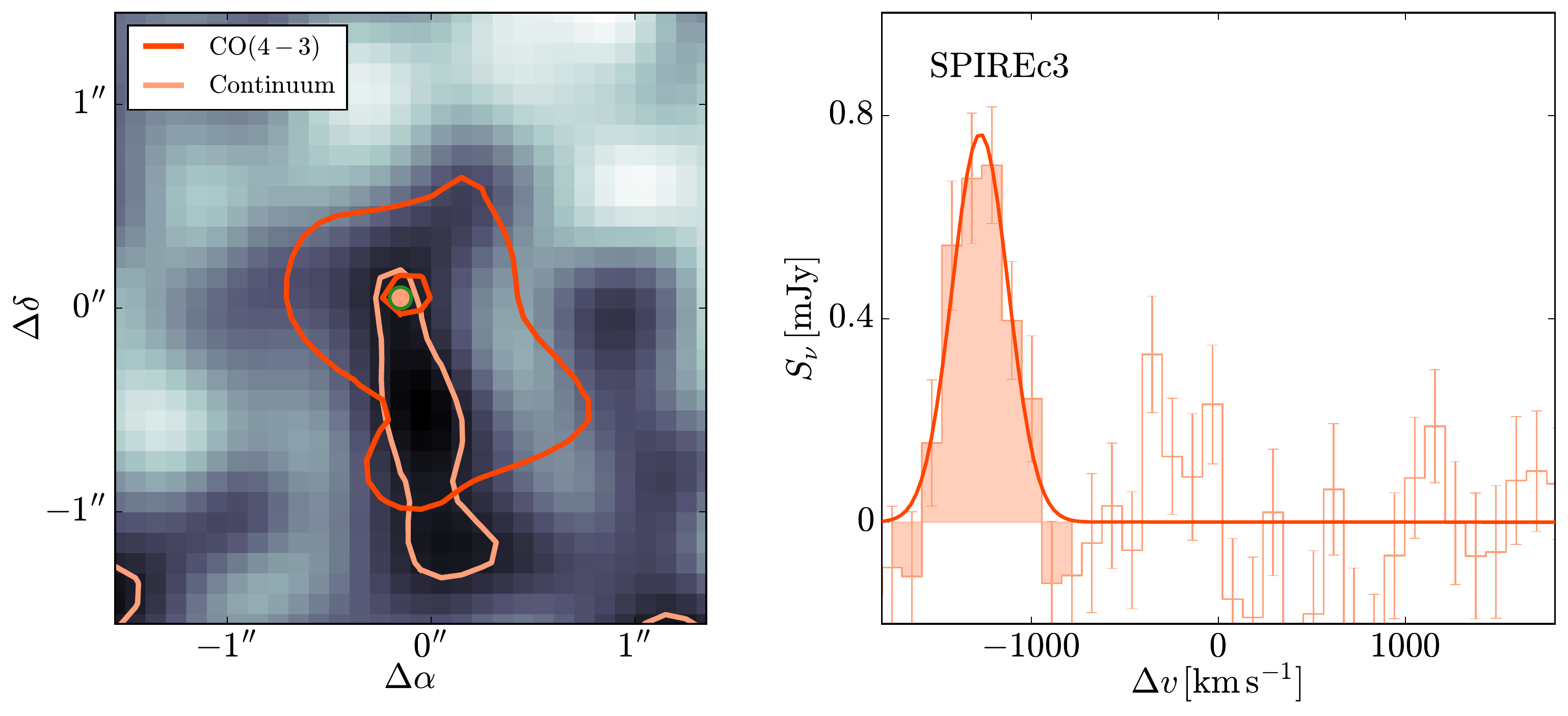}
\end{framed}
\end{subfigure}
\caption{}
\end{figure*}
\renewcommand{\thefigure}{B\arabic{figure}}

\section{High-resolution imaging}
\label{appendix2}

We provide continuum cut-outs of our sources obtained from the high-resolution Band 7 imaging. Line-free channels were determined from our deeper Cycle 5 data by fitting Gaussian profiles to each spectrum (see Section \ref{source_properties}) and stacked. S{\'e}rsic profiles were fit to all sources detected in these images above 3$\sigma$, and half-light radii were estimated from the fits.

The left panels show these stacked images with 2 and 3$\sigma$ countours, then increasing in steps of 3$\sigma$, with positions found in the Cycle 5 data shown as blue points and positions found from the S{\'e}rsic profiles shown as red points. The red bars indicated the sizes of the half-light radii resulting from the S{\'e}rsic profiles, and best-fitting half-light radii and S{\'e}rsic indices are shown in the top left. For source NL1 we provide the best-fitting half-light radius in units of arcseconds (as the redshift of this source cannot be confirmed), otherwise the best-fitting half-light radii are in units of kiloparsecs. The middle panels show our S{\'e}rsic profile models, and the right panels show the residuals. For sources below 3$\sigma$, where did not attempt to fit S{\'e}rsic profiles, we leave the middle panel blank. We also note that source C7 resolves into a complicated structure, possibly a pair of merging galaxies, and we do not attempt to fit a S{\'e}rsic profile and measure a half-light radius for it.

\begin{figure*}
\begin{framed}
\includegraphics[width=\textwidth]{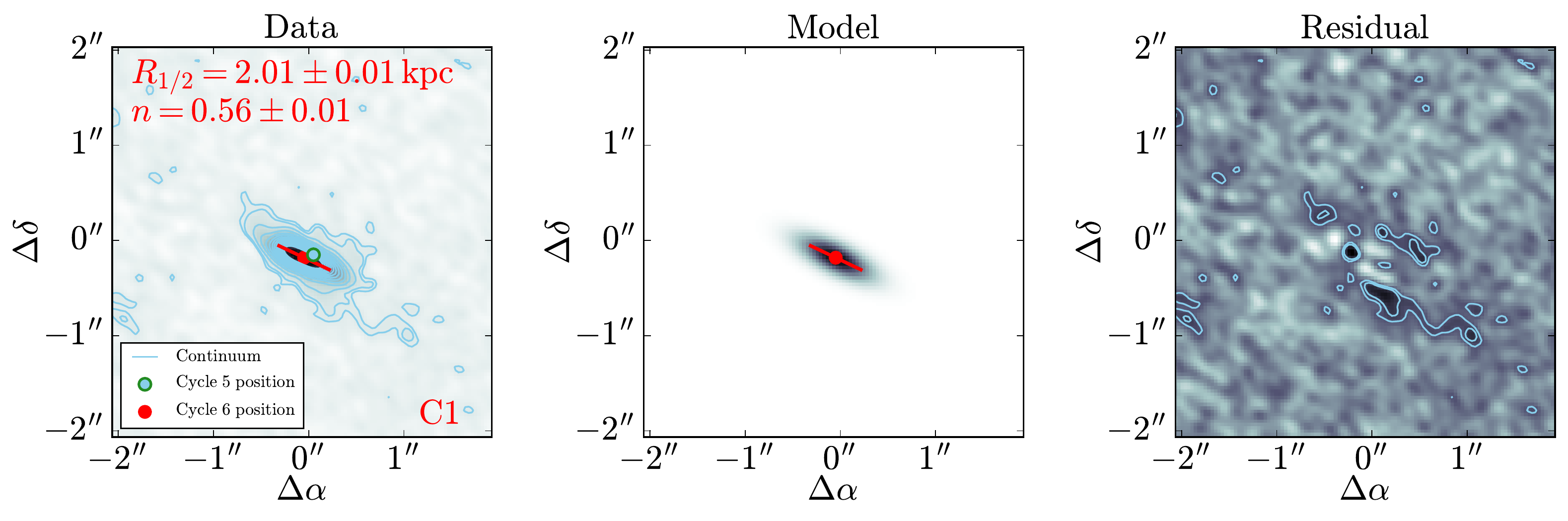}
\end{framed}
\begin{framed}
\includegraphics[width=\textwidth]{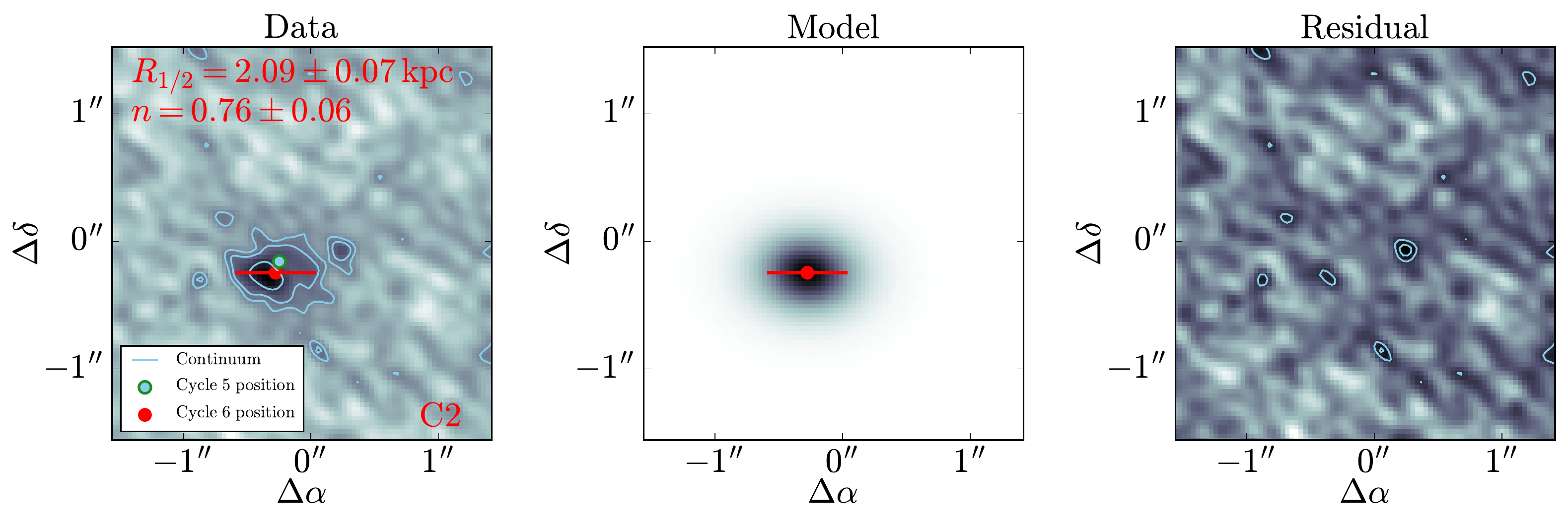}
\end{framed}
\begin{framed}
\includegraphics[width=\textwidth]{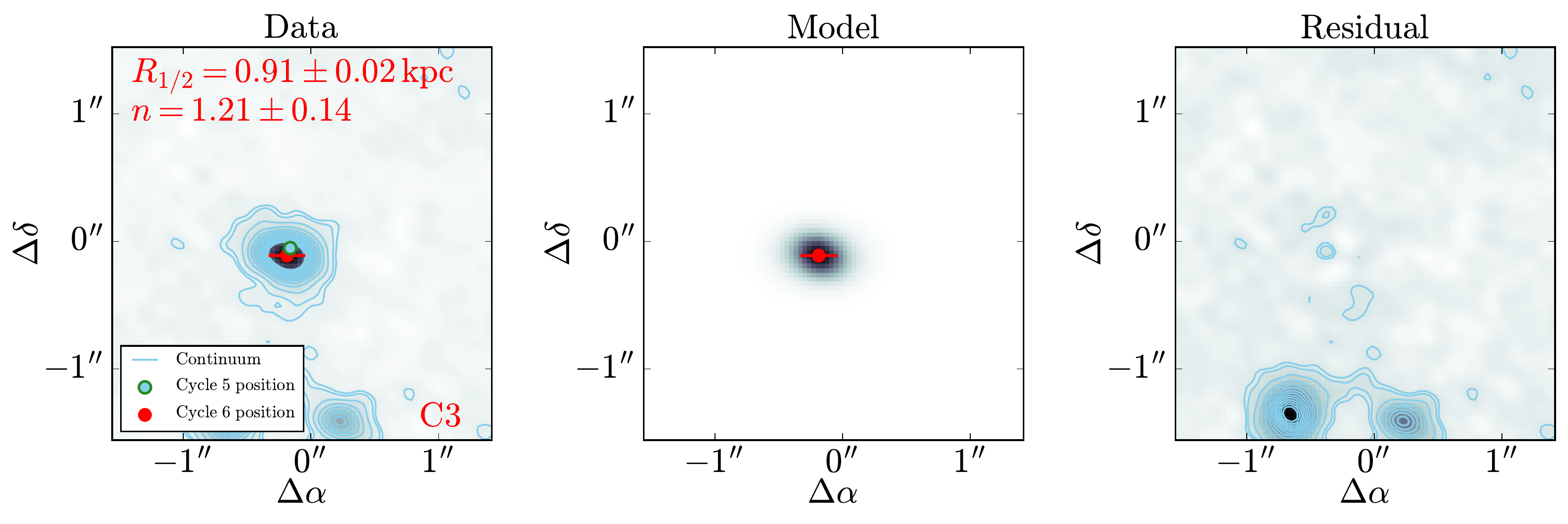}
\end{framed}
\caption{{\it Left:} Continuum images from the high-resolution Cycle 6 data at 850\,$\mu$m. Contours are 2 and 3$\sigma$, then increase in steps of 3$\sigma$. The blue points are positions found in our lower resolution Cycle 5 data, and red points are the centres of the S{\'e}rsic profiles fit to these higher resolution images. The red bars show the lengths of the half-light radii determined from the best-fitting S{\'e}rsic profiles, and best-fitting half-light radii and S{\'e}rsic indices are shown in the top left {\it Middle:} Best-fitting model S{\'e}rsic profiles. Sources undetected above 3$\sigma$ were not fitted, and for these cases we leave this panel blank. {\it Right:} Residuals from the S{\'e}rsic profile fits.}
\label{hires}
\end{figure*}

\renewcommand{\thefigure}{B\arabic{figure} (Cont.)}
\addtocounter{figure}{-1}
\begin{figure*}
\begin{framed}
\includegraphics[width=\textwidth]{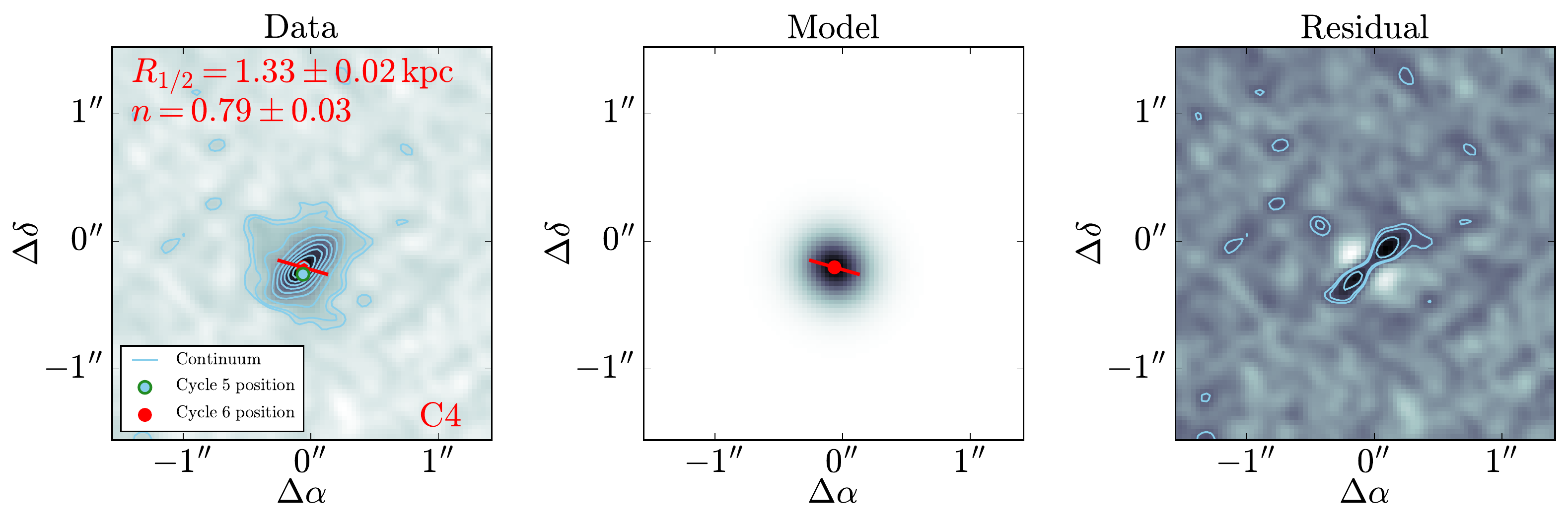}
\end{framed}
\begin{framed}
\includegraphics[width=\textwidth]{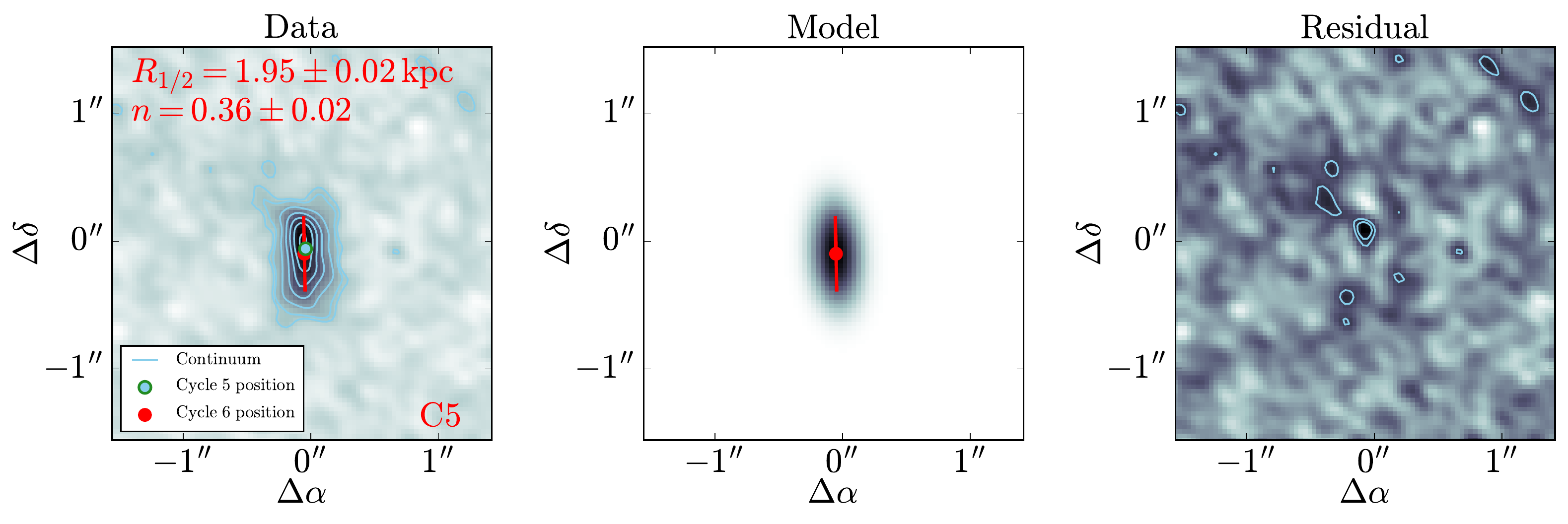}
\end{framed}
\begin{framed}
\includegraphics[width=\textwidth]{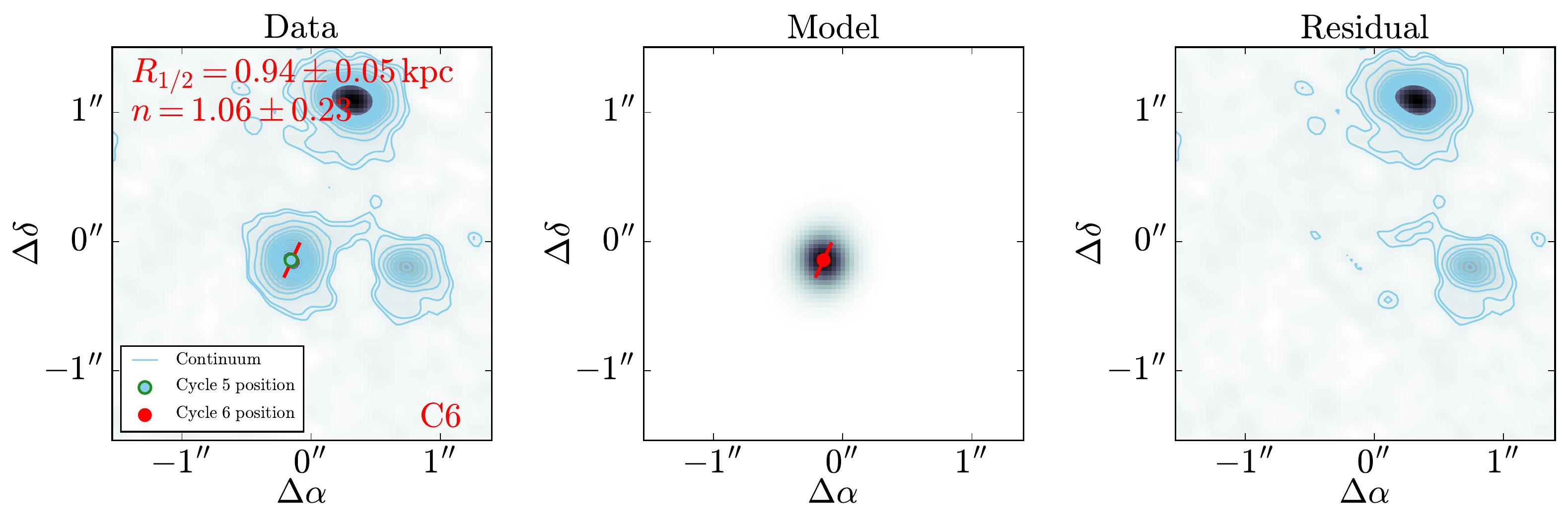}
\end{framed}
\caption{}
\end{figure*}
\renewcommand{\thefigure}{\arabic{figure}}

\renewcommand{\thefigure}{B\arabic{figure} (Cont.)}
\addtocounter{figure}{-1}
\begin{figure*}
\begin{framed}
\includegraphics[width=\textwidth]{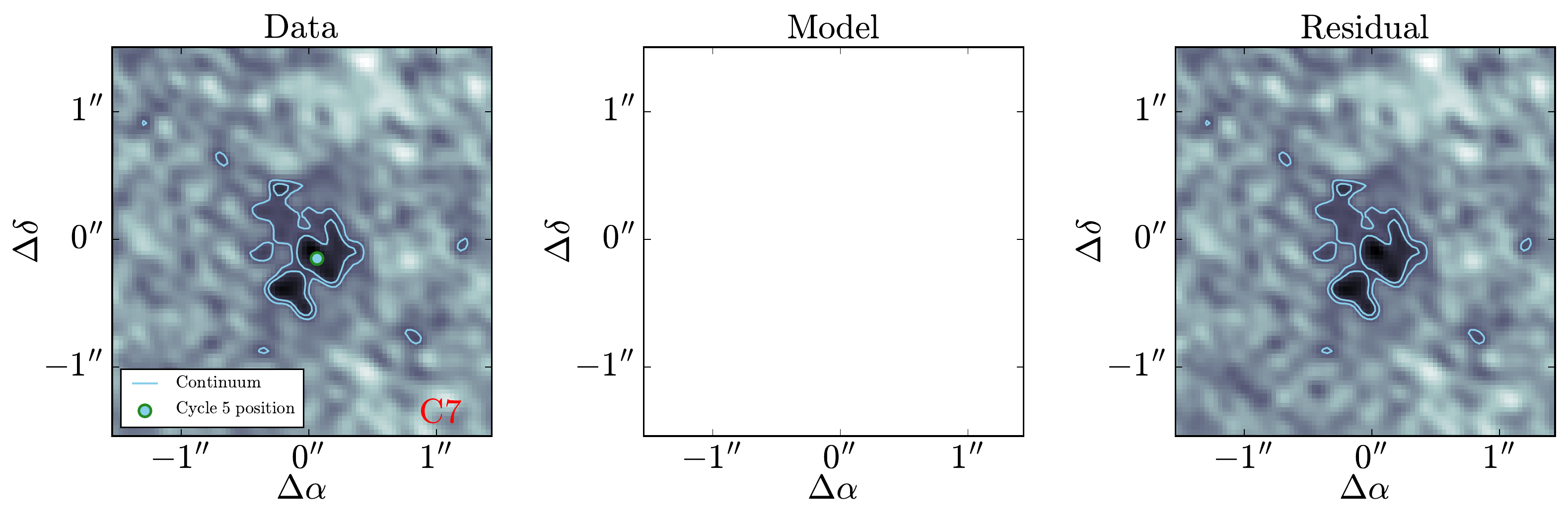}
\end{framed}
\begin{framed}
\includegraphics[width=\textwidth]{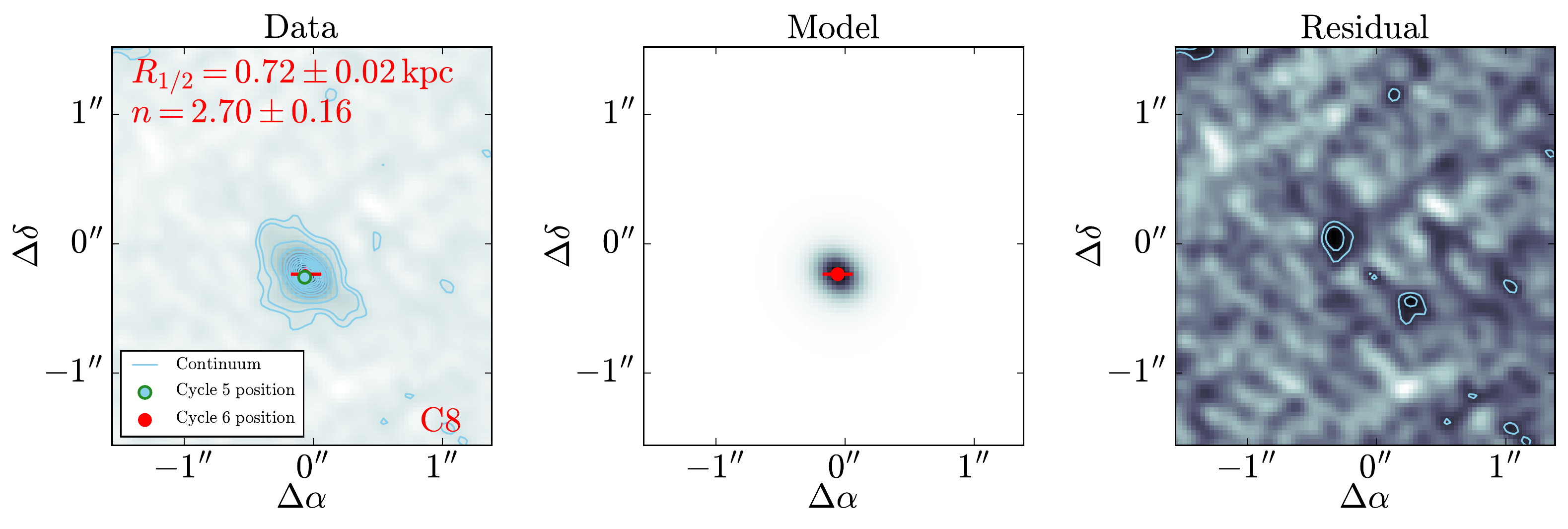}
\end{framed}
\begin{framed}
\includegraphics[width=\textwidth]{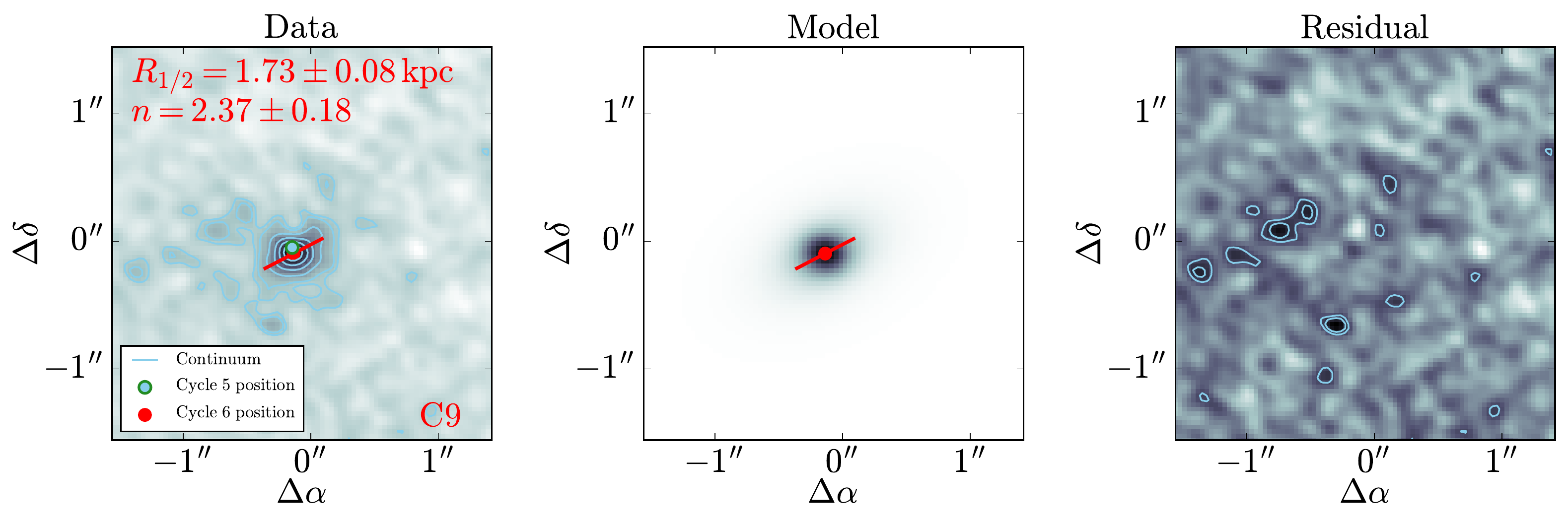}
\end{framed}
\caption{}
\end{figure*}
\renewcommand{\thefigure}{\arabic{figure}}

\renewcommand{\thefigure}{B\arabic{figure} (Cont.)}
\addtocounter{figure}{-1}
\begin{figure*}
\begin{framed}
\includegraphics[width=\textwidth]{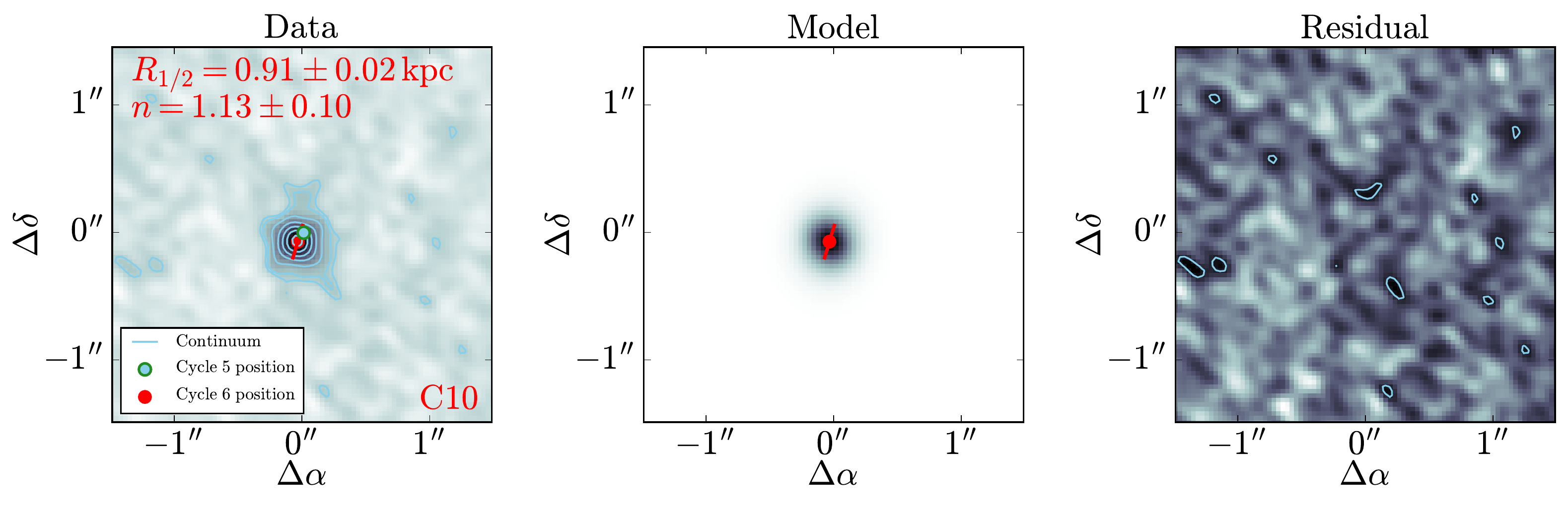}
\end{framed}
\begin{framed}
\includegraphics[width=\textwidth]{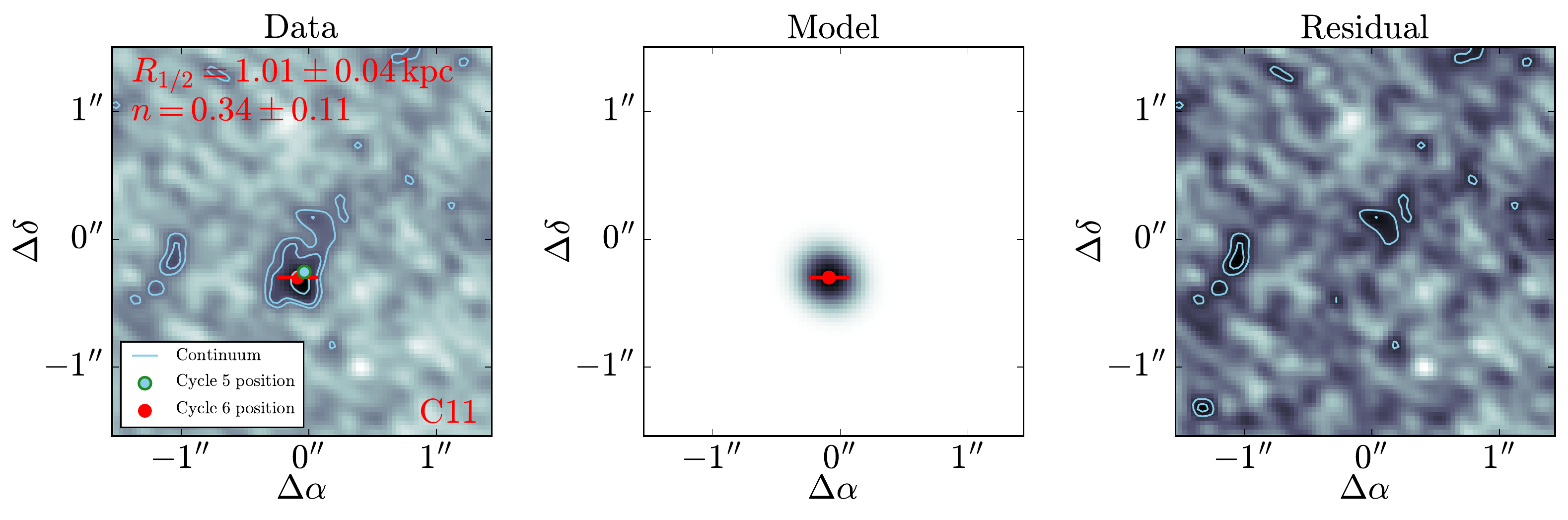}
\end{framed}
\begin{framed}
\includegraphics[width=\textwidth]{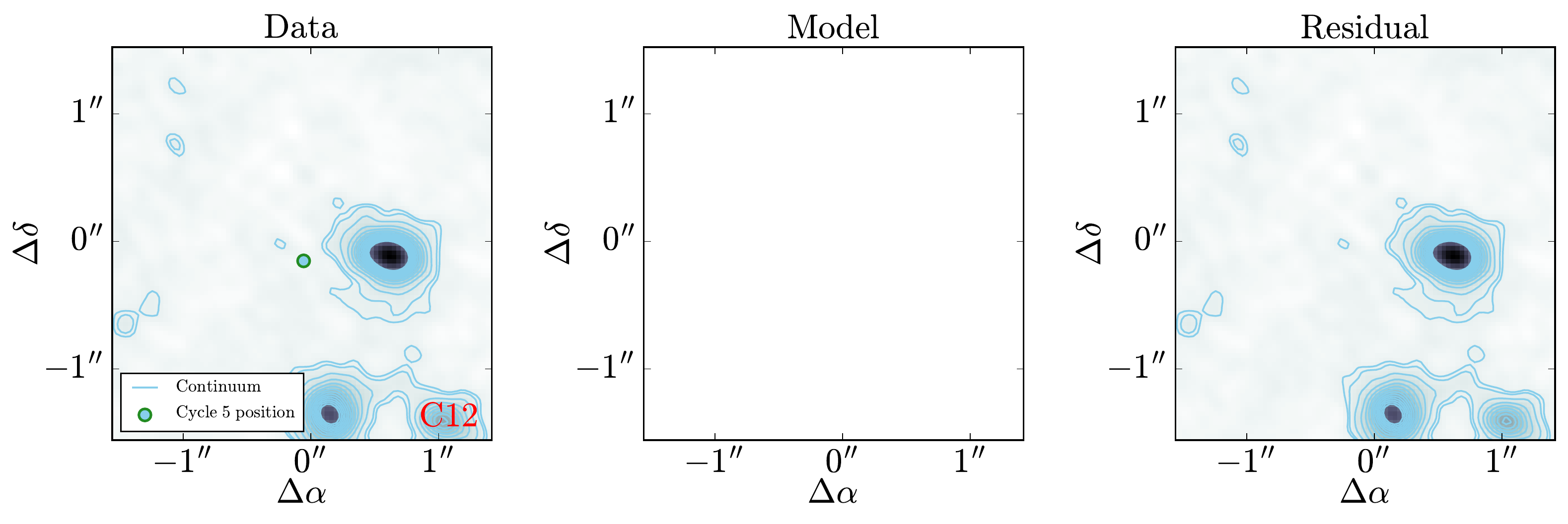}
\end{framed}
\caption{}
\end{figure*}
\renewcommand{\thefigure}{\arabic{figure}}

\renewcommand{\thefigure}{B\arabic{figure} (Cont.)}
\addtocounter{figure}{-1}
\begin{figure*}
\begin{framed}
\includegraphics[width=\textwidth]{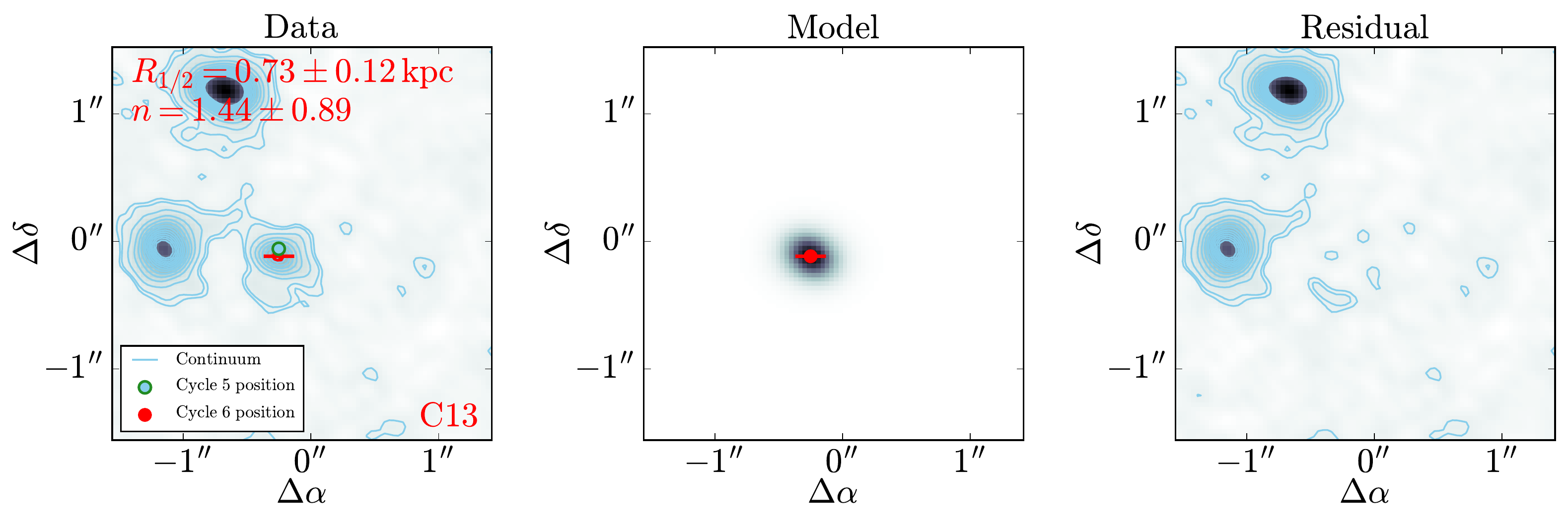}
\end{framed}
\begin{framed}
\includegraphics[width=\textwidth]{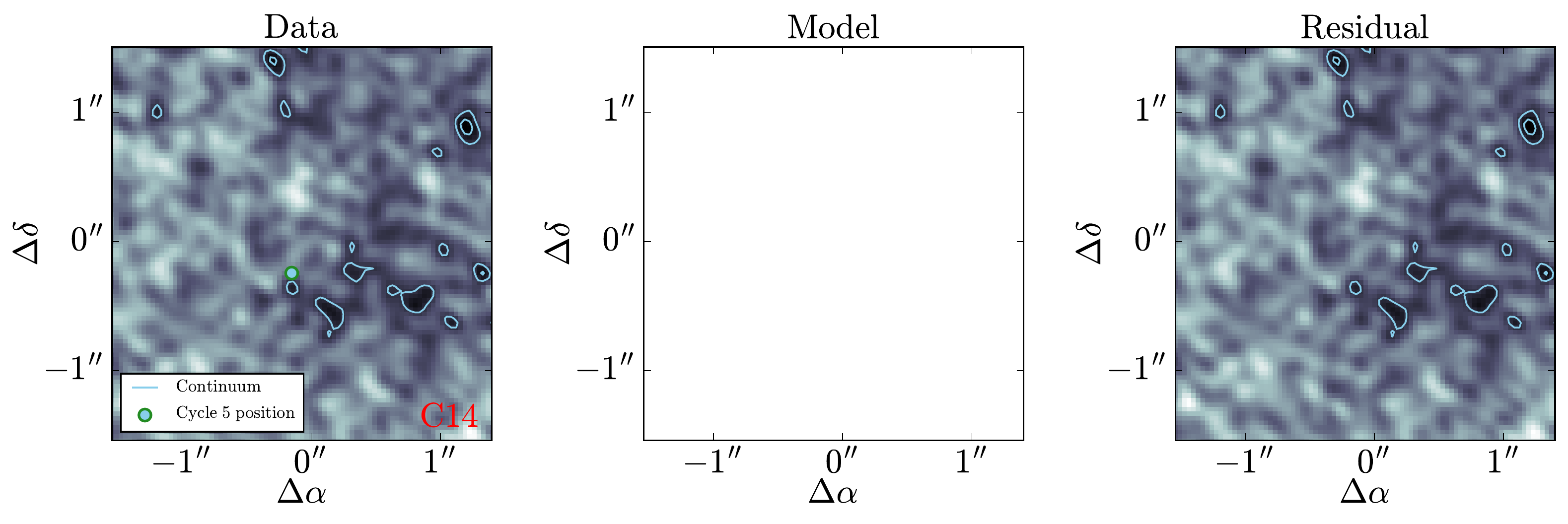}
\end{framed}
\begin{framed}
\includegraphics[width=\textwidth]{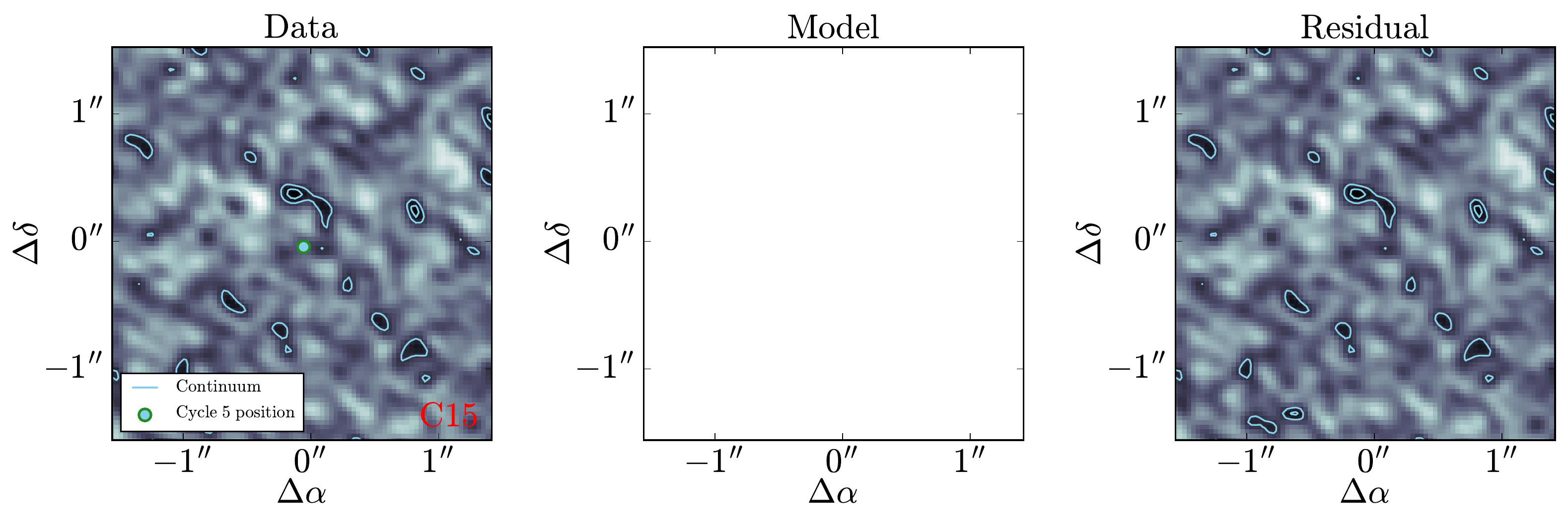}
\end{framed}
\caption{}
\end{figure*}
\renewcommand{\thefigure}{\arabic{figure}}

\renewcommand{\thefigure}{B\arabic{figure} (Cont.)}
\addtocounter{figure}{-1}
\begin{figure*}
\begin{framed}
\includegraphics[width=\textwidth]{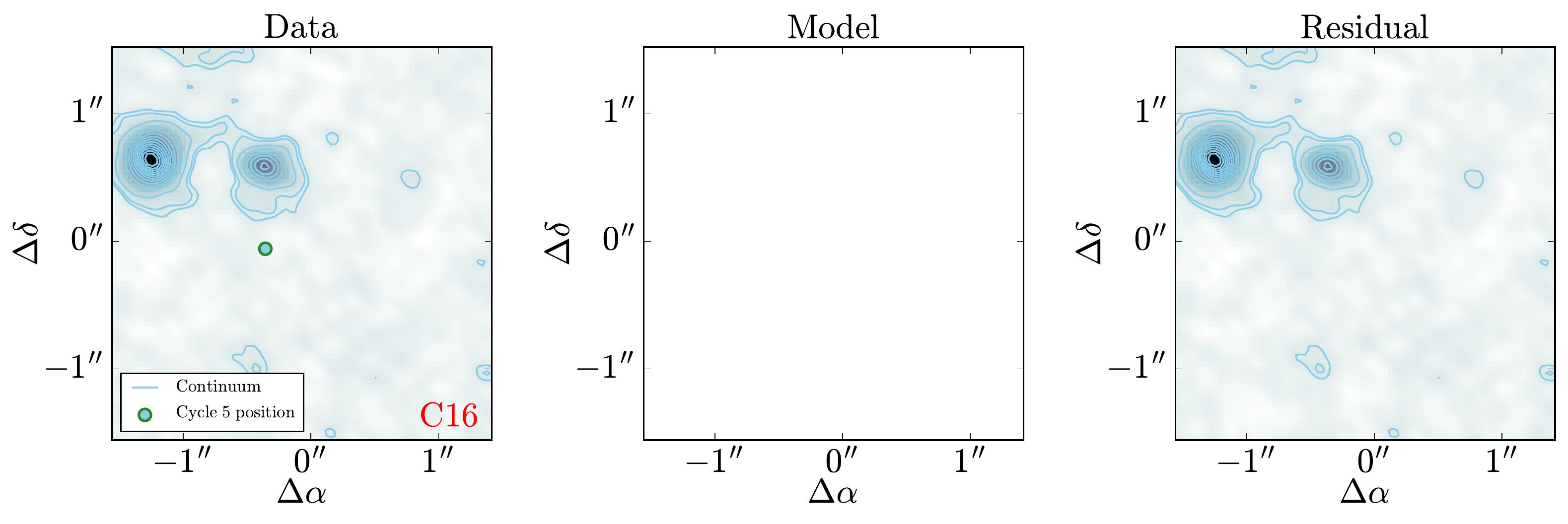}
\end{framed}
\begin{framed}
\includegraphics[width=\textwidth]{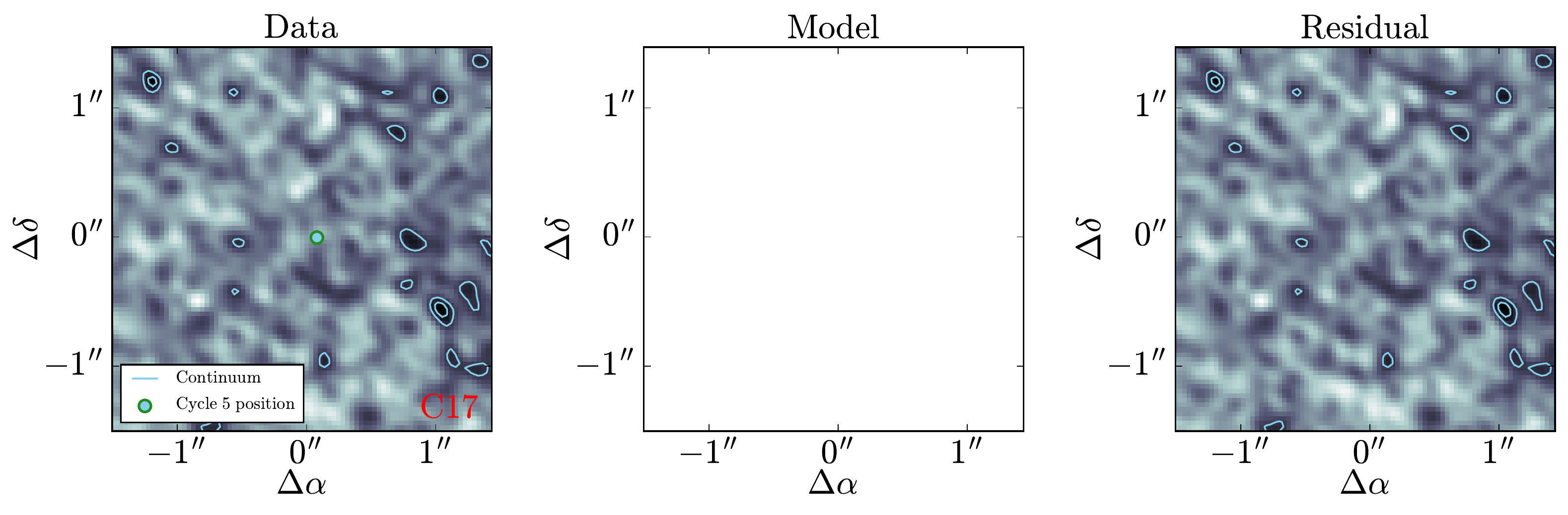}
\end{framed}
\begin{framed}
\includegraphics[width=\textwidth]{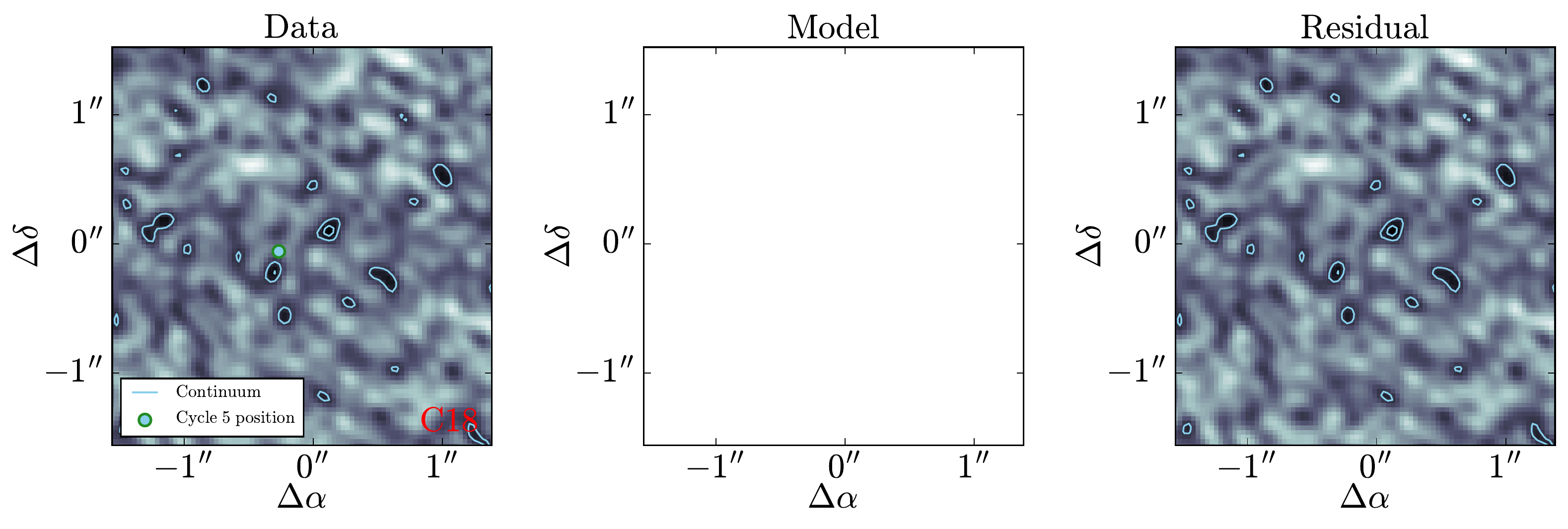}
\end{framed}
\caption{}
\end{figure*}
\renewcommand{\thefigure}{\arabic{figure}}

\renewcommand{\thefigure}{B\arabic{figure} (Cont.)}
\addtocounter{figure}{-1}
\begin{figure*}
\begin{framed}
\includegraphics[width=\textwidth]{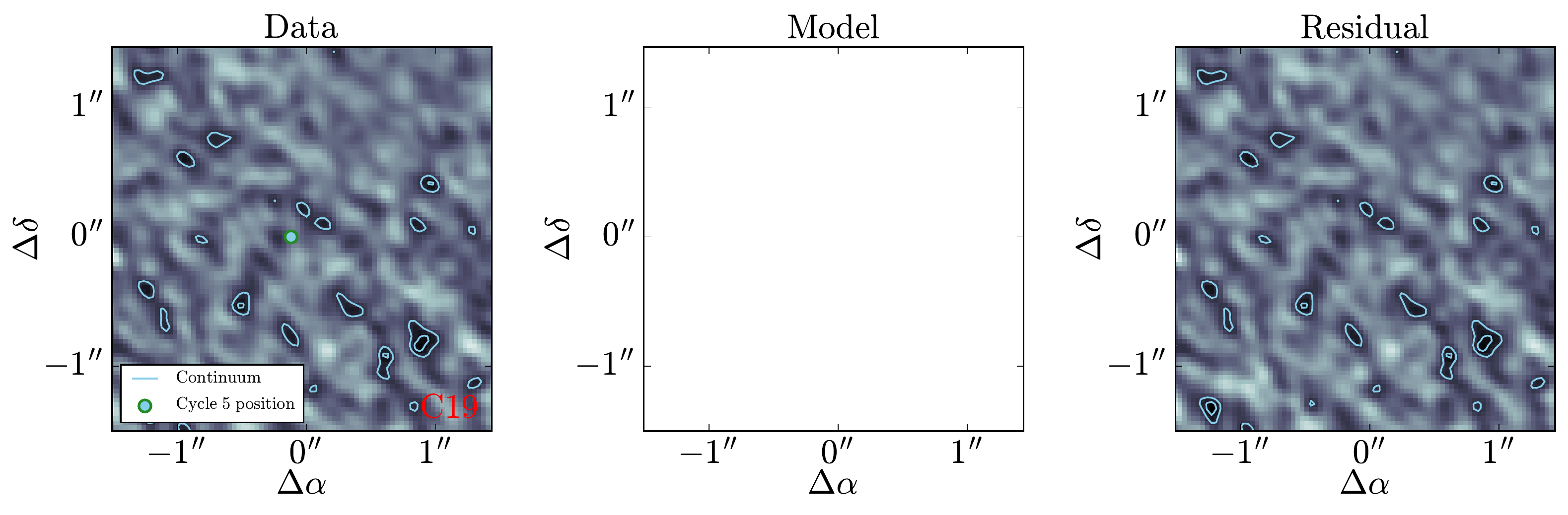}
\end{framed}
\begin{framed}
\includegraphics[width=\textwidth]{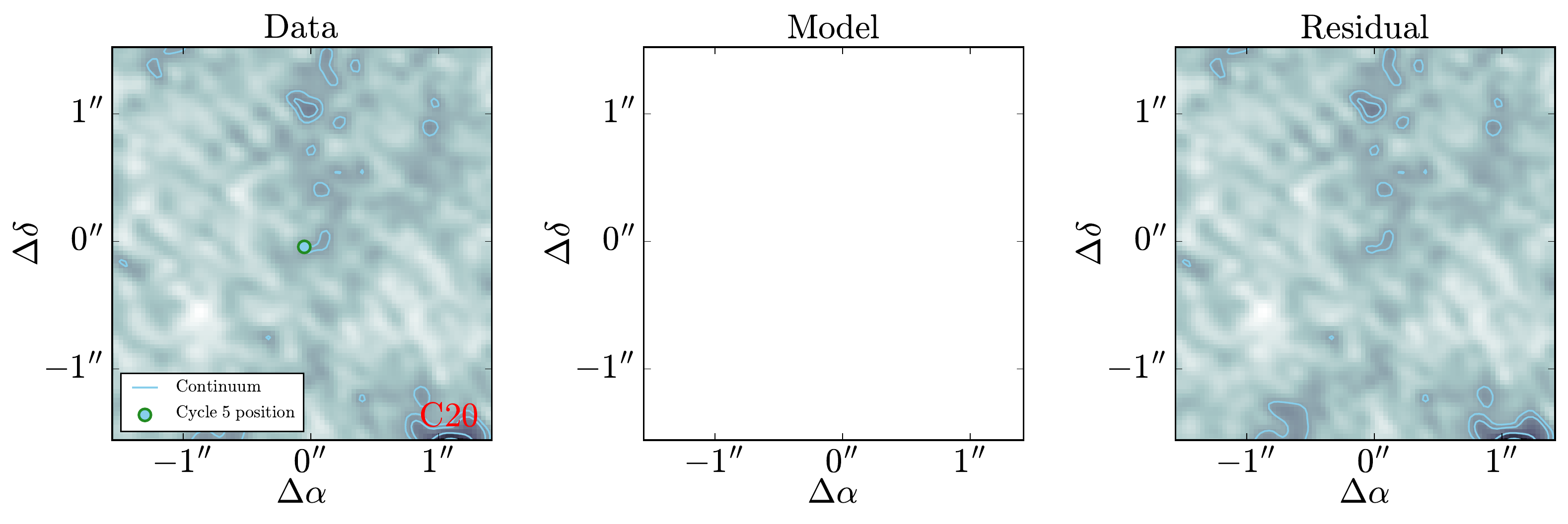}
\end{framed}
\begin{framed}
\includegraphics[width=\textwidth]{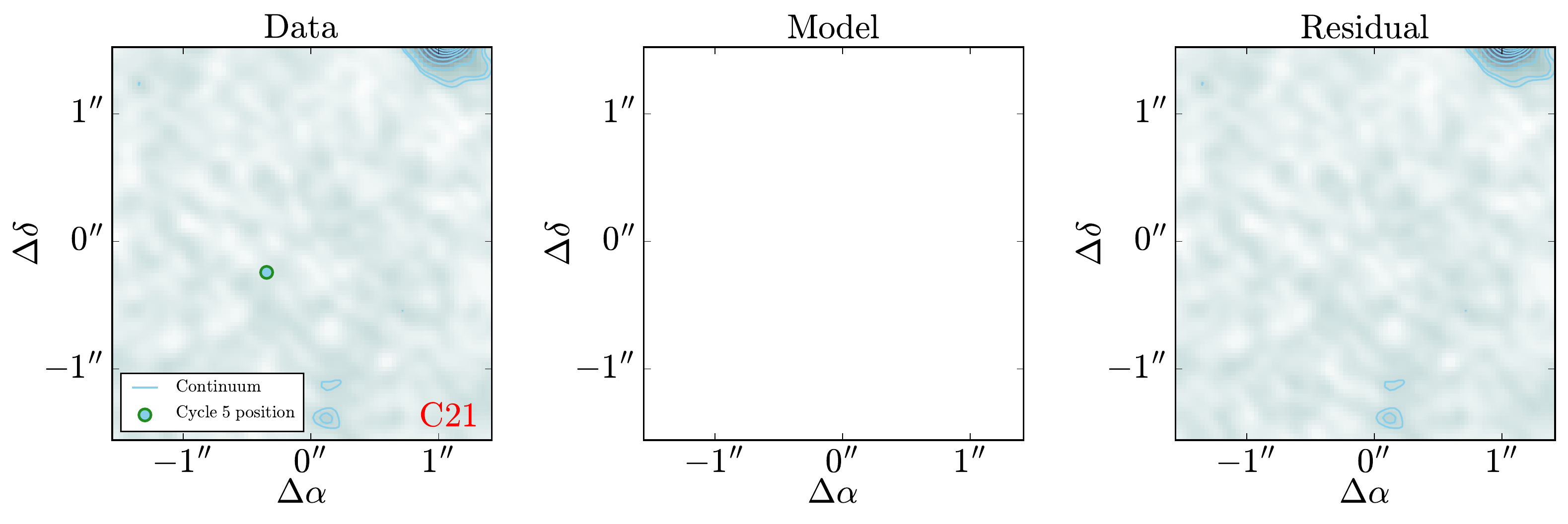}
\end{framed}
\caption{}
\end{figure*}
\renewcommand{\thefigure}{\arabic{figure}}

\renewcommand{\thefigure}{B\arabic{figure} (Cont.)}
\addtocounter{figure}{-1}
\begin{figure*}
\begin{framed}
\includegraphics[width=\textwidth]{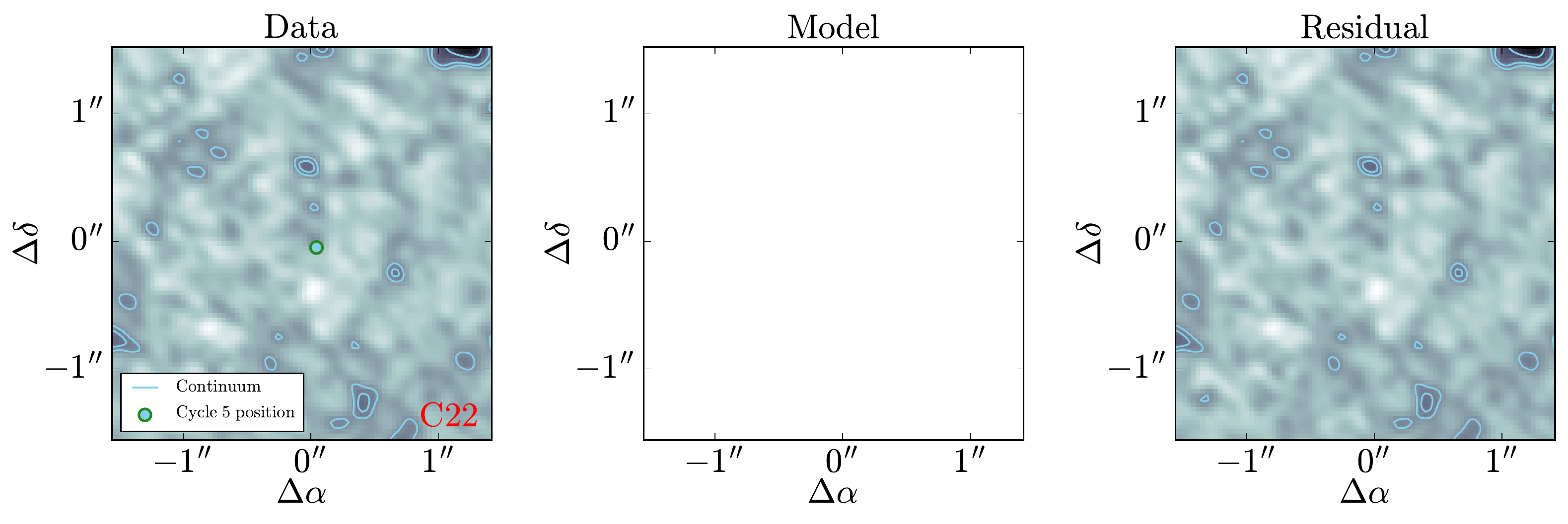}
\end{framed}
\begin{framed}
\includegraphics[width=\textwidth]{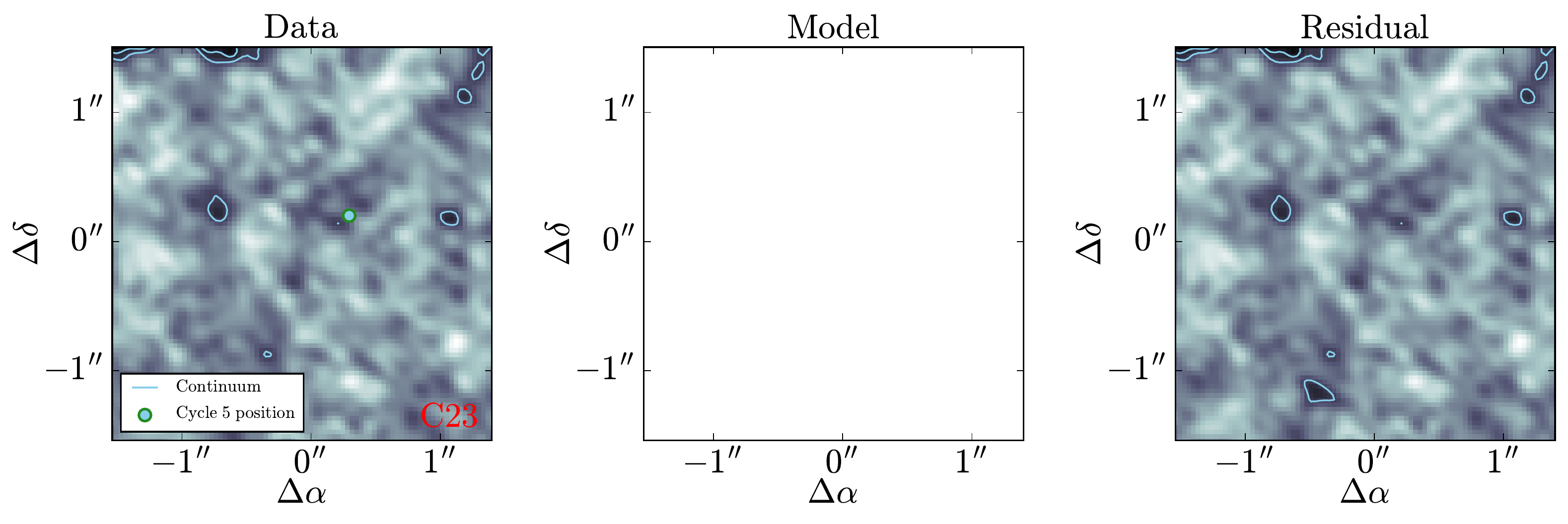}
\end{framed}
\begin{framed}
\includegraphics[width=\textwidth]{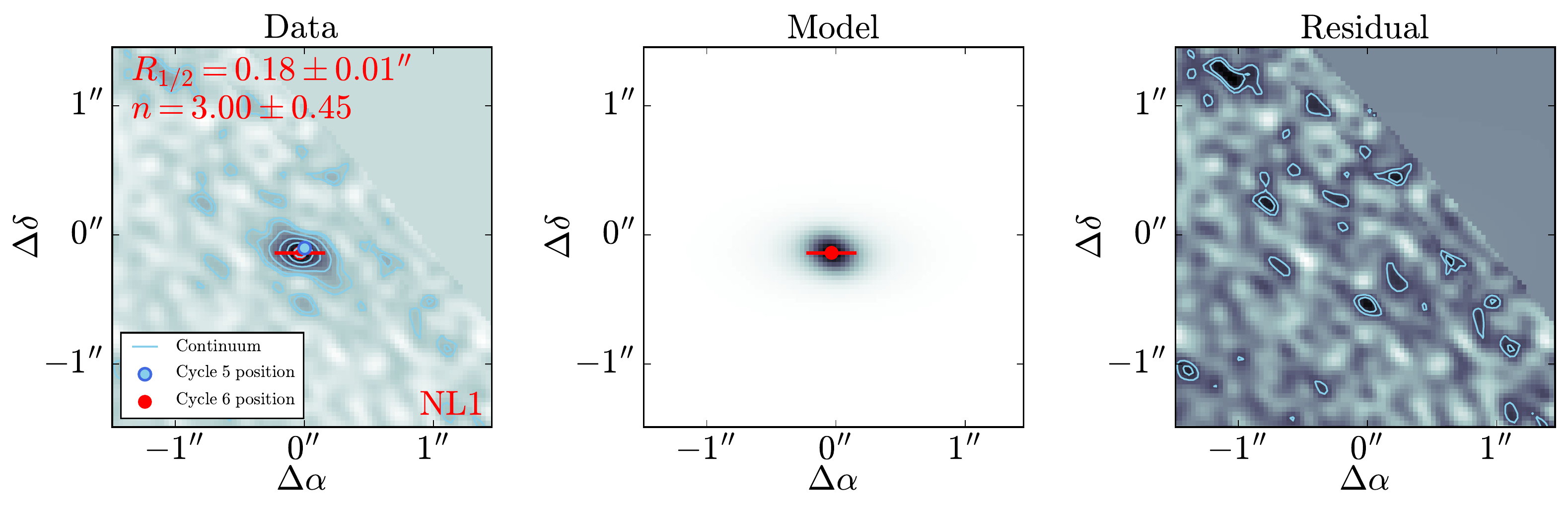}
\end{framed}
\caption{}
\end{figure*}
\renewcommand{\thefigure}{\arabic{figure}}

\renewcommand{\thefigure}{B\arabic{figure} (Cont.)}
\addtocounter{figure}{-1}
\begin{figure*}
\begin{framed}
\includegraphics[width=\textwidth]{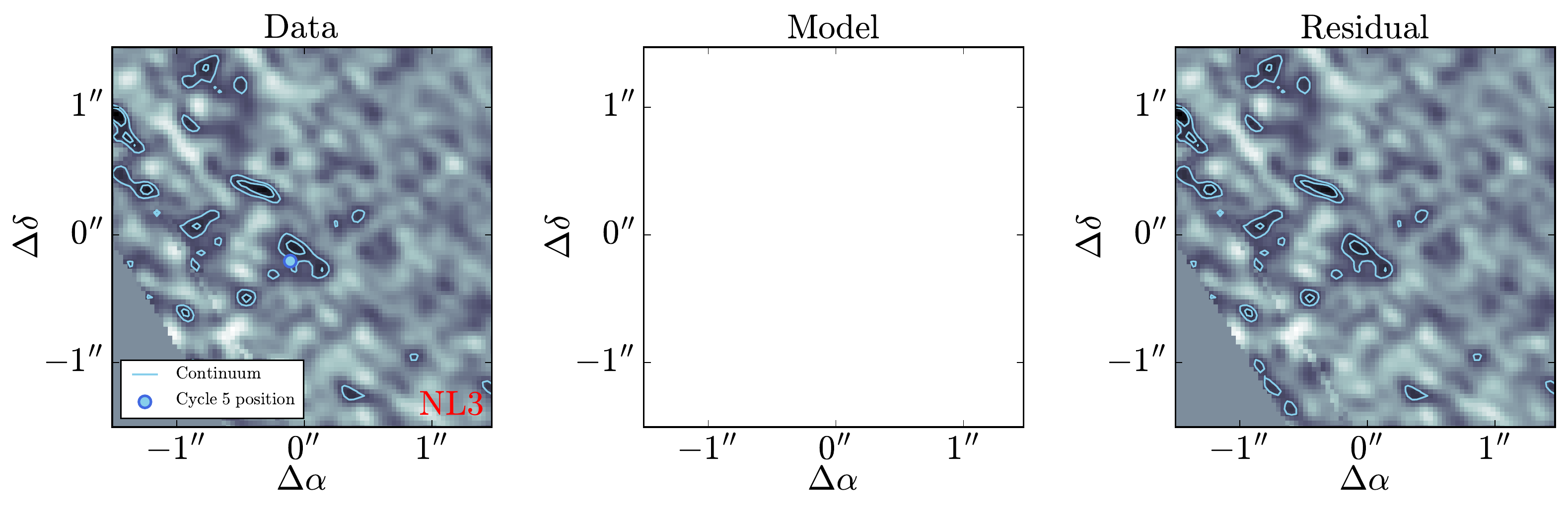}
\end{framed}
\caption{}
\end{figure*}
\renewcommand{\thefigure}{\arabic{figure}}

\end{document}